%% file: thesis.tex
\documentclass[12pt,fleqn,twoside]{report}
\usepackage{amsmath}
\usepackage{ifthen}
\usepackage{amssymb}
\usepackage{varioref}
\usepackage[plainpages=false,pdfpagelabels,hypertexnames=false]{hyperref}
\usepackage[capitalise,noabbrev,nameinlink]{cleveref}
\include{lsp}

\begin{document}
\pagenumbering{roman}

\include{revised-frontpage}

\setcounter{page}{2}
\include{preface}
\include{corrections}

\pagenumbering{arabic}
\setcounter{page}{1}
\include{frontpage}
\tableofcontents
\include{abstract}

\include{declaration}
\include{education}
\include{acknowledgement}
\include{motiv}
\include{asslang}

\include{prglang}

\include{prgsem}

\include{spec}

\include{auxiliary}

\include{specprog}

\include{opass}

\include{wellfound}

\include{ruledes}
\include{exintr}
\include{phil}

\include{bubl}
\include{part}

\include{safety}
\include{dekker}

\include{newsound}

\include{cmpl}
\include{alt}

\addcontentsline{toc}{chapter}{Bibliography}
\bibliographystyle{alpha}
\bibliography{bib/book,bib/article,bib/inproceeding,bib/incollection,bib/phdthesis,bib/techreport,bib/misc,bib/unpublished,bib/focus}
\end{document}

%% file: lsp.tex
\newtheorem{definition}{Definition}
\newtheorem{statement}{Proposition}

\newenvironment{spec}[9]{$\quad\begin{array}{l}\\       
                                 \ #1( #2 )\ #3\\
                                 #4\\

                               \quad\ \ \begin{array}{ll}\\
                                  \npre&#5\\
                                  \nrely&#6\\
                                  \nwait&#7\\
                                  \nguar&#8\\
                                  \neff&#9\\ 
                               \end{array}\\
\\
                          }{\end{array}$}

\newenvironment{tuple}[7]{$\quad\ \begin{array}{ll}\\
                             \nglo&#1\\
                             \naux&#2\\                           
                             \quad\npre&#3\\
                             \quad\nrely&#4\\
                             \quad\nwait&#5\\
                             \quad\nguar&#6\\
                             \quad\neff&#7\\
\\
                         }{\end{array}$}

\newenvironment{oldtuple}[4]{$\quad\ \begin{array}{ll}\\
                             \quad\npre&#1\\
                             \quad\nrely&#2\\
                             \quad\nguar&#3\\
                             \quad\npost&#4\\
\\
                         }{\end{array}$}

\newenvironment{dcl}{\begin{array}{lrcl}}{\end{array}}

\newenvironment{ndcl}[2]{\begin{array}{llrcl}
                        \nglo &#1\\
                        \naux &#2\\}{\end{array}}

\newenvironment{prg}{$\ \quad\begin{array}{l}\\

                           }{\\
                             \end{array}$}

\newenvironment{flist}{$\quad\ \begin{array}{l}\\ 
}{\\

                     \vspace{0mm}    \end{array}$}
										
\newenvironment{nproof}{\noindent{\bf Proof:}}{

                     \vspace{0mm}    \noindent{\bf end of proof}}

\makeatletter
\newcommand{\overleftharpoon}{%
  \mathpalette{\overarrow@\leftharpoonfill@}}
\def\leftharpoonfill@{\arrowfill@\leftharpoonup\relbar\relbar}
\makeatother	

\def\set#1{\{#1\}}

\renewcommand{\and}{\And}
\renewcommand{\emptyset}{\{\}}

\newcommand{\seqof}[1]{{\rm seq\ of\ }#1}
\newcommand{\setof}[1]{{\rm set\ of\ }#1}
\newcommand{\exter}{\mbox{$\stackrel{e}{\rightarrow}$}}
\newcommand{\intern}{\mbox{$\stackrel{i}{\rightarrow}$}}
\newcommand{\aug}[2]{\mbox{$\stackrel{(#1,#2)}{\hookrightarrow}$}}
\newcommand{\arrayof}[2]{\narray\ #1\ \nto\ #2}
\newcommand{\type}[2]{\mbox{$#1=#2$}}
\newcommand{\card}[1]{\mbox{$#1$}}
\newcommand{\satis}{\mbox{\ \underline{{\rm sat}}\ }}

\newcommand{\narray}{{\sf array}}
\newcommand{\nto}{{\sf to}}

\newcommand{\progtran}{\mbox{$\aug{\vartheta}{\alpha}$}}

\newcommand{\nskip}{{\sf skip}}
\newcommand{\nbegin}{{\sf begin}}

\newcommand{\nend}{{\sf end}}
\newcommand{\Or}{\mbox{$\,\vee\,$}}
\newcommand{\nif}{{\sf if}}
\newcommand{\nthen}{{\sf then}}
\newcommand{\nelse}{{\sf else}}
\newcommand{\nfi}{{\sf fi}}
\newcommand{\nwhile}{{\sf while}}
\newcommand{\ndo}{{\sf do}}
\newcommand{\nod}{{\sf od}}
\newcommand{\nawait}{{\sf await}}

\newcommand{\nglo}{{\sf glo}}
\newcommand{\naux}{{\sf aux}}
\newcommand{\lwew}{{\sf ioeo}}
\newcommand{\lwer}{{\sf ioeh}}
\newcommand{\lrew}{{\sf iheo}}
\newcommand{\lrer}{{\sf iheh}}
\newcommand{\loc}{{\sf loc}}
\newcommand{\npre}{{\sf pre}}
\newcommand{\nrely}{{\sf rely}}
\newcommand{\nwait}{{\sf wait}}
\newcommand{\nguar}{{\sf guar}}
\newcommand{\neff}{{\sf eff}}
\newcommand{\npost}{{\sf post}}

\newcommand{\nwf}{{\sf wf}}
\newcommand{\true}{{\sf true}}
\newcommand{\false}{{\sf false}}
\renewcommand{\And}{\mbox{$\,\wedge\,$}}
\newcommand{\wfound}{\mbox{\ \underline{{\rm wf}}\ }}
\newcommand{\Bool}{\ensuremath{{\mathbb B}}}
\newcommand{\Nat}{\ensuremath{{\mathbb N}}}
\newcommand{\Real}{\ensuremath{{\mathbb R}}}
\newcommand{\len}[1]{\ensuremath{{\sf len}\; #1}}
\newcommand{\sconc}{\ensuremath{\curvearrowright}}
\newcommand{\Implies}{\ensuremath{\Rightarrow}}
\newcommand{\Forall}{\ensuremath{\forall}}
\newcommand{\Exists}{\ensuremath{\exists}}

%% file: revised-frontpage.tex
\thispagestyle{empty}
\vspace{2cm}

\begin{center}
\LARGE \bf DEVELOPMENT\\ OF PARALLEL PROGRAMS\\ ON SHARED DATA-STRUCTURES
\end{center}

\begin{center}
\Large \bf REVISED VERSION
\end{center}

\vspace{7cm}

\begin{center}
\small THIS REPORT IS A REVISED VERSION OF UMCS-91-1-1
\end{center}

\vspace{2cm}

\begin{center}\small
Ketil St\o len\\
Oslo, March 2023
\end{center}

%% file: preface.tex
\chapter*{Preface}

Thirty-two years ago, my PhD-thesis was published as Technical Report UMCS-91-1-1 at the computer science department at Manchester University. To save money, I was required to print the thesis on as few pages as possible. This meant small fonts (10 pt), no indentation, narrow page margins, etc., making the thesis unnecessarily hard to read. Over the years, I have also discovered several annoying misprints and minor mistakes. In this revised version these shortcomings have been removed.

Except for the corrections specified in detail on Pages iii-v, I have only changed the formatting of the original report.

Many thanks to Dag Frette Langmyhr, who helped me resolve some Latex-issues.

%% file: corrections.tex
\section*{Corrections}

The known mistakes in {UMCS}-91-1-1 and how they have been corrected in the current version.

\begin{itemize}

\item Page 32, Line $+2$: substitute `of' for `of of'
  
\item Page 32, Line $+2$: substitute `each finite' for `each'
  
\item Page 47, Line ${-10}$: remove `valid'

\item Page 62, Section 11.2.1: substitute `$a\_dcl$' for `$a\mbox{-}dcl$'
  
\item Page 69, Line ${-10}$: substitute `$I$' for `$\overleftharpoon{Inv}\Rightarrow Inv$'
 
\item Page 75:
  \begin{itemize}
    \item Line ${-9}$: substitute `$I$' for `$Dyn\wedge (\overleftharpoon{Inv} \Rightarrow Inv)$'
    \item Line ${-4}$: substitute `$0$' for `$j$'
  \end{itemize}

\item Page 77, Line $+7$: substitute `$\neg{Empty(1)}$' for `$\neg{Empty(l)}$'
  
\item Page 81, Line $+11$: substitute `$\overleftharpoon{TrmS}\wedge \overleftharpoon{Flag}$' for `$\overleftharpoon{TrmS}$'
        
\item Page 82:
   \begin{itemize}
     \item Line $+12$: substitute `$(L\cup\{Min\}=\overleftharpoon{L}\vee L=\overleftharpoon{L})$' for `$L\cup\{Min\}=\overleftharpoon{L}$'
     \item Line $+16$: substitute `$(S\cup\{Max\}=\overleftharpoon{S}\vee S=\overleftharpoon{S})$' for `$S\cup\{Max\}=\overleftharpoon{S}$'
   \end{itemize}

\item Page 83:
	\begin{itemize}
    \item Line $+9$: same as on Page 82, Line $+16$
    \item Line $+14$: same as on Page 82, Line $+12$
  \end{itemize}
  
\item Page 84:
	\begin{itemize}
    \item Line ${-14}$: same as on Page 82, Line $+12$
    \item Line ${-10}$: same as on Page 82, Line $+16$
  \end{itemize}
  
\item Page 86:
  \begin{itemize}
    \item Line $+12$: same as on Page 82 Line $+16$
    \item Line $+17$: same as on Page 82 Line $+12$
  \end{itemize}
	
\item Page 100, Line $+8$: substitute `,' for `.'

\item Page 117: substitute the two occurrences of `$x_n:=r_{x_n}$' by `$x_w:=r_{x_w}$'

\item Page 117: substitute the occurrence of `$1\le k\le n$' by `$1\le k\le w$'

\item Page 117: substitute the occurrence of `$1\le j<k\le n$' in the footnote by `$1\le j<k\le w$'

\item Page 118, Line $+14$: substitute `then
$Rel[R,V,U]$ denotes the assertion' by `then
$Rel[R,V,U]$, $Rel_1[R,V,U]$ and $Rel_2[R,V,U]$ denote respectively the assertions:'

\item Page 118, Line $+15$: substitute the full stop by comma and insert the following

\begin{flist}
(R\wedge (h=3\sconc \overleftharpoon{h}\Or h=2\sconc \overleftharpoon{h})\ \And\\ 
	\qquad\bigwedge_{v\in V} h_{v}=[v]\sconc \overleftharpoon{h_v}\ \wedge\\
  \qquad\bigwedge_{u\in U} h_{u}=[\overleftharpoon{h_u}(1)]\sconc \overleftharpoon{h_u})^*,\\
\\
(R\wedge (h=3\sconc \overleftharpoon{h}\Or h=1\sconc\overleftharpoon{h})\ \And\\
	\qquad\bigwedge_{v\in V} h_{v}=[v]\sconc \overleftharpoon{h_{v}}\ \And\\
  \qquad\bigwedge_{u\in U} h_{u}=[\overleftharpoon{h_u}(1)]\sconc \overleftharpoon{h_u})^*.
\end{flist}

\item Page 118, Line $+17$: substitute `function $Rel$ records any update due to the overall' by `$Rel$ functions record any update due to the relevant'

\item Page 118, Line $+18$: substitute `$Rel$' by `the $Rel$ functions'

\item Page 118, Line ${-12}$: substitute the left-most occurrence of `$::$' by `$\!\!\satis\!\!$'

\item Page 119:

  \begin{itemize}
    \item Line ${-15}$: substitute `$Rel_1[\true,\vartheta,U]$' for
    `$Rel[\true,\vartheta,U]$'
   \item Line ${-14}$: substitute `$Rel_2[\true,\vartheta,U]$'
    for `$Rel[\true,\vartheta,U]$'
  \end{itemize}

\item Page 121:

  \begin{itemize}
  \item Line $+4$: substitute `$Rel_2[\true,\vartheta',U]$' for
    `$Rel[\true,\vartheta',U]$'
  \end{itemize}

\item Page 125:

  \begin{itemize}
    \item Line $-8$: substitute `$Rel_1[\true,\vartheta,U]$' for
    `$Rel[\true,\vartheta,U]$'
    \item Line $-7$: substitute `$Rel_2[\true,\vartheta,U]$'
    for `$Rel[\true,\vartheta,U]$'
  \end{itemize}
  
\item Page 130:

  \begin{itemize}
    \item Line $-8$: substitute
    `$Rel_1[\true,\vartheta\cup\alpha,U]$' for
    `$Rel[\true,\vartheta\cup\alpha,U]$'
    \item Line $-7$: substitute
    `$Rel_2[\true,\vartheta\cup\alpha,U]$' for
    `$Rel[\true,\vartheta\cup\alpha,U]$'
  \end{itemize}
	
	\item Page 149, \cite{jb:BNF}: substitute `conference' for `conferance'
	
	\item Page 151, \cite{dmg:unity}: substitute `Conference' for `Conferance'

 \end{itemize}

\newpage 

\ 

\newpage

%% file: frontpage.tex
\thispagestyle{empty}
\vspace{2cm}

\begin{center}
\LARGE \bf DEVELOPMENT\\ OF PARALLEL PROGRAMS\\ ON SHARED DATA-STRUCTURES
\end{center}

\vspace{7cm}

\begin{center}
\small THIS TECHNICAL REPORT IS A SLIGHTLY MODIFIED VERSION OF \cite{Stoelen90}
\end{center}

\vspace{2cm}

\begin{center}\small
By\\
Ketil St\o len\\
December 1990
\end{center}

%% file: abstract.tex

\addcontentsline{toc}{chapter}{Abstract}
\chapter*{Abstract}

A syntax-directed formal system for the development of totally correct programs with
respect to an unfair shared-state parallel while-language is proposed. The system
can be understood as a compositional reformulation of the Owicki/Gries method for verification of parallel programs.

Auxiliary variables are used both as a specification tool to eliminate undesirable
implementations, and as a verification tool to make it possible to prove that an
already finished program satisfies a particular specification.

Auxiliary variables may be of any sort, and it is up to the user to define the
auxiliary structure he prefers. Moreover, the auxiliary structure is only a part
of the logic. This means that auxiliary variables do not have to be implemented as if
they were ordinary programming variables.

The system is proved sound and relatively complete with respect to an operational
semantics and employed to develop three nontrivial algorithms: the Dining-Philosophers,
the Bubble-Lattice-Sort and the Set-Partition algorithms.

Finally, a related method for the development of (possibly nonterminating) programs
with respect to four properties is described. This approach is then used to develop
Dekker's algorithm.

%% file: declaration.tex
\addcontentsline{toc}{chapter}{Declaration}
\chapter*{Declaration}

No portion of the work referred to in the thesis has been submitted in support of
an application for another degree or qualification of this or any other university
or other institute of learning. 

%% file: education.tex
\addcontentsline{toc}{chapter}{Education}

\chapter*{Education}

Ketil St\o len received a Cand.\ Scient.\ degree
in Computer Science from the University
of Oslo in June 1986. He worked 
at the Norwegian Defense Research Establishment,
Department of System Analysis from October 1986 until July
1987 (part of compulsory military service). 
He has been a postgraduate research 
student
at Manchester University since September 1987.

%% file: acknowledgement.tex
\addcontentsline{toc}{chapter}{Acknowledgements}
\chapter*{Acknowledgements}

First of all I would like to thank my supervisor, 
Professor Cliff B.\ Jones,
for advice, encouragement and for carefully 
reading through and commenting 
on a large number of notes and drafts.

I am also indebted to Xu Qiwen and Wojciech Penczek,
who have read and commented
on several chapters, and to 
Professor Howard Barringer for discussions on 
temporal logic.

A special thanks goes to Jill Jones for 
correcting the English in an early draft of this
thesis. 

Finally, I would like to thank my examiners, Professor Mathai Joseph and Professor
Howard Barringer, for their comments.

Financial support has been received from the 
Norwegian Research Council for
Science and the Humanities.

%% file: motiv.tex
\chapter{Introduction and Summary}

\section{Motivation}

\subsection{Program Development}

In the 1960's it became more and more clear that the existing techniques for 
program development
could not cope with the ever growing size and complexity of software products.
Naur and Floyd proposed the use of logic to formally prove that a program 
had its desired
effect \cite{pn:proof}, \cite{rwf:classic}\footnote{It is
interesting to note that Turing proposed similar strategies 
already in 1949 \cite{at:first}. Naur and Floyd were unaware of this paper.}.  

Hoare reformulated this idea as a set of syntax-directed rules for a simple
programming language \cite{carh:axiom}. This gave rise to what today is called 
Hoare-logic. 

So far
the emphasis was on verifying already finished programs.
In the early 1970's the attention turned towards stepwise program development. The
basic idea was that instead of verifying the correctness of a program first after
it had been finished, such reasoning should be part of the development process and
thereby simplify the argumentation and serve as a guide to the developer.
Dijkstra advocated such views already in 1965 \cite{ewd:step}. 
Wirth, Dahl and Hoare had similar
ideas (see \cite{nw:step}, \cite{ojd:step}, \cite{carh:step}), 
and they were given a formal setting in \cite{rm:reification}, 
\cite{carh:reification}, \cite{cbj:reification}.

\subsection{Concurrency}

Attempts were soon made to employ similar techniques to concurrent programs. In
Hoare-logic a triple of the form

\begin{flist}
\{P\}\ z\ \{Q\},
\end{flist}

\noindent often called a Hoare-triple, is valid if for any initial state which satisfies the 
pre-condition $P$, the program $z$ either ends in an infinite loop or terminates
in a state which satisfies the post-condition $Q$. 

Unfortunately, in the concurrent case such triples are insufficient since they do not
contain information about the other processes running in parallel.
This is why the Owicki/Gries \cite{so:thesis}, \cite{so:await}, \cite{so:res}
extension of Hoare-logic to cover parallel programs
on shared data structures and much of what followed depend upon an 
interference-freedom proof which first can be carried out when the whole program is 
complete. In these approaches programs are developed according to one set of
rules (closely related to the sequential rules of Hoare-logic) down to the most
concrete level, then 
the complete programs (with their proofs) are subject to a non-interference test. 

For large software products this strategy is totally unacceptable, because erroneous
design decisions, taken early in the design process, may remain undetected until
the implementors attempt to provide the final non-interference proof. Since, in
the worst case,  everything
that depends upon such mistakes will have to be thrown away, much
work could be wasted.

The existence of non-interference proofs is not restricted to systems for shared-state
parallel programs.
Several of the early CSP \cite{carh:csp} proof methods, like for example those described
in \cite{gml:csp} and \cite{kra:CSP}, are based on related strategies. 

\subsection{Compositionality}

To facilitate top-down development of programs, a proof method should
satisfy the principle of compositionality (see \cite{wpr:survey}, \cite{jz:CCC},
\cite{rk:comp}):
\begin{itemize}
\item A proof method is compositional if a program's specification can be
verified on the basis of the specifications of its constituent components, 
without knowledge of the
interior program structure of those components.
\end{itemize}
Hoare-logic is compositional, while the Naur/Floyd method \cite{rwf:classic}
is not; nor is any of the methods for concurrent programs discussed above.

Lamport proposed a system \cite{ll:GHL}, \cite{ll:CSP} 
suitable for proving
partial correctness of both shared-state and communication-based parallel
programs. Strictly speaking, the method is compositional. 
However, to prove that a program 
satisfies a specification, it is necessary to show that each of the component programs
satisfies the very same specification. 
This means that program decomposition does
not  reduce  the  specification's complexity and, as pointed out in \cite{ll:comp},
the method actually assumes a context of a
complete program.

A more suitable compositional proof system for CSP was proposed by Misra and Chandy 
in \cite{jm:safety}. Their proof tuples are of the form $r[h]s$, where
$h$ denotes a process, and $r$ and $s$ are assertions which express
respectively what is assumed by the process $h$, and what $h$ establishes under
this assumption. 
The method was later extended in \cite{jm:liveness} to deal with a weak form
of liveness. In the latter approach, the proof tuple is augmented with a third
assertion characterising when the process is guaranteed to progress. 

A compositional proof system 
for CSP was also suggested in \cite{zcc:partial}.

\subsection{Rely- and Guar-Conditions}

Francez and Pnueli were the first to reason about shared-state
parallel programs in terms of assumptions and commitments \cite{nf:thesis}, 
\cite{nf:cyclic}. The basic idea is:
\begin{itemize}
\item If the environment, by which we mean the set of processes running in
parallel with the one in question, fulfills the assumptions, then the actual
process is required to fulfill the commitments.
\end{itemize}

\noindent Jones employed rely- and guar-conditions 
\cite{cbj:thesis}, \cite{cbj:ifip}, \cite{cbj:acm} in a similar style.
However, while the previous approach focused essentially on program verification;
Jones main concern was top-down program development.
His proof tuples are of the form

\begin{flist}
 z\satis (P,R,G,Q),
\end{flist}

\noindent where $ z$ is a program and $(P,R,G,Q)$ is a specification consisting of four
assertions $P$, $R$, $G$ and $Q$.
The pre-condition $P$ and the rely-condition $R$
constitute assumptions that the developer can
make about the environment.
In return the implementation $ z$ must satisfy the 
guar-condition $G$, the post-condition
$Q$, and terminate, when operated in an environment which fulfills
the assumptions.

The pre-condition is intended to characterise a set of states to which the 
implementation is applicable. Any uninterrupted state transition by the environment
is supposed to satisfy the rely-condition, while any atomic state transition
by the implementation must satisfy the guar-condition.
The rely-condition is required to be both reflexive and transitive, while the
guar-condition is only constrained to be reflexive\footnote{In 
\cite{cbj:acm} the guar-condition is also required to be transitive.}.
Finally, the post-condition is required to characterise the overall effect of using
the implementation in such an environment.

This method allows erroneous interference decisions to be spotted and corrected at 
the level
where they are taken. Thus programs can be developed in a top-down style.

Unfortunately, the system cannot deal with synchronisation. Hence,
programs whose algorithms
depend upon some sort of delay-construct cannot be developed.

In \cite{cs:note} programs are specified in a similar style, although the rely-
and guar-conditions are represented as sets of invariants, and the post-condition
is unary, not binary as in \cite{cbj:thesis}. The paper
reformulates and generalises Owicki's approach (in \cite{so:await}) as 
 a system of syntax-directed rules for
a while-language extended with await- and  parallel-constructs. Although, this
method favours top-down development in the sense of \cite{cbj:thesis}, 
it can only be used for the design of partially correct programs.

In \cite{ns:oslo} shared-variable concurrency is 
dealt with in a CSP inspired style. 
The method is compositional, but rather difficult to use 
in practice.

\subsection{Temporal Logic}

Many methods have been suggested for the development of totally correct
programs in a similar way. In most cases they have been based on temporal logic.

The first truly compositional approach to (linear time)
temporal logic was due to Barringer, Kuiper
and Pnueli \cite{hb:preclass}, \cite{hb:class}. They offered an axiomatic 
semantics for a 
shared-state parallel language.
The technique was later modified to deal with a CSP-like language \cite{hb:csp}.
In both cases, they ended up with a very 
general specification language and logic.

\subsection{Compositional State Specifications}

As pointed out in \cite{ews:ref}, an alternative to the use of powerful 
temporal operators is the technique of
conceptual-state specification. In such a method, the behaviour of a process
with respect to a collection of program variables is specified with the help
of a collection of auxiliary variables, whose values serve as an
abstract representation of the internal state  of the process. Auxiliary
variables appearing in a specification are not intended to be implemented.

Approaches in this tradition are described in \cite{bth:monit}, \cite{so:temp} 
and \cite{ll:comp}.
Although these papers employ temporal operators to express liveness
properties, the operators are weaker. Instead, auxiliary 
variables are employed to give the necessary expressive power.

\subsection{Without Temporal Logic}

Several authors have suggested the use
of modal logic to prove termination of sequential programs. 
Burstall \cite{rmb:interm} introduced the so-called intermittent assertions method,
while Knuth applied related strategies on specific examples \cite{dek:art}. 

However, despite the fact that 
the while-construct has a highly `temporal' semantics, the standard 
techniques for proving termination of loops, 
the well-foundedness approach
and the loop method (see \cite{sk:term}),
 do not depend upon temporal logic. 
One may therefore ask, can these methods be extended to deal with shared-state
concurrency? In \cite{dl:ijf} and
\cite{zm:wf} invariance and well-foundedness are used to prove temporal
properties of parallel programs. This is further developed in \cite{zm:wftemp} 
where temporal logic
is applied together with invariance and well-foundedness. However,
these methods are not compositional.

Stark \cite{ews:rely/guar} has proposed
a proof technique where
a program can be proved to satisfy a rely/guarantee specification $R\supset
G$, given that the same program satisfies a finite collection of 
rely/guarantee specifications $R_i\supset G_i$.

In \cite{ba:buchi} Buchi automata are employed to specify temporal
properties. Proof obligations are derived from the automata and 
shown to satisfy an implementation by devising suitable invariant assertions
and variant functions.

Both these methods are a lot more general than our approach. However, neither of
them can be said to provide the conceptual simplicity of Hoare-logic.

\subsection{Other Systems}

The historical overview given above is far from complete. We
believe that the mentioned approaches are amongst the most important, but there are
of course many other interesting and influential papers that could and perhaps
should have been mentioned.

For example nothing has been said about action-based development methods like
Unity (see \cite{kmc:Unity} and \cite{dmg:unity}),
nor has Back's work been discussed \cite{rjrb:atom}, \cite{rjrb:90}. 
Other promising ideas are described in \cite{cl:comm}, 
\cite{rls:interval}, \cite{bm:tempura}, \cite{mb:marktoberdorf},
\cite{pa:object},  \cite{pkp:palogic},
\cite{pkp:assump} and \cite{ll:binary}.

In \cite{km:VIP} VDM is combined with temporal logic. However, this approach is
difficult to evaluate due to lack of worked out examples.

\section{Our Approach}

\subsection{Properties}

This thesis proposes a method for top-down development of totally correct
programs with respect
to an unfair shared-state while-language extended with await- and parallel-constructs. 
The system can be 
interpreted as a compositional reformulation of the Owicki/Gries
approach \cite{so:await}.

Because the programming language is unfair, the method cannot deal with 
programs whose algorithms rely upon busy waiting\footnote{It will later 
be indicated
how similar strategies can be employed to develop totally correct programs 
with respect to
fair programming languages (see page \pageref{fairness:ref}).}. For example, 
the parallel
composition of the two programs

\begin{prg}
\nbegin\\
\qquad \loc\ v;\\
\qquad v:=x;\\
\qquad \nwhile\ v\ \ndo\\
\qquad \qquad v:=x\\
\qquad \nod\\
\nend\\
\end{prg}

\noindent and 

\begin{prg}
x:=\false\\
\end{prg}

\noindent is not guaranteed to terminate, because the latter may be infinitely overtaken by
the former.

Another deficiency with our approach is that it cannot be used to develop 
nonterminating programs\footnote{It will shown below
that the method can be modified to develop (possibly nonterminating) programs with
respect to four properties (see page \pageref{nonterminating:ref}).}. 

\subsection{Lack of Generality}

There are several methods that
can be used to develop totally correct parallel programs with respect to an unfair
programming language, and since 
in most cases they are a lot more general than the one described in this thesis, it may
be argued that our system is of limited interest.

We do not agree with this. Although, it is quite possible to employ
temporal logic in its most general form 
to develop totally correct sequential programs, most
users would prefer to apply ordinary Hoare-logic in the style of for example
Z \cite{jms:Z} or VDM \cite{cbj:VDM2}. 
The reason is that these approaches are designed to deal with the
sequential case only, and they are therefore both simpler to use and easier to 
understand than a formalism powerful enough to deal with concurrency.

The same can be said with respect to the development of terminating programs versus
programs that are not supposed to terminate, and regarding different fairness
constraints.

Our approach can be considered as an attempt to
extend the method in \cite{cbj:thesis} to handle synchronisation, in the same way
as \cite{cbj:thesis} extends VDM to deal with interfering programs.

\subsection{Operational Semantics}

The programming language is given an operational semantics 
in the style of \cite{ka:potcomp}, \cite{gdp:opsem} and \cite{pa:note2}.
A (potential) computation
is defined as a possibly infinite sequence of external and internal 
transitions on configurations. An internal
transition represents a transition by the actual program, while an external
transition is due to the program's environment.
If a computation is finite, then no internal transition is enabled in its final
configuration. The latter constraint ensures that a program will 
always progress given 
that the program stays enabled and is not infinitely overtaken by its environment.
A configuration is blocked if it is disabled and its program component is
different from the empty program.

\subsection{Auxiliary Variables}

Auxiliary variables will be employed for two different purposes:

\begin{itemize}
\item To strengthen a specification to eliminate undesirable implementations.
In this case auxiliary variables are employed as a specification tool; they are used
to characterise a program that has not yet been implemented.

\item To strengthen a specification to make it possible to prove that an already
finished program satisfies a particular specification. Here auxiliary variables
are used as a verification tool, since they are introduced to show that a given
algorithm satisfies a specific property.
\end{itemize}

\noindent In \cite{so:thesis} and \cite{cs:note}, where auxiliary variables are applied 
only as a verification tool, auxiliary variables are first implemented 
as if they were ordinary programming
variables, and then afterwards removed
by a deduction rule specially designed for this purpose.

This is not a very satisfactory method, because in some specifications a
large number of auxiliary variables is needed, and the procedure of first 
implementing them and then removing them is rather tedious.

The method described in this thesis is more closely related to 
\cite{bth:monit}, \cite{ns:oslo},
\cite{carh:cspbook}, where the auxiliary 
structure is only a part of the logic and does not appear in the programs. 
Nevertheless,
although it is possible to define trace-related variables in our
system, auxiliary variables may be of any sort, and it is up to the user to
define the auxiliary structure he prefers.

\subsection{Specified Programs}

The proof tuples of our system will
be called specified programs and are of the form:

\begin{flist}
 z\satis\omega,
\end{flist}

\noindent where $z$ stands for a program and $\omega$ is a specification.

\subsection{Specifications}

A specification is a tuple of the form

\begin{flist}
(\vartheta,\alpha)::(P,R,W,G,E)
\end{flist}

\noindent consisting of two sets of variables $\vartheta$ and $\alpha$, and five
assertions $P$, $R$, $W$, $G$ and $E$.
The glo-set 
$\vartheta$ is the set of global
variables, while the aux-set $\alpha$ is the set of auxiliary variables.

The pre-condition $P$ and the rely-condition $R$ 
describe the assumptions that can be made about
the environment, while the wait-condition $W$, the guar-condition $G$ 
and the eff-condition $E$ 
constrain the implementation.

The pre-, rely- and guar-conditions
are identical to the similarly named assertions in
\cite{cbj:thesis}. The eff-condition corresponds to what 
\cite{cbj:thesis} calls the post-condition and
covers interference both before the first 
internal transition and after the last.
The wait-condition is supposed to characterise the set of states in which 
the implementation may become blocked. The implementation is not allowed to
become blocked inside the body of an await-statement.

\subsection{Convergent Environments}

An environment is called convergent if it can only perform a finite number of
consecutive atomic steps. This means that when a program is executed in a
convergent environment, then no computation has infinitely many external 
transitions unless it also has infinitely many internal transitions.

\subsection{Divergence, Convergence and Deadlock}

Given a convergent environment, an execution of a program will be said to 
diverge if one of its processes
ends in an infinite loop, or the body of one of its await-statements becomes blocked.
Moreover, it will be said to converge if it does not
diverge, and to deadlock if it converges but does not terminate (in which case it
is blocked in its final configuration).

\subsection{Satisfaction --- Ignoring Auxiliary Variables}

A specified program,

\begin{flist}
 z\satis(\vartheta,\emptyset)::(P,R,W,G,E),
\end{flist}

\noindent whose set of auxiliary variables is empty, is valid, 
if
\begin{itemize}
\item $ z$ converges, 
\item any atomic step due to the implementation satisfies
$G$,
\item $z$ can only become blocked in a state which satisfies $W$,
\item the overall effect is characterised by $E$ if $ z$ terminates,
\end{itemize}
whenever
\begin{itemize}
\item the environment is convergent,
\item $ z$ is called in a state which
satisfies $P$, 
\item any uninterrupted state transition by the
environment satisfies $R$.
\end{itemize}

\subsection{Satisfaction --- General Case}

In the general case,

\begin{flist}
 z_1\satis(\vartheta,\alpha)::(P,R,W,G,E)
\end{flist}

\noindent is valid, if there is a program $ z_2$ such that

\begin{flist}
 z_2\satis(\vartheta\cup\alpha,\emptyset)::(P,R,W,G,E)
\end{flist}

\noindent is valid, and $ z_1$ can be obtained from $ z_2$ by removing all auxiliary structure 
with respect to $\alpha$. The auxiliary structure is of course required to satisfy a
number of constraints.

\subsection{Total Correctness}

This means that if

\begin{flist}
z\satis(\vartheta,\alpha)::(P,R,W,G,E)
\end{flist}

\noindent is a valid specified program, and $z$ is executed in a convergent environment
characterised by $P$ and $R$, then $z$ either deadlocks in a state which satisfies
$W$ or terminates in a state such that the overall effect is characterised by
$E$. Thus, $z$ is totally correct with respect to the same specification
if $W$ is equivalent to false.

\subsection{Logic of Specified Programs}

The formal system, which is called LSP (Logic of Specified Programs), consists of 
sixteen decomposition-rules, and a base logic which is a first-order logic with a
second order extension to allow for well-foundedness proofs. 
Only twelve decomposition-rules are needed for top-down
development. The others have been introduced to make it easier to take advantage
of already finished developments.

The basic system, consisting of only twelve rules, 
is called LSP$_B$.

\subsection{Parallel-Rule}

The premises of the parallel-rule must ensure 
that the two component programs 
are compatible with respect to mutual interference.
It should also be clear that the parallel composition of two
programs is only guaranteed to converge
if called in an environment in which
both component programs are guaranteed to 
converge.
Thus, if the component
programs do not deadlock, then the following
rule is sufficient:

\begin{flist}
 z_1 \satis  (\vartheta,\alpha):: (P,R_1,\false,G\And R_2,E_1)\\
 z_2 \satis  (\vartheta,\alpha):: (P,R_2,\false,G\And R_1,E_2)\\
\overline{\{ z_1\parallel  z_2\} \satis  (\vartheta,\alpha):: 
(P,R_1\And R_2,\false,G,E_1\And E_2)}
\end{flist}

\noindent To formulate the general rule, it is enough to observe that $z_1$ is guaranteed
to be released whenever it becomes blocked in a state in which $z_2$ cannot
become blocked or terminate. Similarly, $z_2$ is guaranteed to be released in any state
in which $z_1$ cannot become blocked or terminate. Thus:

\begin{flist}
\neg(W_1\And E_2)\And \neg(W_2\And E_1)\And \neg(W_1\And W_2)\\
 z_1 \satis  (\vartheta,\alpha):: (P,R_1,W\Or W_1,G\And R_2,E_1)\\
 z_2 \satis  (\vartheta,\alpha):: (P,R_2,W\Or W_2,G\And R_1,E_2)\\
\overline{\{ z_1\parallel  z_2\} \satis  (\vartheta,\alpha):: 
(P,R_1\And R_2,W,G,E_1\And E_2)}
\end{flist}

\subsection{Soundness}

LSP is proved sound with respect to the operational semantics; in other words:

\begin{itemize}
\item for any structure $\pi$ and specified program $\psi$, 
\begin{itemize}
\item if $\psi$ is provable in LSP 
given the set of all
base-logic assertions, valid in $\pi$, as axioms, then $\psi$ is
valid in $\pi$.
\end{itemize}
\end{itemize}

\subsection{Relative Completeness}

Under a number of constraints on the assertion language and the set of legal 
structures, the
converse result is also shown:

\begin{itemize}
\item for any structure $\pi$ and specified program $\psi$, 
\begin{itemize}
\item if $\psi$ is valid in $\pi$, then $\psi$ is provable in LSP$_B$ 
given the set of all
base-logic assertions, valid in $\pi$, as axioms.
\end{itemize}
\end{itemize}

\noindent In other words, we can prove any valid specified program $\psi$ with respect
to a structure $\pi$, given an oracle capable of deciding if a base logic
assertion is valid in $\pi$ or not.

\subsection{Compositional Completeness}

The completeness result, together with the fact that our decomposition-rules 
are independent
of the program component's internal structure, implies that LSP$_B$ is 
compositionally complete \cite{jz:CCC}. 
Thus LSP favours top-down development. 

\subsection{Adaptation Completeness}

Unfortunately, our method is less
suited for bottom-up reasoning. It does not satisfy the following
criterion, also called adaptation completeness \cite{jz:CCC}:

\begin{itemize}
\item if the specification of a program, whose internal structure is unknown, implies
another specification for the same program, then the proof system admits a formal
deduction of that fact.
\end{itemize}

\noindent For example, since 

\begin{flist}
 z\satis(\{v\},\emptyset)::
(\true,v=\overleftharpoon{v},\false,v\ge \overleftharpoon{v},v=\overleftharpoon{v}),
\end{flist}

\noindent is valid only if $ z$ leaves $v$ unchanged, it should be possible to deduce

\begin{flist}
 z\satis(\{v\},\emptyset)::(\true,v=\overleftharpoon{v},\false,v=\overleftharpoon{v},v=\overleftharpoon{v}).
\end{flist}

\noindent However, there is no way to do this given the current 
set of rules.

\subsection{Achievements}

A compositional proof system for the deduction of totally correct shared-state
parallel programs with respect to an unfair programming language is presented.

With respect to \cite{so:await}, the main difference is that our system is
compositional, that auxiliary variables can be employed as a specification tool, and
that auxiliary variables are only a part of the logic and do not have to be implemented.

With respect to \cite{cbj:thesis}, the main difference is that our system can deal
with synchronisation, and that auxiliary variables can be used both as specification
and verification tools.

With respect to \cite{ns:oslo}, the main difference is that our system can be used
to prove total correctness and that auxiliary variables can be of any sort.

With respect to \cite{hb:class}, the main difference
is that our system is conceptually simpler because it is designed to deal with
one particular type of parallel programs. 

With respect to \cite{cs:note}, the main difference is that our method can be
used to prove total correctness, that auxiliary variables can be used as a 
specification tool, and that auxiliary variables are only a part of the logic and do
not have to be implemented.

\subsection{Organisation of Thesis}

The thesis is organised as follows. First the assertion language and its
semantics are introduced. 
Secondly, the programming language is defined and given
an operational semantics. 
The next two chapters deal respectively with
specifications and auxiliary variables.

Then specified programs are defined both at the syntactic level and with respect to
the operational semantics. Next, a number of simplifying syntactic operators
are defined, well-foundedness is discussed, the decomposition-rules are introduced, 
and the method is applied on three different examples;
the Bubble-Lattice-Sort, the Dining-Philosophers and the Set-Partition problems.

The approach is thereafter modified to 
allow development of possibly nonterminating programs with respect to four
properties.

Then, LSP is proved sound, and it is shown that LSP$_B$ is relatively
complete.

Finally, alternative approaches and relationships 
with other systems are discussed, and possible extensions are indicated.

%% file: asslang.tex
\chapter{First-Order Language}

\section{Syntax}

\subsection{Restrictions}

Let L be a many-sorted first-order language\index{first-order language} with 
equality. It is assumed
that $\Bool$ is a sort in L, and that L$_P$, the language of
Peano arithmetic on the sort $\Nat$, is contained in L.
Moreover, 
if $\Sigma$ is a sort in L, then $\seqof{\Sigma}$ is a sort in L,
and 

\begin{flist}
\len{\_}:\seqof{\Sigma}\rightarrow\Nat,\\ 
\_(\_):\seqof{\Sigma}\times\Nat\rightarrow \Sigma,\\
{\rm []}:\rightarrow\seqof{\Sigma},\\
{\rm [}\_{\rm ]}:\Sigma\rightarrow \seqof{\Sigma},\\
\_\sconc\_:\seqof{\Sigma}\times\seqof{\Sigma}\rightarrow
\seqof{\Sigma}
\end{flist}

\noindent are  function symbols in L\footnote{Alternatively, we could have introduced the
notion of an acceptable structure (see \cite{ynm:accept}).}.

The notation  $e:\Sigma$  will be used to state that $e$ is an {\em 
\underline{expression}}\index{expression} in L of sort $\Sigma$.

\subsection{Variables}

We will follow \cite{cbj:VDM2} in adding
hooks to variables
when it is necessary to refer to an earlier state (which is not 
necessarily the previous state). This means that, for any {\em \underline{unhooked
variable}}\index{unhooked variable} $v:\Sigma$, there is a {\em 
\underline{hooked variable}}\index{hooked variable} $\overleftharpoon{v}:\Sigma$.

To avoid unnecessary complication, the same variable
cannot be assigned to more than one sort. In other words, if $v_1:\Sigma_1$ and
$v_2:\Sigma_2$, then $\Sigma_1\neq\Sigma_2$ implies $v_1\neq v_2$.

Finally, it is assumed that the same variable cannot be both quantified (bound) 
and unquantified (free)
in the same expression. For example, this means that there is no expression in L
of the form

\begin{flist}
x=y\And \Forall x:x>15.
\end{flist}

\noindent To simplify expressions it is assumed that $\Implies$ has higher priority
than $\And$ and $\Or$, which again have higher priority than $\neg$, which has
higher priority than $=$, which has higher priority than 
all other function symbols in L.

This means that we can write

\begin{flist}
x+y=5\And y=2\Implies x=3,
\end{flist}

\noindent instead of

\begin{flist}
(((x+y)=5)\And (y=2))\Implies (x=3).
\end{flist}

\noindent Expressions of sort $\Bool$ will be called {\em \underline{assertions}}
\index{assertion}.

\section{Semantics}

\subsection{Structures}

The next step is to assign  meanings to expressions in L.

\begin{definition}
A {\em \underline{structure}}\index{structure} $\pi$ is a mapping of 
every sort $\Sigma$ in L, to a nonempty set
of values $\pi(\Sigma)$, and a mapping of every function symbol

\begin{flist}
f:\Sigma_1\times\ldots\times\Sigma_n\rightarrow \Sigma_{n+1}
\end{flist}

\noindent in L, to a total function

\begin{flist}
\pi(f):\pi(\Sigma_1)\times\ldots\times\pi(\Sigma_n)\rightarrow\pi(\Sigma_{n+1}).
\end{flist}

\noindent In particular, it is assumed that 
\begin{itemize}
\item $\pi(\Bool)=\{false,true\}$,
\item $\Nat$ is assigned the set of natural numbers,
\item L$_P$ is given its standard interpretation,
\item for any sort, the equality operator $\_=\_$ is given a standard interpretation,
\item if  $\Sigma$  is assigned
the set $\pi(\Sigma)$, then  $\seqof{\Sigma}$  is assigned the set of all 
finite sequences of the form

\begin{flist}
e_1e_2\ \ldots\ e_n,
\end{flist}

\noindent where for all $1\le j\le n$, $e_j$ is an element of $\pi(\Sigma)$, 
\item if $\pi(s)$ is a finite sequence, then the term  $\len{s}$  denotes the number
of elements in $\pi(s)$, 
\item if
$\pi(n)$ is a natural number greater than 0 and less than or equal to the number 
of elements
in the sequence $\pi(t)$, then the term  $t(n)$  
denotes the $\pi(n)$'th element of $\pi(t)$, 
\item the term  $[]$  denotes the empty sequence,
\item if $\pi(t)$ is an element of $\pi(\Sigma)$, then
the term  $[t]$  denotes the sequence
in $\pi(\seqof{\Sigma})$ consisting of one element $\pi(t)$, 
\item if $\pi(t_1)$ and $\pi(t_2)$ are elements of $\pi(\seqof{\Sigma})$,
then the term  $t_1\sconc t_2$ 
denotes the sequence in $\pi(\seqof{\Sigma})$ consisting of $\pi(t_1)$ concatenated
with $\pi(t_2)$.
\end{itemize}
\end{definition}

\subsection{Valuations and States}

A {\em \underline{valuation}}\index{valuation} $\mho$ over a structure $\pi$, 
is a mapping of all variables in 
L to values in the structure. Obviously, if $v:\Sigma$ 
is a variable in L, then $\mho(v)\in\pi(\Sigma)$. 
Similarly, a {\em \underline{state}}\index{state} is a mapping of
all unhooked variables to values.

Given a set of variables $V$ and two states $s_1$, $s_2$, then

\begin{flist}
s_1\stackrel{V}{=}s_2\index{\stackrel{V}{=}}
\end{flist}

\noindent denotes that for all variables $v\in V$, $s_1(v)=s_2(v)$, while

\begin{flist}
s_1\stackrel{V}{\neq}s_2\index{\stackrel{V}{=}}
\end{flist}

\noindent means that there is a variable $v\in V$, such that $s_1(v)\neq s_2(v)$.

\subsection{Validity}

Given a structure $\pi$, and a valuation $\mho$, 
then the expressions in L can be assigned meanings in the usual way.
In the end, every expression $t:\Sigma$ in L will denote a value
$\pi_{\mho}(t)\in \pi(\Sigma)$.
Given an assertion $A$, then 

\begin{flist}
\models_{\pi}A,
\end{flist}

\noindent if and only if,
for all valuations $\mho$ over $\pi$, 
$\pi_{\mho}(A)=true$; in other words, if and only if 
$A$ is {\em \underline{valid}}\index{valid} in $\pi$. Moreover,

\begin{flist}
\models A,
\end{flist}

\noindent if and only if $A$ is valid in any structure.
Similarly, if $(s_1,s_2)$ is a pair of states, then

\begin{flist}
(s_1,s_2)\models_{\pi} A,
\end{flist}

\noindent if and only if $\pi_{\mho}(A)=true$, where 
for all unhooked variables $v$, $\mho(\overleftharpoon{v})=s_1(v)$ and 
$\mho(v)=s_2(v)$.
The first state $s_1$ may be
omitted if $A$ has no occurrences of hooked variables.

\subsection{Unary and Binary Assertions}

An assertion $A$ defines the set of all pairs of states
$(s_1,s_2)$, such that

\begin{flist}
(s_1,s_2)\models_{\pi} A.
\end{flist}

\noindent If $A$ has no occurrences of hooked variables, it may also be thought of as the
set of all states $s$, such that

\begin{flist}
s\models_{\pi} A.
\end{flist}

\noindent We will use both interpretations below. This means that 
assertions without occurrences
of hooked variables can be given two different interpretations. To indicate
the intended meaning, 
we will distinguish between {\em \underline{binary assertions}}\index{binary assertion}
and {\em \underline{unary assertions}}. When an assertion is binary it 
denotes a set
of pairs of states, and when an assertion is unary it denotes a set
of states.
In other words, an assertion with occurrences of hooked variables is always binary,
while an assertion without occurrences of hooked variables can be both binary
and unary.

\subsection{Reflexivity}

A binary assertion $A$ will be said to be {\em \underline{reflexive}}\index{reflexive}, 
if for all states 
$s$, $(s,s)\models_{\pi} A$.

\subsection{Transitivity}

A binary assertion $A$ will be said to be {\em \underline{transitive}}\index{transitive}, 
if for all states $s_1$,
$s_2$, $s_3$, $(s_1,s_2)\models_{\pi} A$ and $(s_2,s_3)\models_{\pi} A$
implies $(s_1,s_3)\models_{\pi} A$.

\subsection{Respects}

A binary assertion $A$ {\em \underline{respects}} a set of variables $V$, 
if for all states $s_1$, $s_2$,
$(s_1,s_2)\models_{\pi} A$ implies $s_1\stackrel{V}{=}s_2$.

\subsection{Substitution}

Given $m+1$ expressions $t,r_1,\ldots, r_m$ and $m$ distinct variables 
$v_1,\ldots,v_m$, then $t(r_1/v_1,\ldots,r_m/v_m)$ denotes the result of
replacing any occurrence of $v_j$ ($1\le j\le m$) in $t$ with $r_j$.

%% file: prglang.tex
\chapter{Syntax of Programming Language}

\section{Motivation}

The programming language is closely related to the language used in \cite{so:await},
where a traditional while language is extended with a parallel-statement and
an await-construct. 
As pointed out in \cite{so:await}, this is a flexible but primitive
language, so primitive that other methods for synchronisation such as semaphores and
events can be easily implemented using it. In other words, a formal system for
this language can be employed to prove the correctness of programs using other
tools as well.
In our approach there is only one additional construct,
namely a block-statement.

\section{Syntax}

\subsection{BNF}

A {\em \underline{program}}\index{program}
 is a finite, nonempty list of symbols whose context-independent 
syntax can be characterised in the well-known BNF-notation \cite{jb:BNF}:

\begin{itemize}
\item Given that 
$<var>$ denotes an unhooked variable in L, 
$<exp>$ stands for an expression in L without hooks and quantifiers, and
$<tst>$ represents an expression in L of sort $\Bool$ without hooks and quantifiers, 
then any program will be of the form $<pg>$, where
\begin{eqnarray}
<pg>&::=&<sk>|<as>|<bl>|<sc>|<if>|\nonumber\\
    &   &<wd>|<pr>|<aw>\nonumber\\
<sk>&::=&\nskip\nonumber\\
<as>&::=&<var>:=<exp>\nonumber\\
<bl>&::=&\nbegin\ \loc<vl>;<pg>\nend\nonumber\\
<vl>&::=&<var>|<var>,<vl>\nonumber\\
<sc>&::=&<pg>;<pg>\nonumber\\
<if>&::=&\nif <tst> \nthen <pg> \nelse <pg> \nfi\nonumber\\
<wd>&::=&\nwhile <tst> \ndo <pg> \nod\nonumber\\
<pr>&::=&\{<pg>\parallel<pg>\}\nonumber\\
<aw>&::=&\nawait <tst> \ndo <pg> \nod\nonumber
\end{eqnarray}
\end{itemize}

\subsection{Additional Restrictions}

The main structure of a program is characterised above. However, a
syntactically correct program is also required to satisfy four supplementary
constraints, namely:
\begin{itemize}
\item Assignments:

For any assignment-statement ($<as>$), the variable on the left-hand side is of the
same sort as the expression on the right-hand side.

\item Scope of Variables:

The block-statement ($<bl>$) allows us to declare variables. A variable is {\em
\underline{local}}\index{local}
to a program, if it is declared in the program; otherwise it is said to
be {\em \underline{global}}\index{global}. For example,

\begin{prg}
\nbegin\\
\qquad\loc\ x,y;\\
\qquad x:=5+v;\\
\qquad y:=9-x\\
\nend\\
\end{prg}

declares two local variables, namely $x$ and $y$, while $v$ is a global
variable. To avoid complications
due to name clashes, it is required that 
\begin{itemize}
\item the same variable cannot be 
declared more than once in the same program, 
\item a local variable cannot appear outside its block.
\end{itemize}
The first constraint avoids name clashes between local variables, 
while the second ensures that
the set of global variables is disjoint from the set of local variables\footnote{There
are well-known techniques which allow these constraints to be relaxed. See for 
example \cite{jaa:thesis}.}.

\item Initialisation of Local Variables:

To avoid complicating the operational semantics, it is assumed that local 
variables are never read before they have been initialised. This means that

\begin{prg}
\nbegin\\
\qquad \loc\ y;\\
\qquad x:=y\\
\nend\\
\end{prg}

is not a syntactically correct program, because $y$ is not initialised before its
value is (read and) assigned to $x$ (see page \pageref{locinit:lab}).

\item Boolean Tests \label{booltest:text}:

\label{hidden:ref}
To simplify the decomposition rules and the reasoning with auxiliary variables,
it is assumed that for any program 
of the form $\{z_1\parallel
z_2\}$ or of the form $\{z_2\parallel z_1\}$, and any program $z_3$ contained in
$z_1$:
\begin{itemize}
\item if $z_3$ is a while- ($<wd>$) or if-statement ($<if>$), then its 
Boolean test ($<tst>$) can only access variables declared in $z_1$ (in other words,
variables local to $z_1$). 
\end{itemize}

\noindent This constraint does not reduce the number of implementable
algorithms, because if $b$ is a Boolean test with occurrences of global variables,
then

\begin{prg}
\nif\ b\ \nthen\\
\qquad z_1\\
\nelse\\
\qquad z_2\\
\nfi,\\
\end{prg}

\noindent can be rewritten as

\begin{prg}
\nbegin\\
\qquad \loc\ v;\\
\qquad v:=b;\\
\qquad \nif\ v\ \nthen\\
\qquad\qquad  z_1\\
\qquad \nelse\\
\qquad\qquad  z_2\\
\qquad \nfi\\
\nend,\\
\end{prg}

\noindent while a program of the form

\begin{prg}
\nwhile\ b\ \ndo\\
\qquad z\\
\nod,\\
\end{prg}

\noindent can be rewritten as

\begin{prg}
\nbegin\\
\qquad \loc\ v;\\
\qquad v:=b;\\
\qquad \nwhile\ v\ \ndo\\
\qquad\qquad  z;\\
\qquad\qquad v:=b\\
\qquad \nod\\
\nend.\\
\end{prg}

\end{itemize}

This constraint will later be discussed in more detail (see page \pageref{aux:ref}).

\subsection{Further Comments}

Observe that there are no extra constraints on programs that occur in
the bodies of await-statements ($<aw>$). Thus, nested await-statements are legal, and
parallel-statements may occur in the body of await-statements.

Moreover, there is no restriction with respect to the number of variable occurrences
on the right hand side of an assignment-statement.

\subsection{Notation}

In the usual way, a program $z_1$ is a {\em \underline{subprogram}}\index{subprogram} 
of a program $z_2$, if $z_1$
occurs in $z_2$. This means that any program is a subprogram of itself.

For any program $z$, $hid[z]$ \label{hid:ref} 
is the least set consisting of any variable, 
which is declared in $z$ (in other words, is a local variable in $z$) 
or occurs in the Boolean test of an if- or a 
while-statement in $z$.

For any expression or program $t$, $var[t]$ is the least set consisting
of the unhooked version of any free hooked or unhooked variable in 
$t$.

%% file: prgsem.tex
\chapter{Operational Semantics}

\section{Motivation}

The next step is to give a semantics for our programming language. We have 
decided to use an operational model (transition system) 
in the style of \cite{henplot:pap} but
extended to potential computations.
Usually, the meaning of a program in an operational semantics is characterised
by the program's execution sequences. Unfortunately, when dealing with
concurrency in a compositional way, this approach is not satisfactory.

The reason is that other programs running concurrently may interfere, and
this has led several authors, see
\cite{ka:potcomp}, \cite{gdp:opsem}, \cite{pa:note2}, 
\cite{hb:class} and \cite{cs:note},
to suggest that information about the environment should be included in the
execution sequences.
Our approach is in the latter tradition, and what we will call a computation is
actually a potential computation or alternatively a potential execution sequence.

\section{Computations}

\subsection{Configurations}

A {\em \underline{configuration}}\index{configuration} is a pair $< z,s>$, 
where $ z$ is either 
a program or the symbol $\epsilon$\index{$\epsilon$} which will be called the 
{\em \underline{empty program}}\index{empty program}, and
$s$ is a state.

\subsection{Transitions}

An {\em \underline{internal transition}}\index{internal transition} represents a 
transition by the actual
program, while a transition due to the program's environment is called
an {\em \underline{external transition}}\index{external transition}.
We will use two binary relations
on configurations, $\intern$\index{$\intern$} and $\exter$\index{$\extern$}, 
to characterise respectively 
the set of legal internal 
transitions and the set of legal external
transitions
\footnote{It may be argued that this definition could have been simplified by allowing
`programs' of the form $\epsilon; z_1$, $\{\epsilon\parallel  z_1\}$
and $\{ z_1\parallel\epsilon\}$. 
However, we find our approach `cleaner', and
also easier to work with in the definitions and proofs below.}:

\begin{definition}\label{trans:def}
Given a structure $\pi$, let $\exter$ be the binary relation, such that
\begin{itemize}
\item $< z,s_1>\exter < z,s_2>$,
\end{itemize}
and let $\intern$ be the least binary relation, such that either:
\begin{itemize}
\item $<\nskip,s>\intern <\epsilon,s>$,
\item $<v:=r,s>\intern <\epsilon,s(^v_r)>$, where $s(^v_r)$ denotes
the state that
is obtained from $s$, by mapping the variable $v$ to the value of the term $r$,
determined by $\pi$ and $s$, and leaving
all other maplets unchanged,
\item $<\nbegin\ \loc\ v_1,\ldots,v_n; z\ \nend,s>\intern < z,s>$,
\item $< z_1; z_2,s_1>\intern < z_2,s_2>$ if $< z_1,s_1>\intern 
<\epsilon,s_2>$,
\item $< z_1; z_2,s_1>\intern < z_3; z_2,s_2>$ if $< z_1,s_1>\intern 
< z_3,s_2>$ and $ z_3\neq \epsilon$,
\item $<\nif\ b\ \nthen\  z_1\ \nelse\  z_2\ \nfi,s>\intern < z_1,s>$ if $s
\models_{\pi}b$,
\item $<\nif\ b\ \nthen\  z_1\ \nelse\  z_2\ \nfi,s>\intern < z_2,s>$ if $s
\models_{\pi}\neg b$,
\item $<\nwhile\ b\ \ndo\  z_1\ \nod,s>\intern < z_1;\nwhile\ b\ \ndo\  z_1\ 
\nod,s>$ if 
$s\models_{\pi} b$,
\item $<\nwhile\ b\ \ndo\  z_1\ \nod,s>\intern <\epsilon,s>$ if $s\models_{\pi}
\neg b$,
\item $<\{ z_1\parallel  z_2\},s_1>\intern < z_2,s_2>$ if 
$< z_1,s_1>
\intern <\epsilon,s_2>$, 
\item $<\{ z_1\parallel  z_2\},s_1>\intern < z_1,s_2>$ if 
$< z_2,s_1>
\intern <\epsilon,s_2>$, 
\item $<\{ z_1\parallel  z_2\},s_1>\intern <\{ z_3\parallel  z_2\},s_2>$ if 
$< z_1,s_1>
\intern < z_3,s_2>$ and $ z_3\neq \epsilon$,
\item $<\{ z_1\parallel  z_2\},s_1>\intern <\{ z_1\parallel  z_3\},s_2>$ if 
$< z_2,s_1>
\intern < z_3,s_2>$ and $ z_3\neq \epsilon$,
\item $<\nawait\ b\ \ndo\  z_1\ \nod,s_1>\intern < z_n,s_n>$ 
if $s_1\models_{\pi} b$,
and 

\begin{itemize}
\item there is a list of configurations $< z_2,s_2>,
< z_3,s_3>,\ \ldots\ ,< z_{n_
1},s_{n_ 1}>$, such that for all $1<k\le n$,
$< z_{k_ 1},s_{k_ 1}>\intern < z_k,s_k>$ and
$ z_n=\epsilon$, 
\end{itemize}

\item $<\nawait\ b\ \ndo\  z_1\ \nod,s_1>\intern <\nawait\ b\ \ndo\  z_1\ \nod,s_1>$
if $s_1\models_{\pi} b$, and 

\begin{itemize}
\item there is an infinite list of configurations $< z_2,s_2>,
< z_3,s_3>,\ \ldots\ ,< z_n,s_n>,\ \ldots\ $, such that for all $k> 1$, 
$< z_{k_ 1},s_{k_ 1}>\intern< z_k,s_k>$, or
\item there is a finite list of configurations $< z_2,s_2>,
< z_3,s_3>,\ \ldots\ ,< z_n,s_n>$, where $ z_n\neq \epsilon$, there is no
configuration $< z_{n+1},s_{n+1}>$ such that 
$< z_n,s_n>\intern< z_{n+1},s_{n+1}>$, and 
for all $1<k\le n$, 
$< z_{k_ 1},s_{k_ 1}>\intern< z_k,s_k>$.
\end{itemize}

\end{itemize}
\end{definition}

\label{inter:ref}
\noindent The assignment-statement and Boolean tests are here interpreted as atomic. 
We will later discuss this constraint in more detail (see page \pageref{atomic:ref}).

There are two different internal transitions defined for the await-statement. The 
first deals with the case when the await-statement's body terminates. The second
models that the await-statement's body either ends in an infinite loop or becomes
blocked.
This topic will be discussed in greater detail below (see page \pageref{abort:ref}).
The different program constructs
are all deterministic (although the parallel-construct 
has a nondeterministic behaviour). Furthermore, all functions are
assumed to be total.
We will later explain how the system can be extended to handle
both nondeterministic statements (see page \pageref{nondet:ref})
and partial functions (see page \pageref{part:ref}).

\subsection{No Fairness}

No fairness constraint will be assumed. This means that a program can be
infinitely overtaken by its environment. Thus, given that $ z_1$ denotes
the program

\begin{prg}
\nbegin\\
\qquad\loc\ v;\\
\qquad v:=b;\\
\qquad\nwhile\ \neg v\ \ndo\\
\qquad\qquad v:=b\\
\qquad\nod\\
\nend,\\
\end{prg}

\noindent while $ z_2$ represents 

\begin{prg}
b:=\true,\\
\end{prg}

\noindent then the program $\{ z_1\parallel z_2\}$ is not guaranteed to terminate because
$ z_2$ can be infinitely overtaken by $ z_1$.
In other words, our system cannot be used to develop programs whose
algorithms depend upon busy waiting as in this example.

\subsection{Naive Attempt}

As explained above, because there is no fairness constraint, a program can be
infinitely overtaken by its environment. One might therefore think that
it is sufficient to define a computation as a possibly infinite
sequence of the form:

\begin{flist}
< z_1,s_1>\ \stackrel{a_1}{\rightarrow}_{\pi}\ < z_2,s_2>\
\stackrel{a_2}{\rightarrow}_{\pi}\quad 
\ldots\quad\stackrel{a_{k_ 1}}{\rightarrow}_{\pi}\ < z_k,s_k>\
\stackrel{a_{k}}{\rightarrow}_{\pi}\quad\ldots\quad,
\end{flist}

\noindent where each arrow stands for an atomic transition, either internal
or external, and for all $j\ge 1$,
$s_j$ is the state (with respect to the  structure $\pi$) after $j_ 1$
transitions, while
$ z_j$ either equals $\epsilon$, in which case the actual program has terminated,
or is a program describing what is left to be executed.

Unfortunately, this is too weak; for two reasons: we need both a progress property and a
way to constrain the environment's access to hidden variables.

\subsection{Progress}

Nobody doubts that the sequential program

\begin{prg}
x:=1\\
\end{prg}

\noindent (given its usual semantics) eventually will terminate. 
The reason is that any sensible sequential programming
language satisfies the following progress property, 
\begin{itemize}
\item if something can happen then eventually something will happen.
\end{itemize}

\noindent In the concurrent case, with respect to an unfair programming language, a slightly
different progress property is required; namely that if the actual program 
is `enabled' in the current 
configuration, then eventually a transition, either internal or external, will
take place.

Otherwise, sequences of the form

\begin{flist}
<\nskip,s>\\
<\nskip,s_1>\exter<\nskip,s_2>\exter<\nskip,s_3>\exter<\nskip,s_4>\\
<v_1:=t_1;v_2:=t_2,s_1>\exter<v_1:=t_1;v_2:=t_2,s_2>\\
\qquad\intern<v_2:=t_2,s_3>
\end{flist}

\noindent are correct `computations', and no program is totally correct, since this would
allow programs to `stop executing' without being disabled or infinitely overtaken
by the environment.

To give a more accurate description of the progress property, and thereby
define what we mean by a computation, we must first characterise when a configuration
is enabled:

\begin{definition}
A configuration $c_1$ is {\em \underline{enabled}}\index{enabled}
if there is a configuration $c_2$, such that
$c_1\ \intern\ 
c_2$. If a configuration is not enabled, it will be said to be 
{\em \underline{disabled}}\index{disabled}. Finally, a configuration is
{\em \underline{blocked}}\index{blocked}, if it is disabled, 
and its program component is different from
the empty program.
\end{definition}

\noindent The progress property can then be formulated as follows: a computation
is either infinite or the final configuration is disabled.

\subsection{Hidden Variables}

The second reason why the `definition' of a computation given above is
insufficient is that the environment must be constrained from updating hidden
variables.
For example, if $ z_1$ denotes the program

\begin{prg}
\nif\ v=0\ \nthen\\
\qquad\nskip\\
\nfi,\\
\end{prg}

\noindent then there is no program of the form $\{ z_1\parallel z_2\}$, because
$v$ is not declared in $ z_1$ (see page \pageref{hidden:ref}).
To take advantage of this when formulating the decomposition-rules, 
another restriction on computations is needed. 
The following concept is useful:

\begin{definition}
An external transition 

\begin{flist}
< z,s_1>\exter< z,s_2>
\end{flist}

\noindent \underline{{\rm respects}}\index{respect} a set of variables $V$, if and only if 
$s_1\stackrel{V}{=}s_2$.
\end{definition}

\noindent Then, if $ z$ denotes the program

\begin{prg}
\nbegin\\
\qquad\loc\ y;\\
\qquad y:=v;\\
\qquad\nwhile\ x<100\ \ndo\\
\qquad\qquad x:=x+y\\
\qquad\nod\\
\nend,\\
\end{prg}

\noindent it is clear that $hid[ z]=\{x,y\}$ ($hid$ is defined on page \pageref{hid:ref}). 
Since no program running in parallel with $z$
can change the value of $x$ or $y$, the required constraint with respect 
to this example
is of course that
any external transition in a computation of $ z$ must respect
$hid[ z]$.

\subsection{Summary}

A computation can now be defined as below:

\begin{definition}\label{cp:def}
A {\em \underline{computation}}\index{computation} is a possibly
infinite sequence
of the form

\begin{flist}
< z_1,s_1>\ \stackrel{a_1}{\rightarrow}_{\pi}\ < z_2,s_2>\
\stackrel{a_2}{\rightarrow}_{\pi}\quad 
\ldots\quad\stackrel{a_{k_ 1}}{\rightarrow}_{\pi}\ < z_k,s_k>\
\stackrel{a_{k}}{\rightarrow}_{\pi}\quad\ldots\quad,
\end{flist}

\noindent such that
\begin{itemize}
\item $a_j\in\{e,i\}$ and
$< z_j,s_j>\ \stackrel{a_j}{\rightarrow}_{\pi}\ < z_{j+1},s_{j+1}>$, 
\item any external transition respects $hid[ z_1]$, 
\item the sequence is either infinite or the final configuration is disabled.
\end{itemize}
\end{definition}

\subsection{Why Local Variables Must be Initialised}

\label{locinit:lab}One of the constraints on programs is that 
local variables must be initialised before they are
 read. To see why, it is enough to observe that without this constraint and 
given the semantics above, the `program'

\begin{prg}
\nwhile\ v>0\ \ndo\\
\qquad\nbegin\\
\qquad\qquad \loc\ x;\\
\qquad\qquad x:=x_ 1;\\
\qquad\qquad v:=x\\
\qquad\nend\\
\nod\\
\end{prg}

\noindent is guaranteed to terminate. 
The problem is that the operational semantics `remembers' the value of $x$ from
the previous iteration. Thus if the initial value of $x$ is $m$, the loop will
terminate after $m$ iterations.

One way to deal with this is to weaken the second constraint in definition 
\ref{cp:def}, and only constrain the environment to leave local variables
unchanged while they are `active'.

A second alternative is to define some sort of initialisation convention at
semantic level.
However, to avoid complicating the operational semantics, we are instead 
insisting that local variables must be initialised before they are read.

\subsection{Useful Notation}

Given a computation $\sigma$, then $\tau(\sigma)$\index{$\tau(\sigma)$}, 
$\delta(\sigma)$\index{$\delta(\sigma)$} and 
$\lambda(\sigma)$\index{$\lambda(\sigma)$}
are the obvious projection functions to 
sequences of possibly
empty programs, states and transition labels (actions), while 
$\tau(\sigma_j)$\index{$\tau(\sigma_j)$},
$\delta(\sigma_j)$\index{$\delta(\sigma_j)$}, 
$\lambda(\sigma_j)$\index{$\lambda(\sigma_j)$} and $\sigma_j$\index{$\sigma_j$} 
denote respectively 
the $j$'th program, 
the $j$'th state, the $j$'th transition label and the $j$'th configuration. 
Furthermore, $\sigma(j,\ldots,k)$\index{$\sigma(j,\ldots,k)$}
represents

\begin{flist}
<\tau(\sigma_j),\delta(\sigma_j)>
\stackrel{\lambda(\sigma_j)}{\rightarrow}_{\pi}\quad\ldots
\quad\stackrel{\lambda(\sigma_{k_ 1})}{\rightarrow}_{\pi}
<\tau(\sigma_{k}),\delta(\sigma_k)>,
\end{flist}

\noindent while $\sigma(j,\ldots,\infty)$\index{$\sigma(j,\ldots,\infty)$} 
stands for

\begin{flist}
<\tau(\sigma_j),\delta(\sigma_j)>\stackrel{\lambda(\sigma_j)}{\rightarrow}_{\pi}
\quad\ldots
\quad\stackrel{\lambda(\sigma_{k_ 1})}{\rightarrow}_{\pi}
<\tau(\sigma_{k}),\delta(\sigma_k)>\stackrel{\lambda(\sigma_k)}{\rightarrow}_{\pi}
\quad\ldots\quad.
\end{flist}

\noindent If $\sigma$ is infinite, then $len(\sigma)=\infty$, otherwise 
$len(\sigma)$\index{$len(\sigma)$} is
equal to its number of configurations.

Finally, $\sigma_j\stackrel{V}{=}\sigma'_k$\index{$\sigma_j\stackrel{V}{=}\sigma'_k$} 
means that $\tau(\sigma_j)=
\tau(\sigma'_k)$ and $s(\sigma_j)\stackrel{V}{=}s(\sigma'_k)$, while
$\sigma_j\stackrel{V}{\neq}\sigma'_k$\index{$\sigma_j\stackrel{V}{\neq}\sigma'_k$} 
means that $\tau(\sigma_j)\neq
\tau(\sigma'_k)$ or $s(\sigma_j)\stackrel{V}{\neq}s(\sigma'_k)$.

\subsection{Meanings}

As explained above, the meaning of a program is characterised by its set of 
computations. We will use a special notation to refer to this set:

\begin{definition}\label{fcp:def}
Given a structure $\pi$, and a program $z$, let $cp_{\pi}[ z]$ be the 
set of all computations $\sigma$ in $\pi$
such that $\tau(\sigma_1)= z$.
\end{definition}

\subsection{Uninterrupted Transitions}

By an {\em \underline{uninterrupted}}\index{uninterrupted} state transition
by the environment we mean the overall
effect of a finite number of external transitions not interrupted by any internal 
transition. Similarly, an uninterrupted state transition by the implementation is
the overall effect of a finite number of internal transitions not interrupted by
any external transition.

\section{Composition}

\subsection{Motivation}

A computation $\sigma'$ of
a program $ z_1$  will be said to be {\em \underline{compatible}}\index{compatible} 
with a 
computation $\sigma''$ of 
a program $ z_2$, if they
are of the same length, have identical sequences of states, and are not
performing internal transitions at the same time.

For example, let $ z_1$ denote the program

\begin{prg}
v:=v+t_1;z_3,\\
\end{prg}

\noindent where $z_3$ is an atomic statement, and assume that $ z_2$ represents 

\begin{prg}
v:=v+t_2\\
\end{prg}

\noindent then

\begin{flist}
< z_1,s_1>\intern< z_3,s_2>\intern<\epsilon,s_3>
\end{flist}

\noindent is a computation of $ z_1$ which is not compatible with any computation
of $ z_2$, because each finite computation of $ z_2$ has exactly one internal
transition and the computation above has no external transitions.

On the other hand, the computation

\begin{flist}
< z_1,s_1>\intern< z_3,s_2>\exter< z_3,s_3>\exter< z_3,s_4>\intern<\epsilon,s_5>
\end{flist}

\noindent of $ z_1$, and

\begin{flist}
< z_2,s_1>\exter< z_2,s_2>\intern<\epsilon,s_3>\exter<\epsilon,s_4>
\exter<\epsilon,s_5>
\end{flist}

\noindent of $ z_2$, are compatible.
Furthermore, they can be `composed' into a computation 

\begin{flist}
<\{ z_1\parallel z_2\},s_1>\intern<\{ z_3\parallel z_2\},s_2>
\intern< z_3,s_3>\\
\qquad\exter< z_3,s_4>\intern<\epsilon,s_5>
\end{flist}

\noindent of $\{ z_1\parallel z_2\}$, by `composing' the program part of each configuration
pair, and making a transition internal if and only if one of the two
component transitions are internal.

\subsection{Summary}

\begin{definition}
Given a computation $\sigma'$ of $ z_1$ and a computation $\sigma''$ of $ z_2$,
then $\sigma'$ and $\sigma''$ are {\em \underline{compatible}}, also 
written $\sigma'\bullet\sigma''$\index{$\sigma'\bullet\sigma''$}
if and only if
\begin{itemize}
\item $len(\sigma')=len(\sigma'')$,
\item $\delta(\sigma')=\delta(\sigma'')$, 
\item for all $1\le j\le len(\sigma')$, $\lambda(\sigma'_j)=\lambda(\sigma''_j)$ implies
$\lambda(\sigma'_j)=e$.
\end{itemize}
\end{definition}

\noindent By joining together the
compatible computations as described above, the set of computations
representing the parallel composition of two programs can be generated
from the two sets of computations characterising the subprograms.

\section{Decomposition}

\subsection{Motivation}

We have already shown how the execution sequences of two component programs
can be used to determine the computations of their parallel composition. The next
step is to indicate a method for decomposing computations.

Given that $ z_1$ and $ z_2$ denote the same programs as above, then

\begin{flist}
<\{ z_1\parallel z_2\},s_1>\exter<\{ z_1\parallel z_2\},s_2>\intern< z_1,s_3>
\exter < z_1,s_4>\\
\qquad \exter < z_1,s_5>\intern< z_3,s_6>\intern<\epsilon,s_7>
\end{flist}

\noindent is a computation of $\{ z_1\parallel z_2\}$, while

\begin{flist}
< z_1,s_1>\exter< z_1,s_2>\exter< z_1,s_3>\exter < z_1,s_4>\exter < z_1,s_5>\\
\qquad \intern< z_3,s_6>\intern<\epsilon,s_7>
\end{flist}

\noindent is a computation of $ z_1$, and

\begin{flist}
< z_2,s_1>\exter< z_2,s_2>\intern<\epsilon,s_3>
\exter <\epsilon,s_4>\exter <\epsilon,s_5>\\
\qquad \exter<\epsilon,s_6>\exter<\epsilon,s_7>
\end{flist}

\noindent is a computation of $ z_2$. 
In other words, a computation $\sigma$ of 
$\{ z_1\parallel z_2\}$ can be decomposed into two computations of $\sigma$'s length
and with $\sigma$'s sequence of states.
Each configuration in the computation
of $ z_1$ (respectively $ z_2$) gets `what is left of' of $ z_1$ (respectively
$ z_2$) in the corresponding configuration of the initial
computation.
Moreover, an external transition is mapped into two external transitions, while 
an internal transition is split into one internal and one external transition.

\subsection{Summary}

\begin{definition}
Given three computations, $\sigma$, $\sigma'$ and $\sigma''$,
then $\sigma'\bullet\sigma''\leftarrow\sigma$\index{$\sigma'\bullet\sigma''
\leftarrow\sigma$}, 
if and only if 
\begin{itemize}
\item $\sigma'\bullet\sigma''$,
\item $len(\sigma)=len(\sigma')$,
\item $\delta(\sigma)=\delta(\sigma')$, 
\item for all $1\le j\le len(\sigma)$, 
\begin{itemize}
\item $\tau(\sigma'_j)=\epsilon$ implies $\tau(\sigma_j)=\tau(\sigma''_j)$,
\item $\tau(\sigma''_j)=\epsilon$ implies $\tau(\sigma_j)=\tau(\sigma'_j)$,
\item $\tau(\sigma'_j)\neq \epsilon$ and $\tau(\sigma''_j)\neq \epsilon$ implies
      $\tau(\sigma_j)=\{\tau(\sigma'_j)\parallel\tau(\sigma''_j)\}$, 
\item $\lambda(\sigma_j)=e$ if and only if $\lambda(\sigma'_j)=e$ and $\lambda(\sigma''_j)=e$.
\end{itemize}
\end{itemize}
\end{definition}

\section{Composition Proposition}

\begin{statement}
Given a computation $\sigma'$ of $ z_1$ and a computation $\sigma''$ of $ z_2$,
such that $\sigma'\bullet\sigma''$, then there is a computation
$\sigma$ of $\{ z_1\parallel z_2\}$ such that 
$\sigma'\bullet\sigma''\leftarrow\sigma$.
\end{statement}

\begin{nproof} We will first give an algorithm for the construction of $\sigma$.
Let $\tau(\sigma_1)=\{ z_1\parallel z_2\}$, and $\delta(\sigma_1)=\delta(\sigma'_1)$. 
Moreover, iteratively, in ascending order, for all $2\le k\le len(\sigma')$:
\begin{itemize}

\item let $\delta(\sigma_k)=\delta(\sigma'_k)$;

\item if $\lambda(\sigma'_{k_ 1})=\lambda(\sigma''_{k_ 1})=e$ then
\begin{itemize}
\item let $\lambda(\sigma_{k_ 1})=e$
\end{itemize}
\item else
\begin{itemize}
\item let $\lambda(\sigma_{k_ 1})=i$;
\end{itemize}

\item end if;

\item if $\tau(\sigma'_k)=\epsilon$ 
\begin{itemize}
\item let $\tau(\sigma_k)=\tau(\sigma''_k)$
\end{itemize}
\item else if $\tau(\sigma''_k)=\epsilon$
\begin{itemize}
\item let $\tau(\sigma_k)=\tau(\sigma'_k)$
\end{itemize}
\item else 
\begin{itemize}
\item let $\tau(\sigma_k)=\{\tau(\sigma'_k)\parallel\tau(\sigma''_k)\}$
\end{itemize}

\item end if;

\end{itemize}

\noindent It can easily be shown by induction that
each transition step
in $\sigma$ is an element of either $\intern$ or $\exter$. Therefore, since 
\begin{itemize}
\item $len(\sigma)=len(\sigma')=len(\sigma'')$, 
\item $len(\sigma')\neq \infty$ implies that $\sigma'_{len(\sigma')}$ and $\sigma''_{len(\sigma'')}$
are both disabled, which again implies that
$\sigma_{len(\sigma)}$ is disabled, 
\item any external transition in $\sigma$ is also an external transition in both 
$\sigma'$ and $\sigma''$, and must therefore respect  $hid[ z_1]\cup hid[ z_2]$,
\end{itemize}
it follows that $\sigma$ is a computation.

\end{nproof}

\section{Decomposition Proposition}
 
\begin{statement}\label{decomp1:theo}
Given a program $\{ z_1\parallel z_2\}$ and one of its computations $\sigma$,
then there are two computations $\sigma'$ and $\sigma''$ of respectively
$ z_1$ and $ z_2$, such that

\begin{flist}
\sigma'\bullet\sigma''\leftarrow 
\sigma.
\end{flist}

\end{statement}

\begin{nproof} We will first give an algorithm for the 
construction of $\sigma_1$ and $\sigma_2$. Let 
$\tau(\sigma'_1)= z_1$, $\tau(\sigma''_1)= z_2$, and
$\delta(\sigma'_1)=\delta(\sigma''_1)=\delta(\sigma_1)$. Moreover, 
iteratively, in ascending order, for all $2\le k \le len(\sigma)$:

      \begin{itemize}

      \item let $\delta(\sigma'_k)=\delta(\sigma''_k)=\delta(\sigma_k)$;

      \item if $\lambda(\sigma_{k_ 1})=e$, let
            \begin{itemize}
            \item $\tau(\sigma'_k)=\tau(\sigma'_{k_ 1})$;
                  $\tau(\sigma''_k)=\tau(\sigma''_{k_ 1})$;
            \item $\lambda(\sigma'_{k_ 1})=e$;
                  $\lambda(\sigma''_{k_ 1})=e$;
            \end{itemize}

      \item else if $\tau(\sigma'_{k_ 1})=\epsilon$, let
            \begin{itemize}
            \item $\tau(\sigma'_k)=\epsilon$;
                  $\tau(\sigma''_k)=\tau(\sigma_k)$;
            \item $\lambda(\sigma'_{k_ 1})=e$;
                  $\lambda(\sigma''_{k_ 1})=i$;
            \end{itemize}

      \item else if $\tau(\sigma''_{k_ 1})=\epsilon$, let
            \begin{itemize}
            \item $\tau(\sigma'_k)=\tau(\sigma_k)$;
                  $\tau(\sigma''_k)=\epsilon$;
            \item $\lambda(\sigma'_{k_ 1})=i$;
                  $\lambda(\sigma''_{k_ 1})=e$;
            \end{itemize}

      \item else if $\tau(\sigma_k)=\tau(\sigma''_{k_ 1})$, let

            \begin{itemize}

            \item $\tau(\sigma'_k)=\epsilon$;
                  $\tau(\sigma''_k)=\tau(\sigma''_{k_ 1})$;
            \item $\lambda(\sigma'_{k_ 1})=i$;
                  $\lambda(\sigma''_{k_ 1})=e$;

            \end{itemize}

      \item else if $\tau(\sigma_k)=\tau(\sigma'_{k_ 1})$, let

            \begin{itemize}

            \item $\tau(\sigma'_k)=\tau(\sigma'_{k_ 1})$;
                  $\tau(\sigma''_k)=\epsilon$;
            \item $\lambda(\sigma'_{k_ 1})=e$;
                  $\lambda(\sigma''_{k_ 1})=i$;

            \end{itemize}

      \item else if $\tau(\sigma_k)$ is of the form 
             $\{\tau\parallel\tau(\sigma''_{k_ 1})\}$
             and $<\tau(\sigma'_{k_ 1}),s_{k_ 1}>\intern <\tau,s_k>$, let

            \begin{itemize}

            \item $\tau(\sigma'_k)=\tau$;
                  $\tau(\sigma''_k)=\tau(\sigma''_{k_ 1})$;
            \item $\lambda(\sigma'_{k_ 1})=i$;
                  $\lambda(\sigma''_{k_ 1})=e$;

            \end{itemize}

      \item else if $\tau(\sigma_k)$ is of the form 
             $\{\tau(\sigma'_{k_ 1})\parallel\tau\}$
            and $<\tau(\sigma''_{k_ 1}),s_{k_ 1}>\intern <\tau,s_k>$, let

            \begin{itemize}

            \item $\tau(\sigma'_k)=\tau(\sigma'_{k_ 1})$;
                  $\tau(\sigma''_k)=\tau$;
            \item $\lambda(\sigma'_{k_ 1})=e$;
                  $\lambda(\sigma''_{k_ 1})=i$;

            \end{itemize}

          \item end if;

\end{itemize}

\noindent It can easily be shown by induction 
that each transition step in
both $\sigma'$ and $\sigma''$ is an element of either $\intern$ or $\exter$.
Moreover, since 
\begin{itemize}
\item $len(\sigma')=len(\sigma'')=len(\sigma)$, 
\item $len(\sigma)\neq \infty$ and $\sigma_{len(\sigma)}$ is disabled implies
that both $\sigma'_{len(\sigma')}$ and $\sigma''_{len(\sigma'')}$ are disabled,
\item any external transition in $\sigma$ respects $hid[\{ z_1\parallel z_2\}]$,
\end{itemize}
it follows that both $\sigma'$ and $\sigma''$ are computations.

\end{nproof}

%% file: spec.tex
\chapter{Specifications}

\section{Motivation}

The object of this chapter is first of all to characterise what we mean by a
{\em \underline{specification}}\index{specification}
at syntactic level, and secondly to indicate its intended semantics.

\section{Syntax}

\subsection{Tuple}

A specification is a tuple of the form

\begin{flist}
(\vartheta,\alpha)::(P,R,W,G,E).
\end{flist}

\noindent The {\em \underline{pre-condition}}\index{pre-condition} $P$, 
the {\em \underline{rely-condition}}\index{rely-condition} $R$
and the {\em \underline{guar-condition}}\index{guar-condition} $G$
are identical to
 the similarly named 
assertions in \cite{cbj:thesis}, while the {\em 
\underline{eff-condition}}\index{eff-condition}
$E$ corresponds to what 
\cite{cbj:thesis} calls the 
post-condition. Moreover, $W$ will be called 
the {\em \underline{wait-condition}}, while $\vartheta$
and $\alpha$ are finite sets of variables, called respectively the 
{\em \underline{glo-set}}\index{glo:set} and the {\em \underline{aux-set}}
\index{aux:set}.

\subsection{Assertions}

The pre- and wait-conditions are unary assertions. In other words, they are
 without occurrences of
hooked variables. 
The rely-, guar- and eff-conditions are binary assertions and may therefore 
refer to both hooked and
unhooked variables, in which case
the hooked variables refer to the `older' state.

\subsection{Variable Sets}

The glo-set is assumed to contain the global implementable
variables, while the elements of the 
aux-set will be called auxiliary
variables. The auxiliary variables are not 
implementable (with respect to the actual
specification).

Since the elements of the two sets belong to the same syntactical 
category, the sets are required to be disjoint. Moreover, to make sure
that the assertions are only constraining variables from the two sets, the unhooked
version of any free variable that appears in the specification is required to be
an element of one of the two sets.

\section{Semantics}

\subsection{Assumptions}

A specification states a number of assumptions about the environment. 
\begin{itemize}
\item First of all,
it is assumed that the environment can only perform a finite number of consecutive
atomic steps. Such an environment is said to be convergent.
Thus, when a program is executed in a convergent environment,
then no computation has infinitely many external transitions unless it also has 
infinitely many internal transitions.

This is
not a fairness requirement on the programming language, because it 
does not constrain the implementation of a specification. If for example 
a parallel-statement $\{z_1\parallel z_2\}$ 
occurs in the implementation then this assumption does not
influence whether or not $z_1$ is infinitely overtaken by
$z_2$.

Moreover, the assumption can be removed (it is explained how
on page \pageref{conver:ref}). The only disadvantage would be that we no longer
can use our formal system to prove that a program terminates, but only that a program 
terminates when it is not infinitely overtaken by the environment.

\item Another more important assumption  is that the initial state satisfies
the pre-condition.

\item Furthermore, it is also assumed that
any external transition  satisfies the rely-condition.
For example, given the rely-condition 

\begin{flist}
x\le \overleftharpoon{x}\And y=\overleftharpoon{y},
\end{flist}

then it is assumed that the environment will never change the value of $y$. 
Moreover, if the
environment assigns a new value to $x$, then this value
will be less than or equal to the variable's 
previous value.

To make it possible for the implementor to assume that
any uninterrupted interference between two internal transitions
satisfies the rely-condition, 
the rely-condition is required to be 
reflexive because the environment may not interfere at all, and transitive
to cover the case when the environment interferes more than once. This 
constraint has also a simplifying effect on some of the decomposition-rules. (It is
shown on page \pageref{reftran:ref} how the reflexivity and transitivity requirements 
can be removed.) 
\end{itemize}

\subsection{Commitments}

A specification is of course not only stating assumptions about the environment,
but also commitments to the implementation. To simplify their formulation we
will introduce some new notation:
Given a convergent environment, 
a computation $\sigma$ will be said to 
{\em \underline{terminate}}\index{terminate} if $len(\sigma)\neq\infty$ and 
$\tau(\sigma_{len(\sigma)})=\epsilon$, to 
{\em \underline{deadlock}}\index{deadlock}
if $len(\sigma)\neq\infty$ and 
$\tau(\sigma_{len(\sigma)})\neq\epsilon$, to
{\em \underline{diverge}}\index{diverge}
if $len(\sigma)=\infty$, and to {\em \underline{converge}}\index{converge}
if $\sigma$ does not diverge. 

\begin{itemize}
\item Given an environment which satisfies the assumptions, then an implementation
is required to converge.
\item Moreover, the wait-condition is supposed to describe the set of states in which 
the implementation may become blocked. The implementation is not allowed to become
blocked inside the body of an await-statement.
Since a computation which deadlocks is finite (given a convergent
environment), it is enough to insist that the final state of any
deadlocking computation satisfies the 
wait-condition.

\item Any internal transition is required to satisfy the guar-condition.
The guar-condition is constrained to be reflexive because this has a simplifying
effect on some of the decomposition-rules (it is explained
on page \pageref{reftran:ref} how this constraint can be removed). However,
the guar-condition is not required to be transitive. The reason is that it from
time to time 
is necessary to require a certain maximum granularity. For example, if we want
to specify a process which simulates a certain behaviour in a particular
environment. ($Phil(l)$ on page \pageref{phil:ref} is one example.)

As mentioned above, the rely-condition is assumed to characterise
the overall effect of any (finite) 
uninterrupted sequence of
external transitions. 
The guar-condition is closely related to the rely-condition since it determines 
what the environment can rely on with respect to the implementation:
any (finite) uninterrupted sequence of internal transitions is characterised by the 
transitive closure of the guar-condition; in other words, by the
strongest transitive assertions implied by the guar-condition.

\item Finally, the eff-condition 
is intended to characterise the overall 
effect when the implementation terminates.
External transitions both before
the first internal transition and after the last are included. 
This means that given the rely-condition 
$x\ge \overleftharpoon{x}$,
 the strongest eff-condition for
the program 
$\nskip$ is $x\ge \overleftharpoon{x}$.

\end{itemize}

\section{Summary}

The constraints on a specification are restated in a more formal
notation below:

\begin{definition}\label{spec:def}
A specification is a tuple of the form

\begin{flist}
(\vartheta,\alpha)::(P,R,W,G,E),
\end{flist}

\noindent where $R$, $G$ and $E$ are binary assertions, 
$P$ and $W$ are unary assertions, $\vartheta$
and $\alpha$ are finite, disjoint sets of variables, such that:
\begin{itemize}
\item $R$ is reflexive and transitive,
\item $G$ is reflexive,
\item the unhooked version of any free variable occurring in $(P,R,W,G,E)$ is an
element of $\vartheta\cup \alpha$.
\end{itemize}
\end{definition}

\section{An Example}

Assume we want to specify a program that 
reads 10 natural numbers from a variable
$In$ and adds them
to an initially empty bag $Rcv$. Only one number can be stored in $In$ at 
a time, and
it is the environment's job to provide a new number when the old one has been read.

There is no other restriction on when the program may be applied 
given that the constraint
imposed by the rely-condition is satisfied, so it is enough if the 
pre-condition 
constrains $Rcv$ to be empty.

Moreover, the environment is assumed not to change the value of $Rcv$. 
In other words, $Rcv$ will have 10 elements when (and if) the program terminates.
The eff-condition can be used to express this requirement.

To inform the
environment that a number has been read, the program switches on the flag $Flag$. 
The environment signals that
a new number has been loaded into $In$ by changing $Flag$ to false.

It is assumed that
the program updates the variables $Flag$ and $Rcv$
in the same atomic step. Thus the guar-condition must express
that the program 
either leaves the state as it is or changes $Flag$ from false to true and
adds $In$ to $Rcv$ without changing the value of $In$.

Furthermore, it is clear that 
the program will only wait if $Flag$ is true and the size of $Rcv$ is
less than 10, which gives us the wait-condition.

In what way values are loaded into $In$
is unspecified. For example, the environment may change the value of 
$Flag$ any number of times between each addition of a new value to $Rcv$. Thus,
the only constraint on the environment is that it cannot change the value of
$Rcv$.

Therefore, given that $\#Rcv$ denotes the
number of elements in $Rcv$, we end up with the following specification:

\begin{tuple}{\{Rcv,Flag,In\}
            }{\emptyset
            }{Rcv=\{\}
            }{Rcv=\overleftharpoon{Rcv}
            }{Flag\And \#Rcv<10
            }{In=\overleftharpoon{In}\ \And\\
             &((Flag=\overleftharpoon{Flag}\And Rcv=\overleftharpoon{Rcv})\ \Or\\
             &(\neg \overleftharpoon{Flag}\And Flag \And Rcv=\overleftharpoon{Rcv}\cup\{In\}))
            }{\#Rcv=10.}
\end{tuple}

%% file: auxiliary.tex
\chapter{Auxiliary Variables}

\section{Motivation}

Auxiliary variables were early recognised as helpful tools to reason
about concurrency, see \cite{pbh:oper}, \cite{hcl:oper} or 
\cite{mc:corut}. Since then they have occurred in many shapes and forms. 
For example, in CSP related systems, \cite{carh:cspbook}, 
\cite{ns:oslo} or
\cite{pkp:palogic}, auxiliary variables appear
in the logic as nonimplementable traces. 
Another related concept is what \cite{ll:modul} calls state functions.

In
\cite{so:thesis} and \cite{cs:note}, auxiliary variables are first implemented 
as if they were ordinary programming
variables, and then afterwards removed
by a deduction rule specially designed for this purpose.
This is not a very satisfactory method, because in some specifications a
large number of auxiliary variables are needed, and the procedure of first 
implementing them and then removing them is rather tedious.

Our approach is more in the style of \cite{carh:cspbook}, where the auxiliary 
structure is only a part of the logic and does not appear in the programs. 
Nevertheless,
although it is possible to define trace-related variables in our
system, auxiliary variables may be of any sort, and it is up to the user to
define the auxiliary structure he prefers.

We will use auxiliary variables for two different purposes:
\begin{itemize}
\item To strengthen a specification to eliminate undesirable
implementations. In this case auxiliary variables are used as a specification tool; 
they
are employed to characterise a program that has not yet been implemented.
\item To strengthen a specification to make it possible to prove that an
already finished program satisfies a particular specification. Here auxiliary variables
are used as a verification tool, 
since they are introduced to show that a given algorithm
satisfies a specific property.
\end{itemize}

\section{As a Specification Tool}

\subsection{In Isolation}

We will first discuss an example where it is necessary to use auxiliary variables 
as a specification tool.
Assume we want to specify a program that adds a new element to a global
buffer called $\mathit{Buff}$. 
If we are satisfied with an implementation that can only run in
isolation, then this can easily be expressed as follows:

\begin{tuple}{\{\mathit{Buff}\}
            }{\emptyset
            }{\true
            }{\mathit{Buff}=\overleftharpoon{\mathit{Buff}}
            }{\false
            }{\mathit{Buff}=\overleftharpoon{\mathit{Buff}}\Or \mathit{Buff}=[A]\sconc\overleftharpoon{\mathit{Buff}}
            }{\mathit{Buff}=[A]\sconc \overleftharpoon{\mathit{Buff}}}
\end{tuple}                                 

\noindent The pre-condition allows the implementation to be applied in any state.
Moreover, the rely-condition restricts the environment from changing the 
value of $\mathit{Buff}$, thus
we may use the eff-condition to express the desired property. Finally, the 
guar-condition specifies that the concatenation step takes place in isolation, while
the falsity of the wait-condition requires the implementation to terminate.

\subsection{Allowing Interference}

However, if the environment is allowed to make any changes to $\mathit{Buff}$, 
the task of formulating a specification becomes more difficult. Observe that we still
consider the actual concatenation step to be atomic; the only difference 
from above is that
the environment may now interfere immediately before and (or) after
the concatenation takes place.

Thus, since there are
no restrictions on the ways the environment can change $\mathit{Buff}$, and because
external transitions, both
before the first internal transition
and after the last, are included in the eff-condition,
the
eff-condition must allow anything to happen.

This means that the eff-condition is no longer of much use. Thus,
we are left with the guar-condition as our only hope to pin down the intended
meaning. The specification

\begin{tuple}{\{\mathit{Buff}\}
            }{\emptyset
            }{\true
            }{\true
            }{\false
            }{\mathit{Buff}=\overleftharpoon{\mathit{Buff}}\Or \mathit{Buff}=[A]\sconc \overleftharpoon{\mathit{Buff}}
            }{\true}
\end{tuple}                                 

\noindent is almost sufficient. The only problem is that there is no restriction on
the number of times the operation is allowed to add $A$ to $\mathit{Buff}$. Hence, for example,
the statement

\begin{prg}
\nskip\\
\end{prg}

\noindent is one possible implementation, while

\begin{prg}
\mathit{Buff}:=[A]\sconc \mathit{Buff};\\
\mathit{Buff}:=[A]\sconc \mathit{Buff}\\
\end{prg}

\noindent is another correct implementation.

\subsection{Introducing Auxiliary Variables}

One solution is to introduce a Boolean auxiliary variable called
$Done$, and use $Done$ as a flag to indicate whether $A$ has been added to
$\mathit{Buff}$ or not. Then the program can be specified as follows:

\begin{tuple}{\{\mathit{Buff}\}
            }{\{Done\}
            }{\neg Done
            }{Done=\overleftharpoon{Done}
            }{\false
            }{(\mathit{Buff}=\overleftharpoon{\mathit{Buff}}\And Done=\overleftharpoon{Done})\ \Or\\
              & (\mathit{Buff}=[A]\sconc \overleftharpoon{\mathit{Buff}}\And \neg \overleftharpoon{Done}\And Done)
            }{Done}
\end{tuple}                                 

\noindent Since 
\begin{itemize}
\item the environment cannot change the value of $Done$, 
\item $A$ can only be added to $\mathit{Buff}$ in a state where $Done$ is false, 
\item the concatenation transition changes $Done$ from false to true, 
\item the operation is
not allowed to change $Done$ from true to false,
\end{itemize}
the pre- and eff-conditions imply that $A$ is added to $\mathit{Buff}$ once and only once.

\section{As a Verification Tool}

\subsection{Encoding}

We will now try to explain what we mean by using auxiliary variables as a verification
tool. 
At any point during the execution of a program, the set of possible internal
transitions 
is a function of the global state, the local state and the program counter.

In the tradition of \cite{cbj:thesis} specifications will only constrain
the global state, and it is up to the implementor to choose the local
data structure. Thus,
to add auxiliary structure and use this to encode information
about the program counter and the local state into the global state, will 
in many cases be of crucial importance.

\subsection{Too Weak Specifications}

For example, without auxiliary structure,
the strongest possible guar-condition for the program

\begin{prg}
v:=v+1;\\
v:=v+2,\\
\end{prg}

\noindent is 

\begin{flist}
v=\overleftharpoon{v}\Or v=\overleftharpoon{v}+1 \Or v=\overleftharpoon{v}+2.
\end{flist}

\noindent Moreover, 

\begin{tuple}{\{v\}
            }{\emptyset
            }{\true
            }{v\ge \overleftharpoon{v}
            }{\false
            }{v=\overleftharpoon{v}\Or v=\overleftharpoon{v}+1 \Or v=\overleftharpoon{v}+2
            }{v\ge \overleftharpoon{v}+3}
\end{tuple}

\noindent is the specification that gives the best possible 
characterisation given the actual assumptions about the environment. Furthermore, 
this tuple is also
the strongest possible specification of the program

\begin{prg}
v:=v+2;\\
v:=v+1\\ 
\end{prg}

\noindent with respect to the same constraints on the environment.

Let $ z_1$ denote the first program and $ z_2$ the second.
If we constrain the overall environment to leave $v$ unchanged, 
it is clear that the parallel composition of $ z_1$ and $ z_2$
satisfies:

\begin{tuple}{\{v\}
            }{\emptyset
            }{\true
            }{v=\overleftharpoon{v}
            }{\false
            }{\true
            }{v=\overleftharpoon{v}+6.}
\end{tuple}

\noindent Unfortunately, there is no way to deduce this from the information
in the two component
specifications. The problem is of course that the rely-conditions
do not give enough information. 
Thus, unless the expressibility is increased, any
compositional system based on our specifications will be hopelessly
incomplete.

\subsection{Introducing Auxiliary Variables}

One way to increase the expressibility is to introduce auxiliary variables. Let
$p_1$ be a variable that records the overall effect of the updates
 to $v$ in $ z_1$, while $p_2$ characterises the  overall change to $v$ due 
to $ z_2$. 
Clearly, it is required that $ z_2$
has no effect on $p_1$, and also that $ z_1$ cannot change the value of $p_2$. 

The specification of $ z_1$ can then be rewritten as

\begin{tuple}{\{v\}
            }{\{p_1,p_2\}
            }{\true
            }{p_1=\overleftharpoon{p_1}\ \And\\
              &v- \overleftharpoon{v}= p_2- \overleftharpoon{p_2}
            }{\false
            }{p_2=\overleftharpoon{p_2}\ \And\\
              &v- \overleftharpoon{v}= p_1- \overleftharpoon{p_1}
            }{v=\overleftharpoon{v}+3+(p_2- \overleftharpoon{p_2})\And p_1- \overleftharpoon{p_1}=3,}
\end{tuple}

\noindent while $ z_2$ fulfills
               
\begin{tuple}{\{v\}
            }{\{p_1,p_2\}
            }{\true
            }{p_2=\overleftharpoon{p_2}\ \And\\
              &v- \overleftharpoon{v}= p_1- \overleftharpoon{p_1}
            }{\false
            }{p_1=\overleftharpoon{p_1}\ \And\\
              &v- \overleftharpoon{v}= p_2- \overleftharpoon{p_2}
            }{v=\overleftharpoon{v}+(p_1- \overleftharpoon{p_1})+3\And p_2- \overleftharpoon{p_2}=3.}
\end{tuple}

\noindent It follows easily from these two specifications that,
if the overall environment
is restricted from changing $v$, then  
the overall effect of the parallel composition
of $ z_1$ and $ z_2$ is characterised by $v=\overleftharpoon{v}+6$.

In other words, we have given an informal description of how auxiliary
structure can be introduced to prove that an already finished
implementation satisfies a given specification.	

\subsection{Level of Detail}

Not surprisingly, auxiliary variables can be used to specify the global effect
of a program
to any level of detail. For example, 
the specification of $ z_1$ can easily be strengthened to

\begin{tuple}{\{v\}
           }{\{p_1,p_2\}
           }{p_1=0
           }{p_1=\overleftharpoon{p_1}\ \And\\
             &v- \overleftharpoon{v}= p_2- \overleftharpoon{p_2}
           }{\false
           }{p_2=\overleftharpoon{p_2}\ \And\\
              &((p_1=\overleftharpoon{p_1}\And v=\overleftharpoon{v})\ \Or\\
              &(\overleftharpoon{p_1}=0\And p_1=1\And v=\overleftharpoon{v}+1)\ \Or\\
              &(\overleftharpoon{p_1}=1\And p_1=3\And v=\overleftharpoon{v}+2))
           }{v=\overleftharpoon{v}+3+(p_2- \overleftharpoon{p_2})\And p_1=3,}
\end{tuple}

\noindent which not only specifies exactly how the state can be altered by $ z_1$, but also
gives the order of the updates.

%% file: specprog.tex
\chapter{Specified Programs}

\section{Motivation}

A formal system, whose well-formed formulas are 
called {\em \underline{specified programs}}\index{specified program}, 
will be defined below.
The aim of this chapter is to characterise the latter concept; 
both at syntactic level, and 
with respect to the operational semantics. 

Satisfaction will be defined in two steps; first for the situation when the set of
auxiliary variables is empty.
This definition is then extended to characterise satisfaction in the general case.

\section{Syntax}

\subsection{Two Components}

Since the object of our formal system is to prove that a program satisfies a 
specification, a specified program is a tuple of the form

\begin{flist}
z\satis {\omega},
\end{flist}

\noindent where $ z$ is a program and $\omega$ is a specification.
To avoid unnecessary complications when formulating the block-rule, it is
required that none of the program's local variables occurs in the specification.
To ensure that only programming variables are implemented,
any global variable occurring in the program must be included in
the specification's glo-set.

\subsection{Summary}

The constraints are summed up in the definition below:

\begin{definition}\label{specprog:def}
A specified program is a tuple of the form

\begin{flist}
 z\satis(\vartheta,\alpha)::(P,R,W,G,E),
\end{flist}

\noindent where $ z$ is a program and $(\vartheta,\alpha)::(P,R,W,G,E)$ is a specification, 
such that
\begin{itemize}
\item no local variable of $ z$ occurs in $\alpha\cup\vartheta$,
\item any global variable of $ z$ is an element of $\vartheta$.
\end{itemize}
\end{definition}

\noindent The set of all specified programs will be denoted $SP$\index{$SP$}.

\section{Satisfaction --- Ignoring Auxiliary Variables}

\subsection{External}

What does it mean for a program to satisfy a specification when the set
of auxiliary variables is empty? 
First of all, as indicated in the informal description, a specification
invites the implementor to assume a number of things about the environment.
One assumption is that an implementation is 
called only in a state which satisfies the pre-condition. 
Moreover, any uninterrupted
state transition by the environment is supposed to satisfy the rely-condition,
and it is also assumed that the environment is convergent\footnote{This assumption can
be removed. See page \pageref{conver:ref}.}:
Thus, 
we may restrict our attention to computations
characterised by $ext_{\pi}[P,R]$:

\begin{definition}\label{ext:def}
Given a pre-condition $P$, a rely-condition $R$, and a
structure $\pi$, then
$ext_{\pi}[P,R]$ denotes the set of all computations $\sigma$ in $\pi$,
such that:
\begin{itemize}

\item $\delta(\sigma_1)\models_{\pi} P$,

\item for all $1\le j< len(\sigma)$, if $\lambda(\sigma_j)=e$ then
$(\delta(\sigma_j),\delta(\sigma_{j+1}))\models_{\pi} R$,
\item if $len(\sigma)=\infty$, then 
for all $j\ge 1$, there is a $k\ge j$, such that $\lambda(\sigma_k)=i$.
\end{itemize}
\end{definition}

\noindent The need for the first constraint should be obvious. The second guarantees that
any external transition satisfies the rely-condition. Since the rely-condition is
transitive, this means that the rely-condition is satisfied by any uninterrupted
state transition by the environment. Finally, the third condition ensures that 
the environment is convergent.

Clearly, for any program $ z$ and state $s$ 
which satisfies the pre-condition, there are
infinitely many computations 

\begin{flist}
\sigma\in ext_{\pi}[P,R]\cap cp_{\pi}[ z]
\end{flist}

\noindent whose initial configuration is $< z,s>$ ($cp_{\pi}$ is defined on page 
\pageref{fcp:def}).

\subsection{Internal}

We have already described the assumptions the 
implementor can make about the environment. Moreover, for a given structure $\pi$,
the set of all possible computations of a program $z$, with respect to a
convergent environment characterised by a pre-condition $P$ and a rely-condition
$R$, is

\begin{flist}
ext_{\pi}[P,R]\cap cp_{\pi}[ z].
\end{flist}

\noindent The next step is to formulate what the implementor must
provide in return. 
First of
all, the implementation is constrained to converge when operated 
in an environment
which fulfills the assumptions. Secondly, any internal transition is required to
satisfy the guar-condition. 
Thirdly, the implementor must make sure that the implementation can only become
blocked in a state which satisfies the wait-condition, and that
the overall effect satisfies the eff-condition if the implementation
terminates. 
The requirements on the implementation are summed up below:

\begin{definition}\label{int:def}
Given a wait-condition $W$, a guar-condition $G$, an eff-condition $E$, 
and a structure $\pi$, then
$int_{\pi}[W,G,E]$ denotes the set of all computations $\sigma$ in $\pi$,
such that:
\begin{itemize}
\item $len(\sigma)\neq \infty$,

\item for all $1\le j<len(\sigma)$, if $\lambda(\sigma_j)=i$ then
$(\delta(\sigma_j),\delta(\sigma_{j+1}))\models_{\pi} G$, 

\item if $\tau(\sigma_{len(\sigma)})\neq\epsilon$ then 
$\delta(\sigma_{len(\sigma)})\models_{\pi}W$,

\item if $\tau(\sigma_{len(\sigma)})=\epsilon$ then 
$(\delta(\sigma_1),\delta(\sigma_{len(\sigma)}))\models_{\pi}E$.

\end{itemize}
\end{definition}

\noindent The first constraint guarantees that the implementation either terminates or
deadlocks, while the second ensures that
any internal transition satisfies the
guar-condition. The third constraint requires the final state of a deadlocking
computation to satisfy $W$, which means that the implementation can only
become blocked in a state which satisfies the wait-condition.
Finally, the fourth condition makes sure that the overall
effect of a terminating computation satisfies the eff-condition.

\subsection{Summary}

Thus, if
the set of auxiliary variables is empty, satisfaction can be defined as below:

\begin{definition}\label{presatis:def}
Given a specified program $ z\satis(\vartheta,\emptyset)::(P,R,W,G,E)$ 
and a structure $\pi$, then 
\begin{itemize}
\item $\models_{\pi} z\satis(\vartheta,\emptyset) ::(P,R,W,G,E)$ 
\end{itemize}
if and only if 
\begin{itemize}
\item $ext_{\pi}[P,R]\cap cp_{\pi}[ z]\subseteq int_{\pi}[W,G,E]$.
\end{itemize}
\end{definition}

\section{Satisfaction --- General Case}

\subsection{Existence Check}

In program-development systems (like for example \cite{so:await} and \cite{cs:note})
employing the two-step strategy when reasoning
with auxiliary variables,
the object of the 
first step, where auxiliary variables are implemented as if they were ordinary
programming variables, is to prove that such an extension really exists; in other
words,
that the actual program can be extended with auxiliary structure in such a way
that the specification's requirements on both auxiliary variables and 
ordinary programming variables are fulfilled.

\subsection{Removals}

In our approach, this existence requirement is
incorporated in the semantics. 
To characterise it, 
it is necessary to define what we mean by a {\em \underline{removal}}:
\begin{definition}
A removal is an expression of the form

\begin{flist}
z_1\progtran z_2,
\end{flist}

\noindent where $z_1$ and $z_2$ are programs, $\vartheta$ and $\alpha$ are finite,
disjoint (possibly empty) sets of variables such that no element of $\alpha$
occurs in $z_2$, and a variable of $z_2$ is an element of $\vartheta$ if and only
if it is global. (Observe, that this does not mean that $\vartheta$ 
cannot have occurrences
of variables that do not occur in $z_2$.)
\end{definition}
Informally, a removal is valid,
written $\models z_1\progtran z_2$, if $z_1$ can be
obtained from $z_2$ by adding auxiliary structure related to $\vartheta$ and
$\alpha$. 

\subsection{Restrictions}

There are 
of course a number of constraints on the auxiliary structure:
\begin{itemize}
\item First of all, to make sure that the auxiliary structure has no influence on
the algorithm, auxiliary variables must be restricted from occurring in the 
Boolean tests of while-, if- and await-statements. Furthermore, they cannot appear
on the right-hand side of an assignment, unless the variable on the left-hand side
is auxiliary.

\item Moreover, since we want to be able to remove some auxiliary variables from
a specified program without having to remove all the auxiliary variables, 
it is important that
they do not depend upon each other. This means that if an auxiliary variable $v$
occurs on the left-hand side of an assignment-statement, the only auxiliary
variable that may occur on the right-hand side is $v$. 
In other words, to eliminate all occurrences of an auxiliary variable from a program,
it is enough to remove all assignment-statements with this variable on
the left-hand side. However, an assignment
to an auxiliary variable may have any number of programming variables on
the right-hand side.

\item Finally, since auxiliary variables will only be employed to record 
information about state changes and synchronisation, auxiliary variables are only
allowed to be updated in connection with await- and assignment-statements.
\end{itemize}

\subsection{Summary}

What we mean by a valid removal is characterised more formally below:

\begin{definition}\label{removal:def}
Given a removal $z_1\progtran z_2$,
then 

\begin{flist}
\models z_1\stackrel{(\vartheta,\alpha)}{\hookrightarrow} z_2,
\end{flist}

\noindent if and only if  
$z_1$ can be obtained from $z_2$ by substituting
\begin{itemize}
\item a statement of the form

\begin{prg}
\nawait\ \true\ \ndo\\
\qquad a_1:=u_1;\\
\qquad\quad \vdots\\
\qquad a_n:=u_n;\\
\qquad v:=r\\
\nod,\\
\end{prg}

 where for all $1\le j,k\le n$, $var[u_j]\subseteq \vartheta\cup\{a_j\}$, $a_j
\in\alpha$, and $j\neq k$ implies $a_j\neq a_k$,
for each occurrence of an assignment-statement of the form

\begin{prg}
v:=r,\\
\end{prg}

\noindent which does not occur in the body of an await-statement, 
\item a statement
of the form

\begin{prg}
\nawait\ b\ \ndo\\
\qquad z';\\
\qquad a_1:=u_1;\\
\qquad\quad \vdots\\
\qquad a_n:=u_n\\
\nod,\\
\end{prg}

 where $\models z'\progtran z$, for all $1\le j,k\le n$, $var[u_j]\subseteq\vartheta
\cup\{a_j\}$, $a_j\in\alpha$, and $j\neq k$ implies $a_j\neq a_k$, for each 
occurrence of an await-statement of
the form

\begin{prg}
\nawait\ b\ \ndo\\
\qquad z\\
\nod,\\
\end{prg}

which does not occur in the body of another await-statement.
\end{itemize}
\end{definition}

\noindent We will use $\models z_1\hookrightarrow z_2$ to denote that there are sets of variables
$\vartheta$ and $\alpha$, such that $\models z_1\progtran z_2$.

\subsection{Example}

For example, if $z_2$ and $z_1$ denote the programs

\begin{prg}
x:=x+y;\\
\nawait\ b\ \ndo\\
\qquad \nskip\\
\nod;\\
x:=x+y\\
\end{prg}

\noindent and 

\begin{prg}
\nawait\ \true\ \ndo\\
\qquad a_1:=a_1+x;\\
\qquad x:=x+y\\
\nod;\\
\nawait\ b\ \ndo\\
\qquad \nskip;\\
\qquad a_1:=a_1+x\\
\nod;\\
\nawait\ \true\ \ndo\\
\qquad a_2:=x;\\
\qquad x:=x+y\\
\nod,\\
\end{prg}

\noindent then $\models z_1\hookrightarrow z_2$.

\subsection{General Case}

It is now straightforward to extend definition \ref{presatis:def} on page
\pageref{presatis:def} to the general case:

\begin{definition}\label{satis:def}
Given a specified program $ z_1\satis(\vartheta,\alpha) ::(P,R,W,G,E)$ and a structure
$\pi$, then 
\begin{itemize}
\item $\models_{\pi} z_1\satis(\vartheta,\alpha) ::(P,R,W,G,E)$
\end{itemize}
if and only if 
\begin{itemize}
\item there is a program $ z_2$ such that
\begin{itemize}
\item $\models z_2\progtran z_1$, 
\item $ext_{\pi}[P,R]\cap cp_{\pi}[ z_2]\subseteq int_{\pi}[W,G,E]$.
\end{itemize}
\end{itemize}
\end{definition}

\subsection{Auxiliary Form}

The following concept will be useful in the relative completeness proof:

\begin{definition}
\label{auxform:ref}
A specified program of the form

\begin{flist}
z_1\satis(\vartheta,\emptyset)::(P,R,W,G,E)
\end{flist}

\noindent will be said to be of {\em \underline{auxiliary form}}\index{auxiliary form}, 
if there is a program $z_2$ such that $\models z_1\hookrightarrow z_2$.
\end{definition}

\subsection{Uniqueness Proposition}

\begin{statement}
Given that

\begin{flist}
\models z_1\hookrightarrow z_2,\\
\models z_1\hookrightarrow z_3,
\end{flist}

\noindent then $z_2=z_3$.
\end{statement}
\begin{nproof}
Follows easily from definition \ref{removal:def} on page 
\pageref{removal:def}.

\end{nproof}

%% file: opass.tex
\chapter{Syntactic Operators}

\section{Motivation}

By a syntactic operator we mean a
function, which returns an expression of
the first-order language,
given a finite number of expressions as arguments.
Syntactic operators are not themselves symbols of the 
first-order language, nor
may they appear in programs.

The aim of this chapter is to introduce a number of syntactic operators to reduce the
size and complexity of assertions in L.
We will distinguish between two different types of operators:
\begin{itemize}
\item those which can be expressed independently of the sequence concept,
\item those whose expressibility depends upon the existence of finite sequences.
\end{itemize}

\section{Sequence-Independent Operators}

\subsection{Reference}

There are occasions when it is necessary to add hooks to all free unhooked variables
in an expression. To express this, we will follow the convention suggested 
in \cite{cbj:VDM2}, where 
$\overleftharpoon{r}$ denotes
the result of hooking all free unhooked variables in the expression $r$.

\subsection{Identity}

In most cases a program can only change a rather small
subset of the global state. 
Thus, the rely- and guar-conditions will often be of
the form

\begin{flist}
A\And v_1=\overleftharpoon{v_1}\And\ \ldots\ \And v_n=\overleftharpoon{v_n},
\end{flist}

\noindent where $A$ constrains the part of the state that can be changed.

To simplify such binary assertions, we have found it useful
to introduce a syntactic operator $I_{\beta}$, which 
constrains all variables not in the set $\beta$ to remain unchanged. Hence
when it occurs in 
a specified program of the form

\begin{flist}
 z\satis(\vartheta,\alpha) :: (P,R,W,G,E)
\end{flist}

\noindent (or in the context of a glo-set $\vartheta$ and a aux-set $\alpha$),
it denotes the binary assertion

\begin{flist}
v_1=\overleftharpoon{v_1}\And\ \ldots\ \And v_n=\overleftharpoon{v_n},
\end{flist}

\noindent where 

\begin{flist}
(\vartheta\cup\alpha)\setminus\beta=\{v_1,v_2,\ \ldots\ ,v_n\}.
\end{flist}

\noindent This allows
us to shorten the assertion above to

\begin{flist}
A\And I_{\beta}.
\end{flist}

\noindent When the set of changeable variables is empty, we will write $I$ instead of
$I_{\{\}}$.

\subsection{Set Quantification}

To make it easier to deal with sets of variables, we will use

\begin{flist}
\Forall V:A,\\
\Exists V:A,\\
\Forall \overleftharpoon{V}:A,\\
\Exists \overleftharpoon{V}:A,
\end{flist}

\noindent where $V$ is a set of unhooked variables
$\{v_1,\ldots,v_n\}$, to denote respectively

\begin{flist}
\Forall v_1,\ldots,v_n:A,\\
\Exists v_1,\ldots,v_n:A,\\
\Forall \overleftharpoon{v_1},\ldots,\overleftharpoon{v_n}:A,\\
\Exists \overleftharpoon{v_1},\ldots,\overleftharpoon{v_n}:A.
\end{flist}

\subsection{Composition}

Since we are dealing with binary assertions, a 
composition operator on binary assertions, corresponding to relational composition,
is useful. Given two binary assertions $A$ and $B$, 
then for any structure $\pi$, the 
assertion $A|B$ denotes the least binary relation 
$T$ on the set of states, 
such that for all 
states $s_1$, $s_2$, $s_3$:
\begin{itemize}
\item $(s_1,s_2)\models_{\pi} A$ and $(s_2,s_3)\models_{\pi} B$ implies that
      $(s_1,s_3)\in T$.
\end{itemize}

\noindent The proposition below shows that $A|B$ is always expressible in L.

\begin{statement}
Given two binary assertions $A$ and $B$, then $A|B$ is expressible in L.
\end{statement}

\begin{nproof}
Assume the list

\begin{flist}
v_1:\Sigma_1,v_2:\Sigma_2,\ \ldots\ ,v_n:\Sigma_n
\end{flist}

\noindent consists of the unhooked versions of all free variables in $A$ or
$B$, with their respective
sorts. Let

\begin{flist}
v_1':\Sigma_1,v_2':\Sigma_2,\ \ldots\ ,v_n':\Sigma_n
\end{flist}

\noindent be a list of `new' variables. 
Then $A|B$ can be defined to be the assertion:

\begin{flist}
\Exists v_1',\ \ldots\ ,v_n':\\
\qquad\qquad A(v_1'/v_1,\ \ldots\ ,v_n'/v_n)\And B(v_1'/\overleftharpoon{v_1},\ \ldots\ ,v_n'/\overleftharpoon{v_n}).
\end{flist}

\end{nproof}

\subsection{Example}

The sequential composition of the two assertions

\begin{flist}
\overleftharpoon{v_1}=5\And v_1=77\And v_2\ge \overleftharpoon{v_2},\\
\overleftharpoon{v_1}=77\And v_1=5\And v_2=\overleftharpoon{v_2}+5,
\end{flist}

\noindent is for example characterised by

\begin{flist}
\Exists v_1',v_2':\overleftharpoon{v_1}=5\And v_1'=77\And v_2'\ge \overleftharpoon{v_2}\ \And\\
\qquad\qquad v_1'=77\And v_1=5\And v_2=v_2'+5,
\end{flist}

\noindent which simplifies to

\begin{flist}
\Exists v_1',v_2':\overleftharpoon{v_1}=5\And v_1'=77\And v_2'\ge \overleftharpoon{v_2}\ \And\\
\qquad\qquad v_1=5\And v_2=v_2'+5.
\end{flist}

\section{Finite-Sequence Operators}

\subsection{Transitive Closure}

A transitive closure operator on binary assertions is also needed. 
Given a binary assertion $A$, then for any structure $\pi$, 
the assertion $A^{\dagger}$ 
denotes the least
binary relation $T$ on the set of states, such that for all states $s_1$, $s_2$, $s_3$:
\begin{itemize}
\item $(s_1,s_2)\models_{\pi} A$ implies $(s_1,s_2)\in T$, 
\item $(s_1,s_2)\in T$ and $(s_2,s_3)\in T$, implies 
      $(s_1,s_3)\in T$.
\end{itemize}

\noindent The proposition below shows that $A^{\dagger}$ is always expressible in $L$.

\begin{statement}
Given a binary assertion $A$, then $A^{\dagger}$ is expressible in L.
\end{statement}

\begin{nproof}
Assume the list

\begin{flist}
v_1:\Sigma_1,v_2:\Sigma_2,\ \ldots\ ,v_n:\Sigma_n
\end{flist}

\noindent consists of the unhooked versions of all free variables 
in $A$, with their respective sorts. Let

\begin{flist}
h_1:\seqof{\Sigma_1},h_2:\seqof{\Sigma_2},\ \ldots\ ,h_n:\seqof{\Sigma_n}, 
m:\Nat, j:\Nat
\end{flist}

\noindent be a list of `new' variables. Then $A^{\dagger}$ can be defined to be the assertion:

\begin{flist}
\Exists h_1,\ \ldots\ ,h_n,m:\\
\qquad\qquad  \overleftharpoon{v_1}=h_1(1)\And\ \ldots\ \And \overleftharpoon{v_n}=h_n(1)\ \And \\
\qquad\qquad  v_1=h_1(m)\And\ \ldots\ \And v_n=h_n(m)\ \And \\
\qquad\qquad (\Forall j: 1\le j<m\ \Rightarrow\\
\qquad\qquad\qquad A(h_1(j)/\overleftharpoon{v_1},\ \ldots\ ,h_n(j)/\overleftharpoon{v_n},
h_1(j+1)/v_1,\\
\qquad\qquad\qquad\qquad \ \ldots\ ,h_n(j+1)/v_n)).
\end{flist}

\end{nproof}

\subsection{Example}

The transitive closure of the assertion

\begin{flist}
v_1=\overleftharpoon{v_1}+5\And v_2\ge \overleftharpoon{v_2}
\end{flist}

\noindent is for example characterised by 

\begin{flist}
\Exists h_1, h_2,m:\\
\qquad\qquad \overleftharpoon{v_1}=h_1(1)\And \overleftharpoon{v_2}=h_2(1)\ \And\\
\qquad\qquad v_1=h_1(m)\And v_2=h_2(m)\ \And\\
\qquad\qquad (\Forall j: 1\le j<m\ \Implies\\
\qquad\qquad\qquad\qquad h_1(j+1)=h_1(j)+5\And h_2(j+1)\ge h_2(j)).
\end{flist}

\subsection{Transitive and Reflexive Closure}

It is now easy to define a transitive and reflexive closure operator. Given a
binary assertion $A$, then for any structure $\pi$, the assertion $A^*$ 
stands for $(A\Or I)^{\dagger}$.

\subsection{Preservation}

Given a unary assertion $A$ and a binary assertion $B$, 
in the completeness proof it will
often be necessary to define a binary assertion $A^B$ which characterises the least
set of states $T$, such that for all states $s_1$, $s_2$:

\begin{itemize}
\item $s_1\models_{\pi}A$ implies $s_1\in T$,
\item $s_1\in T$ and $(s_1,s_2)\models_{\pi}B$ implies $s_2\in T$.
\end{itemize}

\noindent Even this operator is expressible in L.

\begin{statement}
Given an unary assertion $A$, and a binary assertion $B$, then $A^B$ is expressible
in L.
\end{statement}

\begin{nproof}
Assume the list

\begin{flist}
v_1:\Sigma_1,v_2:\Sigma_2,\ \ldots\ ,v_n:\Sigma_n
\end{flist}

\noindent consists of the unhooked version of all free variables in $A$, with their respective
sorts. Then $A^B$ can be defined to be the assertion

\begin{flist}
\Exists \overleftharpoon{v_1},\ \ldots\ ,\overleftharpoon{v_n}:\overleftharpoon{A}\And B^*.
\end{flist}

\end{nproof}

%% file: wellfound.tex
\chapter{Well-Foundedness}

\section{Motivation}

Many strategies have been proposed to prove termination of loops. In \cite{sk:term}
four different methods are employed on a number of examples. 
Two of them will be discussed here,
namely Floyd's \cite{rwf:classic} well-foundedness approach
and the loop method first suggested in \cite{be:loop}.

The loop method has some obvious advantages.
First of all, the method is easy to use and of
the four methods discussed in \cite{sk:term}, it is recommended as
the one which can most easily be integrated into an
automatic verification system.

Secondly, this approach  has the additional advantage of giving an upper
bound on the number of possible iterations for a specific initial state. In other
words, the method can also be used to estimate the loop's time consumption.

For that reason, some authors
distinguish between proving strong termination, in which case an upper bound
on the number of iterations is required, and proving weak termination when it is
enough that the loop eventually terminates.

\section{Unbounded Nondeterminism}

Unfortunately, with respect to LSP there is generally no upper bound on the
number of iterations for a specific initial state. To see that, 
assume we want to prove that the program $z$:

\begin{prg}
\nwhile\ x\ge 0\ \ndo\\
\qquad y:=\true;\\
\qquad v:=v - 1;\\
\qquad x:=v\\
\nod,\\
\end{prg}

\noindent satisfies the specification

\begin{tuple}{\{v,x,y\}
            }{\emptyset
            }{\true
            }{\overleftharpoon{y}\Implies y\And v=\overleftharpoon{v}
            }{\false
            }{\true
            }{\true.}
\end{tuple}

\noindent Given that the initial state $s$ satisfies $s(x)\ge 0$ and $s(y)$=false,
then although the while-statement is guaranteed eventually to terminate, 
there is no upper bound
on the number of times it may iterate, since between the initial state
and the first execution of $y:=\true$, the environment is free
to update $v$ as it likes. (Remember that since $x\in hid[z]$, no external
transition can change the value of $x$.)

The reason for this is of course that the rely-condition gives rise to unbounded 
nondeterminism (see \cite{ewd:76book}, \cite{kra:ten2} for a detailed
discussion). The program above is `equivalent' to a sequential program, namely

\begin{prg}
y,v:=R(y,v);\\
\nwhile\ x\ge 0\ \ndo\\
\qquad y,v:=R(y,v);\\
\qquad y:=\true;\\
\qquad y,v:=R(y,v);\\
\qquad v:=v - 1;\\
\qquad y,v:=R(y,v);\\
\qquad x:=v;\\
\qquad y,v:=R(y,v);\\
\nod;\\
y,v:=R(y,v)\\
\end{prg}

\noindent where $y,v:=R(y,v)$ is a random assignment constrained by

\begin{flist}
\overleftharpoon{y}\Implies y\And v=\overleftharpoon{v}.
\end{flist}

\section{Well-Foundedness Approach}

We will therefore employ the well-foundedness approach. This means that our system
can only be used to prove weak termination. Given an environment, the basic idea is:
\begin{itemize}
\item if we can show that the loop's body terminates in a state which satisfies 
an unary assertion $A$
for any initial state which satisfies
$A$ and the Boolean test, and $B$ is a binary assertion 
characterising the effect of the loop's body with respect to 
the same environment and the same restrictions on the initial state, then the
loop terminates if and only if, there is no infinite sequence of
states $s_1s_2\ \ldots\ s_n\ \ldots\ $, such that for all $j\ge 1$,

\begin{flist}
(s_j,s_{j+1})\models_{\pi} B.
\end{flist}

\end{itemize}

\noindent This property is a well-known mathematical concept. A number
of authors have proposed (second order) extensions of the first-order logic
to make it possible to express and reason about well-foundedness. See for 
example \cite{ph:wf}, \cite{jwb:wf}, \cite{nf:fairness}, \cite{kra:wf},
\cite{fas:wf}. Unfortunately, none of these approaches consider
unbounded well-foundedness
in the most general case\footnote{In \cite{kra:wf} 
Apt and Plotkin propose a version of
$\mu$-calculus which handles countable (unbounded) nondeterminism. If 
all sorts are restricted to be countable, it should be possible to define a 
similar logic for
our system.}.

We will not attempt to deal with this problem here. 
\label{wf:re} Instead it will be assumed that there is
a conservative (second order) 
extension L$_{wf}$ of L, which, for any binary assertion $A$ in L,
allows us to formulate an assertion, denoted by $\nwf\, A$\index{$\nwf$},
which is valid  in a structure $\pi$
if and only if $A$ is well-founded in $\pi$.

We will use $Tr_{\pi}$ to denote the set of all assertions in L$_{wf}$ which are
valid in $\pi$.

%% file: ruledes.tex
\chapter{Logic of Specified Programs}

\section{Motivation}

\subsection{Formal System}

The object of this chapter is to 
introduce a formal system, called LSP (Logic of Specified Programs),
which consists of 
sixteen {\em \underline{decomposition-rules}}\index{decomposition-rules} 
for specified programs,
axioms and deduction-rules for L$_{wf}$, and all valid removals as axioms.
The decomposition-rules are split into two categories:
\begin{itemize}
\item The  rules needed for top-down development of programs; namely the
consequence-, pre-, access-, skip-, assignment-, block-, sequential-,
if-, while-, parallel-, await- and elimination-rules.

These rules will be called the basic rules, and it will later be shown that
the basic rules are sufficient to prove relative completeness. Thus, 
since the decomposition-rules are compositional, it follows that LSP$_B$ is
{\em \underline{compositionally complete}}\index{compositional complete} 
(see \cite{rk:comp}, \cite{jz:CCC} for a more detailed 
discussion).

\item The rules introduced to simplify the development and 
make it possible to take advantage of already finished designs; namely
the effect-, global-, auxiliary- and introduction-rules.

These rules will be called the adaptation rules. Unfortunately, this 
set of rules is not strong enough
to ensure {\em \underline{adaptation completeness}}\index{adaptation completeness}
(see \cite{jz:CCC})\footnote{It is not difficult to find a stronger
set, but we do not know how to formulate a set which is adaptation complete.},
namely that 
\begin{itemize}
\item if the specification of a program, whose internal structure is unknown, implies
another specification for the same program, then the proof system admits a formal
deduction of that fact.
\end{itemize}

\noindent For example, since 

\begin{flist}
 z\satis(\{v\},\emptyset)::
(\true,v=\overleftharpoon{v},\false,v\ge \overleftharpoon{v},v=\overleftharpoon{v}),
\end{flist}

\noindent is valid only if $ z$ leaves $v$ unchanged, it should be possible to deduce

\begin{flist}
 z\satis(\{v\},\emptyset)::(\true,v=\overleftharpoon{v},\false,v=\overleftharpoon{v},v=\overleftharpoon{v}).
\end{flist}

\noindent However, there is no way to do this given the current 
set of rules.

\end{itemize}

\noindent Axioms and deduction-rules for first-order systems can be found in almost
any text-book on logic, and this will therefore not be discussed here.

\subsection{Decomposition-Rules}

The decomposition-rules are all of the form

\begin{flist}
Prem_1\\
\qquad\vdots\\
\underline{Prem_n}\\
Concl,
\end{flist}

\noindent where the conclusion, $Concl$, is a schema characterising a specified program, 
and the j'th premise, $Prem_j$, is 
a schema denoting either a specified program,
a formula in L$_{wf}$ or a removal.

Moreover, a decomposition-rule should be interpreted as follows:
\begin{itemize}
\item given a structure $\pi$ and $n+1$  matching specified programs
formulas and removals\footnote{$n$ 
premises + conclusion=$n+1$.}, then the conclusion is valid in $\pi$, if each
of the $n$ premises are valid in $\pi$.
\end{itemize}

\subsection{Proofs}

A {\em \underline{proof}}\index{proof} in LSP is a finite tree where 
\begin{itemize}
\item each node represents a removal,
a formula of L$_{wf}$ or a specified program, 
\item all leaves are axioms,
\item a node is either a leaf or can be deduced from its immediate sons by
one of the deduction- or decomposition-rules.
\end{itemize}

\subsection{Depth}

A proof's {\em \underline{depth}}\index{depth} is 0 if the root is a 
formula of L$_{wf}$ or a removal,
1 if the root is a specified program without sons; otherwise $n+1$,
where $n$ is the maximum depth of its immediate subproofs.

\label{depth:ref}

\subsection{Notation}

LSP restricted to the set of basic rules will be denoted LSP$_B$. No proof
in LSP$_B$ has a removal in its tree. This means that
auxiliary variables are only a part of the logic and do 
not have to be implemented.

Given a specified program $\psi$, we will use

\begin{flist}
\vdash \psi
\end{flist}

\noindent to denote that $\psi$ is provable in LSP,

\begin{flist}
\vdash_B \psi
\end{flist}

\noindent to denote that $\psi$ is provable in LSP$_B$, and

\begin{flist}
Tr_{\pi}\vdash \psi,\\
Tr_{\pi}\vdash_B \psi
\end{flist}

\noindent to denote that $\psi$ is provable in LSP, respectively LSP$_B$, given
the assertions in $Tr_{\pi}$ as axioms.

\subsection{Current Set of Rules}

There are many ways to formulate the different decomposition-rules, and we are
not claiming the given set is the best possible for all applications. However,
since this set is both sound and relatively complete, it can always be used
to deduce new more user friendly decomposition-rules.

\section{Consequence-Rule}

The consequence-rule is probably the easiest to understand. 
Basically, the pre- and rely-conditions can be strengthened, because
any program that behaves correctly given 
the original pre- and rely-conditions, will also
behave correctly when they are strengthened.

Similarly, we may weaken the wait-, guar- and eff-conditions, because any program
that satisfies the stronger wait-, guar- and eff-conditions, will also satisfy the new
ones:

\begin{flist}
 z  \satis  (\vartheta,\alpha):: (P_1,R_1,W_1,G_1,E_1)\\
P_2\Implies P_1\\
R_2\Implies R_1\\
W_1\Implies W_2\\
G_1\Implies G_2\\
E_1\Implies E_2\\
\overline{ z \satis(\vartheta,\alpha):: (P_2,R_2,W_2,G_2,E_2)}
\end{flist}

\section{Pre-Rule}

The pre-rule is also straightforward. If the actual program is employed in a state which does not satisfy the pre-condition, there are no constraints on its behaviour. Thus, we may restrict the eff-condition to transitions from states which satisfy the pre-condition:

\begin{flist}
z  \satis(\vartheta,\alpha)::(P,R,W,G,E)\\
\overline{ z  \satis(\vartheta,\alpha)::(P,R,W,G,\overleftharpoon{P}\And E)}
\end{flist}

\section{Access-Rule}

Since for any program $z$, the environment respects $hid[ z ]\cap \vartheta$, a rule which allows us to weaken the rely-condition is also needed:

\begin{flist}
\underline{ z  \satis(\vartheta,\alpha)::(P,R\And v=\overleftharpoon{v},W,G,E)}\\
z  \satis(\vartheta,\alpha)::(P,R,W,G,E)\\
\\
where\ v\in hid[ z ]\cap \vartheta
\end{flist}

\section{Skip-Rule}

The skip-statement generates only one internal transition. Since this transition leaves the state unchanged and since the guar-condition is reflexive, it is clear that any internal transition satifies the guar-condition. Moreover, since each transition due to the environment is assumed to satisfy the rely-condition, which is both reflexive and transitive, it follows that the overall effect ot the skip-statement satifies the rely-condition:

\begin{flist}
\nskip \satis  (\vartheta,\alpha):: (P,R,W,G,R)
\end{flist}

\section{Assignement-Rule}

\subsection{Introduction}

The assignment-rule is more complicated. Although there is only one internal transition, we must allow auxiliary variables to be updated in the same atomic step. The reason is of course that due to the definition of the satisfaction relation, the execution of a statement $z_1$ of the form

\begin{flist}
v:= r
\end{flist}

\noindent corresponds to the execution of a statement $z_2$ of the form

\begin{flist}
\nawait\ \true\ \ndo\\
\qquad a_1:=u_1;\\
\qquad\quad \vdots\\
\qquad a_m:=u_m;\\
\qquad v:=r\\
\nod;
\end{flist}

\noindent where $\models z_2\hookrightarrow z_1$. In other words, $z_1$ may be extended with auxiliary structure in such a way that the auxiliary variables $a_1,\ldots,a_m$ are updated in the same atomic step as $v$.

To simplify the discussion, we will first deal with the situation where the set of auxiliary variables is empty, and thereafter show how the rule can be extended to cover the general case.

\subsection{Without Auxiliary Variables}

Since the assignment-statement is interpreted as atomic in our operational semantics, there is only one internal transition. Clearly, this transition must be shown to satisfy the guar-condition.

Moreover, the assignment can only take place in a state that can be reached from a state which satisfies the pre-condition by a finite number of external transitions, and since the rely-condition is both transitive and reflexive, the following rule is sufficient:

\begin{flist}
\overleftharpoon{P^R}\And v=\overleftharpoon{r}\And I_{\{v\}}
\Rightarrow G\And E\\
\overline{v:=r \satis(\vartheta,\{\}):: (P,R,W,G,R|E|R)}
\end{flist}

\subsection{General Case}

It is now straightforward to extend the rule to handle the general situation. The only real difference is that the premise in this case must also guarantee that the assignment-statement can be extended with auxiliary structure in such a way that the specified changes to both the auxiliary variables and the programming variables will indeed take place:

\begin{flist}
\underline{\overleftharpoon{P^R}\And v=\overleftharpoon{r} \And I_{\{v\}\cup\alpha}\And \bigwedge_{a\in\alpha} a=\overleftharpoon{u_a}\Rightarrow G\And E}\\
v:=r \satis  (\vartheta,\alpha):: (P,R,W,G,R|E|R)\\
\\
{where\ for\ all\ }a\in\alpha, var[u_a]\subseteq \vartheta\cup\{a\}
\end{flist}

\section{Block-Rule}

The block-rule is also easy to understand. When employed in a top-down style,
the set of global programming
variables is extended with the new ones. Moreover, since no
program running in parallel can access these variables, we may restrict the
environment to leave them unchanged:

\begin{flist}
 z  \satis  (\vartheta,\alpha):: (P,R\And \bigwedge_{j=1}^n v_j=\overleftharpoon{v_j},
W,G,E)\\
\overline{\nbegin\ \loc\ v_1,\ \ldots\ ,v_n; z \ \nend \satis  (\vartheta\setminus
\bigcup_{j=1}^n \{v_j\},\alpha):: (P,R,W,G,E)}\\
\end{flist}

\noindent It follows from the 
constraints on specified programs that the variables 

\begin{flist}
v_1,\ \ldots\ ,v_n
\end{flist} 

\noindent do not occur in the specification 

\begin{flist}
(\vartheta\setminus \bigcup_{j=1}^n \{v_j\},\alpha)::(P,R,W,G,E).
\end{flist}

\section{Sequential-Rule}

The sequential-rule is straightforward. Basically,
the first component's eff-condition must imply the second component's pre-condition.
This explains why $P_2$ occurs in 
the first
statement's effect-condition:

\begin{flist}
 z_1 \satis  (\vartheta,\alpha):: (P_1,R,W,G,P_2\And E_1)\\
 z_2 \satis  (\vartheta,\alpha):: (P_2,R,W,G,E_2)\\
\overline{ z_1; z_2 \satis  (\vartheta,\alpha):: (P_1,R,W,G,E_1|E_2)}
\end{flist}

\section{If-Rule}

In this case there is not much to explain. Basically, because
the if-statement's Boolean test has been restricted to have 
no occurrences of variables accessible
by the environment, the truth value of the Boolean test is
maintained by the environment.
Thus, the following rule is sufficient:

\begin{flist}
 z_1 \satis  (\vartheta,\alpha):: (P\And b,R,W,G,E)\\
 z_2 \satis  (\vartheta,\alpha):: (P\And \neg b,R,W,G,E)\\
\overline{\nif\ b\ \nthen\  z_1\ \nelse\  z_2\ \nfi \satis  (\vartheta,\alpha):: (P,R,W,G,E)}
\end{flist}

\section{While-Rule}

In the same way as for the if-rule, interference before
the first evaluation of the Boolean test has no influence on the tests
outcome.
Moreover, the falsity of the Boolean test will be preserved after the
while-statement terminates.
As explained above, we are content with proving weak (conditional) 
termination, thus it is
enough to show that an assertion characterising the effect of the loop's
body in the actual environment is well-founded when considered as a binary
relation on states.
We can therefore formulate a rule fairly similar to the while-rule in \cite{cbj:VDM}
(if we ignore that the latter is not dealing with rely-, wait- and guar-conditions):

\begin{flist}
{\rm\underline{wf}}\ Z\\
 z  \satis  (\vartheta,\alpha):: (P\And b,R,W,G,P\And Z)\\
\overline{\nwhile\ b\ \ndo\  z \ \nod \satis  (\vartheta,\alpha) 
:: (P,R,W,G,(Z^{\dagger}\Or  R)\And \neg b)}
\end{flist}

\noindent The only real difference is that, because of possible interference from the
environment, the $I$ in the
conclusion's eff-condition has been replaced by $R$.

\section{Parallel-Rule}

\subsection{Simple Case}

We will first design a rule for the case when the two component programs do not
deadlock (given their respective environments).
Obviously, the premises must make sure that the 
component programs  
are compatible with respect to mutual interference. 

It should also be clear that the parallel composition of two
programs is only guaranteed to converge if called in an environment in which
both component programs are guaranteed to converge.
Finally, the statements overall effect is also
the overall effect of both component programs:

\begin{flist}
 z_1 \satis  (\vartheta,\alpha):: (P,R_1,\false,G\And R_2,E_1)\\
 z_2 \satis  (\vartheta,\alpha):: (P,R_2,\false,G\And R_1,E_2)\\
\overline{\{ z_1\parallel  z_2\} \satis  (\vartheta,\alpha):: 
(P,R_1\And R_2,\false,G,E_1\And E_2)}
\end{flist}

\subsection{General Case}

To formulate the general rule, it is enough to observe that $z_1$ is guaranteed
to be released whenever it becomes blocked in a state in which $z_2$ cannot
become blocked or terminate. Similarly, $z_2$ is guaranteed to be released in any state
in which $z_1$ cannot become blocked or terminate:

\begin{flist}
\neg(W_1\And E_2)\And \neg(W_2\And E_1)\And \neg(W_1\And W_2)\\
 z_1 \satis  (\vartheta,\alpha):: (P,R_1,W\Or W_1,G\And R_2,E_1)\\
 z_2 \satis  (\vartheta,\alpha):: (P,R_2,W\Or W_2,G\And R_1,E_2)\\
\overline{\{ z_1\parallel  z_2\} \satis  (\vartheta,\alpha):: 
(P,R_1\And R_2,W,G,E_1\And E_2)}
\end{flist}

\noindent One possible alternative is:

\begin{flist}
\neg(W_1\And E_2)\And \neg(W_2\And E_1)\And \neg(W_1\And W_2)\\
 z_1 \satis  (\vartheta,\alpha):: (P,(R\Or G_2)^{\dagger},W\Or W_1,G\And G_1,E_1)\\
\underline{z_2 \satis  (\vartheta,\alpha):: (P,(R\Or G_1)^{\dagger},W\Or W_2,
G\And G_2,E_2)}\\
\{ z_1\parallel  z_2\} \satis  (\vartheta,\alpha):: 
(P,R,W,G,E_1\And E_2)
\end{flist}

\subsection{Generalised Parallel-Rule}

The parallel-rule can also be generalised to deal with more than two processes:

\begin{flist}
\neg(W_j\And \bigwedge_{k=1,k\neq j}^{m} (W_k\Or E_k))_{1\le j\le m}\\
 z_j \satis  (\vartheta,\alpha):: (P,R_j,W\Or W_j,G\And \bigwedge_{k=1,k\neq j}^{m} R_k,
E_j)_{1\le j\le m}\ \\
\overline{\{ z_1\parallel\ \ldots\ \parallel  z_m\} \satis  (\vartheta,\alpha):: 
(P,\bigwedge_{j=1}^m R_j,W,G,\bigwedge_{j=1}^m E_j)}
\end{flist}

\noindent Obviously, $\{ z_1\parallel\ \ldots\ \parallel  z_m\}$ denotes any program that
can be obtained from $\{ z_1\parallel\ \ldots\ \parallel  z_m\}$ by adding
curly brackets.
This rule can be deduced from the rules in LSP$_B$, and it will from now on
be referred to as the generalised parallel-rule.

\section{Await-Rule}

\subsection{Introduction}

The await-rule is closely related to the assignment-rule;
there is only one 
internal transition, and auxiliary variables are allowed to be updated in
the same atomic step. Thus, the
execution of a statement $z_1$ of the form

\begin{prg}
\nawait\ b\ \ndo\\
\qquad z\\
\nod\\
\end{prg}

\noindent corresponds to the execution of a statement $z_2$ of the form

\begin{prg}
\nawait\ b\ \ndo\\
\qquad z';\\
\qquad a_1:=u_1;\\
\qquad\quad\vdots\\
\qquad a_m:=u_m\\
\nod,\\
\end{prg}

\noindent where $\models z_2\hookrightarrow z_1$ and
$\models z'\hookrightarrow z$.

To simplify the discussion, we will first deal with the 
situation where the set of auxiliary variables is empty, and thereafter
show how the rule can be extended to cover the general case.

\subsection{Without Auxiliary Variables}

Clearly, any state which does not satisfy the Boolean test and can be reached from
a state which satisfies the pre-condition by a finite number of
external transitions must satisfy the wait-condition.

The operational semantics guarantees that the environment cannot
interfere with the body of an await-statement. Moreover, the statement's
body is required to terminate for any state which satisfies the Boolean test
and can be reached from a state which satisfies the pre-condition by
a finite number of external transitions. Finally, the
effect of the await-statement's body is required to satisfy 
the overall guar-condition:

\begin{flist}
P^R\And \neg b\Rightarrow W\\
z  \satis  (\vartheta,\{\}):: (P^R\And b,I,\false,\true,G\And E)\\
\overline{\nawait\ b\ \ndo\  z \ \nod \satis  (\vartheta,\{\}):: (P,R,W,G,R|E|R)}
\end{flist}

\subsection{General Case}

It is now straightforward to formulate the rule for the general case:

\begin{flist}
P^R\And \neg b\Rightarrow W\\
E_1|(\bigwedge_{a\in\alpha}a=\overleftharpoon{u_a}\And I_{\alpha})\Implies G\And E_2\\
 z  \satis  (\vartheta,\alpha):: (P^R\And b,I,\false,\true,E_1)\\
\overline{\nawait\ b\ \ndo\  z \ \nod \satis  (\vartheta,\alpha):: (P,R,W,G,R|E_2|R)}\\
\\
{where\ for\ all\ }a\in\alpha, var[u_a]\subseteq \vartheta\cup\{a\}
\end{flist}

\section{Elimination-Rule}

We also need a rule
that allows us to eliminate auxiliary structure from the specification:

\begin{flist}
 z  \satis  (\vartheta,\alpha\cup\{a\})::(P,R,W,G,E)\\
\overline{ z  \satis  (\vartheta,\alpha\setminus\{a\})::(\Exists a:P,
\Forall\overleftharpoon{a}:\Exists a:R,W,G,E)}
\end{flist}

\noindent Since the conclusion is required to be a specified programs, it follows
that  $W$, $R$ and $E$ have no occurrences of $a$.

\section{Effect-Rule}

It is also true that any state reachable from the pre-condition 
can be reached by the transitive closure of the rely- and guar-conditions.
The eff-condition can therefore be strengthened with their closure:

\begin{flist}
 z  \satis  (\vartheta,\alpha):: (P,R,W,G,E)\\
\overline{ z  \satis  (\vartheta,\alpha):: (P,R,W,G,E\And (R\Or G)^{\dagger})}
\end{flist}

\section{Global-Rule}

The global-rule allows us to introduce a new variable. 

\begin{flist}
 z  \satis  (\vartheta\setminus\{v\},\alpha):: (P,R,W,G,E)\\
\overline{ z  \satis  (\vartheta\cup\{v\},\alpha):: 
(P,R,W,G\And v=\overleftharpoon{v},E)}
\end{flist}

\noindent The premise and the constraints on specified programs imply that the
variable does not occur in the actual program. Thus no internal transition
can change its value.

\section{Auxiliary-Rule}

Similarly, the auxiliary-rule allows us to introduce a new auxiliary variable:

\begin{flist}
 z  \satis  (\vartheta,\alpha\setminus\{a\}):: (P,R,W,G,E)\\
\overline{ z  \satis  (\vartheta,\alpha\cup\{a\}):: 
(P,R,W,G\And a=\overleftharpoon{a},E)}
\end{flist}

\section{Introduction-Rule}

Although our logic allows us to reason about auxiliary variables without having
to implement them, the following rule can be useful if we have already developed
a program with respect to a specification and later want to use the same algorithm
in another connection:

\begin{flist}
 z_1\stackrel{(\vartheta\setminus\alpha,\alpha)}{\hookrightarrow} z_2\\
 z_1 \satis  (\vartheta,\emptyset)::(P,R,W,G,E)\\
\overline{ z_2 \satis  (\vartheta\setminus\alpha,\alpha)::(P,R,W,G,E)}
\end{flist}

%% file: exintr.tex
\chapter{Simplifying Notation}

\section{Motivation}

Before applying LSP to the Dining-Philosophers, the Bubble-Lattice-Sort
and the Set-Partition algorithms, we will introduce some simplifying
notation. The new concepts are introduced in a rather informal style. However,
based on the way they are used in  the different examples,
it should not be difficult to
grasp their intended meaning\footnote{The reason why we have not given
a formal semantics is that we are currently experimenting
with this notation, and it is still subject to change.}.

\section{Operations}

\subsection{Main Structure}

First of all, to make specifications more readable, 
a VDM-related \cite{cbj:VDM2} {\em \underline{operation}}\index{operation} 
concept of the form

\begin{spec}{Name}{In}{Out}{\begin{ndcl}{g\_dcl}{a\_dcl}
                            \end{ndcl}
                           }{P
                           }{R
                           }{W
                           }{G
                           }{E}
\end{spec}

\noindent has been found helpful. Not surprisingly, $Name$ is the name of the operation, 
$In$ is the list of input parameters, 
while $Out$ is the list of output parameters.
Moreover, global variables are declared in $g\_dcl$, 
while $a\_dcl$ is used to declare auxiliary variables. Finally,
$P$, $R$, $W$, $G$ and $E$ denote respectively the pre-, rely-,
wait-, guar- and eff-conditions.

\subsection{Variables}

The set of input parameters\index{input parameter} 
is constrained to be disjoint from the set
of output parameters\index{output parameter}. Syntactically the parameters 
are unhooked variables.
Similarly, the set of global variables is required to be disjoint from the
set of auxiliary variables, and no global or auxiliary variable can occur as
a parameter. 

The five conditions may contain occurrences of global variables that are not declared in
$g\_dcl$. However, no such variable is allowed to appear in the final code
implementing the operation. In other words, no global variable may occur in the
implementation unless it is listed in $g\_dcl$.

Auxiliary variables that will be `updated' by the implementation must be listed in
$a\_dcl$.
Variables local to the implementation are not allowed to appear in the
specification.

If a free variable is hooked in the rely-, guar- or eff-conditions, then it is either
auxiliary or global.
If it is clear from $g\_dcl$ and $a\_dcl$ that a particular variable 
cannot be changed by
 the operation, its environment or both, then this will not be restated in
the operation's rely-, guar- and eff-conditions. 

For example, if $v$ is a global variable not
mentioned in $g\_dcl$, then it is clear that any internal transition will satisfy
$v=\overleftharpoon{v}$, but to keep the specifications as simple as possible
this does not have to be restated in
the guar-condition, although it may be added as an extra conjunct when proofs
are undertaken.

\subsection{Observable and Hidden Changes}

In VDM, which covers only non-interfering programs, it is 
indicated in the declaration whether the
operation has write access to a variable, only read access or no access at all.
If an operation has no write access to
a variable, clearly its value will be left unchanged.

Unfortunately, when dealing with concurrency the situation is a lot more complicated.
First of all, other programs running in parallel may interfere. Moreover,
because of the await-statement, there can be  changes to the global
structure due to the implementation 
that are hidden from the environment and vice versa.

If a particular global variable is not updated 
outside await-statements, and the
variable's
value upon entry to an await-statement is always equal to its value when
the await-statement terminates, then changes to this variable are not
observable from the outside.

Therefore, instead of declaring global and auxiliary variables according to whether 
they are writable or only
readable, we
will distinguish variables that can be changed in an observable way from those
that cannot.
Both global variables and auxiliary variables will be declared 
according to the following convention. 
Let $v$ be a variable
of sort $T$, then:

\begin{itemize}
\item $\lwew\ v:T$ --- (internal observe, external observe.)

Means that the operation can change the value of $v$ in such a way that
this can be observed by the environment. Similarly, the operation can observe
changes to $v$ that are due to the environment.

\item $\lwer\ v:T$ --- (internal observe, external hide.) 

Means that the operation can change the value of $v$ in such a way that 
this can be observed by the environment, while any changes to
$v$ due to the environment are hidden from the operation.

\item $\lrew\ v:T$ --- (internal hide, external observe.)

Means that any changes to $v$ due to the operation are hidden from the environment,
while the operation can observe updates to $v$ caused by the environment.

\item $\lrer\ v:T$ --- (internal hide, external hide.)

Means that any changes to $v$ due to the operation are
hidden from the environment, and the 
operation cannot observe any of the environment's changes to $v$.

\end{itemize}

\subsection{Splitting of Objects}

We have also found it convenient to split up objects consisting of more than one
memory location and declare them separately. For example, given a global variable

\begin{flist}
A:\arrayof{\set{1,\ldots,9}}{\Bool},
\end{flist}

\noindent and assuming we are specifying an operation which needs $\lwew$-access to 
$A(8)$, but never accesses the rest of $A$,
then

\begin{flist}
\lwew\ A(8):\Bool,
\end{flist}

\noindent will be the only reference to $A$ in $g\_dcl$.

\section{Assertions Occurring in the Code}

\subsection{Sequential Case}

When proving properties of programs it is often useful to insert assertions into the
code. For example, the sequential program

\begin{prg}
\{\true\}\\
\nwhile\ x>0\ \ndo\\
\qquad\{x>0\}\\
\qquad x:=x - 1\\
\nod\\
\{x\le 0\}\\
\end{prg}

\noindent has three such assertions. The first and last characterise respectively the initial
and final states, while the one in the middle describes the state each time
the program counter is `situated' between the Boolean test and the assignment-statement.

We will insert assertions in a similar style. 
However, because
the assertions may have occurrences of hooked variables, because the environment
may interfere, and because of the way our operational
model is related to a more intuitive interpretation of parallel programs,
it is necessary to
discuss the meaning of such assertions in more detail.

\subsection{Hooking}

Assertions occurring in the code will have occurrences of hooked
variables when this is convenient. The hooked variables are supposed to refer to 
the initial state with respect to the particular piece of code in which the
assertion occur. For example, in the annotated program:

\begin{prg}
\{\true\}\\
x:=x+5;\\
\{x=\overleftharpoon{x}+5\}\\
x:=x+3;\\
\{x=\overleftharpoon{x}+8\}\\
x:=x+2\\
\{x=\overleftharpoon{x}+10\},\\
\end{prg}

\noindent the second assertion states that 
whenever the program counter is situated between the
first and the second assignment-statement, the difference between the current value
of $x$ and the initial value of $x$ is five,
the third assertion states that 
whenever the program counter is situated between the
second and the third assignment-statement, the difference between the current value
of $x$ and the initial value of $x$ is eight, while
the fourth assertion states that whenever the program counter is 
situated after the third assignment-statement, 
the difference between the current value
of $x$ and the initial value of $x$ is ten.

\subsection{Interference}

It is important to realise that assertions inserted into the code are supposed
to be preserved by the actual rely-condition. In the example above it is
implicitly assumed that the environment leaves $x$ unchanged. If we change the
rely-condition to

\begin{flist}
x\ge \overleftharpoon{x},
\end{flist}

\noindent we end up with the following annotated program:

\begin{prg}
\{\true\}\\
x:=x+5;\\
\{x\ge \overleftharpoon{x}+5\}\\
x:=x+3;\\
\{x\ge \overleftharpoon{x}+8\}\\
x:=x+2\\
\{x\ge \overleftharpoon{x}+10\}.\\
\end{prg}

\subsection{Two Interpretations}

\label{overlap:ref}
Await-statements play an important r\^ole in the examples below. Actually, they
can be used in two different ways:
\begin{itemize}
\item First of all, an await-statement can be employed to guarantee 
mutual exclusion in cases where this 
is necessary to ensure the correctness of the actual algorithm. In other words, 
cases when 
the replacement of the await-statement with the await-statement's 
body would result in a
different `behaviour'.
\item Secondly, an await-statement can be applied as an abstraction tool; 
to simplify the development
and verification of an algorithm. In this case, the 
replacement of  the await-statement with
the await-statement's body `has no real effect' on the algorithm's `behaviour'.
The `amount of 
concurrency' is not reduced by using the await-statement as an abstraction
tool. The only disadvantage is the unnecessary 
hiding of variables that are already `hidden'
by some sort of mutual exclusion algorithm in the actual program. 
\end{itemize}

\noindent Some readers may find the second alternative a bit surprising. After all, in the
operational model, the execution
of an await-statement is represented by one internal transition, while the second
option above seems to imply that the await-statement is non-atomic.
Actually, by a similar argument, they might claim that:
\begin{itemize}
\item Since any computation consists of an 
infinite sequence of nonoverlapping external
and internal transitions, there can be no real concurrency at all. In other words, that our
operational model is inadequate.
\end{itemize}

\noindent But this is
a misunderstanding. 
An atomic step should not be interpreted as something
immediate, without extension in time, 
nor should it be understood as a program that must be run in isolation, i.e.\ until
it has terminated no other process is allowed to progress.

Instead, one should consider an atomic step as a program, which, while
it is executing, has some sort
of device to deny other processes access to the memory locations it is
accessing. This way, although the environment may change any of the remaining 
memory locations, no other process can interfere, and the
atomic transition can be reasoned about as a sequential program. 

Moreover, since the
sets of memory locations accessed by two concurrent atomic transitions are
disjoint, it follows that they may just as well be considered to have taken place in 
sequential order without any overlap in time. This motivates our choice of operational
model.

Unfortunately, this distinction between the operational model and a more
intuitive interpretation, where atomic steps may overlap in time, 
can easily lead to confusion when it comes to interpreting
assertions occurring in the code.

If one think in terms of the operational model, which is what we recommend,
assertions occurring in the
code should be interpreted as follows:

\begin{itemize}
\item An assertion occurring
in the code is true
whenever the program counter reaches the assertion's address in the actual program.
\end{itemize}

\noindent This corresponds to how the assertions are interpreted in the
example above. On the other hand, if one insists on thinking in terms of
the more intuitive model, assertions occurring in the code should be interpreted 
as follows:

\begin{itemize}
\item An assertion occurring in the code is true whenever the
program counter reaches the assertion's address, and all the variables 
occurring in the assertion are accessible, i.e.\ not occupied by an atomic transition
currently under execution.
\end{itemize}

\noindent Not surprisingly, the pre-, rely-, wait-, guar- and eff-conditions can be given  
two different interpretations in a similar way.

\section{\texorpdfstring{$Inv$}{Inv} and \texorpdfstring{$Dyn$}{Dyn}}

In the examples we have found it useful to formulate two special assertions;
a {\em \underline{global invariant}}\index{global invariant} denoted by $Inv$ 
and a {\em \underline{dynamic invariant}}\index{dynamic invariant} called 
$Dyn$\footnote{Jones
arrived at a similar conclusion in \cite{cbj:thesis}, although his definition
of a dynamic invariant is slightly different from ours.}.

The global invariant is a unary assertion which is true initially and thereafter
preserved by both the rely- and guar-conditions. 

The dynamic invariant is a binary assertion which is supposed 
to model any finite sequence of consecutive
rely and guar steps. For that reason $Dyn$ is required to be both reflexive 
and transitive. This means that any state transition which satisfies the
rely-condition, will also satisfy $Dyn$. The same is of course also true for
the guar-condition.

Unfortunately, $Inv$ and $Dyn$ are often false inside await-statements, which means
that we cannot automatically interpret

\begin{tuple}{\vartheta
            }{\alpha
            }{P
            }{R
            }{W
            }{G
            }{E}
\end{tuple}

\noindent as

\begin{tuple}{\vartheta
            }{\alpha
            }{P\And Inv
            }{R\And Dyn \And (\overleftharpoon{Inv}\Implies Inv)
            }{W\And Inv
            }{G\And Dyn\And (\overleftharpoon{Inv}\Implies Inv)
            }{E\And Inv\And Dyn,}
\end{tuple}

\noindent since we also want to specify subprograms of await-statements. Instead it will be
stated explicitly in each particular specification whether $Inv$ and
$Dyn$ are valid or not.

It may be argued that $Inv$ and $Dyn$ should have arguments indicating which part
of the global state they affect. We have ignored this because of the large number
of arguments needed in some cases\footnote{The number of arguments could probably have been
reduced if we had introduced records, in the style of for example
VDM, to structure the global
state.}.

%% file: phil.tex
\chapter{Dining-Philosophers}

\section{Task}

\subsection{Mutual Exclusion}

Two statements are {\em \underline{mutually exclusive}}\index{mutually exclusive} if 
they cannot be executed simultaneously. 
Obviously, if the two statements occur in the same process, then they are mutually
exclusive unless one of them is a substatement of the other. 
If the statements
are taken from two different processes, on the other hand, 
it is often difficult to
decide if they
are mutually exclusive or not.

If the sets of memory locations accessed by the two statements overlap, 
one way to
secure mutual exclusion is to place them in the bodies of 
two different await-statements. 
If the sets of memory locations are disjoint, mutual exclusion is normally not 
needed.

Unfortunately, from time to time problems arise where the programmer must do the
synchronisation himself and provide his own
code to secure mutual exclusion. Such a situation will be discussed 
in the example below.

\subsection{Workshop}

Assume $M$ ($M\ge 3$) philosophers are attending a workshop. 
During this workshop each philosopher
is scheduled to eat $Q$ times\footnote{Both $M$ and $Q$ are constants ---
not variables.}. The rest of the time, he is supposed to spend 
thinking or discussing with his colleagues. The food is served on a round table. 
Each philosopher has his own plate, which means that there are $M$ plates.

The dish is spaghetti. Unfortunately the spaghetti is
so long and tangled that a philosopher needs two forks to eat it. 

By some mistake there are only $M$ forks on the table; one between each plate. Moreover,
no philosopher would consider using any other fork than the ones to his immediate
left or right.

It is also assumed that when a philosopher
grabs his forks, he grabs both of them at the same time. In other words, a philosopher 
holds either two forks or no forks at all. 
This means that two neighbours cannot eat at the same time.

Our task is to write a program for each philosopher that provides this synchronisation.
The algorithm presented here is closely related to the one discussed in \cite{so:await}.

\section{Development}

\subsection{Data Structure}

Let

\begin{flist}
\type{D_{(n,m)}}{\set{n,n+1,\ldots,m}}.
\end{flist}

\noindent We will employ one global array

\begin{flist}
Frk:\arrayof{D_{(0,M -  1)}}{D_{(0,2)}},
\end{flist}

\noindent and two auxiliary arrays

\begin{flist}
Num:\arrayof{D_{(0,M -  1)}}{D_{(0,Q)}},\\
Eating:\arrayof{D_{(0,M -  1)}}{D_{(0,1)}}.
\end{flist}

\noindent To
avoid making a special case of the `first' and the `last' philosopher, we will
apply arithmetic modulo $M$ to access the `previous' and the `next' philosopher. 
The symbol $\ominus$ will be used to denote subtraction modulo $M$, while
$\oplus$ stands for addition modulo $M$.

This 
means that we would like the arrays to be defined on 
$D_{(0,M -  1)}$, and that the process $Phil(j -  1)$ simulates the $j$'th philosopher. 
Moreover, at any time, $Frk(j)$ equals the number of forks available to philosopher 
$j+ 1$,
while $Num(j)$ gives the number of times he has eaten.

Finally, since we want to prove that our implementation satisfies the desired mutual
exclusion property, namely that two neighbours cannot eat at the same time,
we will use $Eating(j)$ to indicate if philosopher $j+ 1$
is holding his forks or not. 

If philosopher $j+ 1$ is eating then $Eating(j)=1$, otherwise $Eating(j)=0$. The
reason why the range of $Eating$ is $D_{(0,1)}$ and not $\Bool$, is that this 
allows us to formulate the following invariant

\begin{flist}
\forall{j\in D_{(0,M -  1)}}{Frk(j)=2 -  (Eating(j\oplus 1)+Eating(j\ominus 1))}.
\end{flist}

\noindent Moreover, since a philosopher can only eat if he holds both his forks, we will
also insist on the truth of

\begin{flist}
\forall{j\in D_{(0,M -  1)}}{Eating(j)=1\Implies Frk(j)=2}.
\end{flist}

\noindent From now on, $Inv$ will denote the conjunction of these two assertions.

\subsection{Main Program}

This means that the main program will be of the form:

\begin{prg}
\{\true\}\\
\\
\nbegin\\
\qquad\loc\ Frk;\\
\qquad Init();\\
\\
\qquad \{\forall{j\in D_{(0,M -  1)}}{Eating(j)=0\And Num(j)=0\And Inv}\}\\
\\
\qquad \{Phil(0)\parallel Phil(1)\parallel \ \ldots\ \parallel Phil(M -  1)\}\\
\nend\\
\\
\{\forall{j\in D_{(0,M -  1)}}{Eating(j)=0}\And Num(j)=Q\And Inv\}.\\
\end{prg}

\subsection{Specification of Processes}

Since $Phil(l)$ only waits if  $Phil(l\ominus 1)$ or $Phil(l\oplus 1)$ are
eating, it follows that:
\label{phil:ref}

\begin{spec}{Phil}{l:D_{(0,M -  1)}}{}{\begin{ndcl}{
                              \lwew & Frk(l\ominus 1):D_{(0,2)},\\
                               &     & Frk(l\oplus 1):D_{(0,2)},\\
                               &\lrew & Frk(l):D_{(0,2)},}{
                               \lwer & Eating(l):D_{(0,1)},\\
                               &\lwer & Num(l):D_{(0,Q)}}
                            \end{ndcl}
                          }{Eating(l)=0\And Num(l)=0\And Inv
                          }{\overleftharpoon{Inv}\Implies Inv
                          }{(Eating(l\ominus 1)=1\Or
                            Eating(l\oplus 1)=1)\And  Inv
                          }{((Eating(l)=\overleftharpoon{Eating}(l)\And Num(l)=\overleftharpoon{Num}(l))\ \Or\\
                           &(\overleftharpoon{Eating}(l)=0\And Eating(l)=1\And Num(l)=\overleftharpoon{Num}(l))
                             \ \Or\\
                           & (\overleftharpoon{Eating}(l)=1\And Eating(l)=0\And 
                                Num(l)=\overleftharpoon{Num}(l)+1))\ \And\\
                           & (\overleftharpoon{Inv}\Implies Inv)
                          }{Eating(l)=0\And Num(l)=Q\And Inv.}
\end{spec}

\subsection{Correctness Proof}

We will now show that the parallel composition of these $M$ processes has the disered
effect.
Firstly, observe that the constraints imposed on the global memory locations by the 
declarations in the respective processes are satisfied. 
Secondly, since it is clear that

\begin{flist}
\vdash Eating(l\oplus 1)=1\And Inv\Implies\\
\qquad \neg((Eating(l\oplus 1)=0\And Num(l+1)=Q\And Inv)\ \Or \\
\qquad\qquad            ((Eating(l)=1\Or Eating(l\oplus 2)=1)\And Inv)),
\end{flist}

\begin{flist}
\vdash Eating(l\ominus 1)=1\And Inv\Implies\\
\qquad \neg((Eating(l\ominus 1)=0\And Num(l+1)=Q\And Inv)\ \Or \\
\qquad\qquad            ((Eating(l)=1\Or Eating(l\ominus 2)=1)\And Inv)),
\end{flist}

\noindent it follows that whenever $Phil(l)$ is blocked, then there is at least one other
process which is enabled.
Thus, it can be deduced by the consequence- and generalised parallel-rules 
that the parallel-statement satisfies:

\begin{tuple}{\{Frk\}
            }{\{Num,Eating\}
            }{\forall{j\in D_{(0,M -  1)}}{Eating(j)=0\And Num(j)=0}\And Inv
            }{I
            }{\false
            }{\overleftharpoon{Inv}\Implies Inv
            }{\forall{j\in D_{(0,M -  1)}}{Eating(j)=0\And Num(j)=Q}
                            \And Inv.}
\end{tuple}

\subsection{Process Decomposition}
 
The final step is to implement $Phil(l)$. Assume that $Eat(l)$ and $Study(l)$ are
operations whose guar-condition implies that

\begin{flist}
Num=\overleftharpoon{Num}\And Frk=\overleftharpoon{Frk}\And Eating=\overleftharpoon{Eating}.
\end{flist}

\noindent Then, since a philosopher will only have to wait
if one of his two neighbours is eating, it follows that $Phil(l)$ should be
of the following form:

\begin{prg}
\{Num(l)=0\And Eating(l)=0\And Inv\}\\
\\
\nbegin\\
\qquad\loc\ j;\\
\qquad j:=0;\\
\qquad \nwhile\ j<Q\ \ndo\\
\\
       \qquad\qquad \{Num(l)<Q\And Eating(l)=0\And Inv\}\\
\\
       \qquad\qquad \nawait\ Frk(l)=2\ \ndo\ GrabFrks(l)\ \nod;\\
\\  
     \qquad\qquad \{Num(l)<Q\And Eating(l)=1\ \And\\
      \qquad\qquad\qquad              Frk(l\ominus 1)<2\And Frk(l\oplus 1)<2\And Inv\}\\
\\
       \qquad\qquad Eat(l);\\
\\
       \qquad\qquad \{Num(l)<Q\And Eating(l)=1\ \And\\
        \qquad\qquad\qquad          Frk(l\ominus 1)<2\And Frk(l\oplus 1)<2\And Inv\}\\
\\
       \qquad\qquad \nawait\ \true\ \ndo\ DropFrks(l)\ \nod;\\
\\
       \qquad\qquad \{Num(l)\le Q\And Eating(l)=0\ \And Inv\}\\
\\
       \qquad\qquad Think(l);\\
       \qquad\qquad j:=j+1;\\
\\
       \qquad\qquad \{Num(l)\le Q\And Eating(l)=0 \And Inv\}\\ 
\\
\qquad \nod\\
\nend\\
\\
\{Num(l)=Q\And Eating(l)=0 \And Inv\}.\\
\end{prg}

\subsection{Implementing Atomic Statements}

The final step is to implement the atomic statements $GrabFrks(l)$ and
$DropFrks(l)$. The first one must satisfy:

\begin{tuple}{\{Frk\}
            }{\{Num,Eating\}
            }{Num(l)<Q\And Eating(l)=0\And Frk(l)=2\And Inv
            }{I
            }{\false
            }{\true
            }{Num(l)<Q\And Eating(l)=1\And Frk(l)=2\ \And\\
               &Frk(l\ominus 1)=\overleftharpoon{Frk}(l\ominus 1) -  1\ \And\\ 
               &Frk(l\oplus 1)=\overleftharpoon{Frk}(l\oplus 1)  -  1\And Inv.}
\end{tuple}

\noindent Thus,

\begin{prg}
          Frk(l\ominus 1):=Frk(l\ominus 1) -  1;\\
          Frk(l\oplus 1):=Frk(l\oplus 1) -  1\\
\end{prg}

\noindent is a valid implementation. Similarly, $DropFrks(l)$ requires that

\begin{tuple}{\{Frk\}
            }{\{Num,Eating\}
            }{Num(l)<Q\And Eating(l)=1\And Frk(l)=2\ \And \\
               &Frk(l\ominus 1)<2\And Frk(l\oplus 1)<2\And Inv
            }{I
            }{\false
            }{\true
            }{Num(l)<Q\And Eating(l)=0\And Frk(l)=2\ \And \\
              & Frk(l\ominus 1)=\overleftharpoon{Frk}(l\ominus 1)+1\ \And\\ 
							& Frk(l\oplus 1)=\overleftharpoon{Frk}(l\oplus 1)+1\And Inv,}
\end{tuple}

\noindent which is satisfied by

\begin{prg}
          Frk(l\ominus 1):=Frk(l\ominus 1)+1;\\
          Frk(l\oplus 1):=Frk(l\oplus 1)+1.\\
\end{prg}

%% file: bubl.tex
\chapter{Bubble-Lattice-Sort}

\section{Task}

We will now employ LSP to develop Bubble-Lattice-Sort, a sorting algorithm which
relies heavily upon synchronisation between $M+1$ processes\footnote{$M$ is
a constant --- not a variable.}. 
A number of proof techniques are used to verify a related 
 algorithm in \cite{hb:lncs}.

Let

\begin{flist}
\type{D_{(n,m)}}{\set{n,n+1,\ldots,m}.}
\end{flist}

\noindent Our job is to implement the following operation specification

\begin{spec}{Sort}{}{}{\begin{dcl}
                            \lwer &B&:&\arrayof{D_{(1,M)}}{Value}\\
                       \end{dcl}
                       }{\true
                       }{\true
                       }{\false
                       }{\true
                       }{\forall{j\in D_{(1,M -  1)}}\cdot{B(j)\ge B(j+1)}\ \And \\
                         &\forall{v\in Value}\cdot\\
												 &\qquad {\card{\set{j\in D_{(1,M)}|\overleftharpoon{B}(j)=v}}=
                                             \card{\set{j\in D_{(1,M)}|B(j)=v}}}.}
\end{spec}

\noindent The operation $Sort$ sorts $B$ into decreasing order, given that the
environment does not change the value of $B$ 
(in a way observable to $Sort$) while
$Sort$ is being executed\footnote{Observe that the rely-condition actually is 
`equivalent' to $B=\overleftharpoon{B}$
since the $\lwer$-declaration is constraining $B$ from being changed (in an
observable way) by the
environment.}. Moreover, $Sort$ does not rely 
upon any help from the environment
to terminate.

\section{Development}

\subsection{Division into Processes}

There are of
course many ways to implement this operation. As already indicated, we will use 
an algorithm called
Bubble-Lattice-Sort\footnote{The origin of this problem is a description given
by T.\ C.\ Chen to C.\ B.\ Jones (in a {\em Heueriger}!) of how lattices of
magnetic bubbles can be used to sort values in linear time. Chen's claim, in 1975, was 
that there was no algebra in which this could be verified.}, 
which is related to the well-known 
sequential Bubble-Sort algorithm. 

Bubble-Lattice-Sort consists of $M+1$ processes:

\begin{flist}
\{Send()\parallel Bubble(1)\parallel\ \ldots\ \parallel Bubble(M)\}.
\end{flist}

\noindent $Send()$ reads the next value of $B$ to be bubbled (from left to right) and
passes it on to $Bubble(1)$, which, when it has received two values, will
transfer the smallest to Bubble(2) (or one of them if they are equal). This is
thereafter repeated each time $Bubble(1)$ receives a value from $Send()$.

$Bubble(2)$ feeds $Bubble(3)$ in a similar style and so on.
The algorithm terminates when (for the first time) $Bubble(M)$ receives a value 
from $Bubble(M -  1)$ .

The sending and receiving is synchronised in such a way that no process is `storing'
more than two values at a time.

\subsection{More Data Structure}

To implement the algorithm we will employ two global arrays

\begin{flist}
 Feed:\arrayof{D_{(1,M+1)}}{Value},\\
 Empty:\arrayof{D_{(1,M+1)}}{\Bool}.
\end{flist}

\noindent The following auxiliary array will also be useful

\begin{flist}
 Num:\arrayof{D_{(0,M+1)}}{D_{(0,M)}}.
\end{flist}

\noindent $Send()$ can only access the memory locations $Empty(1)$, $Feed(1)$, $Num(0)$ and $B$.
$Bubble(l)$, on the other hand,
is restricted to accessing $Empty(l)$, $Empty(l+1)$, $Feed(l)$, $Feed(l+1)$,
$Num(l)$ and $B(l)$.

$Bubble(j)$ is ready to receive
a value when $Empty(j)$ is true, otherwise $Empty(j)$ is false. When $Empty(j)$ is false
and $Empty(j+1)$ is true, then no other process will try to access any of the memory
locations accessible by $Bubble(j)$. 

$Num(j)$ is equal to the number of values bubbled by $Bubble(j)$ (the number
of values read by $Send()$ if $j=0$), while
$Feed(j)$ will be used by $Bubble(j -  1)$ to pass on a new
value to $Bubble(j)$ (from $Send()$ to $Bubble(1)$ if $j=1$).

A value passed on to $Bubble(j)$ will be sent directly to $Bubble(j+1)$,
if the value is less than or equal to
$B(j)$, otherwise the value stored in $B(j)$ will be passed on to $Bubble(j+1)$, and
the new value will be assigned to $B(j)$. Moreover, because the internal steps of one
particular bubbling may be hidden completely inside an await-statement, 
we can insist on the truth of

\begin{flist}
\forall j\in D_{(1,M -  1)} \cdot(Num(j+1)\neq 0\Rightarrow B(j)\ge B(j+1))\ \And\\
\qquad (\neg Empty(j+1)\Rightarrow Feed(j+1)\le B(j)).
\end{flist}

\noindent In other words, this
condition can be used as an invariant.

Moreover, to ensure that the different processes stay in step, we would 
also like 

\begin{flist}
(Empty(1)\Rightarrow Num(0)=Num(1))\ \And\\
(\neg Empty(1)\Implies Num(0)=Num(1)+1)\ \And\\
\forall j\in D_{(1,M)}\cdot (\neg Empty(j+1)\Implies Num(j+1)+2=Num(j))\ \And\\
    \qquad(Empty(j+1)\Implies Num(j+1)+1=Num(j)\ \Or\\
		\qquad\qquad Num(j)=Num(j+1)=0)
\end{flist}

\noindent to be an invariant. From now on, we will use $Inv$ to denote 
the conjunction of these two conditions.

The reason why $M+1$ occurs in the domains of the new arrays, is that we
then avoid making a special case of implementing $Bubble(M)$. 
Furthermore,
this also has a simplifying effect on some of the proofs.

\subsection{Dynamic Invariant}

The next step is to characterise the way in which the $M+1$ processes may interact.
Since any externally observable state change by any of the processes is
assumed to maintain the invariant,
this has to some extent already been done. What remains 
is to make sure that the bag of values occurring in 
$B$ is preserved. The condition below
is sufficient:

\begin{flist}
\forall v\in Value\cdot \\
\qquad \card{\set{j\in D_{(1,M)}|\overleftharpoon{B}(j)=v\And \overleftharpoon{Num}(j)>0}}\ +\\
\qquad\qquad\card{\set{j\in D_{(1,M)}|\overleftharpoon{Feed}(j)=v\And \neg \overleftharpoon{Empty}(j)}}\ +\\
\qquad\qquad\card{\set{j\in D_{(1,M)}|\overleftharpoon{B}(j)=v\And j>\overleftharpoon{Num}(0)}}=\\
\qquad \card{\set{j\in D_{(1,M)}|B(j)=v\And Num(j)>0}}\ +\\
\qquad\qquad\card{\set{j\in D_{(1,M)}|Feed(j)=v\And \neg Empty(j)}}\ +\\
\qquad\qquad\card{\set{j\in D_{(1,M)}|B(j)=v\And j>Num(0)}}.
\end{flist}

\noindent The reason why this constraint is not included in the
invariant $Inv$ is of course that it refers to a previous state. Instead, it
can be defined as a dynamic invariant, and 
we will use $Dyn$ to denote this formula. In other
words, in any specification where we need to state the condition
above, we will instead write $Dyn$.

\subsection{Main Structure}

This means that $Sort$ will be of the form:

\begin{prg}
\{\true\}\\
\\
\loc\ Feed:\arrayof{D_{(1,M+1)}}{Value};\\
\loc\ Empty:\arrayof{D_{(1,M+1)}}{\Bool};\\
Init();\\
\\
\{Num(0)=0\And Inv\And Dyn\}\\
\\
\{Send()\parallel Bubble(1)\parallel\ \ldots\ \parallel Bubble(M)\}\\
\\
\{Num(0)=M\And (\forall{j\in D_{(1,M+1)}}\cdot{Num(j)=M -  j+1})\ \And\\
\qquad Inv\And Dyn\}.\\
\end{prg}

\noindent Observe that 

\begin{flist}
\vdash Num(0)=0\And Inv\Implies\forall{j\in D_{(1,M)}}\cdot{Num(j)=0\And Empty(j)}.
\end{flist}

\subsection{Specification of Processes}

Based on the discussion above, it is now straightforward to specify
the component processes:

\begin{spec}{Send}{}{}{\begin{ndcl}{\lwew &Empty(1)&:&\Bool,\\
                                    &\lwer      &Feed(1)&:&Value,
                                  }{\lwer &Num(0)&:&D_{(0,M)}}
                       \end{ndcl}
                       }{Num(0)=0\And Inv
                       }{Dyn\And (\overleftharpoon{Inv}\Implies Inv)
                       }{Num(0)<M\And \neg Empty(1)\And Inv
                       }{Dyn\And (\overleftharpoon{Inv}\Implies Inv)
                       }{Num(0)=M\And Dyn\And Inv,}
\end{spec}

\begin{spec}{Bubble}{l:D_{(1,M)}}{}{\begin{ndcl}{\lwew &Empty(l)&:&\Bool,\\
                                     &          &Empty(l+1)&:&\Bool,\\
                                     & \lrew      &Feed(l)&:&Value,\\
                                     &  \lwer  &Feed(l+1)&:&Value,\\
                                        &       &B(l)&:&Value,
                                       }{\lwer &Num(l)&:&D_{(0,M)}}
                            \end{ndcl}
                       }{Num(l)=0\And Inv
                       }{Dyn\And (\overleftharpoon{Inv}\Implies Inv)
                       }{Num(l)<M -  l+1\And (Empty(l)\Or 
                         \neg Empty(l+1))\And Inv
                       }{Dyn\And (\overleftharpoon{Inv}\Implies Inv)
                       }{Num(l)=M -  l+1\And Dyn\And Inv,}
\end{spec}

\subsection{Correctness Proof}

The next step is to prove that the parallel composition of 
these $M+1$ processes has
the desired effect. 
Firstly, observe that the constraints imposed on the global memory locations 
by the declarations in the respective operations are satisfied.
Secondly, since

\begin{flist}
\vdash Num(0)<M\And \neg Empty(1)\And Inv\ \And\\
\ ((Num(1)=M\And Dyn \And Inv)\ \Or\\
\ (Num(1)<M\And (Empty(1)\Or \neg Empty(2))\And Inv))\Implies\\
\qquad Num(1)<M\And \neg Empty(2)\And Inv
\end{flist}

\begin{flist}
\vdash \forall l\in D_{(1,M -  1)}\cdot Num(l)<M -  l+1\And \neg 
Empty(l+1)\And Inv\ \And\\
\ ((Num(l+1)=M -  l\And Dyn\And Inv)\ \Or\\
\ (Num(l+1)<M -  l\And (Empty(l+1)\Or \neg Empty(l+2))\And Inv))\Implies\\
\qquad Num(l+1)<M -  l\And \neg Empty(l+2)\And Inv
\end{flist}

\begin{flist}
\vdash \neg(Num(M)<1\And \neg Empty(M+1)\And Inv),
\end{flist}

\noindent it follows that whenever $Send()$ becomes blocked, there is at least one other
process which is enabled. In a similar way it can be shown that for all $1\le l\le M$,
$Bubble(l)$ cannot become blocked in a state in which no other process is enabled.
Thus, it follows by the consequence-
and generalised parallel-rules that the parallel-statement satisfies:

\begin{tuple}{\{B,Empty,Feed\}
            }{\{Num\}
            }{Num(0)=0\And Inv
            }{I
            }{\false
            }{Dyn\And (\overleftharpoon{Inv}\Implies Inv)
            }{\forall{j\in D_{(1,M)}}\cdot {Num(j)=M -  j+1}\And Num(0)=M\ \And\\
              &Dyn\And Inv.}
\end{tuple}

\noindent But then, since it is clear that

\begin{flist}
\vdash \overleftharpoon{Num}(0)=0\And \overleftharpoon{Inv}\ \And\\
\ \forall{j\in\set{1,\ldots,M}}\cdot{Num(j)=M -  j+1}\ \And\\
\qquad Num(0)=M\And Dyn\And Inv\ \Implies\\
\qquad \qquad\forall j\in D_{(1,M -  1)} \cdot B(j)\ge B(j+1)\ \And \\
\qquad \qquad\forall v\in Value\cdot \\
\qquad \qquad\qquad\card{\set{j\in D_{(1,M)}|\overleftharpoon{B}(j)=v}}=
                                             \card{\set{j\in D_{(1,M)}|B(j)=v}},
\end{flist}

\noindent we have proved that the parallel composition of the $M+1$ processes has the desired
effect. Moreover, the proof is only relying upon assumptions about the environment
allowed by the specification
of $Sort$.

\subsection{Process Decomposition}

The final step is to implement the specified operations.
As should be pretty obvious by now, we will have
to rely upon the
await-statement as an abstraction tool. Let us postpone implementation of the
await-statement's bodies, and first design the synchronisation structure.

Since $Send()$ is supposed to read $M$ different values in $B$, and $Bubble(1)$ is
ready to receive a value when $Empty(1)$ is true, it follows that $Send()$ 
should be of the form:

\begin{prg}
             \{Num(0)=0\And Inv\}\\
\\
             \nbegin\\
             \qquad \loc\ j;\\
             \qquad j:=0;\\
             \qquad\nwhile\ j< M\ \ndo\\
\\
                  \qquad \qquad\{Num(0)<M\And Inv\And Dyn\}\\
\\
                  \qquad\qquad j:=j+1;\\
                  \qquad \qquad \nawait\ Empty(1)\ \ndo\ S(j)\ \nod\\
             \qquad\nod\\
             \nend\\
\\
             \{Num(0)=M\And Inv\And Dyn\}.\\
\end{prg}

\noindent Moreover, since $Bubble(l)$ will receive $M -  l+1$ values, 
$Empty(l)$ is false if
and only if an unread value is stored in $Feed(l)$, and $Bubble(l+1)$ is ready
to receive a new value if $Empty(l+1)$ is true, it follows that $Bubble(l)$ should be
of the form:

\begin{prg}
             \{Num(l)=0\And Inv\}\\
\\
             \nawait\ \neg Empty(l)\ \ndo\ Fb(l)\ \nod;\\
             \nbegin\\
             \qquad\loc\ j;\\
             \qquad j:=1;\\
             \qquad\nwhile\ j< M -  l+1\ \ndo\\
\\
             \qquad\qquad\{Num(l)<M -  l+1\And Inv\And Dyn\}\\
\\
                  \qquad\qquad\nawait\ \neg Empty(l)\And Empty(l+1)\ \ndo\ Rb(l)\ \nod;\\
                  \qquad\qquad j:=j+1\\
             \qquad\nod\\
             \nend\\
\\
             \{Num(l)=M -  l+1\And Inv\And Dyn\}.\\
\end{prg}

\subsection{Implementing Atomic Statements}

To finish the implementation of $Send()$, it is necessary to substitute code for
$S(j)$ and verify that the final product has the desired properties.
The latter follows easily if $S(j)$ satisfies:

\begin{tuple}{\{B,Empty,Feed\}
            }{\{Num\}
            }{Num(0)<M\And Empty(1)\And Inv}{I}{\false}{\true}{Num(0)=\overleftharpoon{Num}(0)+1\And 
\neg Empty(1)\And Inv\And Dyn.}
\end{tuple}

\noindent Thus,

\begin{prg}
       Feed(1):=B(j);\\
       Empty(1):=\false\\
\end{prg}

\noindent is all that is required. 

Similarly, to finish the code of $Bubble(l)$ we must
implement $Fb(l)$ and $Rb(l)$ and make sure the 
specified properties are satisfied.
Again, if $Fb(l)$ satisfies:

\begin{tuple}{\{B,Empty,Feed\}
            }{\{Num\}
            }{Num(l)=0\And \neg Empty(l)\And Inv}{I}{\false}{\true}{
Num(l)=1\And Empty(l)\And Inv\And Dyn,}
\end{tuple}

\noindent and $Rb(l)$ satisfies:

\begin{tuple}{\{B,Empty,Feed\}
            }{\{Num\}
            }{Num(l)<M -  l+1\And \neg Empty(l)\And Empty(l+1)\And Inv}{I}{\false}{\true}{
Num(l)=\overleftharpoon{Num}(l)+1\And Empty(l)\And \neg Empty(l+1)\ \And\\
& Inv\And Dyn,}
\end{tuple}

\noindent it is clear that the implementation is correct, so

\begin{prg}
          B(l):=Feed(l);\\
          Empty(l):=\true,\\
\end{prg}

\noindent and 

\begin{prg}
          Feed(l+1):=\min{\set{Feed(l),B(l)}};\\
          B(l):=\max{\set{Feed(l),B(l)}};\\
          Empty(l):=\true;\\
          Empty(l+1):=\false\\
\end{prg}

\noindent are what is missing.

%% file: part.tex
\chapter{Set-Partition}

\section{Task}

Given two non-empty, disjoint sets of integers, $S$ and $L$;
our task is to develop a program which terminates in a state where
the maximum element of $S$ is less than the minimum element of $L$.
The sizes of the two sets must remain unchanged. Moreover, after termination,
the union of $S$
and $L$ is required to equal the union of their initial values.

This informal specification can be translated into a more mathematical
notation:

\begin{spec}{SetPart}{}{}{\begin{ndcl}{\lwer& S&:&\setof{\Nat},\\
                                       &    & L&:&\setof{\Nat}}
                                      {}
                          \end{ndcl}
                        }{S\cap L=\emptyset\And S\not=\emptyset\And L\not=\emptyset
                        }{\true
                        }{\false
                        }{\true
                        }{max(S)<min(L)\And \#S=\#\overleftharpoon{S}\ \And\\
                          &\#L=\#\overleftharpoon{L}\And S\cup L=\overleftharpoon{S}\cup \overleftharpoon{L}.}
\end{spec}

\noindent The declarations of $S$ and $L$ allow us to assume that the environment
will leave $S$ and $L$ unchanged.
Moreover, from the wait-condition it follows that $SetPart$ is 
required to
terminate. Finally, the guar-condition allows us to
change $S$ and $L$ as we like. The rest should be clear from the informal specification.

\section{Development}

\subsection{Algorithm}

Our implementation is inspired by \cite{ewd:607}.
In \cite{hb:lncs} a number of methods are used to verify a related algorithm\footnote{
The algorithm used in \cite{hb:lncs}, where verification methods 
for both CSP and shared-state programs are compared, is understandly a direct 
translation from CSP. It may be argued that to make it easier to compare our
approach with those discussed by Barringer, we should have developed the
same algorithm.

However, as pointed out in \cite{hb:lncs}, the shared state version may deadlock, 
and since we 
want to prove total correctness, a modification is necessary. Moreover, the algorithm
in \cite{hb:lncs} employs two flags, one for each CSP-channel. We decided to
replace these two flags with one, because this results in a simpler and more
natural development and therefore gives a more correct impression of what can be
achieved in LSP with respect to this particular synchronisation problem.}.

The algorithm employs two processes
called respectively $Small$ and $Large$.
The basic idea is as follows:
\begin{itemize}
\item The process $Small$ starts by finding the maximum element of $S$. This integer
is sent on to 
$Large$ and then subtracted from $S$.

The task of $Large$ is to add the received integer to $L$, 
and thereafter send the minimum element
of $L$ (which by then contains the integer just received from $Small$)
back to $Small$ and remove it from $L$. 

The process $Small$ adds the element sent from $Large$ to $S$. Then, if the maximum of
$S$ equals the integer just received from $Large$, it follows that the maximum
of $S$ is less than the minimum of $L$ and the algorithm terminates. Otherwise,
the whole procedure is repeated.

Since the difference between the maximum of $S$ and the minimum of $L$ is decreased
at each iteration, it follows that the program will eventually terminate.
\end{itemize}

\subsection{Data Structure}

The variables $Max$ and $Min$ simulate respectively `the channel' from $Small$ to 
$Large$ and `the channel' from $Large$ to $Small$. 

To secure that the two processes stay in step, the Boolean variable $Flag$ 
is introduced. 
When $Small$ switches on $Flag$, it means that $Large$ may
read the next value from $Max$, and when $Large$ makes $Flag$ false, it signals that
$Min$ is ready to be read by $Small$.

The adding, finding the maximum and sending section of $Small$ is mutually 
exclusive with
the adding, finding the minimum and sending section of $Large$. 

The only thing the process $Small$ is allowed to do while $Flag$ is true, is
to remove from $S$ the integer it just sent to $Large$. Similarly, when $Flag$ is
false, $Large$ is only allowed to remove the element it just sent to $Small$.

\subsection{First Decomposition Step}

Our implementation will be of the form:

\begin{prg}
\nbegin\\
\qquad \loc\ Max,Min,Flag;\\
\qquad Init();\\
\qquad \{Small()\parallel Large()\}\\
\nend.\\
\end{prg}

\noindent The task of $Init$ is of course to initialise the local state.

\subsection{Specifying \texorpdfstring{$Init$}{Init}}

To make it easier to formulate and reason about properties
satisfied by the concurrent part of our implementation,
we will use $Init$ to simulate the first iteration of the algorithm; in other
words, to perform
the first interchange of values. This means that:

\begin{spec}{Init}{}{}{\begin{ndcl}{\lwer& S&:&\setof{\Nat},\\
                                    &     & L&:&\setof{\Nat},\\
                                    &     & Min&:&\Nat,\\
                                    &     & Max&:&\Nat,\\
                                    &     & Flag&:&\Bool
                                   }{}
                        \end{ndcl}
            }{S\cap L=\emptyset\And S\not=\emptyset\And L\not=\emptyset
            }{\true
            }{\false
            }{\true
            }{\#S=\#\overleftharpoon{S}\ \And\\
             &\#L=\#\overleftharpoon{L}\ \And\\
             &S\cup L\cup \{Max,Min\}=\overleftharpoon{S}\cup \overleftharpoon{L}\ \And\\
             &\neg Flag\And Min\le Max\ \And\\
             &Max= max(S)\And Min<min(L)\ \And\\
             &S\cap L=\emptyset.}
\end{spec}

\noindent Basically, this operation simulates `the sending' of one element in both directions. 
Thus, the next process to transfer a value is $Small$, which explains the restriction 
on $Flag$. Moreover, $Small$ has already determined the `new' maximum of $S$.

\subsection{Final Implementation of \texorpdfstring{$Init$}{Init}}

The implementation of $Init$ is not very challenging. The program below is obviously 
sufficient:

\begin{prg}
Max:=max(S);\\
L:=L\cup \{Max\};\\
S:=S -  \{Max\};\\
Min:=min(L);\\
S:=S\cup \{Min\};\\
L:=L -  \{Min\};\\
Flag:=\false;\\
Max:=max(S).
\end{prg}

\subsection{Invariant}

We will now characterise a few properties that will be invariantly true for
the concurrent part of the implementation. 
Since for both processes the previously sent element is 
removed before the actual process starts to look for a 
new integer to send, it is clear that:

\begin{flist}
S\cap L\subseteq \{Min,Max\}.
\end{flist}

\noindent Moreover, because $Large$ will return the integer just received if the maximum of $S$
is less than the minimum of $L$, it is also true that:

\begin{flist}
Min\le Max.
\end{flist}

\noindent We will use $Inv$ to denote the conjunction of these 
two assertions. $Inv$ is obviously implied by the eff-condition of $Init$.

\subsection{Dynamic Invariant}

To ensure maintenance of the original integers we will require that any
state transition must satisfy:

\begin{flist}
S\cup L\cup\{Max,Min\}=\overleftharpoon{S}\cup \overleftharpoon{L}\cup \{\overleftharpoon{Max},\overleftharpoon{Min}\}.
\end{flist}

\noindent This is of course not enough on its own; however, if we 
insist that the conjunction of the eff-conditions of
the two processes implies that $\{Max,Min\}\subseteq S\cup L$,
it follows easily from
the eff-condition of $Init$ that the desired maintenance property is satisfied by
the overall program.

Moreover, since the first interchange of elements has already taken place in $Init$,
it is clear that any transition by either $Small$ or $Large$ will
satisfy:

\begin{flist}
Max\le \overleftharpoon{Max}\And
Min\ge \overleftharpoon{Min}.
\end{flist}

\noindent To prove freedom from deadlock an
auxiliary Boolean variable $TrmS$ is needed. The idea is that $TrmS$ is switched on
when $Small$ leaves its critical section for the last time, i.e.\ in the case that 
$Max$ 
equals $Min$. To show that $Flag$ is true 
when $Small$ has terminated, and that $TrmS$ is
true when $Large$ has terminated, we must insist that:

\begin{flist}
\overleftharpoon{TrmS}\And \overleftharpoon{Flag}\Implies TrmS\And Flag.
\end{flist}

\noindent From now on we will use $Dyn$ to denote the conjunction of these three assertions.

\subsection{Freedom from Deadlock}

The process $Small$ can only become blocked if it wants to enter its critical section, i.e.\ if
$Flag$ is true and $TrmS$ is still false. Hence, $Small$ will
only wait in a state which satisfies:

\begin{flist}
Flag\And \neg TrmS\And Inv.
\end{flist}

\noindent Similarly, it is clear that $Large$ will only be held back in a state characterised 
by:

\begin{flist}
\neg Flag\And Inv.
\end{flist}

\noindent The conjunction of these two assertions is obviously inconsistent. Moreover, if we
insist that the eff-condition of $Small$ implies that $Flag$ is true, while
the eff-condition of $Large$ implies that $TrmS$ is switched on, it follows that the parallel
composition of $Small$ and $Large$
will never deadlock.

\subsection{Specifying \texorpdfstring{$Small$}{Small}}

From the discussion above it is clear that $Small$ does not need write access to $L$ and
$Min$. Similarly, $Large$ will never have to change the value of $S$, $Max$
and $TrmS$. 

To secure mutual exclusion the environment must maintain the falsity of $Flag$, while
$Small$ in return must guarantee never to make $Flag$ false. 

Moreover, the only possible change of state due to the environment while $Flag$ is
false is that $Min$ is removed from from $L$. Similarly, the only thing $Small$ is
allowed to do while $Flag$ is true is to remove $Max$ from $S$.

Furthermore, to prove that the number of elements in $S$, when $Small$ terminates, 
equals the set's
initial size, any internal transition must satisfy:

\begin{flist}
\neg\overleftharpoon{Flag}\And Flag\Implies Max=Min\Or (Max\not\in L\And Min\in S),
\end{flist}

\noindent and for similar reasons when an integer is sent in the other direction, it 
is necessary that any external transition satisfies:

\begin{flist}
\overleftharpoon{Flag}\And \neg Flag\Implies Max=Min\Or (Min\not\in S\And Max\in L).
\end{flist}

\noindent Finally, to ensure that $TrmS$ is switched on if and only if $Small$ has left its
critical section for the last time, any internal transition must satisfy:

\begin{flist}
\neg \overleftharpoon{TrmS}\And TrmS\Leftrightarrow \neg \overleftharpoon{Flag}\And Flag\And Max=Min.
\end{flist}

\noindent Thus, in a more formal notation:

\begin{spec}{Small}{}{}{\begin{ndcl}{
                           \lwew& Flag&:&\Bool,\\
                           &\lwer& S&:&\setof{\Nat},\\
                           &     & Max&:&\Nat,\\
                           &\lrew& L&:&\setof{\Nat},\\
                           &     & Min&:&\Nat\\
                                   }{\lwer& TrmS&:&\Bool
                       }\end{ndcl}
                                   }{Max= max(S)\And \neg Flag\And Max\not\in L\ \And\\
                                    & \neg TrmS\And Inv
                                   }{(\neg \overleftharpoon{Flag}\Implies \neg Flag
                                     \And Min=\overleftharpoon{Min}\And (L\cup\{Min\}=\overleftharpoon{L}\Or L=\overleftharpoon{L}))\ \And\\
                                     &(\overleftharpoon{Flag}\And \neg Flag\Implies 
                                      Max=Min\Or (Min\not\in S\And Max\in L))\ \And\\
                                     &Dyn\And (\overleftharpoon{Inv}\Implies Inv)
                                   }{Flag \And \neg TrmS\And Inv
                                   }{( \overleftharpoon{Flag}\Implies 
                                      Flag \And Max=\overleftharpoon{Max}\And (S\cup\{Max\}=\overleftharpoon{S}\Or S=\overleftharpoon{S}))\ 
                                       \And\\
                                     &(\neg\overleftharpoon{Flag}\And Flag\Implies Max=Min\Or 
                                       (Max\not\in L\And Min\in S))\ \And\\
                                     &(\neg \overleftharpoon{TrmS}\And TrmS\Leftrightarrow 
                                      \neg \overleftharpoon{Flag}\And Flag\And Max=Min)\ \And\\
                                     &Dyn\And (\overleftharpoon{Inv}\Implies Inv)
                                   }{\#S=\#\overleftharpoon{S}\And Max=max(S)\ \And\\
                                     &Max=Min\And Flag\ \And\\
                                     &Dyn\And Inv.}
\end{spec}

\subsection{Specifying \texorpdfstring{$Large$}{Large}}

The specification of $Large$ is very similar:

\begin{spec}{Large}{}{}{\begin{ndcl}{
                           \lwew& Flag&:&\Bool,\\
                           &\lwer& L&:&\setof{\Nat},\\
                           &     & Min&:&\Nat,\\
                           &\lrew& S&:&\setof{\Nat},\\
                           &     & Max&:&\Nat\\
                                   }{
                           \lrew& TrmS&:&\Bool
                                   }\end{ndcl}
                                   }{Min< min(L)\And Max\not\in L\And\neg TrmS\And 
                                     \neg Flag\And Inv
                                   }{( \overleftharpoon{Flag}\Implies 
                                      Flag \And Max=\overleftharpoon{Max}\And 
                                         (S\cup\{Max\}=\overleftharpoon{S}\Or S=\overleftharpoon{S}))\ \And\\
                                     &(\neg\overleftharpoon{Flag}\And Flag\Implies Max=Min\Or 
                                       (Max\not\in L\And Min\in S))\ \And\\
                                     &(\neg \overleftharpoon{TrmS}\And TrmS\Leftrightarrow 
                                      \neg \overleftharpoon{Flag}\And Flag\And Max=Min)\ \And\\
                                     &Dyn\And (\overleftharpoon{Inv}\Implies Inv)
                                   }{\neg Flag\And Inv
                                   }{(\neg \overleftharpoon{Flag}\Implies \neg Flag
                                     \And Min=\overleftharpoon{Min}\And (L\cup\{Min\}=\overleftharpoon{L}\Or L=\overleftharpoon{L}))\ \And\\
                                     &(\overleftharpoon{Flag}\And \neg Flag\Implies 
                                      Max=Min\Or (Min\not\in S\And Max\in L))\ \And\\
                                     &Dyn\And (\overleftharpoon{Inv}\Implies Inv)
                                   }{\#L=\#\overleftharpoon{L}\And Min<min(L)\ \And\\
                                     &Max=Min\And TrmS\ \And Dyn\And Inv.}
\end{spec}

\subsection{Parallel Composition}

Since the wait-conditions of both operations are inconsistent with the
other operation's wait- and eff-conditions, it follows by the parallel-
and consequence-rules
that the concurrent part of our implementation 
satisfies:

\begin{tuple}{\{L,S,Flag,Min,Max\}
            }{\{TrmS\}
            }{Max= max(S)\And \neg Flag\And \neg TrmS\And Max\not\in L\ \And\\
             & Min< min(L)\And Inv
            }{I
            }{\false
            }{\true
            }{\#L=\#\overleftharpoon{L}\And Min<min(L)\ \And\\
              &\#S=\#\overleftharpoon{S}\And Max=max(S)\ \And\\
              &Max=Min\And Dyn\And Inv,}
\end{tuple}

\noindent which together with $Init$ gives the desired overall effect.

\subsection{Decomposing \texorpdfstring{$Small$}{Small}}

How can we best decompose $Small$? Obviously, the while-construct is needed. One 
possible strategy is the following:

\begin{prg}
\nbegin\\
\qquad\loc\ V_S;\\
\qquad V_S:=(Max\not=Min);\\
\qquad \nwhile\ V_S\ \ndo\\
\qquad\qquad Sml();\\
\qquad\qquad V_S:=(Max\not=Min)\\
\qquad \nod;\\
\qquad Flag:=\true\\
\nend.\\
\end{prg}

\noindent The obvious termination expression is:

\begin{flist}
Max -  Min.
\end{flist}

\noindent Thus, since

\begin{flist}
\overleftharpoon{Min}<\overleftharpoon{Max}\And (Max<\overleftharpoon{Max}\Or Max=Min)\ \And\\ 
Dyn\Implies Max -  Min< \overleftharpoon{Max} -  \overleftharpoon{Min},
\end{flist}

\noindent and

\begin{flist}
Inv\Implies Max -  Min\ge 0,
\end{flist}

\noindent it follows that the loop terminates and that the specification of $Small$ is
satisfied, if we can prove that $Sml$ is characterised by:

\begin{spec}{Sml}{}{}{\begin{ndcl}{
                           \lwew& Flag&:&\Bool,\\
                           &\lwer& S&:&\setof{\Nat},\\
                           &     & Max&:&\Nat,\\
                           &\lrew& L&:&\setof{\Nat},\\
                           &     & Min&:&\Nat\\
                                   }{\lwer& TrmS&:&\Bool
                       }\end{ndcl}
            }{Max= max(S)\And \neg Flag\ \And\\
						  &\neg TrmS\And Min<Max\And Max\not\in L 
              \And Inv
                                   }{(\neg \overleftharpoon{Flag}\Implies \neg Flag
                                     \And Min=\overleftharpoon{Min}\And (L\cup\{Min\}= \overleftharpoon{L}\Or L=\overleftharpoon{L}))\ \And\\
                                     &(\overleftharpoon{Flag}\And \neg Flag\Implies 
                                      Max=Min\Or (Min\not\in S\And Max\in L))\ \And\\
                                     &Dyn\And (\overleftharpoon{Inv}\Implies Inv)
                                   }{Flag \And \neg TrmS\And Inv
                                   }{( \overleftharpoon{Flag}\Implies 
                                      Flag \And Max=\overleftharpoon{Max}\And (S\cup\{Max\}=\overleftharpoon{S}\Or S=\overleftharpoon{S}))\ 
                                       \And\\
                                     &(\neg\overleftharpoon{Flag}\And Flag\Implies Max=Min\Or 
                                       (Max\not\in L\And Min\in S))\ \And\\
                                     &(\neg \overleftharpoon{TrmS}\And TrmS\Leftrightarrow 
                                      \neg \overleftharpoon{Flag}\And Flag\And Max=Min)\ \And\\
                                     &Dyn\And (\overleftharpoon{Inv}\Implies Inv)
            }{\#S=\#\overleftharpoon{S}\And Max=max(S)\ \And\\
              & \neg Flag\And \neg TrmS\ \And\\
              &((Max<\overleftharpoon{Max}\And Max\not\in L)\Or Max=Min)\ \And\\
              &Dyn\And Inv.}
\end{spec}

\subsection{Final Implementation of \texorpdfstring{$Sml$}{Sml}}

It can be shown that this property is satisfied by the program below:

\begin{prg}
\{Max=max(S)\And Min<Max\And Max\not\in L\ \And\\
\qquad \neg TrmS
\And \neg Flag\And Dyn\And Inv\}\\
\\
Flag:=\true;\\
\\
\{Max=max(S)\And \#S=\#\overleftharpoon{S}\And Max=\overleftharpoon{Max}\And\neg TrmS\ \And\\
 \qquad (\neg Flag\Implies Max=Min\Or (Min\not\in S\And Max\in L))\And Dyn\And Inv\}\\
\\
S:=S\setminus \set{Max};\\
\\
\{(S\not=\emptyset\Implies Max>max(S))\And Max=\overleftharpoon{Max}\And \#S=\#\overleftharpoon{S} -  1\ \And\\
\qquad (\neg Flag\Implies Max=Min\Or (Min\not\in S\And Max\in L))\ \And\\
\qquad \neg TrmS\And Dyn\And Inv\}\\
\\
\nawait\ \neg Flag\ \ndo\\
\qquad \nskip\\
\nod;\\
\\
\{(S\not=\emptyset\Implies Max>max(S))\And \#S=\#\overleftharpoon{S} -  1\ \And\\ 
\qquad (Max=Min\Or (Min\not\in S\And Max\in L))\ \And\\
\qquad Max=\overleftharpoon{Max}\And\neg TrmS\And Dyn\And\neg Flag\And Inv\}\\
\\
S:=S\cup\{Min\};\\
\\
\{((Max>max(S)\And Max\in L)\Or (Max=max(S)\And Max=Min))\ \And\\
\qquad \#S=\#\overleftharpoon{S}\And Max=\overleftharpoon{Max}\ \And\\
\qquad  Min\in S\And \neg TrmS\And Dyn\And\neg Flag\And Inv\}\\
\\
Max:=max(S);\\
\\
\{Max=max(S)\And ((Max<\overleftharpoon{Max}\And Max\not\in L)\Or Max=Min)\ \And\\
\qquad \#S=\#\overleftharpoon{S}\And\neg TrmS\And Dyn\And \neg Flag\And Inv\}.\\
\end{prg}

\subsection{Decomposing \texorpdfstring{$Large$}{Large}}

What remains is to decompose $Large$. The main structure is given below:

\begin{prg}
\nbegin\\
\qquad\loc\ V_L;\\
\qquad\nawait\ Flag\ \ndo\\
\qquad\qquad \nskip\\
\qquad\nod;\\
\qquad V_L:=(Max\not=Min);\\
\qquad \nwhile\ V_L\ \ndo\\
\qquad\qquad Lrg();\\
\qquad\qquad V_L:=(Max\not=Min)\\
\qquad \nod\\
\nend.\\
\end{prg}

\noindent Again, the termination expression is 

\begin{flist}
Max -  Min.
\end{flist}

\noindent Moreover, since

\begin{flist}
\overleftharpoon{Min}<\overleftharpoon{Max}\And Min>\overleftharpoon{Min}\And Dyn\Implies Max -  
Min<\overleftharpoon{Max} -  \overleftharpoon{Min},
\end{flist}

\noindent and

\begin{flist}
Inv\Implies Max -  Min\ge 0,
\end{flist}

\noindent it is enough to show that $Lrg$ satisfies:

\begin{spec}{Lrg}{}{}{\begin{ndcl}{
                           \lwew& Flag&:&\Bool,\\
                           &\lwer& L&:&\setof{\Nat},\\
                           &     & Min&:&\Nat,\\
                           &\lrew& S&:&\setof{\Nat},\\
                           &     & Max&:&\Nat\\
                                   }{
                           \lrew& TrmS&:&\Bool
                     }\end{ndcl}
            }{Min< min(L)\And Max\not\in L\And Min\in S\And Flag\ \And\\ 
						  &\neg TrmS\And Min<Max\And Inv
                                   }{( \overleftharpoon{Flag}\Implies 
                                      Flag \And Max=\overleftharpoon{Max}\And (S\cup\{Max\}=\overleftharpoon{S}\Or S=\overleftharpoon{S}))\ \And\\
                                     &(\neg\overleftharpoon{Flag}\And Flag\Implies Max=Min\Or 
                                       (Max\not\in L\And Min\in S))\ \And\\
                                     &(\neg \overleftharpoon{TrmS}\And TrmS\Leftrightarrow 
                                      \neg \overleftharpoon{Flag}\And Flag\And Max=Min)\ \And\\
                                     &Dyn\And (\overleftharpoon{Inv}\Implies Inv)
                                   }{\neg Flag\And Inv
                                   }{(\neg \overleftharpoon{Flag}\Implies \neg Flag
                                     \And Min=\overleftharpoon{Min}\And (L\cup\{Min\}=\overleftharpoon{L}\Or L=\overleftharpoon{L}))\ \And\\
                                     &(\overleftharpoon{Flag}\And \neg Flag\Implies 
                                      Max=Min\Or (Min\not\in S\And Max\in L))\ \And\\
                                     &Dyn\And (\overleftharpoon{Inv}\Implies Inv)
            }{\#L=\#\overleftharpoon{L}\And Min<min(L)\And Flag\And Min>\overleftharpoon{Min}\ \And\\
              &(Max=Min\Or Max\not\in L)\And (Max=Min\Leftrightarrow TrmS)\ \And\\
              &Dyn\And Inv.}
\end{spec}

\subsection{Final Implementation of \texorpdfstring{$Lrg$}{Lrg}}

\begin{prg}
\{Min<min(L)\And Max\not\in L\And Min\in S\And Min<Max\ \And\\
\qquad Flag\And \neg TrmS\And Dyn \And Inv\}\\
\\
L:=L\cup\{Max\};\\
\\
\{\#L=\#\overleftharpoon{L}+1\And Flag\And \neg TrmS\And Min=\overleftharpoon{Min}\And Min\in S\ \And\\
\qquad Min<min(L)\And Max\in L\And Dyn\And Inv\}\\
\\
Min:=min(L);\\
\\
\{Min=min(L)\And\#L=\#\overleftharpoon{L}+1\And Flag\And \neg TrmS\ \And\\
\qquad (Max=Min\Or Min\not\in S)\And Min>\overleftharpoon{Min}\And Dyn\And Inv\}\\
\\
Flag:=\false;\\
\\
\{Min=min(L)\And \#L=\#\overleftharpoon{L}+1\ \And\\ 
\qquad Min>\overleftharpoon{Min}\And (Flag\Implies (Max=Min\Leftrightarrow TermS)) \And\\
\qquad (Flag\Implies Max=Min\Or (Max\not\in L\And Min\in S))\And Dyn\And Inv\}\\
\\
L:=L - \{Min\};\\
\\
\{Min< min(L)\And\#L=\#\overleftharpoon{L}\ \And\\
\qquad Min>\overleftharpoon{Min}\And (Flag\Implies (Max=Min\Leftrightarrow TermS)) \And\\
\qquad (Flag\Implies Max=Min\Or (Max\not\in L\And Min\in S))\And Dyn\And Inv\}\\
\\
\nawait\ Flag\ \ndo\\
\qquad\nskip\\
\nod;\\
\\
\{Min< min(L)\And \#L=\#\overleftharpoon{L}\ \And\\
\qquad Min>\overleftharpoon{Min}\And Flag\And (Max=Min\Or (Max\not\in L\And Min\in S))\ \And\\
\qquad  (Max=Min\Leftrightarrow TrmS)\And Dyn\And Inv\}.\\
\end{prg}

%% file: safety.tex
\chapter{Modified System}

\section{Motivation}

\label{nonterminating:ref}
LSP can only be used to develop programs which are intended to converge. There is
no obvious way the system can be modified to handle more general liveness
(see \cite{ba:liveness})
constraints in
a similar style. However, if we restrict ourselves
to the following four characteristics:
\begin{itemize}
\item the overall effect --- if the implementation terminates,
\item the effect of any atomic step due to the implementation,
\item the set of states in which the implementation can become blocked,
\item that the body of an await-statement terminates whenever it is executed,
\end{itemize}
LSP can easily be adapted. LSP$_S$,
which is what the new system is called, depends upon LSP to prove termination of
await-bodies. The rest of the system is described below.

\section{Modifications}

\subsection{Specified Programs}

A specified program is of the form

\begin{flist}
z\satis[\vartheta,\alpha]::[P,R,W,G,E].
\end{flist}

\noindent Syntactically, the only difference from above is that square brackets are used 
instead of curly brackets. Both specifications and specified programs are 
required to satisfy the same constraints as earlier. 
However, their interpretations are different.

\subsection{Assumptions}

The pre-condition is still assumed to denote a set of initial states to
which the implementation is applicable, while the rely-condition as before is
supposed to characterise any uninterrupted state transition by the 
environment.

\noindent Thus, the pre- and rely-conditions constitute assumptions which the developer can make
about the environment. 

\subsection{Commitments}

Any state in which the implementation can become blocked is required to
satisfy the wait-condition. However, the implementation is not allowed to become
blocked inside the body of an await-statement. Moreover, any internal transition 
is constrained to satisfy the guar-condition, while the
overall effect, if the implementation terminates, must satisfy the
eff-condition. 

\subsection{Satisfaction}

This means that the definition of satisfaction (see page \pageref{satis:def})
can be carried over from earlier,
given that $ext_{\pi}$ and $int_{\pi}$ (see pages \pageref{ext:def} and
\pageref{int:def}) are redefined as below:

\begin{definition}
Given a pre-condition $P$, a rely-condition $R$, and a
structure $\pi$, then
$ext_{\pi}[P,R]$ denotes the set of all computations $\sigma$ in $\pi$,
such that:
\begin{itemize}

\item $\delta(\sigma_1)\models_{\pi} P$,

\item for all $1\le j< len(\sigma)$, if $\lambda(\sigma_j)=e$ then
$(\delta(\sigma_j),\delta(\sigma_{j+1}))\models_{\pi} R$.
\end{itemize}
\end{definition}

\begin{definition}
Given a wait-condition $W$, a guar-condition $G$, 
an eff-condition $E$, and a structure $\pi$, then
$int_{\pi}[W,G,E]$ denotes the set of all computations $\sigma$ in $\pi$,
such that:
\begin{itemize}
\item for all $1\le j\le len(\sigma)$, if $\sigma_j$ is blocked
then $\delta(\sigma_j)\models_{\pi} W$,
\item for all $1\le j<len(\sigma)$, if $\lambda(\sigma_j)=i$ then
$(\delta(\sigma_j),\delta(\sigma_{j+1}))\models_{\pi} G$ and $\tau(\sigma_j)\neq\tau(\sigma_{j+1})$, 
\item for all $1\le j\le len(\sigma)$, if $\tau(\sigma_j)=\epsilon$  then 
$(\delta(\sigma_1),\delta(\sigma_j))\models_{\pi}E$.

\end{itemize}
\end{definition}

\noindent The second conjunct of $int$'s second condition restricts the bodies 
of await-statements to terminate. (Remember that the only internal
transition which leaves the program component unchanged is the one which
models that the execution of an await-statement's body either ends in an infinite
loop or becomes blocked. See page \pageref{inter:ref}.)

\subsection{Decomposition-Rules}

With two exceptions the decomposition-rules are identical to the 
decomposition-rules for LSP (although
it is of course necessary to substitute square brackets for curly brackets).
The first exception is the while-rule. Since the statement is no longer required to
terminate, the first premise may be removed:

\begin{flist}
z\satis  [\vartheta,\alpha]:: [P\And b,R,W,G,P\And Z]\\
\overline{\nwhile\ b\ \ndo\  z \ \nod \satis  [\vartheta,\alpha] 
:: [P,R,W,G,(Z^{\dagger}\Or  R)\And \neg b]}
\end{flist}

\noindent Secondly, since it is necessary to use LSP to prove total correctness of the 
await-statement's body, the await-rule is of the form:

\begin{flist}
P^R\And \neg b\Rightarrow W\\
E_1|(\bigwedge_{a\in\alpha}a=\overleftharpoon{u_a}\And I_{\alpha})\Implies G\And E_2\\
 z  \satis  (\vartheta,\alpha):: (P^R\And b,I,\false,\true,E_1)\\
\overline{\nawait\ b\ \ndo\  z \ \nod \satis  [\vartheta,\alpha]:: [P,R,W,G,R|E_2|R]}\\
\\
{where\ for\ all\ }a\in\alpha, var[u_a]\subseteq \vartheta\cup\{a\}
\end{flist}

\subsection{Soundness and Completeness}

LSP$_S$ is sound and satisfies that same relative-completeness criterion
as LSP. The proofs are straightforward modifications of the proofs for LSP.

\section{Advantages}

\subsection{Always Enabled}

LSP$_S$ allows us to prove that a program is always enabled. If for example

\begin{flist}
\vdash_S z_1\satis[\vartheta,\alpha] :: [\true,\true,b,\true,\false], \\
\vdash_S z_2\satis[\vartheta,\alpha] :: 
[\true,\true,\neg b,\true,\false],
\end{flist}

\noindent we may deduce that

\begin{flist}
\vdash_S \{z_1\parallel z_2\}\satis[\vartheta,\alpha] :: [\true,\true,\false,
\true,\false],
\end{flist}

\noindent which means that $\{z_1\parallel z_2\}$ will never become blocked.

\subsection{Global Invariants}

When developing nonterminating programs it is often useful to state a global
invariant. As should be clear from the previous examples, LSP$_S$ is
well suited to deal with invariants. If for example

\begin{flist}
\vdash_S z\satis [\vartheta,\alpha]::[P,R,W,G,E],
\end{flist}

\noindent then to prove that the assertion $T$ is a global invariant, it is enough to show that

\begin{flist}
\vdash P\Implies T,\\
\vdash \overleftharpoon{T}\And (G\Or R)\Implies T.
\end{flist}

%% file: dekker.tex
\chapter{Dekker's Algorithm}

\section{Task}

We will now use LSP$_S$ to develop 
a program with respect to the four properties discussed above.
Consider the following problem:
\begin{itemize}
\item Two processes $P(0)$ and $P(1)$ are executing in an infinite loop. Both processes
consist of two sections; a critical section and a uncritical section. The executions
of the two critical sections are not allowed to overlap. Our job is to find an 
implementation.
\end{itemize}
\noindent Given that the sets of memory locations accessed by the critical sections are 
not disjoint,
then this can easily be achieved by placing the two critical sections inside the 
bodies of two await-statements.
Thus, to increase the challenge, we will add an extra constraint:
\begin{itemize}
\item Each process can only `hide' at most one memory location at the same time.
\end{itemize}

\noindent Exactly what the critical and uncritical sections are doing is not known. However,
we will assume that their access is restricted to a global data structure $Glb$ of
an unspecified sort $T$.

\section{Development}

\subsection{Data Structure}

We will employ Dekker's algorithm \cite{ewd:dekker}
to deal with this mutual exclusion problem.
Let

\begin{flist}
\type{D_{(0,1)}}{\set{0,1}.}
\end{flist}

\noindent To do the basic synchronisation we will employ an array $Ok$, defined with $D_{(0,1)}$ 
as domain and $\Bool$ as range, and a variable $Turn$ of sort $D_{(0,1)}$. These memory
locations will only be accessed by the processes $P(0)$ and $P(1)$. Thus we will
restrict our attention to a program of the form:

\begin{prg}
\nbegin\\
 \qquad \loc\ Ok,Turn;\\
 \qquad \{P(0)\parallel P(1)\}\\
\nend.\\
\end{prg}

\noindent To simplify the presentation, we will use arithmetics modulo 2;
$\ominus$ denotes subtraction modulo 2, while $\oplus$ stands for 
addition modulo 2.

\subsection{Invariant}

To prove mutual exclusion it is useful to introduce an auxiliary
array $Crit$ that maps $D_{(0,1)}$ to $\Bool$;
$Crit(l)$ is true whenever the process $P(l)$ is inside its critical section. Thus 

\begin{flist}
\neg (Crit(l)\And Crit(l\oplus 1))
\end{flist}

\noindent is an invariant.

Whenever the process $P(l)$ is ready to let
the other process $P(l\oplus 1)$ enter its critical section, 
then $Ok(l)$ is true. To secure mutual
exclusion we must therefore insist that

\begin{flist}
Crit(l)\Implies \neg Ok(l).
\end{flist}

\noindent We will use $Inv$ to denote the conjunction of these two assertions.

\subsection{Main Structure}

The process $P(l)$ can therefore be structured as below:

\begin{prg}
 \{Ok(l)\And \neg Crit(l)\And Inv\}\\
\\
 \nwhile\ \true\ \ndo\\
\\
       \qquad \{Ok(l)\And \neg Crit(l)\And Inv\}\\
\\
       \qquad GetAcc(l);\\
\\
       \qquad \{\neg Ok(l)\And Crit(l)\And \neg Crit(l\oplus 1)\And Inv\}\\
\\
       \qquad DoCrit(l);\\
\\
       \qquad \{\neg Ok(l)\And Crit(l)\And \neg Crit(l\oplus 1)\And Inv\}\\
\\
       \qquad RlsAcc(l);\\
\\
       \qquad \{Ok(l)\And \neg Crit(l)\And Inv\}\\
\\
       \qquad DoUnCrit(l)\\
\\
       \qquad \{Ok(l)\And \neg Crit(l)\And Inv\}\\
\\
 \nod\\
\\
\{\false\}.\\
\end{prg}

\noindent Obviously, $DoCrit(l)$ and $DoUnCrit(l)$ are the critical and uncritical sections.
Moreover, the object of $GetAcc(l)$ is to find the right moment to let the process 
$P(l)$ enter its critical section, while $RlsAcc(l)$ makes it possible for the 
process $P(l\oplus 1)$ to start on its critical section.

\subsection{Guar-Condition}

The critical and uncritical sections are assumed to leave the synchronisation structure
unchanged, thus the atomic state changes of $DoCrit(l)$ and $DoUnCrit(l)$ must imply:

\begin{flist}
 Ok=\overleftharpoon{Ok}\And Turn=\overleftharpoon{Turn}\And Crit=\overleftharpoon{Crit}.
\end{flist}

\noindent Furthermore, because the only task of $GetAcc(l)$ and $RlsAcc(l)$ is to secure mutual
exclusion, their internal transitions must leave $Glb$ unchanged.

Since the process $P(l)$ is only allowed to enter its critical section if
$Ok(l\oplus 1)$ is true, and because $P(l)$ cannot change the value of $Ok(l\oplus 1)$, 
it follows that any state transition due to $P(l)$ must satisfy:

\begin{flist}
\neg \overleftharpoon{Crit}(l)\And Crit(l)\Implies Ok(l\oplus 1).
\end{flist}

\noindent To avoid  one process infinitely overtaking the other, $Turn$ will be
used to provide fairness in cases where both processes want to enter
their respective critical sections. 

If  both processes want
to enter their critical sections and $Turn=l$, then $P(l)$ 
is the last process to have completed its critical section, while
$Turn=l\oplus 1$ implies that $P(l\oplus 1)$ is the most recent process to have
finished its critical section. Thus, it is enough if $P(l)$ updates $Turn$
once per iteration, namely immediately after it has finished its critical
section. Furthermore, it is clear that
the guar-condition of $P(l)$ must imply:

\begin{flist}
\overleftharpoon{Turn}=l\Implies Turn=l.
\end{flist}

\subsection{Wait-Condition}

If both processes want to enter their 
critical sections, in which case 

\begin{flist}
\neg Ok(l)\And \neg Ok(l\oplus 1)
\end{flist}

\noindent is true, we can use $Turn$ to hold back the process that last finished its 
critical section, namely by changing its $Ok$ location to true, and let it wait until
the other process changes the value of $Turn$. This means that the process $P(l)$
may have to wait in a state which satisfies:

\begin{flist}
Ok(l)\And Turn=l.
\end{flist}

\noindent Moreover, the process $P(l)$ may also have to be held back if it wants to enter its
critical section in a state that satisfies:

\begin{flist}
Turn=l\oplus 1\And \neg Ok(l)\And \neg Ok(l\oplus 1).
\end{flist}

\noindent In this case, the 
process $P(l\oplus 1)$ has not had enough time to switch on its $Ok$ flag. 

\subsection{Specification}

This leaves us with the following specification of the process $P(l)$, ($0\le l\le 1$):

\begin{spec}{P}{l:D_{(0,1)}}{}{\begin{ndcl}{
                              \lwew & Turn&:&D_{(0,1)},\\
                              &\lwew & Glb&:&T,\\
                              &\lwer & Ok(l)&:&\Bool,\\
                              &\lrew & Ok(l+1)&:&\Bool\\
                               }{\lwer & Crit(l)&:&\Bool,\\
                                 &\lrew & Crit(l+1)&:&\Bool
                      }\end{ndcl}
                      }{\neg Crit(l)\And Ok(l)\And Inv
                      }{(\overleftharpoon{Turn}=l\oplus 1\Rightarrow Turn=l\oplus 1)\ \And\\
                      & (\neg \overleftharpoon{Crit}(l\oplus 1)\And Crit(l\oplus 1)\Implies Ok(l))\ 
                                                                         \And\\
                      & (\overleftharpoon{Inv}\Implies Inv)
                      }{(Ok(l)\And Turn=l)\ \Or\\ 
                      & (\neg Ok(l)\And Turn=l\oplus 1
                         \And \neg Ok(l\oplus 1))
                      }{(\overleftharpoon{Turn}=l\Rightarrow Turn=l)\ \And\\
                      & (\neg \overleftharpoon{Crit}(l)\And Crit(l)\Implies Ok(l\oplus 1))\ \And\\
                      & (\overleftharpoon{Inv}\Implies Inv)
                      }{\false.}
\end{spec}

\subsection{Composition Proof}

The obvious question at this stage is: May the processes $P(0)$ and $P(1)$ be
composed in parallel? The answer is --- Yes! Since

\begin{flist}
\vdash\neg (((Ok(l)\And Turn=l)\ \Or\\ 
\qquad\qquad (\neg Ok(l)\And Turn=l\oplus 1\And \neg Ok(l\oplus 1)))\ \And\\
\qquad ((Ok(l\oplus 1)\And Turn=l\oplus 1)\ \Or\\ 
\qquad\qquad (\neg Ok(l\oplus 1)\And Turn=l\And \neg Ok(l)))),
\end{flist}

\noindent it follows by the consequence- and parallel-rules
that $\{P(0)\parallel P(1)\}$ satisfies:

\begin{tuple}{\{Turn,Ok,Glb\}
            }{\{Crit\}
            }{\neg Crit(l)\And \neg Crit(l\oplus 1)\And Ok(l)\And Ok(l\oplus 1)\And Inv
            }{I
            }{\false
            }{\true
            }{\false.}
\end{tuple}

\subsection{Get Access}

The next step is to implement $GetAcc(l)$. From the earlier discussion
it follows that an implementation of $GetAcc(l)$ must
satisfy:

\begin{tuple}{\{Turn,Ok,Glb\}
            }{\{Crit\}
            }{Ok(l)\And Inv
            }{(\overleftharpoon{Turn}=l\oplus 1\Rightarrow Turn=l\oplus 1)\ \And\\
              & (\neg \overleftharpoon{Crit}(l\oplus 1)\And Crit(l\oplus 1)\Implies Ok(l))\ 
                                                                         \And\\
              & (\overleftharpoon{Inv}\Implies Inv)
            }{(Ok(l)\And Turn=l)\ \Or\\ 
              & (\neg Ok(l)\And Turn=l\oplus 1\And 
                 \neg Ok(l\oplus 1))
            }{(\overleftharpoon{Turn}=l\Rightarrow Turn=l)\ \And\\
              & (\neg \overleftharpoon{Crit}(l)\And Crit(l)\Implies Ok(l\oplus 1))\ \And\\
              & Glb=\overleftharpoon{Glb}\And (\overleftharpoon{Inv}\Implies Inv)
            }{Crit(l)\And Inv.}
\end{tuple}

\noindent Thus, the following is a correct implementation:

\begin{prg}
\nbegin\\
       \qquad \loc\ V_l;\\
\\
       \qquad \{Ok(l)\And \neg Crit(l)\And Inv\}\\
\\
       \qquad Ok(l):=\false;\\
       \qquad V_l:=Ok(l\oplus 1);\\
\\
       \qquad \{\neg Ok(l)\And (Crit(l)\Leftrightarrow V_l)\And Inv\}\\
\\
       \qquad \nif\ \neg V_l\ \nthen\\
\\
                   \qquad\qquad\{\neg Ok(l)\And \neg Crit(l)\And Inv\}\\
\\
                   \qquad\qquad \nif\ Turn=l\ \nthen\\
                                \qquad\qquad\qquad Ok(l):=\true;\\
\\
                                \qquad\qquad\qquad\{Ok(l)\And \neg Crit(l)\And Inv\}\\
\\
                                \qquad\qquad\qquad \nawait\ Turn=l\oplus 1\ \ndo\ \nskip\ \nod;\\
\\  
                              \qquad\qquad\qquad \{Ok(l)\And \neg Crit(l)\And 
                                                          Turn=l\oplus 1\And Inv\}\\
\\
                                \qquad\qquad\qquad Ok(l):=\false;\\
                   \qquad\qquad \nfi;\\
\\
                   \qquad\qquad \{\neg Ok(l)\And \neg Crit(l)\And Turn=l\oplus 1
                                  \And Inv\}\\
\\
                   \qquad\qquad \nawait\ Ok(l\oplus 1)\ \ndo\ \nskip\ \nod\\
\\
                   \qquad\qquad \{\neg Ok(l)\And Crit(l)\And \neg Crit(l\oplus 1)
                                  \And Turn=l\oplus 1\And Inv\}\\
\\
                   \qquad  \nfi\\
\\
                   \qquad \{\neg Ok(l)\And Crit(l)\And \neg Crit(l\oplus 1)\And Inv\}\\
\nend.\\
\end{prg}

\subsection{Release Access}

What remains is to implement $RlsAcc$. In other words, to find a program that
satisfies:

\begin{tuple}{\{Turn,Ok,Glb\}
            }{\{Crit\}
            }{Crit(l)\And Inv
            }{(\overleftharpoon{Turn}=l\oplus 1\Rightarrow Turn=l\oplus 1)\ \And\\
              & (\neg \overleftharpoon{Crit}(l\oplus 1)\And Crit(l\oplus 1)\Implies Ok(l))\ 
                                                                         \And\\
              & (\overleftharpoon{Inv}\Implies Inv)
            }{(Ok(l)\And Turn=l)\ \Or\\ 
              & (\neg Ok(l)\And Turn=l\oplus 1\And \neg Ok(l\oplus 1))
            }{(\overleftharpoon{Turn}=l\Rightarrow Turn=l)\ \And\\
              & (\neg \overleftharpoon{Crit}(l)\And Crit(l)\Implies Ok(l\oplus 1))\ \And\\
              & Glb=\overleftharpoon{Glb}\And (\overleftharpoon{Inv}\Implies Inv)
            }{Ok(l)\And \neg Crit(l)\And Inv.}
\end{tuple}

\noindent This is not very difficult:

\begin{prg}
\{\neg Ok(l)\And Crit(l)\And \neg Crit(l\oplus 1)\And Inv\}\\
\\
 Ok(l):=\true\\
\\
\{Ok(l)\And \neg Crit(l)\And Inv\}.
\end{prg}

%% file: newsound.tex
\chapter{Soundness}

\section{Parallel-Decomposition Proposition}

The object of this chapter is prove that LSP is sound. This will be shown by
induction on the depth (see page \ref{depth:ref}) of a LSP proof. 
The following proposition,
which characterises a decomposition condition, will be useful:

\begin{statement}\label{decomp2:theo}
Given that
\begin{eqnarray}
&&\models_{\pi} \neg(W_1\And E_2)\And 
                \neg(W_2\And E_1)\And \neg(W_1\And W_2),
\label{ftheo:two}\\
&&ext_{\pi}[P,R_1]\cap cp_{\pi}[z_1]\subseteq 
int_{\pi}[W\Or W_1,G\And R_2,E_1],\label{ftheo:three}\\
&&ext_{\pi}[P,R_2]\cap cp_{\pi}[z_2]\subseteq int_{\pi}[W\Or W_2,G\And R_1,E_2],
\label{ftheo:four}\\
&&\{ z_1\parallel z_2\}\satis(\vartheta,\emptyset)
 :: (P,R_1\And R_2,W,G,E_1\And E_2)\in SP,\nonumber
\end{eqnarray}
then 
\begin{eqnarray}
&&ext_{\pi}[P,R_1\And R_2]\cap cp_{\pi}[\{z_1\parallel z_2\}]\subseteq 
int_{\pi}[W,G,E_1\And E_2].\label{ftheo:six}
\end{eqnarray}
\end{statement}

\begin{nproof}
Let
\begin{eqnarray}
&&\sigma\in ext_{\pi}[P,R_1\And R_2]\cap 
cp_{\pi}[\{ z_1\parallel  z_2\}]
,\label{ftheo:five}
\end{eqnarray}
it follows from proposition \ref{decomp1:theo} on page \pageref{decomp1:theo}
that there are two computations
$\sigma'\in cp_{\pi}[ z_1]$ and 
$\sigma''\in cp_{\pi}[ z_2]$, 
such that $\sigma'\bullet\sigma''\leftarrow \sigma$. We will first show that
$\sigma'\in ext_{\pi}[P,R_1]$
and $\sigma''\in ext_{\pi}[P,R_2]$; in other words, 
that both computations satisfy their respective instances of the three
conditions in definition \ref{ext:def} on page \pageref{ext:def}.

Since \eqref{ftheo:five} implies that $\delta(\sigma_1)\models_{\pi} P$, and
since by definition $\delta(\sigma_1)=\delta(\sigma'_1)=\delta(\sigma''_1)$, 
it follows that $\delta(\sigma'_1)\models_{\pi} P$ 
and $\delta(\sigma''_1)\models_{\pi} P$. Hence, they both satisfy the first condition.

To prove that they also
satisfy the second condition, let:

\begin{itemize} 

\item $max(\sigma')=len(\sigma')$, if for all $1\le k\le len(\sigma')$, there is a 
      $\sigma'''\in ext_{\pi}[P,R_1]\cap cp_{\pi}[z_1]$, which satisfies
      \[\sigma'(1,\ldots,k)=\sigma'''(1,\ldots,k).\]
      Otherwise, $max(\sigma')=m$, where $m$ is the maximum
      natural number such that there is a
      $\sigma'''\in ext_{\pi}[P,R_1]\cap cp_{\pi}[z_1]$, which satisfies
      \[\sigma'(1,\ldots,m)=\sigma'''(1,\ldots,m).\]

\item $max(\sigma'')=len(\sigma'')$, if for all $1\le k\le len(\sigma'')$, there is a 
      $\sigma'''\in ext_{\pi}[P,R_2]\cap cp_{\pi}[z_2]$, which satisfies
      \[\sigma''(1,\ldots,k)=\sigma'''(1,\ldots,k).\]
      Otherwise, $max(\sigma'')=m$, where $m$ is the maximum
      natural number such that there is a
      $\sigma'''\in ext_{\pi}[P,R_2]\cap cp_{\pi}[z_2]$, which satisfies
      \[\sigma''(1,\ldots,m)=\sigma'''(1,\ldots,m).\]
\end{itemize}

\noindent If $max(\sigma')=len(\sigma')$, then it is obvious that $\sigma'$ satisfies
the second condition. The same is of course true for
$\sigma''$ if $max(\sigma'')=len(\sigma'')$.

Assume that  $max(\sigma')\neq len(\sigma')$ or $max(\sigma'')\neq  len(\sigma'')$.
We will show that this leads to a contradiction. There are two cases:

\begin{itemize}

\item $m=max(\sigma')=max(\sigma'')$: Since the constraints imposed on the environment
have no effect on the number of possible internal transitions in a given 
configuration, and since
and $\sigma'\bullet\sigma''\leftarrow\sigma$, it is clear that
\[\lambda(\sigma'_m)=\lambda(\sigma''_m)=\lambda(\sigma_m)=e.\]
But then, $(\delta(\sigma_m),\delta(\sigma_{m+1}))\models_{\pi}\neg(R_1\And R_2)$ 
which  contradicts \eqref{ftheo:five}.

\item $max(\sigma')\neq max(\sigma'')$: Without loss of generality, it may be assumed
that 
\[m=max(\sigma')<max(\sigma'').\] 
Moreover, since the constraints imposed on the environment 
have no effect on the number of possible internal transitions in a given 
configuration, there are
two possibilities:
\begin{itemize}
\item $\lambda(\sigma'_m)=\lambda(\sigma''_m)=e$: This leads to a contradiction by
an argument similar to the one above.
\item $\lambda(\sigma'_m)=e$ and $\lambda(\sigma''_m)=i$: Then there is a 
$\sigma'''\in ext_{\pi}[P,R_2]\cap cp_{\pi}[z_2]$ 
such that $\sigma''(1,\ldots,m+1)=\sigma'''(1,\ldots,m+1)$. From \eqref{ftheo:four} 
it follows that $\sigma'''\in int_{\pi}[W\Or W_2,G\And R_1,E_2]$, which
implies that 
\[(\delta(\sigma''_m),\delta(\sigma''_{m+1}))\models_{\pi} G\And R_1.\] 
Then, since $\sigma'\bullet\sigma''\leftarrow\sigma$, it is also true that 
\[(\delta(\sigma'_{m}),\delta(\sigma'_{m+1}))\models_{\pi}
G\And R_1,\] which again implies that  
\[(\delta(\sigma'_{m}),\delta(\sigma'_{m+1}))
\models_{\pi} R_1.\] 
But, this
contradicts that $max(\sigma')=m$.

\end{itemize}

\end{itemize}

\noindent Thus, both $\sigma'$ and $\sigma''$ satisfy the second condition in definition
\ref{ext:def} on page \pageref{ext:def}. 
To see that they also satisfy the final constraint,
assume that $\sigma$ diverges. This means that for any $j\ge 1$, 
there is a $k\ge j$, such that 
$\lambda(\sigma_k)=i$. But then $\sigma'\bullet\sigma''\leftarrow
\sigma$ implies that for any $j\ge 1$, there is a $k\ge j$, 
such that $\lambda(\sigma_k')=i$ or
$\lambda(\sigma_k'')=i$. Thus, 
$\sigma'\in ext_{\pi}[P,R_1]$ or 
$\sigma''\in ext_{\pi}[P,R_2]$, which contradicts \eqref{ftheo:three}
or \eqref{ftheo:four}. This means that both $\sigma'$ and $\sigma''$ are finite, 
and it is clear
that
\begin{eqnarray}
&&\sigma'\in ext_{\pi}[P,R_1]\cap cp_{\pi}[z_1],\nonumber\\
&&\sigma''\in ext_{\pi}[P,R_2]\cap cp_{\pi}[z_2].\nonumber
\end{eqnarray}
But then
\begin{eqnarray}
&&\sigma'\in int_{\pi}[W\Or W_1,G\Or R_2,E_1],\nonumber\\
&&\sigma''\in int_{\pi}[W\Or W_2,G\Or R_1,E_2]\nonumber
\end{eqnarray}
follows from \eqref{ftheo:three} and \eqref{ftheo:four}.
Hence, \eqref{ftheo:two} implies that
\begin{eqnarray}
&&\sigma\in int_{\pi}[W,G,E_1\And E_2].\nonumber
\end{eqnarray}
This proves \eqref{ftheo:six}.

\end{nproof}

\section{Soundness Proposition}

\begin{statement}
For any structure $\pi$ and specified program $\psi$, if
\begin{eqnarray}
&&Tr_{\pi}\vdash \psi\nonumber
\end{eqnarray}
then
\begin{eqnarray}
&&\models_{\pi} \psi.\nonumber
\end{eqnarray}
\end{statement}

\begin{nproof}
We will show the proposition by induction on the proof's depth (see page 
\pageref{depth:ref}).
If the proof's depth is one, there are only two possibilities:

\begin{itemize}

\item Skip-Rule:

Assume that we have a proof of depth 1 one whose root
\begin{eqnarray}
&& Tr_{\pi}\vdash  \nskip\satis(\vartheta,\alpha) :: (P,R,W,G,R)\nonumber
\end{eqnarray}
is deduced by the skip-rule. Since 
\begin{eqnarray}
&&\models \nskip\progtran \nskip,\nonumber
\end{eqnarray}
it is enough to show that
\begin{eqnarray}
&&ext_{\pi}[P,R]\cap cp_{\pi}[\nskip]\subseteq int_{\pi}[W,G,R].\label{skip:ref}
\end{eqnarray}
Let
\begin{eqnarray}
&&\sigma\in ext_{\pi}[P,R]\cap cp_{\pi}[ \nskip],\nonumber
\end{eqnarray}
then $\sigma$ is of the form
\begin{eqnarray}
&&< \nskip,s_1>\exter\ \ldots\ \exter< \nskip,s_j>\intern
      <\epsilon,s_{j}>\exter\ \ldots\ \exter<\epsilon,s_n>,\nonumber
\end{eqnarray}
in which case the reflexivity and 
transitivity of $R$, and the reflexivity of $G$ imply that
$\sigma\in int_{\pi}[W,G,R]$. This proves \eqref{skip:ref}.

\item Assignment-Rule:

Assume $ z$ is of the form 
\begin{eqnarray}
&& v:=r,\nonumber
\end{eqnarray}
and that we have a proof of depth 1 whose root
\begin{eqnarray}
&& Tr_{\pi}\vdash  z\satis(\vartheta,\alpha) ::(P,R,W,G,R|E|R)
\nonumber
\end{eqnarray}
is deduced by the assignment-rule. This means that there are expressions
$u_1,\ldots,u_{card(\alpha)}$ such that
\begin{eqnarray}
&& \models_{\pi} \overleftharpoon{P^R}\And v=\overleftharpoon{r}\And I_{\{v\}\cup\alpha}\And 
\bigwedge_{j=1}^{card(\alpha)} a_j=\overleftharpoon{u_j}\Implies G\And E, \label{as:one}\\
&&for\ all\ 1\le j,k\le card(\alpha),\nonumber \\
&&\qquad var[u_j]\in\vartheta\cup\{a_j\},\nonumber \\
&&\qquad a_j\in\alpha, \ and\ j\neq k\ implies\
a_j\neq a_k.\label{as:oneb}
\end{eqnarray}

Let $ z'$ denote the program
\begin{eqnarray}
&&\nawait\ \true\ \ndo\nonumber\\
&&\qquad a_1:=u_1;\nonumber\\
&&\qquad\quad\vdots\nonumber\\
&&\qquad a_{card(\alpha)}:=u_{card(\alpha)};\nonumber\\
&&\qquad v:=r\nonumber\\
&&\nod,\nonumber
\end{eqnarray}
then it follows from \eqref{as:oneb} that
\begin{eqnarray}
&& \models z'\progtran z,\nonumber
\end{eqnarray}
and it is enough to show that
\begin{eqnarray}
&& ext_{\pi}[P,R]\cap cp_{\pi}[z']\subseteq int_{\pi}[W,G,R|E|R].
\label{as:two}
\end{eqnarray}
Let
\begin{eqnarray}
&&\sigma\in ext_{\pi}[P,R]\cap cp_{\pi}[ z'],\nonumber
\end{eqnarray}
then $\sigma$ is of the form
\begin{eqnarray}
&&< z',s_1>\exter\ \ldots\ \exter< z',s_{j -  1}>\intern<\epsilon,s_{j}>
\exter\ \ldots\ \exter<\epsilon,s_{j+m}>,\nonumber
\end{eqnarray}
where the internal transition is characterised by 
\begin{eqnarray}
&&\overleftharpoon{P^R}\And v=\overleftharpoon{r}\And I_{\{v\}\cup\alpha}
\And \bigwedge_{j=1}^{card(\alpha)} a_j=\overleftharpoon{u_j}. \nonumber
\end{eqnarray}
Hence, \eqref{as:one} and the reflexivity and transitivity
of $R$ imply that 
\begin{eqnarray}
&&\sigma\in
int_{\pi}[W,G,R|E|R],\nonumber
\end{eqnarray}
which again implies \eqref{as:two}.

\end{itemize}

\noindent This means that the proposition is true for proofs of depth one. 
Assume the proposition is true for proofs of depth less than or equal to $n -  1$.
We will show that the proposition is true for proofs of depth $n$. 
There are fourteen cases:

\begin{itemize}

\item Consequence-Rule:

Assume we have a proof of depth $n$, whose root
\begin{eqnarray}
&&Tr_{\pi}\vdash z\satis(\vartheta,\alpha) ::  
(P_2,R_2,W_2,G_2,E_2)\nonumber
\end{eqnarray}
is deduced from
\begin{eqnarray}
&&Tr_{\pi}\vdash  z\satis(\vartheta,\alpha) ::  
(P_1,R_1,W_1,G_1,E_1),\nonumber
\end{eqnarray}
by the consequence-rule. This means that
\begin{eqnarray}
&&\models_{\pi} P_2\Implies P_1,\label{co:one}\\
&&\models_{\pi} R_2\Implies R_1,\label{co:two}\\
&&\models_{\pi} W_1\Implies W_2,\label{co:three}\\
&&\models_{\pi} G_1\Implies G_2,\label{co:four}\\
&&\models_{\pi} E_1\Implies E_2.\label{co:five}
\end{eqnarray}
Moreover, the induction hypothesis implies that
\begin{eqnarray}
&&\models_{\pi}  z\satis(\vartheta,\alpha) ::  
(P_1,R_1,W_1,G_1,E_1),\nonumber
\end{eqnarray}
which means that there is a program
$ z'$ such that 
\begin{eqnarray}
&&\models z'\progtran z,\nonumber\\
&&ext_{\pi}[P_1,R_1]\cap cp_{\pi}[z']\subseteq int_{\pi}[W_1,G_1,E_1].\label{co:six}
\end{eqnarray}
Thus, 
it is enough to show that
\begin{eqnarray}
&&ext_{\pi}[P_2,R_2]\cap cp_{\pi}[z']\subseteq int_{\pi}[W_2,G_2,E_2],\nonumber
\end{eqnarray}
which follows from \eqref{co:six}, since
\eqref{co:one} and \eqref{co:two} imply that
\begin{eqnarray}
&&ext_{\pi}[P_2,R_2]\subseteq ext_{\pi}[P_1,R_1],\nonumber
\end{eqnarray}
and \eqref{co:three}, \eqref{co:four} and \eqref{co:five} imply that
\begin{eqnarray}
&&int_{\pi}[W_1,G_1,E_1]\subseteq 
int_{\pi}[W_2,G_2,E_2].\nonumber
\end{eqnarray}

\item Pre-Rule: 

Assume we have a proof of depth $n$, whose root
\begin{eqnarray}
&&Tr_{\pi}\vdash z\satis(\vartheta,\alpha) ::  (P,R,W,G,
\overleftharpoon{P}\And E)\nonumber
\end{eqnarray}
is deduced from
\begin{eqnarray}
&&Tr_{\pi}\vdash z\satis(\vartheta,\alpha) ::  (P,R,W,G,E)\nonumber,
\end{eqnarray}
by the pre-rule. Then the induction hypothesis implies that
\begin{eqnarray}
&&\models_{\pi}  z\satis(\vartheta,\alpha) ::  (P,R,W,G,E)\nonumber,
\end{eqnarray}
which means that there is a program
$ z'$ such that 
\begin{eqnarray}
&&\models  z'\progtran z,\nonumber\\
&&ext_{\pi}[P,R]\cap cp_{\pi}[z']\subseteq int_{\pi}[W,G,E].
\nonumber
\end{eqnarray}
Thus, it is enough to show that
\begin{eqnarray}
&&ext_{\pi}[P,R]\cap cp_{\pi}[z']\subseteq int_{\pi}[W,G,\overleftharpoon{P}\And E]\label{pre:sev}.
\end{eqnarray}

Let
\begin{eqnarray}
&&\sigma\in ext_{\pi}[P,R]\cap cp_{\pi}[ z'].\nonumber
\end{eqnarray}

Since
\begin{eqnarray}
&&int_{\pi}[W,G,\overleftharpoon{P}\And E]=
\{\sigma\in int_{\pi}[W,G,E]|\delta(\sigma_1)\models_{\pi} P\},
\nonumber
\end{eqnarray}
and since by definition $\delta(\sigma_1)\models_{\pi} P$, it follows that
\begin{eqnarray}
&& \sigma\in int_{\pi}[W,G,\overleftharpoon{P}\And E].\nonumber
\end{eqnarray}
In other words, \eqref{pre:sev} has been shown.

\item Access-Rule:

Assume we have a proof of depth $n$, whose root 
\begin{eqnarray}
&& Tr_{\pi}\vdash z\satis(\vartheta,\alpha) ::(P,R,W,G,E)\nonumber
\end{eqnarray}
is deduced from
\begin{eqnarray}
&& Tr_{\pi}\vdash z\satis(\vartheta,\alpha) ::(P,R\And v=\overleftharpoon{v},W,G,E)\nonumber
\end{eqnarray}
by the access-rule. This means that
\begin{eqnarray}
&&v\in hid[z]\cap\vartheta.
\nonumber
\end{eqnarray}
Moreover, the induction hypothesis implies that
\begin{eqnarray}
&& \models_{\pi} z\satis(\vartheta,\alpha) :: (P,R\And v=\overleftharpoon{v},W,G,E)\nonumber
\end{eqnarray}
This means that there is a program $z'$ such that
\begin{eqnarray}
&&\models z'\progtran z,\nonumber\\
&& ext_{\pi}[P,R\And v=\overleftharpoon{v}]\cap cp_{\pi}[z']\subseteq int_{\pi}[W,G,E]
\label{ac:four}.
\end{eqnarray}
Thus, it is enough to show that
\begin{eqnarray}
&&ext_{\pi}[P,R]\cap cp_{\pi}[z']\subseteq int_{\pi}[W,G,E]\label{ac:five}.
\end{eqnarray}
Let 
\begin{eqnarray}
&&\sigma\in ext_{\pi}[P,R]\cap cp_{\pi}[z'].\nonumber
\end{eqnarray}
Since $v\in hid[z']$ and the environment is required to respect $hid[z']$, 
it is clear that
\begin{eqnarray}
&&\sigma\in ext_{\pi}[P,R\And v=\overleftharpoon{v}]\cap cp_{\pi}[z'],\nonumber
\end{eqnarray}
in which case \eqref{ac:four} implies that
\begin{eqnarray}
&&\sigma\in int_{\pi}[W,G,E],\nonumber
\end{eqnarray}
which proves \eqref{ac:five}.

\item Block-Rule:

Given that $ z_1$ is of the form 
\begin{eqnarray}
&&\nbegin\ \loc\ v_1,\ \ldots\ ,v_m; z_2\ \nend,\nonumber
\end{eqnarray}
and assume we have a proof of depth $n$, whose root
\begin{eqnarray}
&&Tr_{\pi}\vdash z_1\satis(\vartheta\setminus\bigcup_{j=1}^m\{v_j\},\alpha) ::  (P,R,W,G,E)\nonumber
\end{eqnarray}
is deduced from
\begin{eqnarray}
&&Tr_{\pi}\vdash z_2\satis(\vartheta,\alpha) ::  (P,R\And\bigwedge_{j=1}^m v_j=\overleftharpoon{v_j},W,G,E),\nonumber
\end{eqnarray}
by the block-rule. Then the induction hypothesis implies that
\begin{eqnarray}
&&\models_{\pi} z_2\satis(\vartheta,\alpha) ::  (P,R\And\bigwedge_{j=1}^m v_j=\overleftharpoon{v_j},W,G,E),\nonumber
\end{eqnarray}
which means that there is a program
$ z_2'$ such that
\begin{eqnarray}
&&\models z_2'\progtran z_2,\label{bl:two}\\
&&ext_{\pi}[P,R\And\bigwedge_{j=1}^m v_j=\overleftharpoon{v_j}]\cap cp_{\pi}[z_2']\subseteq 
int_{\pi}[W,G,E].
\label{bl:four}
\end{eqnarray}
Moreover, if $ z_1'$ denotes the program
\begin{eqnarray}
&&\nbegin\ \loc\ v_1,\ \ldots\ ,v_m; z_2'\ \nend\nonumber
\end{eqnarray}
it follows from \eqref{bl:two} that
\begin{eqnarray}
&&\models z_1'\progtran z_1.\nonumber
\end{eqnarray}
Thus,   
it is enough to show that
\begin{eqnarray}
&&ext_{\pi}[P,R]\cap cp_{\pi}[z_1']\subseteq int_{\pi}[W,G,E].\label{bl:five}
\end{eqnarray}
Observe that the constraints on a specified program imply that $v_1,\ldots,v_m$
cannot occur in $P,R,W,G$ or $E$. Let
\begin{eqnarray}
&&\sigma\in ext_{\pi}[P,R]\cap cp_{\pi}[ z_1'].\nonumber
\end{eqnarray}
Since any external transition in $\sigma$ must respect
$\{v_1,\ \ldots\ ,v_n\}$, it follows that $\sigma$ is of
the form
\begin{eqnarray}
&&< z_1',s_1>\exter\ \ldots\ \exter< z_1',s_j>\intern\sigma',\nonumber
\end{eqnarray}
where 
\begin{eqnarray}
&&\sigma'\in ext_{\pi}[P^R,R\And\bigwedge_{j=1}^m v_j=\overleftharpoon{v_j}]\cap cp_{\pi}[ z_2'].
\nonumber
\end{eqnarray}
Moreover, since  $\delta(\sigma'_1)=s_j$, it is also clear that
\begin{eqnarray}
&&< z_2',s_1>\exter\ \ldots\ \exter< z_2',s_j>\exter\sigma'\nonumber
\end{eqnarray}
is an element of  $ext_{\pi}[P,R\And \bigwedge_{j=1}^m v_j=\overleftharpoon{v_j}]\cap cp_{\pi}[ z_2']$.
Thus, it follows from \eqref{bl:four} that $\sigma\in int_{\pi}[W,G,E]$. 
This proves
\eqref{bl:five}.

\item Sequential-Rule: 

Given that $ z_1$ is of the form 
\begin{eqnarray}
&& z_3; z_2, \nonumber
\end{eqnarray}
and assume we have a proof of depth $n$, whose root
\begin{eqnarray}
&&Tr_{\pi}\vdash z_1\satis(\vartheta,\alpha) ::  (P_1,R,W,G,E_1|E_2)\nonumber
\end{eqnarray}
is deduced from
\begin{eqnarray}
&&Tr_{\pi}\vdash z_3\satis(\vartheta,\alpha) ::  (P_1,R,W,G,P_2\And E_1),
\nonumber\\
&&Tr_{\pi}\vdash z_2\satis(\vartheta,\alpha) ::  (P_2,R,W,G,E_2),\nonumber
\end{eqnarray}
by the sequential-rule. Then the induction hypothesis implies that
\begin{eqnarray}
&&\models_{\pi} z_3\satis(\vartheta,\alpha) ::  (P_1,R,W,G,P_2\And E_1), 
\nonumber\\
&&\models_{\pi} z_2\satis(\vartheta,\alpha) ::  (P_2,R,W,G,E_2),\nonumber
\end{eqnarray}
which means that there are two programs
$ z_3'$ and $ z_2'$ such that 
\begin{eqnarray}
&&\models z_3'\progtran z_3,\label{sc:two}\\
&&\models z_2'\progtran z_2,\label{sc:three}\\
&&ext_{\pi}[P_1,R]\cap cp_{\pi}[z_3']\subseteq int_{\pi}[W,G,P_2\And E_1], 
\label{sc:four}\\
&&ext_{\pi}[P_2,R]\cap cp_{\pi}[z_2']\subseteq int_{\pi}[W,G,E_2].\label{sc:five}
\end{eqnarray}
Moreover, if $ z_1'$ denotes the program
\begin{eqnarray}
&& z_3'; z_2',\nonumber
\end{eqnarray}
it follows from \eqref{sc:two} and \eqref{sc:three} 
that
\begin{eqnarray}
&&\models z_1'\progtran z_1.\nonumber
\end{eqnarray}
Thus, it  
it is enough to show that
\begin{eqnarray}
&&ext_{\pi}[P_1,R]\cap cp_{\pi}[z_1']\subseteq int_{\pi}[W,G,E_1|E_2]\label{sc:sev}.
\end{eqnarray}

Let
\begin{eqnarray}
&&\sigma\in ext_{\pi}[P_1,R]\cap cp_{\pi}[ z_1'].\nonumber
\end{eqnarray}
It is clear from \eqref{sc:four} and \eqref{sc:five} 
that $\sigma$ converges.
If $\sigma$ deadlocks before $z_3'$ has terminated, it follows from \eqref{sc:four}
that $\sigma\in int_{\pi}[W,G,P_2\And E_1]$, in which case it is also true that
$\sigma\in int_{\pi}[W,G,E_1|E_2]$.

On the other hand, if $\sigma$ terminates or deadlocks after $z_3'$ has terminated,
it follows that $\sigma$ is of the form
\begin{eqnarray}
&&< z_3'; z_2',s_1>\stackrel{a_1}{\rightarrow}\quad\ldots\quad
\stackrel{a_{k -  2}}{\rightarrow}< z_{k -  1}'; z_2',s_{k -  1}>
\stackrel{a_{k -  1}}{\rightarrow}< z_2',s_k>\
\stackrel{a_k}{\rightarrow}\ \sigma'\nonumber
\end{eqnarray}
where 
\begin{eqnarray}
&&< z_2,s_k>\stackrel{a_k}{
\rightarrow}\sigma'\in ext_{\pi}[P_2,R]\cap
cp_{\pi}[ z_2'],\nonumber
\end{eqnarray}
and there is a $\sigma''\in ext_{\pi}[P_1,R]\cap
cp_{\pi}[ z_3']$ of the form
\begin{eqnarray}
&&< z_3',s_1>\stackrel{a_1}{\rightarrow}\quad\ldots\quad
\stackrel{a_{k -  2}}{\rightarrow}< z_{k -  1}',s_{k -  1}>
\stackrel{a_{k -  1}}{\rightarrow}<\epsilon,s_k>.\nonumber
\end{eqnarray}
Thus, \eqref{sc:four} and
\eqref{sc:five} imply that
$\sigma\in int_{\pi}[W,G,E_1|E_2]$, which proves \eqref{sc:sev}.

\item If-Rule:

Given that $ z_1$ is of the form 
\begin{eqnarray}
&&\nif\ b\ \nthen\  z_2\ \nelse\  z_3\ \nfi, \nonumber
\end{eqnarray}
and assume we have a proof of depth $n$, whose root
\begin{eqnarray}
&&Tr_{\pi}\vdash z_1\satis(\vartheta,\alpha) ::  (P,R,W,G,E)\nonumber
\end{eqnarray}
is deduced from
\begin{eqnarray}
&&Tr_{\pi}\vdash z_2\satis(\vartheta,\alpha) ::  (P\and b,R,W,G,E),\nonumber\\
&&Tr_{\pi}\vdash z_3\satis(\vartheta,\alpha) ::  (P\and \neg b,R,W,G,E),\nonumber
\end{eqnarray}
by the if-rule. Then the induction hypothesis implies that
\begin{eqnarray}
&&\models_{\pi} z_2\satis(\vartheta,\alpha) ::  (P\and b,R,W,G,E), \nonumber\\
&&\models_{\pi} z_3\satis(\vartheta,\alpha) ::  (P\and \neg b,R,W,G,E),\nonumber
\end{eqnarray}
which means that there are two programs
$ z_2'$ and $ z_3'$ such that 
\begin{eqnarray}
&&\models z_2'\progtran z_2,\label{if:two}\\
&&\models z_3'\progtran z_3,\label{if:three}\\
&&ext_{\pi}[P\and b,R]\cap cp_{\pi}[z_2]\subseteq int_{\pi}[W,G,E], 
\label{if:four}\\
&&ext_{\pi}[P\and \neg b,R]\cap cp_{\pi}[z_3]\subseteq int_{\pi}[W,G,E].
\label{if:five}
\end{eqnarray}
Moreover, if $ z_1'$ denotes the program
\begin{eqnarray}
&&\nif\ b\ \nthen\  z_2'\ \nelse\  z_3'\ \nfi,\nonumber
\end{eqnarray}
it follows from \eqref{if:two} and \eqref{if:three} that
\begin{eqnarray}
&&\models z_1'\progtran z_1.\nonumber
\end{eqnarray}
Thus, it  
it is enough to show that
\begin{eqnarray}
&&ext_{\pi}[P,R]\cap cp_{\pi}[z_1]\subseteq int_{\pi}[W,G,E]\label{if:sev}.
\end{eqnarray}

Let
\begin{eqnarray}
&&\sigma\in ext_{\pi}[P,R]\cap cp_{\pi}[ z_1'].\nonumber
\end{eqnarray}
It is clear
that $\sigma$ is of the form
\begin{eqnarray}
&&< z_1',s_1>\exter\quad\ldots\quad\exter< z_1',s_j>\intern\ \sigma'\nonumber
\end{eqnarray}
where 
\begin{eqnarray}
&&\sigma'\in (ext_{\pi}[P^R\And b,R]\cap cp_{\pi}[ z_2'])
\cup(ext_{\pi}[P^R\And\neg b,R]\cap cp_{\pi}[ z_3']).\nonumber
\end{eqnarray}

Since any external
transition must respect $hid[ z_1']$, it follows that no
external transition can change the truth value of $b$. 
This means that even
\begin{eqnarray}
&&<z_2',s_1>\exter\quad \ldots\quad \exter<z_2',s_j>\exter\ 
\sigma'\nonumber
\end{eqnarray}
is an element of
\begin{eqnarray}
&&(ext_{\pi}[P\And b,R]\cap cp_{\pi}[ z_2'])
\cup(ext_{\pi}[P\And\neg b,R]\cap cp_{\pi}[ z_3']).\nonumber
\end{eqnarray}
Thus, the reflexivity of $G$ together with \eqref{if:four} and \eqref{if:five} 
imply that $\sigma\in int_{\pi}[W,G,E]$, which
again implies \eqref{if:sev}.

\item While-Rule:

Given that $ z_1$ is of the form 
\begin{eqnarray}
&&\nwhile\ b\ \ndo\  z_2\ \nod, \nonumber
\end{eqnarray}
and assume we have a proof of depth $n$, whose root
\begin{eqnarray}
&&Tr_{\pi}\vdash z_1\satis(\vartheta,\alpha) ::  (P,R,W,G,(Z^{\dagger}\Or R)\And \neg b)\nonumber
\end{eqnarray}
is deduced from
\begin{eqnarray}
&&Tr_{\pi}\vdash z_2\satis(\vartheta,\alpha) ::  (P\and b,R,W,G,P\And Z)\nonumber,
\end{eqnarray}
by the while-rule. This means that
\begin{eqnarray}
&&\models_{\pi} \wfound\ Z.\label{wd:one}
\end{eqnarray}
Moreover,
the induction hypothesis implies that
\begin{eqnarray}
&&\models_{\pi}  z_2\satis(\vartheta,\alpha) ::  (P\and b,R,W,G,P\And Z)\nonumber,
\end{eqnarray}
which means that there is a program
$ z_2'$ such that 
\begin{eqnarray}
&&\models z_2'\progtran z_2,\label{wd:two}\\
&&ext_{\pi}[P\and b,R]\cap cp_{\pi}[z_2']\subseteq int_{\pi}[W,G,P\And Z].
\label{wd:four}
\end{eqnarray}
Moreover, if $ z_1'$ denotes the program
\begin{eqnarray}
&&\nwhile\ b\ \ndo\  z_2'\ \nod, \nonumber
\end{eqnarray}
it follows from \eqref{wd:two} that
\begin{eqnarray}
&&\models z_1'\progtran z_1.\nonumber
\end{eqnarray}
Thus, it  
it is enough to show that
\begin{eqnarray}
&&ext_{\pi}[P,R]\cap cp_{\pi}[z_1']\subseteq int_{\pi}[W,G,(Z^{\dagger}\Or R)\And 
\neg b].\label{wd:sev}
\end{eqnarray}

Let
\begin{eqnarray}
&&\sigma\in ext_{\pi}[P,R]\cap cp_{\pi}[ z_1'].\nonumber
\end{eqnarray}
There are three cases to consider:
\begin{itemize}
\item Assume that $\sigma$ diverges:

This means there is an infinite sequence of natural numbers
\[n_1n_2n_3n_4n_5n_6n_7\quad \ldots\quad  n_kn_{k+1}\quad\ldots\]
such that:
\begin{itemize}
\item for all $j$, $j<n_1$ implies that 
$\lambda(\sigma_j)=e$,
\item for all $j$,
\begin{itemize}
\item $n_j<n_{j+1}$,
\item $\tau(\sigma_{n_j})=\tau(\sigma_{n_1})$,
\item $\lambda(\sigma_{n_j})=i,$
\item $\tau(\sigma_{n_{j}+1})= z_2';\tau(\sigma_{n_1})$, 
\item $n_j<k<n_{j+1}$ and $\lambda(\sigma_k)=i$ imply 
$\tau(\sigma_k)\neq \tau(\sigma_{n_1})$.
\end{itemize}
\end{itemize}
In other words, $\sigma(n_j,\ldots,n_{j+1})$ characterises the 
$j$-th iteration with its interference.

Since the environment cannot change the truth value of $b$, it is clear that
for any computation $\sigma'\in ext_{\pi}[P,R]\cap cp_{\pi}[z_1']$ of
the form
\begin{eqnarray}
&&<z_1',s_1>\exter\ \ldots\ \exter<z_1',s_m>\intern\sigma'', \nonumber
\end{eqnarray}
where the $m -  1$ first transitions are due to the environment, 
there is a computation
$\sigma'''\in ext_{\pi}[P,R]\cap cp_{\pi}[z_1']$ of
the form
\begin{eqnarray}
&&<z_1',s_1>\intern<z_2',s_1>\exter\ \ldots\ \exter<z_2',s_m>\exter\sigma''. \nonumber
\end{eqnarray}
Thus, we may assume that $n_1=1$, in which case it follows
that $\delta(\sigma_{n_1})\models_{\pi} P$, which together with
\eqref{wd:four} imply that for all $j\ge 1$, $\delta(\sigma_{n_j})
\models_{\pi} P$, and it is also clear that for all $j$,
\begin{eqnarray}
&&(\delta(\sigma_{n_j}),\delta(\sigma_{n_{j+1}}))\models_{\pi} Z.\nonumber
\end{eqnarray}
This contradicts
\eqref{wd:one}. In other words, the statement converges.

\item Assume that $\sigma$ deadlocks:

This means that 
there is a finite sequence of natural numbers
\[\quad n_1n_2n_3n_4n_5n_6n_7\quad \ldots\quad n_{m -  1}n_m,\]
such that 
\begin{itemize}
\item for all $j$, $j<n_1$ implies $\lambda(\sigma_j)=e$,
\item for all $j\le m$:
\begin{itemize}
\item $n_j<n_{j+1}$,
\item $\tau(\sigma_{n_j})=\tau(\sigma_{n_1})$,
\item $\lambda(\sigma_{n_j})=i$,
\item $\tau(\sigma_{n_{j}+1})= z_2';\tau(\sigma_{n_1})$, 
\item $n_j<k<n_{j+1}$ and
$\lambda(\sigma_k)=i$ imply $\tau(\sigma_k)\neq \tau(\sigma_{n_1})$,
\end{itemize}
\item $n_m<k<len(\sigma)$ and
$\lambda(\sigma_k)=i$ implies $\tau(\sigma_k)\neq \tau(\sigma_{n_1})$.
\end{itemize}
In other words, the loop iterates $m -  1$ times before it deadlocks. In the same way
as above, we may assume that $n_1=1$, in which case it follows 
that $\delta(\sigma_{n_1})\models_{\pi} P$, which together with
\eqref{wd:four} imply that for all $1\le j\le m$, $\delta(\sigma_{n_j})
\models_{\pi} P$. Hence, it follows from \eqref{wd:four} that 
$\delta(len(\sigma))\models_{\pi} W$. Thus, $\sigma\in int_{\pi}[W,G,(Z^{\dagger}\Or R)
\And \neg b]$. 

\item Assume that $\sigma$ terminates:

 This means that
there is a finite sequence of natural numbers
\[\quad n_1n_2n_3n_4n_5n_6n_7\quad \ldots\quad n_{m -  1}n_m,\]
such that 
\begin{itemize}
\item for all $j$, $j<n_1$ or $j>n_m$ implies $\lambda(\sigma_j)=e$,
\item for all $j< m$:
\begin{itemize}
\item $n_j<n_{j+1}$,
\item $\tau(\sigma_{n_j})=\tau(\sigma_{n_1})$,
\item $\lambda(\sigma_{n_j})=i$,
\item $\tau(\sigma_{n_{j}+1})= z_2';\tau(\sigma_{n_1})$, 
\item $n_j<k<n_{j+1}$ and
$\lambda(\sigma_k)=i$ implies $\tau(\sigma_k)\neq \tau(\sigma_{n_1})$,
\end{itemize}
\item $\lambda(\sigma_{n_m})=i$,
\item $\tau(\sigma_{n_m})=\tau(\sigma_{n_1})$, 
\item $\tau(\sigma_{n_{m}+1})=\epsilon$.
\end{itemize}

In other words, the loop iterates $m -  1$ times before it terminates. 
In the same way as above, we may assume that $n_1=1$, in which case
it follows that $\delta(\sigma_1)\models_{\pi} P$, which together with 
\eqref{wd:four} 
imply
that for all $1\le j\le m$, $\delta(\sigma_{n_j})\models_{\pi} P$. 

Moreover, 
\eqref{wd:four} and the transitivity of $Z^{\dagger}$, 
imply that 

\begin{flist}
(\delta(\sigma_1),\delta(\sigma_{n_{m}}))\models_{\pi} Z^{\dagger}.
\end{flist}

\noindent Therefore, since $Z$ covers any interference 
after the last internal transition, and since any external 
transition must respect $hid[ z_1]$ and hence cannot change the truth value
of $b$, it is clear that:
\begin{eqnarray} 
&&(\delta(\sigma_1),\delta(\sigma_{len(\sigma)}))\models_{\pi} 
Z^{\dagger}\And \neg b. \nonumber
\end{eqnarray}

On the other
hand, if $m=1$, then:
\begin{eqnarray}
&&(\delta(\sigma_1),\delta(\sigma_{len(\sigma)}))\models_{\pi} R\And \neg b.\nonumber
\end{eqnarray}

That each internal transition satisfies $G$
follows from
\eqref{wd:four} and the reflexivity of $G$. Again, it is clear that
$\sigma\in int_{\pi}[W,G,(Z^{\dagger}\Or R)\And \neg b]$.
\end{itemize}

This proves \eqref{wd:sev}.

\item Parallel-Rule:

Given that $ z_1$ is of the form 
\begin{eqnarray}
&&\{ z_2\parallel z_3\}, \nonumber
\end{eqnarray}
and assume we have a proof of depth $n$, whose root
\begin{eqnarray}
&&Tr_{\pi}\vdash z_1\satis(\vartheta,\alpha) ::  (P,R_1\And 
R_2,W,G,E_1\And E_2)\nonumber
\end{eqnarray}
is deduced from
\begin{eqnarray}
&&Tr_{\pi}\vdash z_2\satis(\vartheta,\alpha) ::  (P,R_1,W\Or W_1,G\And R_2,E_1),
\nonumber\\
&&Tr_{\pi}\vdash z_3\satis(\vartheta,\alpha) ::  (P,R_2,W\Or W_2,G\And R_1,E_2),\nonumber
\end{eqnarray}
by the parallel-rule. This means that
\begin{eqnarray}
&&\models_{\pi} \neg(W_1\And E_2)\And \neg(W_2\And E_1)\And \neg(W_1\And W_2).
\label{pa:ten}
\end{eqnarray}
Moreover, the induction hypothesis  
implies that
\begin{eqnarray}
&&\models_{\pi} z_2\satis(\vartheta,\alpha) ::  (P,R_1,W\Or W_1,G\And R_2,E_1),
\nonumber\\
&&\models_{\pi} z_3\satis(\vartheta,\alpha) ::  (P,R_2,W\Or W_2,G\And R_1,E_2),\nonumber
\end{eqnarray}
which means that there are two programs
$ z_2'$ and $ z_3'$ such that 
\begin{eqnarray}
&&\models z_2'\progtran z_2,\label{pa:two}\\
&&\models z_3'\progtran z_3,\label{pa:three}\\
&&ext_{\pi}[P,R_1]\cap cp_{\pi}[z_2']\subseteq int_{\pi}[W\Or 
W_1,G\And R_2,E_1], 
\label{pa:four}\\
&&ext_{\pi}[P,R_2]\cap cp_{\pi}[z_3']\subseteq int_{\pi}[W\Or W_2,
G\And R_1,E_2].\label{pa:five}
\end{eqnarray}
Moreover, if $ z_1'$ denotes the program
\begin{eqnarray}
&&\{ z_2'\parallel z_3'\},\nonumber
\end{eqnarray}
it follows from \eqref{pa:two} and \eqref{pa:three} that
\begin{eqnarray}
&&\models z_1'\progtran z_1.\nonumber
\end{eqnarray}
Thus,  
it is enough to show that
\begin{eqnarray}
&&ext_{\pi}[P,R_1\And R_2]\cap cp_{\pi}[z_1']
\subseteq int_{\pi}[W,G,E_1\And E_2]\nonumber,
\end{eqnarray}
which follows from \eqref{pa:ten}, \eqref{pa:four} and \eqref{pa:five} by 
proposition \ref{decomp2:theo} on page \pageref{decomp2:theo}.

\item Await-Rule:

Given that $ z_1$ is of the form 
\begin{eqnarray}
&&\nawait\ b\ \ndo\  z_2\ \nod, \nonumber
\end{eqnarray}
and assume we have a proof of depth $n$, whose root
\begin{eqnarray}
&&Tr_{\pi}\vdash z_1\satis(\vartheta,\alpha) ::  (P,R,W,G,R|E_2|R)\nonumber
\end{eqnarray}
is deduced from
\begin{eqnarray}
&&Tr_{\pi}\vdash z_2\satis(\vartheta,\alpha) ::  (P^R\and b,I,\false,\true,E_1)\nonumber,
\end{eqnarray}
by the await-rule. This means that there are expressions $u_1,\ldots, u_{card(\alpha)}$
such that
\begin{eqnarray}
&&\models_{\pi} P^R\And\neg b\Implies W,\label{aw:one}\\
&&\models_{\pi} E_1|(\bigwedge_{j=1}^{card(\alpha)} a_j=\overleftharpoon{u_j}\And I_{\alpha})
\Implies G\And E_2,\label{aw:onea}\\
&&for\ all\ 1\le j,k\le card(\alpha),\nonumber\\ 
&&\qquad var[u_j]\in\vartheta\cup\{a_j\}, a_j\in\alpha,\ and\nonumber\\
&&\qquad j\neq k\ implies\ a_j\neq a_k. \label{aw:oneb}
\end{eqnarray}
Moreover, the induction hypothesis implies that
\begin{eqnarray}
&&\models_{\pi}  z_2\satis(\vartheta,\alpha) ::  
(P^R\and b,I,\false,\true,E_1)\nonumber,
\end{eqnarray}
which means that there is a program
$ z_2'$ such that 
\begin{eqnarray}
&&\models z_2'\progtran z_2,\label{aw:two}\\
&&ext_{\pi}[P^R\and b,I]\cap cp_{\pi}[z_2']\subseteq int_{\pi}[\false,\true,E_1].
\label{aw:four}
\end{eqnarray}
Moreover, if $ z_1'$ denotes the program
\begin{eqnarray}
&&\nawait\ b\ \ndo\nonumber\\
&&\qquad z_2';\nonumber\\
&&\qquad a_1:=u_1;\nonumber\\
&&\qquad\quad\vdots\nonumber\\
&&\qquad a_{card(\alpha)}:=u_{card(\alpha)}\nonumber\\
&&\nod,\nonumber
\end{eqnarray}
it follows from \eqref{aw:oneb} and \eqref{aw:two} that
\begin{eqnarray}
&&\models z_1'\progtran z_1.\nonumber
\end{eqnarray}
Thus, it  
it is enough to show that
\begin{eqnarray}
&&ext_{\pi}[P,R]\cap cp_{\pi}[z_1']\subseteq int_{\pi}[W,G,R|E_2|R]\label{aw:sev}.
\end{eqnarray}

Let
\begin{eqnarray}
&&\sigma\in ext_{\pi}[P,R]\cap cp_{\pi}[ z_1'].\nonumber
\end{eqnarray}

It is clear from \eqref{aw:four} that $\sigma$ converges.
If $\sigma$ deadlocks, \eqref{aw:one} implies that
 $\sigma$ is of the form
\begin{eqnarray}
&&< z_1',s_1>\exter\quad
 \ldots\quad\exter< z_1',s_l>\exter\quad\ldots\quad\exter< z_1',s_j>,\nonumber
\end{eqnarray}
where $s_j\models_{\pi}W$. Otherwise, $\sigma$ is of the form
\begin{eqnarray}
&&< z_1',s_1>\exter\ \ldots\ \exter< z_1',s_j>\intern<\epsilon,s_{j+1}>\exter\ 
\ldots\ \exter<\epsilon,s_k>,\nonumber
\end{eqnarray}
in which case \eqref{aw:onea} and the transitivity and reflexivity of $R$
imply that
$\sigma\in int_{\pi}[W,G,R|E_2|R]$. 
Thus, \eqref{aw:sev} has
been proved.

\item Elimination-Rule:

Assume we have a proof of depth $n$, whose root
\begin{eqnarray}
&&Tr_{\pi}\vdash z\satis(\vartheta,\alpha\setminus\{a\})
 :: (\Exists a:P,
\Forall \overleftharpoon{a}:\Exists a:R,W,G,E)\nonumber
\end{eqnarray}
is deduced from
\begin{eqnarray}
&&Tr_{\pi}\vdash z\satis(\vartheta,\alpha\cup\{a\}) ::  (P,R,W,G,E)\nonumber
\end{eqnarray}
by the elimination-rule. Then the induction hypothesis implies that
\begin{eqnarray}
&&\models_{\pi} z\satis(\vartheta,\alpha\cup\{a\}) ::  (P,R,W,G,E),\nonumber
\end{eqnarray}
which means that there is a program
$ z'$ such that
\begin{eqnarray}
&&\models z'\stackrel{(\vartheta,\alpha\cup\{a\})}{\hookrightarrow} z,\label{el:two}\\
&&ext_{\pi}[P,R]\cap cp_{\pi}[z']\subseteq int_{\pi}[W,G,E].
\label{el:four}
\end{eqnarray}
Moreover, it follows from \eqref{el:two} that there is a program $ z''$, which can be
got from $z'$ by removing all occurrences of assignments to $a$, such that
\begin{eqnarray}
&&\models z''\progtran
 z.\nonumber
\end{eqnarray}
Thus, it is enough to show that
\begin{eqnarray}
&&ext_{\pi}[\Exists a:P,\Forall \overleftharpoon{a}:\Exists a:R]\cap cp_{\pi}[z'']\subseteq 
int_{\pi}[W,G,E]\label{el:seven}.
\end{eqnarray}

Let 
\begin{eqnarray}
&&\sigma\in ext_{\pi}[\Exists a:P,\Forall \overleftharpoon{a}:\Exists a:R]
\cap cp_{\pi}[ z''].\nonumber
\end{eqnarray}

But then, because of the constraints on the pre- and rely-conditions, we can easily 
construct a computation
\begin{eqnarray}
&&\sigma'\in ext_{\pi}[P,R]
\cap cp_{\pi}[ z'],\nonumber
\end{eqnarray}
such that for all $1\le j\le len(\sigma)$, 
\begin{eqnarray}
&&\delta(\sigma_j)\stackrel{\vartheta\cup(\alpha\setminus\{a\})}{=}
\delta(\sigma'_j).\nonumber
\end{eqnarray}
It follows from \eqref{el:four} 
that $\sigma'\in int_{\pi}[W,G,E]$, in which case it also true that
$\sigma\in int_{\pi}[W,G,E]$, 
since $W$, $G$ and $E$ are only constraining the
elements of $\vartheta\cup(\alpha\setminus\{a\})$. 
Thus, \eqref{el:seven} has been proved.

\item Effect-Rule: 

Assume we have a proof of depth $n$, whose root
\begin{eqnarray}
&&Tr_{\pi}\vdash z\satis(\vartheta,\alpha) ::  (P,R,W,G,
(R\Or G)^{\dagger}\And E)\nonumber
\end{eqnarray}
is deduced from
\begin{eqnarray}
&&Tr_{\pi}\vdash z\satis(\vartheta,\alpha) ::  (P,R,W,G,E)\nonumber,
\end{eqnarray}
by the effect-rule. Then the induction hypothesis implies that
\begin{eqnarray}
&&\models_{\pi}  z\satis(\vartheta,\alpha) ::  (P,R,W,G,E)\nonumber,
\end{eqnarray}
which means that there is a program
$ z'$ such that 
\begin{eqnarray}
&&\models  z'\progtran z, \nonumber\\
&&ext_{\pi}[P,R]\cap cp_{\pi}[z']\subseteq int_{\pi}[W,G,E]
\label{ef:four}.
\end{eqnarray}
Thus, it is enough to show that
\begin{eqnarray}
&&ext_{\pi}[P,R]\cap cp_{\pi}[z']\subseteq int_{\pi}[W,G,(R\Or G)^{\dagger}\And E]
\label{ef:sev}.
\end{eqnarray}

Let
\begin{eqnarray}
&&\sigma\in ext_{\pi}[P,R]\cap cp_{\pi}[ z'].\nonumber
\end{eqnarray}
Since any external transition in $\sigma$ satisfies $R$ and \eqref{ef:four}
implies that any internal transition satisfies $G$, it is clear that
\begin{eqnarray}
&& \sigma\in int_{\pi}[W,G,
(R\Or G)^{\dagger}\And E].\nonumber
\end{eqnarray}
Thus, \eqref{ef:sev} has been shown.

\item Global-Rule:

Assume we have a proof of depth $n$, whose root
\begin{eqnarray}
&&Tr_{\pi}\vdash z\satis(\vartheta\cup\{v\},\alpha) ::  
(P,R,W,G\And v=\overleftharpoon{v},E)\nonumber
\end{eqnarray}
is deduced from
\begin{eqnarray}
&&Tr_{\pi}\vdash z\satis(\vartheta\setminus\{v\},\alpha) ::  
(P,R,W,G,E)\nonumber
\end{eqnarray}
by the global-rule.
The induction hypothesis implies that
\begin{eqnarray}
&&\models_{\pi}  z\satis(\vartheta\setminus\{v\},\alpha) ::  
(P,R,W,G,E),\nonumber
\end{eqnarray}
which means that there is a program
$ z'$ such that 
\begin{eqnarray}
&&\models z'\progtran z,\nonumber\\
&&ext_{\pi}[P,R]\cap cp_{\pi}[z']\subseteq int_{\pi}[W,G,E].\label{var:two}
\end{eqnarray}
Thus it is enough to show that
\begin{eqnarray}
&&ext_{\pi}[P,R]\cap cp_{\pi}[z']\subseteq int_{\pi}[W,G\And v=\overleftharpoon{v},E].\label{var:three}
\end{eqnarray}

Let
\begin{eqnarray}
&&\sigma\in ext_{\pi}[P,R]\cap cp_{\pi}[ z'],\nonumber
\end{eqnarray}
then it follows from \eqref{var:two} that $\sigma\in int_{\pi}[W,G,E]$. Moreover, since
the constraints on specified programs imply that the variable $v$
occur in neither $ z'$ nor $G$, it is also clear that 
$\sigma\in int_{\pi}[W,G\And v=\overleftharpoon{v},E]$. 
This proves \eqref{var:three}.

\item Auxiliary-Rule:

Assume we have a proof of depth $n$, whose root
\begin{eqnarray}
&&Tr_{\pi}\vdash z\satis(\vartheta,\alpha\cup\{a\}) ::  
(P,R,W,G\And a=\overleftharpoon{a},E)\nonumber
\end{eqnarray}
is deduced from
\begin{eqnarray}
&&Tr_{\pi}\vdash z\satis(\vartheta,\alpha\setminus\{a\}) ::  
(P,R,W,G,E)\nonumber
\end{eqnarray}
by the auxiliary-rule.
The induction hypothesis implies that
\begin{eqnarray}
&&\models_{\pi}  z\satis(\vartheta,\alpha\setminus\{a\}) ::  
(P,R,W,G,E),\nonumber
\end{eqnarray}
which means that there is a program
$ z'$ such that 
\begin{eqnarray}
&&\models z'\progtran z,\nonumber\\
&&ext_{\pi}[P,R]\cap cp_{\pi}[z']\subseteq int_{\pi}[W,G,E].\label{aux:two}
\end{eqnarray}
Thus it is enough to show that
\begin{eqnarray}
&&ext_{\pi}[P,R]\cap cp_{\pi}[z']\subseteq int_{\pi}[W,G\And a=\overleftharpoon{a},E].\label{aux:three}
\end{eqnarray}

Let
\begin{eqnarray}
&&\sigma\in ext_{\pi}[P,R]\cap cp_{\pi}[ z'],\nonumber
\end{eqnarray}
then it follows from \eqref{aux:two} that $\sigma\in int_{\pi}[W,G,E]$. Moreover, since
the constraints on specified programs imply that the variable $a$
occur in neither  $ z'$ nor $G$, it is also clear that 
$\sigma\in int_{\pi}[W,G\And a=\overleftharpoon{a},E]$. 
This proves \eqref{aux:three}.

\item Introduction-Rule:

Assume we have a proof of depth $n$, whose root 
\begin{eqnarray}
&& Tr_{\pi}\vdash z_2\satis(\vartheta\setminus\alpha,\alpha) ::  (P,R,W,G,E)\nonumber
\end{eqnarray}
is deduced from
\begin{eqnarray}
&& Tr_{\pi}\vdash z_1\satis(\vartheta\cup\alpha,\emptyset) ::  (P,R,W,G,E)\nonumber
\end{eqnarray}
by the introduction-rule. This means that
\begin{eqnarray}
&&\models z_1\progtran z_2.
\label{in:one}
\end{eqnarray}
Moreover, the induction hypothesis implies that
\begin{eqnarray}
&& ext_{\pi}[P,R]\cap cp_{\pi}[z_1]\subseteq int_{\pi}[W,G,E]\label{in:two}
\end{eqnarray}
Clearly, \eqref{in:one} and \eqref{in:two} imply
\begin{eqnarray}
&&\models_{\pi} z_2\satis(\vartheta\setminus\alpha,\alpha) ::  (P,R,W,G,E).\nonumber
\end{eqnarray}
\end{itemize}

\end{nproof}

%% file: cmpl.tex
\chapter{Relative Completeness}

\section{Motivation}

\subsection{Completeness}

We have already proved that the decomposition-rules in LSP are sound.
This means that for any structure
$\pi$ and specified program $\psi$, if

\begin{flist}
Tr_{\pi}\vdash \psi
\end{flist}

\noindent then

\begin{flist}
\models_{\pi}\psi.
\end{flist}

\noindent The topic of this chapter is to discuss the converse question: in what way 
can we characterise the applicability of LSP, and in particular the 
applicability of LSP$_B$?

\subsection{Incompleteness of LSP}

Not surprisingly, because of the dependence upon first-order logic, LSP is 
incomplete. To see this, it is enough to observe that since L$_P$, the language of
Peano arithmetic, is contained in L$_{wf}$, and given its standard interpretation
in any structure $\pi$, the set $Tr_{\pi}$ is not recursively
ennumerable \cite{jrs:logic}. Moreover, since for any unary expression $E$,

\begin{flist}
\models_{\pi} E
\end{flist}

\noindent if and only if

\begin{flist}
\models_{\pi}\nskip\satis(var[E],\emptyset):: (\true,\true,\false,\true,E),
\end{flist}

\noindent it is clear that for any structure $\pi$, the number of valid specified programs 
is not recursively ennumerable either. Thus, since for any formal system the set
of provable formulas is recursively ennumerable, it follows that
LSP is incomplete.

This means that the best we can hope for is to prove {\em
\underline{relative completeness}}
(see \cite{sac:comp}, \cite{mw:incomplet}, 
\cite{dh:lncs}, \cite{kra:ten1} for a more detailed discussion); namely that
for any structure
$\pi$ and specified program $\psi$, if

\begin{flist}
\models_{\pi}\psi
\end{flist}

\noindent then

\begin{flist}
Tr_{\pi}\vdash \psi.
\end{flist}

\noindent It will be shown below that LSP$_B$ satisfies this criterion. 
The proof depends upon the assumption that for any structure $\pi$ and 
binary assertion $A$ in L, 
there is an assertion $A'$ in L$_{wf}$, which is valid in $\pi$,
if and only if $A$ is well-founded in $\pi$ (see page \pageref{wf:re}).

\subsection{Structure of Relative Completeness Proof}

The relative completeness proof is split into
four main sections. 
Firstly,
three new semantic concepts are introduced --- the strongest eff-, wait- and
guar-relations characterising respectively the 
strongest eff-, wait- and guar-conditions for a specified program of auxiliary
form (see page \pageref{auxform:ref}).

Secondly, we define a function which transforms a certain type of 
specified program into another specified
program, called its historic form. Moreover, it is shown that a specified program
of historic form has some very useful properties.

The historic form function is then employed to prove
that we can always express the strongest
eff-, wait- and guar-conditions for a specified program of auxiliary form.

Finally, due to this result and the properties satisfied by a specified program
of historic form, relative completeness is shown by 
structural induction on the program 
component.

\section{Construction of New Assertions}

\subsection{Decomposition}

In the relative-completeness proof it will often be necessary to construct 
new assertions. For example, 
to decompose the specified program 

\begin{flist}
 z_1; z_2\satis(\vartheta,\alpha):: (P,R,W,G,E),
\end{flist}

\noindent we must find three new assertions $P_1$, $E_1$, $E_2$ such that

\begin{flist}
 z_1\satis(\vartheta,\alpha):: (P,R,W,G,E_1),\\
 z_2\satis(\vartheta,\alpha):: (P_1,R,W,G,E_2),
\end{flist}

\noindent are specified programs, and

\begin{flist}
\models_{\pi} (E_1\Implies P_1)\And (E_1|E_2\Implies E).
\end{flist}

\noindent This ensures that

\begin{flist}
Tr_{\pi}\vdash_B z_1; z_2\satis(\vartheta,\alpha):: (P,R,W,G,E)
\end{flist}

\noindent is deducible from

\begin{flist}
Tr_{\pi}\vdash_B z_1\satis(\vartheta,\alpha):: (P,R,W,G,E_1),\\
Tr_{\pi}\vdash_B z_2\satis(\vartheta,\alpha):: (P_1,R,W,G,E_2)
\end{flist}

\noindent by the consequence- and sequential-rules. 

To simplify the
arguments we have found it useful to introduce three new concepts at the semantic level
called respectively the {\em \underline{strongest eff-relation}}, 
the {\em \underline{strongest wait-relation}}
and the {\em \underline{strongest guar-relation}}. 
They are all defined
with respect to a specified program of auxiliary form.

\subsection{Closed Relations}

Let $z$ denote the program

\begin{prg}
\nbegin\\
\qquad \loc\ y;\\
\qquad y:=1;\\
\qquad \nwhile\ y\le 10\ \ndo\\
\qquad\qquad x:=x+y;\\
\qquad\qquad y:=y+1\\
\qquad \nod\\
\nend,\\
\end{prg}

\noindent and assume we want to characterise the strongest eff-condition $E$, such that

\begin{flist}
\models_{\pi} z\satis(\{x\},\emptyset):: (\true,x=\overleftharpoon{x},\false,\true,E).
\end{flist}

\noindent Clearly, for any computation

\begin{flist}
\sigma\in ext_{\pi}[\true,x=\overleftharpoon{x}]\cap cp_{\pi}[ z],
\end{flist}

\noindent it must be true that

\begin{flist}
(\delta(\sigma_1),\delta(\sigma_{len(\sigma)}))\models_{\pi} E.
\end{flist}

\noindent Since

\begin{flist}
\delta(\sigma_{len(\sigma)})(x)=\delta(\sigma_1)(x)+55,\\
\delta(\sigma_{len(\sigma)})(y)=\delta(\sigma_1)(y)+11,
\end{flist}

\noindent the assertion

\begin{flist}
x=\overleftharpoon{x}+55\And y=\overleftharpoon{y}+11
\end{flist}

\noindent satisfies this criterion.
Nevertheless, if we substitute this formula for $E$ in the specified program above,
the result is no longer a specified program. The reason is that $y$ is a local variable
and therefore restricted from occurring in the specification. 
In other words, $E$ is required to be `closed' with respect to $\{x\}$, in which
case

\begin{flist}
x=\overleftharpoon{x}+55
\end{flist}

\noindent is the strongest eff-condition. This motivates the following two 
definitions:

\begin{definition}
A unary relation $L$ on states is \underline{{\em closed}} 
with respect to a set of variables
$V$, if and only if for all states $s_1$ and $s_2$,
\begin{itemize}
\item $s_1\in L$ and $s_1\stackrel{V}{=}s_2$ imply $s_2\in L$.
\end{itemize}
\end{definition}

\begin{definition}
A binary relation $L$ on states is \underline{{\em closed}} with respect to a 
set of variables 
$V$, if and only if for all states $s_1, s_2, s_3, s_4$,
\begin{itemize}
\item $(s_1,s_2)\in L$, $s_1\stackrel{V}{=}s_3$, and
$s_2\stackrel{V}{=}s_4$ imply $(s_3,s_4)\in L$.
\end{itemize}
\end{definition}

\subsection{Strongest Eff-Relation}

The strongest eff-relation is supposed to model the semantic equivalent of the
strongest eff-condition.
Given a specified program 

\begin{flist}
z\satis(\vartheta,\emptyset)::(P,R,W,G,E)
\end{flist}

\noindent of auxiliary form, we will use $ef_{\pi}[ z,\vartheta,P,R]$ to denote its strongest
eff-relation.

\begin{definition}\label{effect:def}
Given a specified program $z\satis(\vartheta,\emptyset)::(P,R,W,G,E)$ of
auxiliary form,
let $ef_{\pi}[z,\vartheta,P,R]$ be
the 
least binary relation on states, closed with respect to $\vartheta$, such that for
any terminating computation $\sigma\in ext_{\pi}[P,R]\cap 
cp_{\pi}[ z]$,
\begin{itemize}
\item $(\delta(\sigma_1),\delta(\sigma_{len(\sigma)}))\in ef_{\pi}[ z,\vartheta,P,R]$.
\end{itemize}

\end{definition}

\subsection{Strongest Wait-Relation}

Similarly, the strongest wait-relation is intended to characterise the semantic
equivalent of the strongest wait-condition. We will use 
$wa_{\pi}[ z,\vartheta,P,R]$ to denote the strongest wait-condition with respect
to a specified program $z\satis(\vartheta,\emptyset)::(P,R,W,G,E)$ of
auxiliary form.

\begin{definition}\label{wait:def}
Given a specified program $z\satis(\vartheta,\emptyset)::(P,R,W,G,E)$ of
auxiliary form, let
$wa_{\pi}[ z,\vartheta,P,R]$ be the least unary relation on states, closed with
respect to $\vartheta$, such that for any deadlocking computation
$\sigma\in ext_{\pi}[P,R]\cap cp_{\pi}[ z]$, 
\begin{itemize}
\item $\delta(\sigma_{len(\sigma)})\in wa_{\pi}[ z,\vartheta,P,R]$.
\end{itemize}
\end{definition}

\subsection{Strongest Guar-Relation}

The strongest guar-relation is the semantic
equivalent of the strongest guar-condition.
For any specified program $z\satis(\vartheta,\emptyset)::(P,R,W,G,E)$
of auxiliary form, we will use  
$gu_{\pi}[ z,\vartheta,P,R]$ to denote its strongest guar-relation.

\begin{definition}\label{guar:def}
Given a specified program $z\satis(\vartheta,\emptyset)::(P,R,W,G,E)$ of
auxiliary form, let
$gu_{\pi}[ z,\vartheta,P,R]$ be the least binary, reflexive relation on states, 
closed with
respect to $\vartheta$, such that for any 
computation $\sigma\in ext_{\pi}[P,R]\cap cp_{\pi}[ z]$ and $1\le l<len(\sigma)$,

\begin{itemize}
\item if $\lambda(\sigma_l)=i$  
then $(\delta(\sigma_l),\delta(\sigma_{l+1}))\in gu_{\pi}[ z,\vartheta,P,R]$.
\end{itemize}

\end{definition}

\subsection{Conversion Function}

We will later prove that for any specified program

\begin{flist}
z\satis(\vartheta,\emptyset)::(P,R,W,G,E)
\end{flist}

\noindent of auxiliary form, 
the strongest eff-, wait- and guar-relations can always be expressed in L. 
In other words,
we can always find two binary assertions $A_1, A_2$, and one unary assertion
$A_3$, such that for all states $s_1,s_2$:

\begin{itemize}
\item $(s_1,s_2)\models_{\pi} A_1$ if and only if $(s_1,s_2)\in 
ef_{\pi}[ z,\vartheta,P,R]$,
\item $(s_1,s_2)\models_{\pi} A_2$ if and only if $(s_1,s_2)\in 
gu_{\pi}[ z,\vartheta,P,R]$,
\item $s_1\models_{\pi} A_3$ if and only if $s_1\in 
wa_{\pi}[ z,\vartheta,P,R]$.
\end{itemize}

\noindent To make it easy to convert a strongest eff-, wait- or guar-relation
into an assertion, we have found it helpful to define a specific conversion
function $\wp$, which, when applied to a strongest eff-, wait- or 
guar-relation, returns a corresponding assertion in L.

\section{Historic Form}

\subsection{Enrichment of the Global State}

At any stage during the execution of a program, the set of possible internal transitions
is a function of the current configuration.
Since specifications are restricted to constrain only the global state transitions, the 
provability of many programs depends upon the possibility to enrich the global 
state with auxiliary variables,
and use them to encode information about the local state and the program counter 
into the specification. 
Let for example $ z$ denote the program

\begin{prg}
x:=x+1\\
\end{prg}

\noindent then 

\begin{flist}
\wp(gu_{\pi}[ z,\{x\},\true,x\ge \overleftharpoon{x}])\quad =\quad x=\overleftharpoon{x}\Or x=\overleftharpoon{x}+1,\\
\wp(wa_{\pi}[ z,\{x\},\true,x\ge \overleftharpoon{x}])\quad =\quad \false, \\
\wp(ef_{\pi}[ z,\{x\},\true,x\ge \overleftharpoon{x}])\quad =\quad x>\overleftharpoon{x}.
\end{flist}

\noindent Thus, if

\begin{flist}
\models_{\pi} z\satis(\{x\},\emptyset):: (\true,x\ge \overleftharpoon{x},W,G,E),
\end{flist}

\noindent it follows that

\begin{flist}
\models_{\pi} x=\overleftharpoon{x}\Or x=\overleftharpoon{x}+1\Implies G, \\
\models_{\pi} \false\Implies W,\\
\models_{\pi} x> \overleftharpoon{x}\Implies E.
\end{flist}

\noindent Furthermore, it is clear that

\begin{flist}
\models_{\pi}\{ z\parallel z\}\satis(\{x\},\emptyset)
::\\
\qquad (\true,x=\overleftharpoon{x},\false,x=\overleftharpoon{x}\Or x=\overleftharpoon{x}+1,
x=\overleftharpoon{x}+2).
\end{flist}

\noindent Unfortunately, without taking advantage of auxiliary variables, this is not provable in
LSP. The closest we can get is

\begin{flist}
Tr_{\pi}\vdash\{ z\parallel z\}\satis(\{x\},\emptyset)::\\
\qquad (\true,x=\overleftharpoon{x},\false,x=\overleftharpoon{x}\Or x=\overleftharpoon{x}+1,
x> \overleftharpoon{x}).
\end{flist}

\noindent Thus, LSP without the auxiliary-variable rules is incomplete not only because the 
first-order logic is incomplete.
The problem is not that the strongest 
guar-, wait- and eff-conditions cannot
be expressed in $L$, because, as we have seen in the example above, 
they can,
but that the global state is not rich enough.

\subsection{History Variables}

To make it easy to deal with auxiliary structure in the completeness proof,
we have found it useful to introduce a function which transforms a
 specified
program of auxiliary form

\begin{flist}
\{ z_1\parallel z_2\}\satis(\vartheta,\emptyset):: (P,R,W,G,E),
\end{flist}

\noindent whose program component has a parallel-statement as main construct,
into another specified program, called its {\em \underline{historic
form}}, 
which is sufficiently expressive
for our purposes. 

The basic idea is to introduce auxiliary history variables; one 
for each variable in $\vartheta$, to record
the use of the global state, one for each local variable to record 
changes to the local state, and one history variable to record if an update
was due to the overall environment, due to $z_1$ or due to $ z_2$.
For example, if $z$ denotes the program

\begin{flist}
 \nawait\ \true\ \ndo\ x:=x+1\ \nod,
\end{flist}

\noindent then the specified program

\begin{flist}
\{z\parallel z\}\satis(\{x\},\emptyset):: (\true,x=\overleftharpoon{x},
\false,\true,x=\overleftharpoon{x}+2)
\end{flist}

\noindent is transformed into a specified program of the form

\begin{flist}
\{ z'\parallel z''\}\satis(\vartheta',\emptyset):: (P',R',\false,\true,x=\overleftharpoon{x}+2),
\end{flist}

\noindent where 

\begin{itemize}
\item $\vartheta'$ is a set $\{x,h_x,h\}$ such that $h$ is of sort $\seqof{\Nat}$,
      and if $x$ is of sort $\Sigma$,
      then $h_x$ is of sort $\seqof{\Sigma}$,
\item $ z'$ and $ z''$ denote respectively

      \begin{prg}
      \nawait\ \true\ \ndo\\
      \qquad h_x:=[h_x(1)+1]\sconc h_x;\\
      \qquad h:=[1]\sconc h;\\
      \qquad x:=x+1\\
      \nod,\\
      \end{prg}

      \noindent and

      \begin{prg}
      \nawait\ \true\ \ndo\\
      \qquad h_x:=[h_x(1)+1]\sconc h_x;\\
      \qquad h:=[2]\sconc h;\\
      \qquad x:=x+1\\
      \nod,\\
      \end{prg}

\item $P'$ represents

      \begin{flist}
      h_x=[x]\And h=[3],
      \end{flist}

\item while $R'$ stands for

      \begin{flist}
      (x=\overleftharpoon{x}\And h=[3]\sconc \overleftharpoon{h}\And h_{x}=[x]\sconc \overleftharpoon{h_{x}})^*.
      \end{flist}

\end{itemize}
\noindent Clearly, we can now use
$h$ to determine if the $n$'th state change was due to $ z'$, in which case
$h(\len{h} -  n)=1$, to $ z''$, in which case $h(\len{h} -  n)=2$, or 
to the environment, in which
case $h(\len{h} -  n)=3$. 

Moreover, if $m\circ h$ stands for the number of occurrences of $m$ in $h$, and

\begin{flist}
G_1\quad =\quad 1\circ \overleftharpoon{h}=0\And x=\overleftharpoon{x}+1\And h_x=[x]\sconc \overleftharpoon{h_x}
                                 \And h=[1]\sconc\overleftharpoon{h},\\
G_2\quad =\quad 2\circ \overleftharpoon{h}=0\And x=\overleftharpoon{x}+1\And h_x=[x]\sconc \overleftharpoon{h_x}
                                 \And h=[2]\sconc\overleftharpoon{h},\\
R_1\quad =\quad (R'\Or G_2)^{\dagger},\\
R_2\quad=\quad (R'\Or G_1)^{\dagger},\\
E_1\quad =\quad 1\circ h=1\And((2\circ h=0\And x=\overleftharpoon{x}+1)\ \Or\\
\qquad\qquad\qquad (2\circ h=1\And x=\overleftharpoon{x}+2)),\\
E_2\quad =\quad 2\circ h=1\And((1\circ h=0\And x=\overleftharpoon{x}+1)\ \Or\\
\qquad\qquad\qquad (1\circ h=1\And x=\overleftharpoon{x}+2)),
\end{flist}

\noindent it follows by the assignment-rule that

\begin{flist}
Tr_{\pi}\vdash z'\satis(\vartheta',\emptyset):: (P',R_1,\false,\true\And R_2,E_1),\\
Tr_{\pi}\vdash z''\satis(\vartheta',\emptyset):: (P',R_2,\false,\true\And R_1,E_2),
\end{flist}

\noindent which by the parallel-rule give

\begin{flist}
Tr_{\pi}\vdash\{ z'\parallel z''\}\satis(\vartheta',\emptyset)
        :: (P',R_1\And R_2,\false,\true,E_1\And E_2),
\end{flist}

\noindent in which case 

\begin{flist}
Tr_{\pi}\vdash\{z\parallel z\}\satis(\{x\},\emptyset)
        :: (\true,x=\overleftharpoon{x},\false,\true,x=\overleftharpoon{x}+2)
\end{flist}

\noindent follows by the elimination-, introduction- and consequence-rules.

\subsection{Component Functions}

To simplify the definition of the historic-form function, we will first
define four component functions: $\hbar$ which generates the
set of history variables, $\Im$ which extends a program with 
auxiliary structure, $Pre$ which initialises the auxiliary structure, and
$Rel$ which constrains the external updates of the auxiliary structure.

\begin{itemize}
\item If $V$ is an ordered set of $m$ variables $\{v_1,\ \ldots\ ,v_m\}$,
then $\hbar[V]$ denotes an ordered set of $m+1$ variables

\begin{flist}
\{h_{v_1},\ \ldots\ ,h_{v_m},h\},
\end{flist}

\noindent disjoint from $V$, such that
$h$ is of sort $\seqof{\Nat}$, and for all $1\le j\le m$, 
if $v_j$ is of sort $\Sigma_{v_j}$, then $h_{v_j}$ is of sort $\seqof{\Sigma_{v_j}}$.
\end{itemize}

\noindent In other words, $\hbar$ is a function which returns a new history variable of
appropriate sort for each element of the argument, plus a new history
variable to record the origin of updates.

\begin{itemize}
\item If $z$ is a program, $V$ is an ordered set of $m$ variables 
$\{v_1,\ \ldots\ ,v_m\}$, $\hbar[V]=\{h_{v_1},\ \ldots\ ,h_{v_m},h\}$ and $n\in\{1,2\}$,
then $\Im[ z,V,n]$ 
denotes the program that can be obtained from $ z$ by replacing any await-statement
of the form

\begin{prg}
\nawait\ b\ \ndo\\
\qquad x_1:=r_{x_1};\\
\qquad\quad \vdots\\
\qquad x_w:=r_{x_w}\\
\nod,\\
\end{prg}

\noindent not contained in the body of an await-statement\footnote{Since the historic-form
function is only defined for specified programs of auxiliary form, this means that
the Boolean test b is of the form $\true$ and for all $1\le j<k\le w$, 
$x_j\not=x_k$.}, with
an await-statement of the form

\begin{prg}
\nawait\ \true\ \ndo\\
\qquad h_{v_1}:=[t_{v_1}]\sconc h_{v_1};\\
\qquad\quad\vdots\\
\qquad h_{v_m}:=[t_{v_m}]\sconc h_{v_m};\\
\qquad h:=[n]\sconc h;\\
\qquad x_1:=r_{x_1};\\
\qquad\quad \vdots\\
\qquad x_w:=r_{x_w}\\
\nod,\\
\\
where\ for\ all\ 1\le j\le m,\\
\quad t_{v_j}=r_{x_k}(h_{v_j}(1)/x_k)\ if\ there\ is\ a\ 1\le k\le w\ such\ that\ v_j=x_k,\\
\quad and\ t_{v_j}=h_{v_j}(1)\ otherwise,\\
\end{prg}

\noindent and by replacing any other await-statement of the form

\begin{prg}
\nawait\ b\ \ndo\\
\qquad z'\\
\nod,\\
\end{prg}

\noindent not contained in the body of an await-statement, with an await-statement of the form

\begin{prg}
\nawait\ b\ \ndo\\
\qquad \Im[z',V,n];\\
\qquad  h_{v_1}:=[h_{v_1}(1)]\sconc h_{v_1};\\
\qquad \quad\vdots\\
\qquad h_{v_m}:=[h_{v_m}(1)]\sconc h_{v_m};\\
\qquad h:=[n]\sconc h\\
\nod.\\
\end{prg}
\end{itemize}

\noindent This means that the history variables are updated only in connection with 
await-statements. The reason is first of all that only
assignment- and await-statements can access the global state. Secondly, since 
the historic-form
function is defined only for specified programs of auxiliary form,
assignments to local variables can only take place inside await-statements.

Observe, that no
variable $v_1$ in $V$ is allowed to occur on the right-hand side of a  
`new' assignment-statement unless there is an assignment-statement in 
$z$ of the form
$v_2:=r$, where $v_1\neq v_2$ and $v_1$ occurs in $r$. Thus, the new auxiliary
structure does not depend upon the auxiliary structure of $z$.
	
\begin{itemize}
\item If $P$ is an unary assertion, 
$V\cup U$ is an ordered set of $m$
variables $\{v_1,\ \ldots\ ,v_m\}$, $V\cap U=\emptyset$ and 
$\hbar[V\cup U]=\{h_{v_1},\ \ldots\ ,h_{v_m},h\}$,
then $Pre[P,V,U]$ denotes the assertion

\begin{flist}
P\And h=[3]\And \bigwedge_{v\in V} h_{v}=[v]\And 
\bigwedge_{u\in U} \len{h_u}=1.
\end{flist}
\end{itemize}
\noindent The object of the function $Pre$ is to initialise the history variables. The
set $V$ contains the global variables, while the set of local variables is denoted by
$U$. 

\begin{itemize}
\item If $R$ is a binary assertion, $V\cup U$ is an ordered set of $m$ variables
$\{v_1,\ \ldots\ ,v_m\}$,
$V\cap U=\emptyset$ and $\hbar[V\cup U]=\{h_{v_1},\ \ldots\ ,h_{v_m},h\}$, then
$Rel[R,V,U]$, $Rel_1[R,V,U]$ and $Rel_2[R,V,U]$ denote respectively the assertions:

\begin{flist}
(R\And h=[3]\sconc \overleftharpoon{h}\And
  \bigwedge_{v\in V}h_{v}=[v]\sconc \overleftharpoon{h_{v}}\ \And \\
  \qquad\bigwedge_{u\in U}h_{u}=[\overleftharpoon{h_{u}}(1)]\sconc \overleftharpoon{h_{u}})^*,\\
\\
(R\wedge (h=3\sconc \overleftharpoon{h}\Or h=2\sconc \overleftharpoon{h})\ \And\\ 
	\qquad\bigwedge_{v\in V} h_{v}=[v]\sconc \overleftharpoon{h_v}\ \wedge\\
  \qquad\bigwedge_{u\in U} h_{u}=[\overleftharpoon{h_u}(1)]\sconc \overleftharpoon{h_u})^*,\\
\\
(R\wedge (h=3\sconc \overleftharpoon{h}\Or h=1\sconc\overleftharpoon{h})\ \And\\
	\qquad\bigwedge_{v\in V} h_{v}=[v]\sconc \overleftharpoon{h_{v}}\ \And\\
  \qquad\bigwedge_{u\in U} h_{u}=[\overleftharpoon{h_u}(1)]\sconc \overleftharpoon{h_u})^*.
\end{flist}

\end{itemize}

\noindent Again, the
set $V$ contains the global variables, while the set of local variables is denoted by
$U$. Clearly, the $Rel$ functions record any update due to the relevant environment.
(Remember that the environment is restricted from updating local variables.)
Observe, that $\Im$, $Pre$ and the $Rel$ functions together ensure 
that the value of any global variable and any `active' local variable always 
equals the first element of its history variable.

\subsection{Historic-Form Function}

The historic form function can then be defined:

\begin{definition}
Given a specified program of auxiliary form

\begin{flist}
\{ z_1\parallel z_2\}\satis(\vartheta,\emptyset):: (P,R,W,G,E),
\end{flist}

\noindent whose program component has a parallel-statement as main construct, then

\begin{flist}
\{\Im[ z_1,V,1]\parallel\Im[ z_2,V,2]\}\satis
(\vartheta\cup \hbar[V],\emptyset):: \\
\qquad (Pre[P,\vartheta,U],Rel[R,\vartheta,U],W,G,E),
\end{flist}

\noindent where $U=var[\{z_1\parallel z_2\}]\setminus \vartheta$ and 
$V=\vartheta\cup U$,
is its historic form.
\end{definition}

\subsection{Historic-Form Proposition}

To show that the historic-form function has the desired properties, we will
prove four propositions. The first one is rather trivial. Basically, it
asserts that the result of applying the historic-form function is a specified program,
that the historic-form function preserves validity, and that the history variables
introduced by the historic-form function are used in such a way that they do not
depend upon auxiliary structure already introduced.

\begin{statement}\label{exform:theo}
Given a specified program 

\begin{flist}
z\satis(\vartheta,\emptyset)::\psi
\end{flist}

\noindent of auxiliary form such that $z$'s main construct is a 
parallel-statement, and assume
that

\begin{flist}
z'\satis(\vartheta\cup\vartheta',\emptyset)::\psi',
\end{flist}

\noindent where $\vartheta\cap\vartheta'=\emptyset$, is its historic form, then
\begin{itemize}
\item $z'\satis(\vartheta\cup\vartheta',\emptyset)::\psi'$ 
is a specified program,
\item $\models_{\pi} z\satis(\vartheta,\emptyset)::\psi$ 
implies $\models_{\pi} z'\satis(\vartheta\cup\vartheta',\emptyset)::\psi'$,
\item $\models z\stackrel{(\vartheta\setminus\alpha,\alpha)}{\hookrightarrow}  
z''$ implies
$\models z'\stackrel{(\vartheta\setminus\alpha,
\alpha\cup\vartheta')}{\hookrightarrow} z''$.
\end{itemize}
\end{statement}

\begin{nproof}
The first result follows from the fact that $Rel[R,\vartheta,U]$ by definition
is both reflexive and transitive, and that any new variable introduced by 
$\Im$, $Pre$ or $Rel$ is an element of $\vartheta'$ and is distinct from any
local variable in $z$.

Moreover, since

\begin{flist}
\models_{\pi} Pre[P,\vartheta,U]\Implies P,\\
\models_{\pi} Rel[R,\vartheta,U]\Implies R
\end{flist}

\noindent the elements of $\vartheta'$ do not occur in 
$W$, $G$ and $E$, and the elements of $\vartheta'$ has no influence on the
elements of $\vartheta$,
the second result is also correct.

The third result follows from the fact that no variable $v_1$ in $\vartheta$ is
allowed to occur on the right-hand side of a `new' assignment-statement unless there
is an assignment-statement in $z$ of the form $v_2:=r$,
where $v_1\neq v_2$ and $v_1$ occurs in $r$.

\end{nproof}

\subsection{Decomposition Propositions}

The object of the three remaining propositions is to show that the historic-form 
function gives us the 
necessary expressive power to decompose any valid specified program whose
program component has a parallel-statement as main construct. Let

\begin{flist}
\{z_1\parallel z_2\}\satis(\vartheta,\emptyset)::(P,R,W,G,E)
\end{flist}

\noindent be a specified program of auxiliary form, such that $R$ respects

\begin{flist}
hid[\{z_1\parallel z_2\}]\cap \vartheta
\end{flist}

\noindent and

\begin{flist}
U=var[\{z_1\parallel z_2\}]\setminus\vartheta.
\end{flist}

\noindent Moreover, let

\begin{flist}
\{z_1'\parallel z_2'\}\satis(\vartheta',\emptyset)::(P',R',W,G,E)
\end{flist}

\noindent be its historic form, and  assume there are assertions 
$G_1$ and $G_2$ such
that

\begin{flist}
G_1=\wp(gu_{\pi}[z_1',\vartheta', Pre[\true,\vartheta,U],
Rel_2[\true,\vartheta,U]]),\\
G_2=\wp(gu_{\pi}[z_2',\vartheta', Pre[\true,\vartheta,U],
Rel_1[\true,\vartheta,U]]),\\
R_1=(R'\Or G_2)^{\dagger},\\
R_2=(R'\Or G_1)^{\dagger}.
\end{flist}

\noindent Clearly, $G_1$ defines $z_1'$'s set of possible 
state changes as a function of the history,
while $G_2$ characterises a similar relationship with respect to $z_2'$.

\begin{statement}\label{histtwo:theo}
Given the context above, and two computations 

\begin{flist}
\sigma\in ext_{\pi}[P',R_1]\cap cp_{\pi}[z_1'],\\
\sigma'\in ext_{\pi}[P',R_2]\cap cp_{\pi}[z_2'],
\end{flist}

\noindent $1\le j\le len(\sigma)$ and $1\le k\le len(\sigma')$,
such that $\delta(\sigma_{j})\stackrel{\vartheta'}{=}\delta(\sigma'_{k})$, 
then there are computations

\begin{flist}
\sigma''\in ext_{\pi}[P',R_1]\cap cp_{\pi}[z_1'],\\
\sigma'''\in ext_{\pi}[P',R_2]\cap cp_{\pi}[z_2']
\end{flist}

\noindent and $1\le l\le \{len(\sigma''),len(\sigma''')\}$, such that 
$\sigma''(1,\ldots,l)\bullet\sigma'''(1,\ldots,l)$,
$\sigma_j\stackrel{\vartheta'}{=}\sigma''_l$ and $\sigma'_k
\stackrel{\vartheta'}{=}\sigma'''_l$.
\end{statement}

\begin{nproof} Assume that any external transition in $\sigma$ satisfies
$R'$ or $G_2$. $\sigma$ can easily be transformed into such a computation
by splitting external transitions, if this is not the case. Due to the information
stored in the history variables no external transition of $\sigma$ can satisfy both
$R'$ and $G_2$ unless it leaves $\vartheta'$ unchanged. 
Similarly, we may assume that any external transition in $\sigma'$
satisfies $R'$ or $G_1$.

The proof is by induction on $j$. The base case $j=1$ is
trivial. Assume the proposition is correct for $j\le n$. We will prove that
the proposition is correct for $j=n+1$. There are  four  cases to consider:
\begin{itemize}
\item If there is an $1\le l<j$ such that 
$\lambda(\sigma_l)=e$ and $\delta(\sigma_l)
\stackrel{\vartheta'}{=}\delta(\sigma_{l+1})$:
\begin{itemize}

\item This
means that there is a computation

\begin{flist}
\sigma''''\in ext_{\pi}[P',R_1]\cap cp_{\pi}[z_1'],
\end{flist}

such that $\sigma''''_{j -  1}\stackrel{\vartheta'}{=}\sigma_j$. Thus,
the proposition follows by the induction hypothesis.
\end{itemize}

\item Else if there is an $1\le l<j$ such that $\lambda(\sigma_l)=i$, 
$\delta(\sigma_l)=
\delta(\sigma_{l+1})$ and for all $l<m<j$, $\lambda(\sigma_l)=e$:
\begin{itemize}
\item Since the internal transition is independent of the global state, and
no external transition can change the local state,
it follows that

\begin{flist}
\sigma(1,\ldots,l)\ \exter<\tau(\sigma_l),\delta(\sigma_{l+2})>\exter\ \ldots\\ 
\qquad \exter<\tau(\sigma_l),\delta(\sigma_{j})>\intern\sigma_j
\end{flist}

is a computation in $ext_{\pi}[P',R_1]\cap cp_{\pi}[z_1']$, in 
which case the proposition follows easily by the induction hypothesis.
\end{itemize}

\item Else if there is an $1\le l<j$, such that for all $l\le m<j$,
$\lambda(\sigma_m)=i$, and $l=1$ or $\lambda(\sigma_{l -  1})=e$:
\begin{itemize}
\item Due to the information stored in the history variables it follows 
that there
is a $1\le p<k$, such that $\delta(\sigma_l)\stackrel{\vartheta'}{=}
\delta(\sigma'_p)$. Since the value of any 
active local variable is determined by the first value of 
its history variable, the proposition follows easily by the
induction hypothesis.
\end{itemize}

\item Else:
\begin{itemize}
\item Due to the information stored in the history variables it 
follows that there are
$1\le l<j$ and $1\le m<k$ such that $\delta(\sigma_l)
\stackrel{\vartheta'}{=}
\delta(\sigma'_{m})$ and $(\delta(\sigma_l),
\delta(\sigma_j))\models_{\pi} R'$ and
$(\delta(\sigma'_m),\delta(\sigma'_k))\models_{\pi} R'$, in which 
case the proposition
follows by the induction hypothesis.
\end{itemize}
\end{itemize}
\end{nproof}

\begin{statement}\label{histthree:theo}
Given the context above, a computation

\begin{flist}
\sigma\in ext_{\pi}[P',R_1]\cap cp[z_1']
\end{flist}

\noindent and $1\le j\le len(\sigma)$, then there are computations

\begin{flist}
\sigma'\in ext_{\pi}[P',R_1]\cap cp[z_1'],\\
\sigma''\in ext_{\pi}[P',R_2]\cap cp[z_2'],
\end{flist}

\noindent and $1\le k\le min\{len(\sigma'),len(\sigma'')\}$, such that 
$\sigma'(1,\ldots,k)\bullet\sigma''(1,\ldots,k)$ and $\sigma_j\stackrel{\vartheta'}{=}
\sigma'_k$.
\end{statement}

\begin{nproof} In the same way as above, we may assume that any external
transition in $\sigma$ satisfies $R'$ or $G_2$.

The proof is by induction on $j$. The base case $j=1$ is trivial. Assume the
proposition is correct for all $j\le n$. We will prove that the proposition is correct
for $j=n+1$. There are three cases to consider:
\begin{itemize}
\item If $\lambda(\sigma_n)=i$:
\begin{itemize}
\item The proposition follows by the induction hypotheses, since the value of any 
active local variable is determined by the first value of
its history variable.
\end{itemize}
\item Else if $(\delta(\sigma_n),\delta(\sigma_{n+1}))\models_{\pi} R'$:
\begin{itemize}
\item Follows easily by the induction hypothesis, since $R'$ respects

\begin{flist}
hid[\{z_1'\parallel z_2'\}]\cap \vartheta'.
\end{flist}

\end{itemize}
\item Else:
\begin{itemize}
\item This means that $\lambda(\sigma_n)=e$ and 
$(\delta(\sigma_n),\delta(\sigma_{n+1}))\models_{\pi} G_2$. By definition, there
is a computation

\begin{flist}
\sigma'\in ext_{\pi}[Pre[\true,\vartheta',U],
Rel_2[\true,\vartheta',U]]\cap cp_{\pi}[z_2']
\end{flist}

and $1\le l\le len(\sigma')$,
such that $\delta(\sigma_l')\stackrel{\vartheta'}{=}\delta(\sigma_{n+1})$. But
then, due to the information stored in the history variables, there is
a computation

\begin{flist}
\sigma''\in ext_{\pi}[P',R_2]\cap cp_{\pi}[z_2'],
\end{flist}

and $1\le l\le len(\sigma'')$, such that $\delta(\sigma_l'')
\stackrel{\vartheta'}{=}\delta(\sigma_{n+1})$,
in which case the desired result follows from proposition \ref{histtwo:theo} on
page \pageref{histtwo:theo}.
\end{itemize}

\end{itemize}
\end{nproof}

\begin{statement}\label{histfour:theo}
Given the context above, if there is a diverging computation

\begin{flist}
\sigma\in ext_{\pi}[P',R_1]\cap cp[z_1'],
\end{flist}

\noindent then there is a diverging computation

\begin{flist}
\sigma'\in ext_{\pi}[P',R']\cap cp[\{z_1'\parallel z_2'\}].
\end{flist}
\end{statement}
\begin{nproof}
There are three cases to consider:
\begin{itemize}
\item There is a $j\ge 1$, such that for all $k\ge j$, 
$\lambda(\sigma_k)=e$ implies 

\begin{flist}
(\delta(\sigma_k),\delta(\sigma_{k+1}))\models_{\pi} R':
\end{flist}

\begin{itemize}
\item Proposition \ref{histthree:theo} on page \pageref{histthree:theo} implies that
there are 

\begin{flist}
\sigma'\in ext_{\pi}[P',R_1]\cap cp_{\pi}[z_1'],\\
\sigma''\in ext_{\pi}[P',R_2]\cap cp_{\pi}[z_2']
\end{flist}

and $1\le k\le min\{len(\sigma'),len(\sigma'')\}$, such that 

\begin{flist}
\sigma'(1,\ldots,k)\bullet\sigma''(1,\ldots,k)
\end{flist}

\noindent and
$\sigma'_k\stackrel{\vartheta'}{=}\sigma_j$. Since the value of any 
active local variable is determined by the first value of its
history variable, and since $R'$ respects
$hid[\{z_1'\parallel z_2'\}]\cap \vartheta'$, it is straightforward to construct an 
infinite computation

\begin{flist}
\sigma'\in ext_{\pi}[P',R']\cap cp_{\pi}[\{z_1'\parallel z_2'\}].
\end{flist}

\end{itemize}

\item There is a $j\ge 1$, such that for all $k\ge j$, 
$\lambda(\sigma_k)=i$ implies $\delta(\sigma_k)=\delta(\sigma_{k+1})$:

Since after the first $j -  1$ transitions no internal transition depends upon
the global state, we can easily construct an 
infinite computation

\begin{flist}
\sigma'\in ext_{\pi}[P',R']\cap cp_{\pi}[\{z_1'\parallel z_2'\}]
\end{flist}

by an argument similar to the one above.

\item Else:

\begin{itemize}
\item Since any internal transition which leaves the state unchanged can be moved,
and any external transition which leaves $\vartheta'$ unchanged can be removed, we
may assume that there is an infinite sequence of natural numbers 

\begin{flist}
n_1<m_1<n_2<m_2<\ \ldots\ <n_k<m_k<\ \ldots\ 
\end{flist}

such that for all $j,k$:

\begin{flist}
\delta(\sigma_{n_j})\stackrel{\vartheta'}{\neq}\delta(\sigma_{m_j}),\\
\delta(\sigma_{m_j})\stackrel{\vartheta'}{\neq}\delta(\sigma_{n_{j+1}}),\\
n_j\le k<m_j\Implies (\delta(\sigma_k),\delta(\sigma_{k+1}))\models_{\pi} G_1\Or R',\\
m_j\le k<n_{j+1}\Implies (\delta(\sigma_k),\delta(\sigma_{k+1}))
\models_{\pi} G_2.
\end{flist}

For any $j\ge 1$, let 
 $T_j$ be the set of all tuples of the form $(\sigma',\sigma'',k)$ such that
there are 

\begin{flist}
\sigma'\in ext_{\pi}[P',R_1]\cap cp_{\pi}[z_1'],\\
\sigma''\in ext_{\pi}[P',R_2]\cap cp_{\pi}[z_2']
\end{flist}

and $1\le k\le min\{len(\sigma'),len(\sigma'')\}$ which satisfy
$\sigma'(1,\ldots,k)\bullet\sigma''(1,\ldots,k)$ and  $\delta(\sigma_{m_j})
\stackrel{\vartheta'}{=}\delta(\sigma'_k)$.

Clearly, for all $1\le p<j$, if $(\sigma',\sigma'',k)\in T_j$, then there is an
$l$ such that $(\sigma',\sigma'',l)\in T_p$. 

Moreover, since the value of any active local variable is determined by the first
value of its history variable, 
if $(\sigma',\sigma'',k), (\sigma''',\sigma'''',p)\in T_j$,
$\tau(\sigma'_k)=\tau(\sigma'''_p)$,
$\tau(\sigma''_k)=\tau(\sigma''''_p)$, and there are $o>j$, extensions
$\sigma'_e$ of $\sigma'(1,\ldots,k)$, $\sigma''_e$ of $\sigma''(1,\ldots,k)$ and $k_e$,
such that $(\sigma'_e,\sigma''_e,k_e)\in T_o$, then there are extensions
$\sigma'''_e$ of $\sigma'''(1,\ldots,k)$ and $\sigma''''_e$ of 
$\sigma''''(1,\ldots,k)$ and $p_e$
such that $(\sigma'''_e,\sigma''''_e,p_e)\in T_o$.

Let $Z_j=\{(\tau(\sigma'_k),\tau(\sigma''_k))|(\sigma',\sigma'',k)\in T_j\}$. 
It follows from
proposition \ref{histthree:theo} on page \pageref{histthree:theo} that $Z_j$ is
nonempty, and from the definition of a computation that $Z_j$ is finite.

But then, since a computation is not
required to satisfy any fairness constraint, we can easily construct an
infinite computation

\begin{flist}
\sigma'\in ext_{\pi}[P',R']\cap cp_{\pi}[\{z_1'\parallel z_2'\}].
\end{flist}

\end{itemize}
\end{itemize}
\end{nproof}

\section{Expressiveness Proposition}

The next proposition shows that we can always express the strongest wait-, guar-
and eff-relations.

\begin{statement}\label{express:theo}
Given a specified program 
\begin{eqnarray}
&&z\satis(\vartheta,\emptyset)::(P,R,W,G,E)\nonumber
\end{eqnarray}
of auxiliary form, then 
$ef_{\pi}[ z,\vartheta,P,R]$, $wa_{\pi}[ z,\vartheta,P,R]$ and
$gu_{\pi}[ z,\vartheta,P,R]$ are expressible in L.
\end{statement}

\begin{nproof} 
For any specified program of auxiliary form

\begin{flist}
z\satis(\vartheta,\emptyset)::(P,R,W,G,E), 
\end{flist}

\noindent there is a unique program $z'$
such that $\models z\hookrightarrow z'$. 
We will prove the above proposition by 
structural induction on this program. 
Since the environment respects $hid[z]\cap\vartheta$, it follows that
for all $v\in hid[z]\cap \vartheta$:
\begin{eqnarray}
&&ef_{\pi}[ z,\vartheta,P,R]=ef_{\pi}[ z,\vartheta,P,R\And v=\overleftharpoon{v}],\nonumber\\
&&wa_{\pi}[ z,\vartheta,P,R]=wa_{\pi}[ z,\vartheta,P,R\And v=\overleftharpoon{v}],\nonumber\\
&&gu_{\pi}[ z,\vartheta,P,R]=gu_{\pi}[ z,\vartheta,P,R\And v=\overleftharpoon{v}].\nonumber
\end{eqnarray}
In each case below it is therefore assumed that $R$ respects $hid[z]\cap \vartheta$.
The base-cases are the skip-
and assignment-statements.

\begin{itemize}
\item Skip: Given a specified program 
\begin{eqnarray}
&&z_1\satis(\vartheta,\emptyset)::(P,R,W,G,E)\nonumber
\end{eqnarray}
of auxiliary form, and a program $z_1'$ of the form 
\begin{eqnarray}
&&\nskip,\nonumber
\end{eqnarray}
such that $\models z_1\hookrightarrow z_1'$. Clearly, 
$ef_{\pi}[ z_1,\vartheta,P,R]$ is characterised by
\begin{eqnarray}
&&\overleftharpoon{P}\And R,\nonumber
\end{eqnarray}
$wa_{\pi}[ z_1,\vartheta,P,R]$ is characterised by
\begin{eqnarray}
&&\false,\nonumber
\end{eqnarray}
while $gu_{\pi}[ z_1,\vartheta,P,R]$ is characterised by
\begin{eqnarray}
&&I.\nonumber
\end{eqnarray}

\item Assignment: Given a specified program 
\begin{eqnarray}
&&z_1\satis(\vartheta,\emptyset)::(P,R,W,G,E)\nonumber
\end{eqnarray}
of auxiliary form, a program $z_1'$ of the form 
\begin{eqnarray}
&&v:=r,\nonumber
\end{eqnarray}
such that
$\models z_1 \hookrightarrow z_1'$, and assume $\alpha$ characterises the set of 
auxiliary variables in $z_1$. 
This means that $z_1$ is of the form
\begin{eqnarray}
&&\nawait\ \true\ \ndo\nonumber\\
&&\qquad a_1:=u_1;\nonumber\\
&&\qquad\quad\vdots\nonumber\\
&&\qquad a_m:=u_m;\nonumber\\
&&\qquad v:=r\nonumber\\
&&\nod,\nonumber
\end{eqnarray}
where for all $1\le j,k\le m$, $var[u_j]\subseteq (\vartheta\setminus\alpha)
\cup\{a_j\}$,
$j\neq k\Implies a_j\neq a_k$ and
$a_j\in \alpha$.
Clearly, $ef_{\pi}[ z_1,\vartheta,P,R]$ is characterised by
\begin{eqnarray}
&&\overleftharpoon{P}\And R|(v=\overleftharpoon{r}\And I_{\{v\}\cup\alpha}\And \bigwedge_{j=1}^m a_j=\overleftharpoon{u_j})|R,
\nonumber
\end{eqnarray}
$wa_{\pi}[ z_1,\vartheta,P,R]$ is characterised by
\begin{eqnarray}
&&\false,\nonumber
\end{eqnarray}
while $gu_{\pi}[ z_1,\vartheta,P,R]$ is characterised by
\begin{eqnarray}
&& \overleftharpoon{P^R}\And v=\overleftharpoon{r}\And I_{\{v\}\cup\alpha}\And \bigwedge_{j=1}^m a_j=\overleftharpoon{u_j}.\nonumber
\end{eqnarray}
\end{itemize}

\begin{itemize}
\item Block: Given a specified program 
\begin{eqnarray}
&&z_1\satis(\vartheta,\emptyset)::(P,R,W,G,E)\nonumber
\end{eqnarray}
of auxiliary form, and a program $z_1'$ of the form 
\begin{eqnarray}
&&\nbegin\ \loc\ v_1,\ \ldots\ ,v_m; z_2'\ \nend,\nonumber
\end{eqnarray}
such that $\models z_1 \hookrightarrow z_1'$.
This means that $z_1$ is of the form 
\begin{eqnarray}
&&\nbegin\ \loc\ v_1,\ \ldots\ ,v_m; z_2\ \nend,\nonumber
\end{eqnarray}
where 
$\models z_2\hookrightarrow z_2'$.
Thus, by the induction hypothesis it follows that 
$ef_{\pi}[ z_1,\vartheta,P,R]$ 
is characterised by
\begin{eqnarray}
&&\Exists \overleftharpoon{v_1},\ \ldots\ ,\overleftharpoon{v_m},v_1,\ \ldots\ ,v_m:\nonumber\\
&&\qquad \wp(ef_{\pi}[ z_2,\vartheta\cup\bigcup_{j=1}^m\{v_j\},P,
R\And\bigwedge_{j=1}^m v_j=\overleftharpoon{v_j}]),\nonumber
\end{eqnarray}
$wa_{\pi}[ z_1,\vartheta,P,R]$ is characterised by
\begin{eqnarray}
&&\Exists v_1,\ \ldots\ ,v_m:
\wp(wa_{\pi}[ z_2,\vartheta\cup\bigcup_{j=1}^m\{v_j\},P,R\And
\bigwedge_{j=1}^m v_j=\overleftharpoon{v_j}]),\nonumber
\end{eqnarray}
while $gu_{\pi}[ z_1,\vartheta,P,R]$ is characterised by
\begin{eqnarray}
&& \Exists \overleftharpoon{v_1},\ \ldots\ , \overleftharpoon{v_m}, v_1,\ \ldots\ , v_m:\nonumber\\
&&\qquad\wp(gu_{\pi}[ z_2,\vartheta\cup\bigcup_{j=1}^m\{v_j\},P,
R\And\bigwedge_{j=1}^m v_j=\overleftharpoon{v_j}])\nonumber
\end{eqnarray}

\item Composition: Given a specified program 
\begin{eqnarray}
&&z_1\satis(\vartheta,\emptyset)::(P,R,W,G,E)\nonumber
\end{eqnarray}
of auxiliary form, and a program $z_1'$ of the form 
\begin{eqnarray}
&&z_2'; z_3',\nonumber
\end{eqnarray}
such that
$\models z_1\hookrightarrow z_1'$.
This means that $z_1$ is of the form 
\begin{eqnarray}
&&z_2; z_3,\nonumber
\end{eqnarray}
where $\models z_2\hookrightarrow z_2'$ and $\models z_3\hookrightarrow z_3'$.
Thus, by the induction hypothesis it follows that 
$ef_{\pi}[ z_1,\vartheta,P,R]$ is characterised by
\begin{eqnarray}
&&\wp(ef_{\pi}[ z_2,\vartheta,P,R])|\nonumber\\
&&\qquad\wp(ef_{\pi}[ z_3,\vartheta,
\Exists \overleftharpoon{\vartheta}:
\wp(ef_{\pi}[ z_2,\vartheta,P,R]),R]),\nonumber
\end{eqnarray}
$wa_{\pi}[ z_1,\vartheta,P,R]$ is characterised by
\begin{eqnarray}
&&\wp(wa_{\pi}[ z_2,\vartheta,P,R])\Or\nonumber\\
&&\qquad\wp(wa_{\pi}[ z_3,\vartheta,
\Exists \overleftharpoon{\vartheta}:
\wp(ef_{\pi}[ z_2,\vartheta,P,R]),R]),\nonumber
\end{eqnarray}
while $gu_{\pi}[ z_1,\vartheta,P,R]$ is characterised by
\begin{eqnarray}
&& \wp(gu_{\pi}[ z_2,\vartheta,P,R])\Or\nonumber\\
&&\qquad\wp(gu_{\pi}[ z_3,\vartheta,
\Exists \overleftharpoon{\vartheta}:
\wp(ef_{\pi}[ z_2,\vartheta,P,R]),R]).\nonumber
\end{eqnarray}

\item If: Given a specified program 
\begin{eqnarray}
&&z_1\satis(\vartheta,\emptyset)::(P,R,W,G,E)\nonumber
\end{eqnarray}
of auxiliary form, and a program $z_1'$ of the form 
\begin{eqnarray}
&&\nif\ b\ \nthen\  z_2'\ \nelse\  z_3'\ \nfi,\nonumber
\end{eqnarray}
such that
$\models z_1\hookrightarrow z_1'$.
This means that $z_1$ is of the form 
\begin{eqnarray}
&&\nif\ b\ \nthen\  z_2\ \nelse\  z_3\ \nfi,\nonumber
\end{eqnarray}
where 
$\models z_2\hookrightarrow z_2'$ and $\models z_3\hookrightarrow z_3'$.
Thus, by the induction hypothesis it follows that 
$ef_{\pi}[ z_1,\vartheta,P,R]$ is characterised by
\begin{eqnarray}
&&\wp(ef_{\pi}[ z_2,\vartheta,P\And \neg b,R])\Or \wp(ef_{\pi}[ z_3,\vartheta,
P\And b,R]),\nonumber
\end{eqnarray}
$wa_{\pi}[ z_1,\vartheta,P,R]$ is characterised by
\begin{eqnarray}
&&\wp(wa_{\pi}[ z_2,\vartheta,P\And \neg b,R])\Or \wp(wa_{\pi}[ z_3,
\vartheta,P\And b,R]),\nonumber
\end{eqnarray}
while $gu_{\pi}[ z_1,\vartheta,P,R]$ is characterised by
\begin{eqnarray}
&&\wp(gu_{\pi}[ z_2,\vartheta,P\And \neg b,R])\Or 
\wp(gu_{\pi}[ z_3,\vartheta,P\And b,R]).\nonumber
\end{eqnarray}

\item While: Given a specified program 
\begin{eqnarray}
&&z_1\satis(\vartheta,\emptyset)::(P,R,W,G,E)\nonumber
\end{eqnarray}
of auxiliary form, and a program $z_1'$ of the form 
\begin{eqnarray}
&&\nwhile\ b\ \ndo\  z_2'\ \nod,\nonumber
\end{eqnarray}
such that $\models z_1\hookrightarrow z_1'$.
This means that $z_1$ is of the form 
\begin{eqnarray}
&&\nwhile\ b\ \ndo\  z_2\ \nod,\nonumber
\end{eqnarray}
where $\models z_2\hookrightarrow z_2'$.
Thus, by the induction hypothesis, if $Z$ denotes
\begin{eqnarray}
&&\wp(ef_{\pi}[ z_2,\vartheta,b,R]), \nonumber
\end{eqnarray}
it follows that
$ef_{\pi}[ z_1,\vartheta,P,R]$ is characterised by
\begin{eqnarray}
&& \overleftharpoon{P}\And (Z^{\dagger}\Or R)\And \neg b,\nonumber
\end{eqnarray}
$wa_{\pi}[ z_1,\vartheta,P,R]$ is characterised by
\begin{eqnarray}
&& \wp(wa_{\pi}[ z_2,\vartheta,P^{Z}\And b,R]),\nonumber
\end{eqnarray}
while $gu_{\pi}[ z_1,P,R]$ is characterised by
\begin{eqnarray}
&& \wp(gu_{\pi}[ z_2,\vartheta,P^{Z}\And b,R]).\nonumber
\end{eqnarray}

\item Parallel: Given a specified program 
\begin{eqnarray}
&&z_1\satis(\vartheta,\emptyset)::(P,R,W,G,E)\nonumber
\end{eqnarray}
of auxiliary form, and a program $z_1'$ of the form 
\begin{eqnarray}
&&\{z_2'\parallel z_3'\},\nonumber
\end{eqnarray}
such that
$\models z_1\hookrightarrow z_1'$.
This means that $z_1$ is of the form 
\begin{eqnarray}
&&\{z_2\parallel z_3\},\nonumber
\end{eqnarray}
where 
$\models z_2\hookrightarrow z_2'$ and $\models z_3\hookrightarrow z_3'$.

Let 
\begin{eqnarray}
&&z_1''\satis(\vartheta'',\emptyset)::(P'',R'',W,G,E)\nonumber
\end{eqnarray}
be the historic form. This means that $z_1''$ is of the form
\begin{eqnarray}
&&\{z_2''\parallel z_3''\},\nonumber
\end{eqnarray}
where $\models z_2''\hookrightarrow z_2'$ and $\models z_3''\hookrightarrow z_3'$.
Let $U=var[z_1]\setminus\vartheta$, then
by the induction hypothesis it follows 
that there are assertions such that
\begin{eqnarray}
&& G_1\quad=\quad \wp(gu_{\pi}[ z_2'',\vartheta'',Pre[\true,\vartheta,U],
   Rel_1[\true,\vartheta,U]]),\nonumber\\
&& G_2\quad=\quad \wp(gu_{\pi}[ z_3'',\vartheta'',Pre[\true,\vartheta,U],
   Rel_2[\true,\vartheta,U]]),\nonumber\\
&& R_1\quad=\quad (G_2\Or R'')^{\dagger},\nonumber\\
&& R_2\quad=\quad (G_1\Or R'')^{\dagger},\nonumber
\end{eqnarray}
Hence, the induction hypothesis, proposition \ref{histtwo:theo} on page
\pageref{histtwo:theo} and proposition \ref{histthree:theo} on page 
\pageref{histthree:theo} imply that
$ef_{\pi}[ z_1,\vartheta,P,R]$ is characterised by
\begin{eqnarray}
&&\Exists \overleftharpoon{\hbar[\vartheta]},\hbar[\vartheta]:\nonumber\\
&&\qquad \wp(ef_{\pi}[ z_2'',\vartheta'',P'',R_1])\And
\wp(ef_{\pi}[ z_3'',\vartheta'',P'',R_2]),
\nonumber
\end{eqnarray}
$wa_{\pi}[ z_1,\vartheta,P,R]$ is characterised by
\begin{eqnarray}
&&\Exists \hbar[\vartheta]:\nonumber\\
&&\qquad(\wp(wa_{\pi}[ z_2'',\vartheta'',P'',R_1])\And 
\wp(wa_{\pi}[ z_3'',\vartheta'',P'',R_2]))\ \Or\nonumber\\
&&\qquad(\wp(wa_{\pi}[ z_2'',\vartheta'',P'',R_1])\And 
\Exists\overleftharpoon{\vartheta''}:\wp(ef_{\pi}[ z_3'',\vartheta'',P'',R_2]))\ \Or\nonumber\\
&&\qquad(\wp(wa_{\pi}[ z_3'',\vartheta'',P'',R_2])\And 
\Exists\overleftharpoon{\vartheta''}:\wp(ef_{\pi}[ z_2'',\vartheta'',P'',R_1]))\nonumber
\end{eqnarray}
while $gu_{\pi}[ z_1,\vartheta,P,R]$ is characterised by
\begin{eqnarray}
&&\Exists \overleftharpoon{\hbar[\vartheta]}, \hbar[\vartheta]:\nonumber\\
&&\qquad \wp(gu_{\pi}[ z_2'',\vartheta'',P'',R_1])\Or 
\wp(gu_{\pi}[ z_3'',\vartheta'',P'',R_2]).
\nonumber
\end{eqnarray}

\item Await: Given a specified program 
\begin{eqnarray}
&&z_1\satis(\vartheta,\emptyset)::(P,R,W,G,E)\nonumber
\end{eqnarray}
of auxiliary form, a program $z_1'$ of the form 
\begin{eqnarray}
&&\nawait\ b\ \ndo\  z_2'\ \nod,\nonumber
\end{eqnarray}
such that $\models z_1\hookrightarrow z_1'$, and assume that
$\alpha$ is the set of auxiliary variables in $z_1$. This means that 
$z_1$ is of the form
\begin{eqnarray}
&&\nawait\ b\ \true\nonumber\\
&&\qquad z_2;\nonumber\\
&&\qquad a_1:=u_1;\nonumber\\
&&\qquad\quad\vdots\nonumber\\
&&\qquad a_m:=u_m\nonumber\\
&&\nod,\nonumber
\end{eqnarray}
where for all $1\le j,k\le m$, $var[u_j]\subseteq 
(\vartheta\setminus\alpha)\cup\{a_j\}$,
$a_j\in \alpha$, $j\neq k\Implies a_j\neq a_k$ and $\models z_2\hookrightarrow z_2'$.

By the induction hypothesis it follows that 
$ef_{\pi}[ z_1,\vartheta,P,R]$ is characterised by
\begin{eqnarray}
&& \overleftharpoon{P}\And R|\wp(ef_{\pi}[ z_2,\vartheta,P^R\And b,I])|(I_{\alpha}\And 
\bigwedge_{j=1}^m a_j=\overleftharpoon{u_j})|R,\nonumber
\end{eqnarray}
$wa_{\pi}[ z_1,\vartheta,P,R]$ is characterised by
\begin{eqnarray}
&&P^R\And \neg b,\nonumber
\end{eqnarray}
while $gu_{\pi}[ z_1,\vartheta,P,R]$ is characterised by
\begin{eqnarray}
&& \wp(ef_{\pi}[ z_2,\vartheta,P^R\And b,I])|(I_{\alpha}\And 
\bigwedge_{j=1}^m a_j=\overleftharpoon{u_j}).\nonumber
\end{eqnarray}

\end{itemize}

\end{nproof}

\section{Relative-Completeness Propositions}

\begin{statement}\label{basiccomp:theo}
Given that 
\begin{itemize}
\item $\models_{\pi} z\satis(\vartheta,\alpha)::(P,R,W,G,E)$,
\item $R$ respects $\vartheta\cap hid[ z]$,
\end{itemize}
then
\begin{itemize}
\item $Tr_{\pi}\vdash_B z\satis(\vartheta,\alpha)::(P,R,W,G,E)$.
\end{itemize}
\end{statement}

\begin{nproof} We will prove the proposition by structural induction on $z$.
The base-cases are the skip- and assignment-statements.

\begin{itemize}
\item Skip: Assume that 
\begin{eqnarray}
&&\models_{\pi}  \nskip\satis(\vartheta,\alpha):: (P,R,W,G,E)\nonumber.
\end{eqnarray}
The skip- and  pre-rules imply that
\begin{eqnarray}
&&Tr_{\pi}\vdash_B \nskip\satis(\vartheta,\alpha):: 
(P,R,W,G,\overleftharpoon{P}\And R).\label{cskip:two}
\end{eqnarray}
Moreover, since 
\begin{eqnarray}
&&ef_{\pi}[\nskip,\vartheta\cup\alpha,P,R]=\overleftharpoon{P}\And R,\nonumber
\end{eqnarray}
it follows that
\begin{eqnarray}
&&\models_{\pi}\overleftharpoon{P}\And R\Rightarrow E,\nonumber
\end{eqnarray}
in which case, \ref{cskip:two} and the consequence-rule give
\begin{eqnarray}
&&Tr_{\pi}\vdash_B \nskip\satis(\vartheta,\alpha):: (P,R,W,G,E).\nonumber
\end{eqnarray}

\item Assignment: Assume that $ z_1$ is of the form
\begin{eqnarray}
&&v:=r,\nonumber
\end{eqnarray}
and that
\begin{eqnarray}
&&\models_{\pi}  z_1\satis(\vartheta,\bigcup_{j=1}^m\{a_j\}):: (P,R,W,G,E).\nonumber
\end{eqnarray}

This means that there is a program $z_1'$ of the form
\begin{eqnarray}
&&\nawait\ \true\ \ndo\nonumber\\
&&\qquad a_1:=u_1;\nonumber\\
&&\qquad\quad\vdots\nonumber\\
&&\qquad a_m:=u_m;\nonumber\\
&&\qquad v:=r\nonumber\\
&&\nod\nonumber
\end{eqnarray}
such that
\begin{eqnarray}
&&\models  z_1'\stackrel{(\vartheta,\bigcup_{j=1}^m\{a_j\})}{\hookrightarrow} z_1,
\nonumber\\
&&\models_{\pi} z_1'\satis(\vartheta\cup\bigcup_{j=1}^m\{a_j\},\emptyset)::
(P,R,W,G,E),\label{cass:two}
\end{eqnarray}
where for all $1\le j,k\le m$, $var[u_j]\subseteq \vartheta\cup\{a_j\}$
and $j\neq k\Implies a_j\neq a_k$.

Moreover, since
\begin{eqnarray}
&&ef_{\pi}[ z_1',\vartheta\cup\bigcup_{j=1}^m\{a_j\},P,R]=\nonumber\\
&&\qquad\qquad \overleftharpoon{P}
\And R|(v=\overleftharpoon{r}\And I_{\{v\}\cup\bigcup_{j=1}^m\{a_j\}}
\And \bigwedge_{j=1}^m a_j=\overleftharpoon{u_j})|R,\nonumber\\
&& gu_{\pi}[ z_1',\vartheta\cup\bigcup_{j=1}^m\{a_j\},P,R]=\nonumber\\
&&\qquad\qquad \overleftharpoon{P^R}\And v=\overleftharpoon{r}
\And I_{\{v\}\cup\bigcup_{j=1}^m\{a_j\}}\And \bigwedge_{j=1}^m a_j=\overleftharpoon{u_j}
,\nonumber
\end{eqnarray}
it follows from \ref{cass:two} that there is an assertion $E_1$ such that
\begin{eqnarray}
&&\models_{\pi} \overleftharpoon{P^R}\And v=\overleftharpoon{r}\And I_{\{v\}\cup\bigcup_{j=1}^m\{a_j\}}\And 
\bigwedge_{j=1}^m a_j=\overleftharpoon{u_j}\Rightarrow G\And E_1,\nonumber\\
&&\models_{\pi}\overleftharpoon{P}\And R|E_1|R\Implies E,\nonumber
\end{eqnarray}
in which case 
\begin{eqnarray}
&&Tr_{\pi}\vdash_B  z_1\satis(\vartheta,\bigcup_{j=1}^m\{a_j\}):: (P,R,W,G,E).\nonumber
\end{eqnarray}
follows by the assignment-, consequence- and pre-rules.

\item Block:  Assume that $ z_1$ is of the form 
\begin{eqnarray}
&&\nbegin\ \loc\ v_1,\ \ldots\ ,v_m; z_2\ \nend,\nonumber
\end{eqnarray}
and that
\begin{eqnarray}
&&\models_{\pi} z_1\satis(\vartheta,\alpha):: (P,R,W,G,E).\nonumber
\end{eqnarray}

Since $\bigcup_{j=1}^m\{v_j\}\subseteq hid[z_1]$, it is clear that
\begin{eqnarray}
&&\models_{\pi} z_2\satis(\vartheta\cup\bigcup_{j=1}^m\{v_j\},\alpha):: 
(P,R\And\bigwedge_{j=1}^m v_j=\overleftharpoon{v_j},W,G,E).\nonumber
\end{eqnarray}
Moreover, since $R\And \bigwedge_{j=1}^m v_j=\overleftharpoon{v_j}$ respects $\vartheta\cap hid[z_2]$,
the induction hypothesis implies that
\begin{eqnarray}
&&Tr_{\pi}\vdash_B z_2\satis(\vartheta\cup \bigcup_{j=1}^m\{v_j\},\alpha):: 
(P,R\And\bigwedge_{j=1}^m v_j=\overleftharpoon{v_j},W,G,E),\nonumber
\end{eqnarray}
in which case 
\begin{eqnarray}
&&Tr_{\pi}\vdash_B z_1\satis(\vartheta,\alpha):: 
(P,R,W,G,E),\nonumber
\end{eqnarray}
follows by the block-rule.

\item Sequential: Assume that $z_1$ is of the form 
\begin{eqnarray}
&& z_2; z_3,\nonumber
\end{eqnarray}
and that
\begin{eqnarray}
&&\models_{\pi} z_1\satis(\vartheta,\alpha):: (P,R,W,G,E).\nonumber
\end{eqnarray}
This means that there is a program $z_1'$ of the form
\begin{eqnarray}
&& z_2';z_3',\nonumber
\end{eqnarray}
such that
\begin{eqnarray}
&& \models z_1'\progtran z_1,\nonumber\\
&& \models_{\pi}z_1'\satis(\vartheta\cup\alpha,\emptyset)::
(P,R,W,G,E).\label{cscomp:two}
\end{eqnarray}

It follows from proposition \ref{express:theo} on page \pageref{express:theo}
that there are assertions such that
\begin{eqnarray}
&& E_1=\wp(ef_{\pi}[ z_2',\vartheta\cup\alpha,P,R]),\nonumber\\
&& P_2=\Exists \overleftharpoon{\vartheta\cup\alpha}:\overleftharpoon{P}\And E_1,\nonumber\\
&& E_2=\wp(ef_{\pi}[ z_3',\vartheta\cup\alpha,P_2,R]),\nonumber
\end{eqnarray}
in which case \ref{cscomp:two} and
the induction hypothesis imply
\begin{eqnarray}
&&Tr_{\pi}\vdash_B z_2\satis(\vartheta,\alpha):: (P,R,W,G,E_1\And P_2),
\nonumber\\
&&Tr_{\pi}\vdash_B z_3\satis(\vartheta,\alpha):: (P_2,R,W,G,E_2).\nonumber
\end{eqnarray}
It is also clear that
\begin{eqnarray}
&&\models_{\pi}E_1|E_2\Rightarrow E\nonumber,\nonumber
\end{eqnarray}
in which case
\begin{eqnarray}
&&Tr_{\pi}\vdash_B z_1\satis(\vartheta,\emptyset):: (P,R,W,G,E),\nonumber
\end{eqnarray}
follows by the sequential- and consequence-rules.

\item If: Assume that $ z_1$ is of the form 
\begin{eqnarray}
&&\nif\ b\ \nthen\  z_2\ \nelse\  z_3\ \nfi,\nonumber
\end{eqnarray}
and that
\begin{eqnarray}
&&\models_{\pi} z_1\satis(\vartheta,\alpha):: (P,R,W,G,E).\nonumber
\end{eqnarray}
Then it is clear that
\begin{eqnarray}
&&\models_{\pi} z_2\satis(\vartheta,\alpha):: (P\And b,R,W,G,E), \nonumber\\
&&\models_{\pi} z_3\satis(\vartheta,\alpha):: (P\And\neg b,R,W,G,E), 
\nonumber
\end{eqnarray}
and it follows from the induction hypothesis that
\begin{eqnarray}
&&Tr_{\pi}\vdash_B z_2\satis(\vartheta,\alpha):: (P\And b,R,W,G,E), 
\nonumber\\
&&Tr_{\pi}\vdash_B z_3\satis(\vartheta,\alpha):: (P\And \neg b,R,W,G,E), \nonumber
\end{eqnarray}
which means that 
\begin{eqnarray}
&&Tr_{\pi}\vdash_B z_1\satis(\vartheta,\alpha):: (P,R,W,G,E)\nonumber
\end{eqnarray}
can be deduced by the if-rule.

\item While: Assume that $ z_1$ is of the form 
\begin{eqnarray}
&&\nwhile\ b\ \ndo\  z_2\ \nod,\nonumber
\end{eqnarray}
and that
\begin{eqnarray}
&&\models_{\pi} z_1\satis(\vartheta,\alpha):: (P,R,W,G,E).\nonumber
\end{eqnarray}

This means that there is a program $z_1'$ of the form
\begin{eqnarray}
&&\nwhile\ b\ \ndo\ z_2'\ \nod,\nonumber
\end{eqnarray}
such that
\begin{eqnarray}
&&\models  z_1'\progtran z_1,\nonumber\\
&&\models_{\pi} z_1'\satis(\vartheta\cup \alpha,\emptyset)::
(P,R,W,G,E).\label{cwhile:two}
\end{eqnarray}

It follows from proposition \ref{express:theo} on page \pageref{express:theo}
that there are assertions such that
\begin{eqnarray}
&&P'=\Exists \overleftharpoon{\vartheta\cup \alpha}:\overleftharpoon{P}\And \wp(ef_{\pi}[
 z_2',\vartheta\cup\alpha,b,R])^*,\nonumber\\
&&Z=\wp(ef_{\pi}[ z_2',\vartheta\cup\alpha,P'\And b,R]). \nonumber
\end{eqnarray}
Assume $Z$ is not well-founded. But then, there is a diverging computation
\begin{eqnarray}
&&\sigma\in ext_{\pi}[P,R]\cap cp_{\pi}[ z_1'].\nonumber
\end{eqnarray}
This contradicts \ref{cwhile:two}. Thus,
\begin{eqnarray}
&&\models_{\pi}\underline{\rm wf}\ Z. \label{cwhile:four}
\end{eqnarray}
Moreover, \ref{cwhile:two} and the induction hypothesis give
\begin{eqnarray}
&&Tr_{\pi}\vdash_B z_2\satis(\vartheta,\alpha):: (P',R,W,G,P'\And Z).
\nonumber
\end{eqnarray}
But then,
\begin{eqnarray}
&& Tr_{\pi}\vdash_B z_1\satis(\vartheta,\alpha)
:: (P',R,W,G,(Z^{\dagger}\Or R)\And \neg b)\nonumber
\end{eqnarray}
follows from \ref{cwhile:four} by the while-rule. The consequence- 
and pre-rules give
\begin{eqnarray}
&& Tr_{\pi}\vdash_B z_1\satis(\vartheta,\alpha):: 
(P,R,W,G,\overleftharpoon{P}\And (Z^{\dagger}\Or R)\And \neg b).\nonumber
\end{eqnarray}
Furthermore, it follows easily that
\begin{eqnarray}
&& \models_{\pi} \overleftharpoon{P}\And (Z^{\dagger}\Or R)\And \neg b\Implies E,\nonumber
\end{eqnarray}
which by the consequence-rule gives
\begin{eqnarray}
&& Tr_{\pi}\vdash_B z_1\satis(\vartheta,\alpha):: (P,R,W,G,E).\nonumber
\end{eqnarray}

\item Parallel: Assume that $ z_1$ is of the form
\begin{eqnarray}
&&\{ z_2\parallel z_3\}\nonumber
\end{eqnarray}
and that
\begin{eqnarray}
&&\models_{\pi} z_1\satis(\vartheta,\alpha):: (P,R,W,G,E).\nonumber
\end{eqnarray}

This means that there is a program $z_1'$ of the form
\begin{eqnarray}
&&\{ z_2'\parallel z_3'\}\nonumber
\end{eqnarray}
such that
\begin{eqnarray}
&& \models z_1'\progtran z_1,\nonumber\\
&& \models_{\pi} z_1'\satis(\vartheta\cup\alpha,\emptyset):: (P,R,W,G,E).\nonumber
\end{eqnarray}
This specified program is of auxiliary form, which means that it has
a historic form
\begin{eqnarray}
&& \models_{\pi} z_1''\satis(\vartheta\cup\alpha'',\emptyset):: (P'',R'',W,G,E)
\label{cpar:one}
\end{eqnarray}
where $z_1''$ is of the form
\begin{eqnarray}
&&\{ z_2''\parallel z_3''\}\nonumber
\end{eqnarray}
and 
\begin{eqnarray}
&& \models z_1''\stackrel{(\vartheta,\alpha'')}{\hookrightarrow} z_1.\nonumber
\end{eqnarray}

Let $U=var[z_1]\setminus\vartheta$,
it follows from proposition \ref{express:theo} on page \pageref{express:theo}
that there are assertions
such that
\begin{eqnarray}
&& G_1=\wp(gu_{\pi}[ z_2'',\vartheta\cup\alpha'',Pre[\true,\vartheta\cup\alpha,U],
Rel_1[\true,\vartheta\cup\alpha,U]),\nonumber\\
&& G_2=\wp(gu_{\pi}[ z_3'',\vartheta\cup\alpha'',Pre[\true,\vartheta\cup\alpha,U],
Rel_2[\true,\vartheta\cup\alpha,U]),\nonumber\\
&& R_1=(G_2\Or R'')^{\dagger},\nonumber\\
&& R_2=(G_1\Or R'')^{\dagger},\nonumber\\
&& E_1=\wp(ef_{\pi}[ z_2'',\vartheta\cup\alpha'',P'',R_1]),\nonumber\\
&& E_2=\wp(ef_{\pi}[ z_3'',\vartheta\cup\alpha'',P'',R_2]),\nonumber\\
&& W_1=\wp(wa_{\pi}[ z_2'',\vartheta\cup\alpha'',P'',R_1])\And \neg W,\nonumber\\
&& W_2=\wp(wa_{\pi}[ z_3'',\vartheta\cup\alpha'',P'',R_2])\And \neg W.\nonumber
\end{eqnarray}

It follows easily from \ref{cpar:one} and proposition \ref{histtwo:theo} on page 
\pageref{histtwo:theo} that
\begin{eqnarray}
&&\models_{\pi}\neg(W_1\And W_2)\And \neg(W_1\And E_2)\And
\neg(W_2\And E_1).\label{cpar:two}
\end{eqnarray}

Moreover, \ref{cpar:one} and proposition \ref{histfour:theo} on page
\pageref{histfour:theo} imply that any computation
\begin{eqnarray}
&&\sigma\in ext_{\pi}[P'',R_1]\cap cp_{\pi}[ z_2'']\nonumber
\end{eqnarray}
either deadlocks or terminates.
The same is of course true for any computation
\begin{eqnarray}
&&\sigma'\in ext_{\pi}[P'',R_2]\cap cp_{\pi}[ z_3''].\nonumber
\end{eqnarray}
By proposition \ref{histthree:theo} on page \pageref{histthree:theo}
and the induction hypothesis we get
\begin{eqnarray}
&&Tr_{\pi}\vdash_B z_2\satis(\vartheta,\alpha''):: 
(P'',R_1,W\Or W_1,G\And R_2,E_1),\nonumber\\
&&Tr_{\pi}\vdash_B z_3\satis(\vartheta,\alpha''):: 
(P'',R_2,W\Or W_2,G\And R_1,E_2),
\nonumber
\end{eqnarray}
which together with \ref{cpar:two} and the parallel-rule give
\begin{eqnarray}
&&Tr_{\pi}\vdash_B  z_1\satis(\vartheta,\alpha''):: 
(P'',R_1\And R_2,W,G,E_1\And E_2).\nonumber
\end{eqnarray}
Moreover, since
\begin{eqnarray}
&&\models_{\pi}R''\Rightarrow R_1\And R_2, \nonumber\\
&&\models_{\pi}E_1\And E_2\Rightarrow E,\nonumber
\end{eqnarray}
it follows by the consequence-rule that
\begin{eqnarray}
&&Tr_{\pi}\vdash_B  z_1\satis(\vartheta,\alpha'')::(P'',R'',W,G,E),\nonumber
\end{eqnarray}
in which case
\begin{eqnarray}
&&Tr_{\pi}\vdash_B  z_1\satis(\vartheta,\alpha)::(P,R,W,G,E),\nonumber
\end{eqnarray}
can be deduced by the elimination- and consequence-rules.

\item Await:

Assume that $ z_1$ is of the form
\begin{eqnarray}
&&\nawait\ b\ \ndo\  z_2\ \nod,\nonumber
\end{eqnarray}
and that
\begin{eqnarray}
&&\models_{\pi} z_1\satis(\vartheta,\bigcup_{j=1}^m\{a_j\}):: (P,R,W,G,E).\nonumber
\end{eqnarray}

This means that there is a program $z_1'$ of the form 
\begin{eqnarray}
&&\nawait\ b\ \ndo\nonumber\\
&&\qquad  z_2';\nonumber\\
&&\qquad a_1:=u_1;\nonumber\\
&&\qquad\quad\vdots\nonumber\\
&&\qquad a_m:=u_m\nonumber\\
&&\nod\nonumber
\end{eqnarray}
such that
\begin{eqnarray}
&&\models z_1'\stackrel{(\vartheta,\bigcup_{j=1}^m\{a_j\})}{\hookrightarrow} z_1,\nonumber\\
&&\models_{\pi} z_1'\satis(\vartheta\cup\bigcup_{j=1}^m\{a_j\},\emptyset):: 
(P,R,W,G,E),\label{cawait:ten}
\end{eqnarray}
where for all $1\le j,k\le m$, $var[u_j]\subseteq \vartheta\cup\{a_j\}$,
$j\neq k\Implies a_j\neq a_k$ and $\models z_2'\hookrightarrow z_2$.

It follows from proposition \ref{express:theo} on page \pageref{express:theo}
that there is an  assertion such that
\begin{eqnarray}
&&E_1=\wp(ef_{\pi}
[ z_2',\vartheta\cup\bigcup_{j=1}^m\{a_j\},P^R\And b,I]),\nonumber
\end{eqnarray}
Moreover, \ref{cawait:ten} implies that
\begin{eqnarray}
&&\models_{\pi}P^R\And \neg b\Rightarrow W,\label{cawait:two}
\end{eqnarray}
and together with the induction hypothesis also that
\begin{eqnarray}
&&Tr_{\pi}\vdash_B z_2\satis(\vartheta,\bigcup_{j=1}^m\{a_j\})::(P^R\And b,I,\false,\true,E_1).\label{cawait:fifteen}
\end{eqnarray}
Moreover, it is clear from \ref{cawait:ten} that there is an assertion $E_2$
such that
\begin{eqnarray}
&&\models_{\pi}E_1|(\bigwedge_{j=1}^m a_j=\overleftharpoon{u_j}\And I_{\bigcup_{j=1}^m\{a_j\}})
\Implies G\And E_2,\nonumber\\
&&\models_{\pi} \overleftharpoon{P}\And R|E_2|R\Implies E,
\label{cawait:twenty}
\end{eqnarray}
in which case
\begin{eqnarray}
&&Tr_{\pi}\vdash_B z_1\satis(\vartheta,\bigcup_{j=1}^m\{a_j\}):: (P,R,W,G,E)\nonumber
\end{eqnarray}
follows from \ref{cawait:two}, \ref{cawait:fifteen}, \ref{cawait:twenty}
by the await-, pre- and consequence-rules.
\end{itemize}

\end{nproof}

\begin{statement}
Given that
\begin{eqnarray}
&&\models_{\pi} z\satis(\vartheta,\alpha)::(P,R,W,G,E)\nonumber
\end{eqnarray}
then
\begin{eqnarray}
&&Tr_{\pi}\vdash_B z\satis(\vartheta,\alpha)::(P,R,W,G,E).\nonumber
\end{eqnarray}
\end{statement}

\begin{nproof}
Assume that
\begin{eqnarray}
&&\models_{\pi} z\satis(\vartheta,\alpha):: 
(P,R,W,G,E).\nonumber
\end{eqnarray}
It follows easily that
\begin{eqnarray}
&&\models_{\pi} z\satis(\vartheta,\alpha):: 
(P,R\And \bigwedge_{j=1}^m v_j=\overleftharpoon{v_j},W,G,E),\nonumber
\end{eqnarray}
where $\vartheta\cap hid[ z]=\bigcup_{j=1}^m\{v_j\}$.
But then, it is clear from proposition \ref{basiccomp:theo} on page \pageref{basiccomp:theo} that
\begin{eqnarray}
&&Tr_{\pi}\vdash_B z\satis(\vartheta,\alpha):: 
(P,R\And \bigwedge_{j=1}^m v_j=\overleftharpoon{v_j},W,G,E),\nonumber
\end{eqnarray}
in which case it follows by the access-rule that
\begin{eqnarray}
&&Tr_{\pi}\vdash_B z\satis(\vartheta,\alpha):: 
(P,R,W,G,E).\nonumber
\end{eqnarray}

\end{nproof}

%% file: alt.tex
\chapter{Discussion}

\section{Motivation}

The object of this chapter is first of all 
to motivate some of the design decisions
taken at different points above. It will be explained how things could have been done 
differently, and alternative approaches will be discussed and compared with the one 
chosen.

The reason why this has been postponed until now is that it is first at this stage
the
reader has an overview of the whole system. Thus it is only now that the reader
can fully evaluate and understand the consequences of the alternative
approaches.

It is also the object of this chapter to indicate weaknesses and areas for further 
research, and furthermore to compare LSP with some of the most closely 
related methods known from the literature.

\section{Without Hooked Variables}

\subsection{Two Representations}

In LSP the rely-, guar- and eff-conditions are binary assertions. This means that 
they may have free occurrences of hooked variables.

However, our approach
does not depend upon the use of binary assertions. LSP 
can be transformed into a logic where all assertions are unary. The choice between the
two representations can be seen as a matter of taste, although we believe that the 
binary assertions in many cases result in shorter and more readable
specifications.

\subsection{Unary Eff-Conditions}

Since the pre-condition characterises the initial state, and since the use of 
auxiliary variables is allowed, the eff-condition could just as well be unary. The
post-condition in Hoare-logic is for example unary.

One
advantage of such a modification is a simpler sequential-rule. 
Unfortunately, this change has the
opposite effect on some of the other rules; for example the while-rule. A unary
eff-condition would also have a complicating effect on certain specifications. 
Consider for example

\begin{flist}
(\{S\},\emptyset)::(\true,I,\false,\true,\#S=\#\overleftharpoon{S}+1),
\end{flist}

\noindent which restricts an implementation to increase the number of elements in
the set $S$ by one. If the eff-condition is constrained to be unary, it is necessary
to introduce an auxiliary variable to write an equivalent specification:

\begin{flist}
(\{S\},\{m\})::(\#S=m,I,\false,\true,\#S=m+1).
\end{flist}

\subsection{Rely and Guar as Assertion Sets}

In \cite{cs:note} rely- and guar-conditions are represented as sets of assertions.
Given a structure $\pi$, the basic idea is that for any set of
assertions $\Delta$, there is
a set of state changes that are invariant with respect to it, namely the set of
all pairs of states $(s_1,s_2)$, such that for all assertions $A\in \Delta$:

\begin{itemize}
\item $s_1\models_{\pi} A$ implies $s_2\models_{\pi} A$.
\end{itemize}

\noindent To compare this approach with ours, consider

\begin{flist}
x=\overleftharpoon{x}\And y\ge \overleftharpoon{y},
\end{flist}

\noindent and assume the variables $x$ and $y$ are respectively of the sorts $\Nat$ and $\Real$.
To express 
the first conjunct in the notation of \cite{cs:note},
it is enough to constrain the environment to maintain
any formula of the form $x=n$, while the set of
all assertions of the form $y\ge r$ characterises the second. 
Thus, the binary assertion above can
be transformed into the following set of unary assertions:

\begin{flist}
\{x=n|n\in\Nat\}\cup \{y\ge r|r\in\Real\}.
\end{flist}

\section{Separate Pre-Conditions}

It may be argued that because the eff-condition is binary, a separate 
pre-condition is not 
really needed (see \cite{ecrh:pre}). 
The reason is of course that the eff-condition can be redefined
to constrain both the set of initial states and the overall effect. Given such a
formalism we may write

\begin{flist}
(\vartheta,\alpha)::(R,W,G,\overleftharpoon{P}\And E)
\end{flist}

\noindent instead of

\begin{flist}
(\vartheta,\alpha)::(P,R,W,G,E).
\end{flist}

\noindent Nevertheless, we have decided to use five assertions.
There
are two reasons for that:
\begin{itemize}
\item A specification with a separate pre-condition is
more readable.
\item The pre-condition is a useful tool when formulating the program 
decomposition rules. Alternatively, we could have introduced a new 
syntactic operator
on assertions to characterise the domain of the eff-condition. This is for example the
view taken in Z \cite{jms:Z}.
\end{itemize}

\section{Scope of Eff-Conditions}

\subsection{Three Alternatives}

In LSP the eff-condition not only characterises the overall 
effect of the specified program, but
also any change due to the environment both 
before the first internal transition and after the last. This means that interference 
both before the implementation starts up and after it has terminated is included
in the eff-condition. A similar view was taken
in \cite{cbj:thesis}.
However, there are some obvious alternatives to this:
\begin{itemize}
\item The eff-condition characterises the overall state transition from
the initial state to immediately after the last internal transition.

\item The eff-condition characterises the overall state transition from 
immediately before the first internal transition to the final state. (Remember that
a terminating program has only finite computations with respect to a
convergent environment.)

\item The eff-condition characterises the overall state transition from
immediately before the first internal transition to immediately after the
last internal transition.
\end{itemize}

\noindent In the first case the eff-condition is only required to be true
immediately after the implementation terminates and does not have to be preserved by
the environment.

In the second case no interference before the first internal transition 
is included in the eff-condition, while interference after the last is.

In the third only interference 
which occurs after the first internal transition and before the last 
influences the eff-condition.

\subsection{Stirling}

The first position is taken by Stirling in \cite{cs:note}. 
However, his approach is not
completely consistent, because the rely-condition of a program whose
main construct is the parallel-statement is required to 
preserve the post-condition.
For example, it is possible to prove that the program
$b:=\false$ satisfies:

\begin{oldtuple}{a\And b
            }{\{b,\neg b,b\Or c\Implies a\}
            }{\{c,\neg c,b\Or c\Implies a\}
            }{a\And \neg b,}
\end{oldtuple}

\noindent and that the program $c:=\false$ satisfies:

\begin{oldtuple}{a\And c
            }{\{c,\neg c,b\Or c\Implies a\}
            }{\{b,\neg b,b\Or c\Implies a\}
            }{a\And \neg c.}
\end{oldtuple}

\noindent However, because Stirling's parallel-rule 
insists that the two post-conditions are preserved by their 
respective rely-conditions, it is as far as we can see only possible to prove that
$\{b:=\false\parallel c:=\false\}$ satisfies

\begin{oldtuple}{a\And b\And c
            }{\{c,\neg c,b,\neg b,b\Or c\Implies a\}
            }{\emptyset
            }{\neg b\And \neg c,}
\end{oldtuple}

\noindent and not that the same program satisfies

\begin{oldtuple}{a\And b\And c
            }{\{c,\neg c,b,\neg b,b\Or c\Implies a\}
            }{\emptyset
            }{\neg b\And \neg c\And a,}
\end{oldtuple}

\noindent which is what one would have expected. 
To achieve this a more sophisticated parallel-rule is needed. 

\subsection{First Alternative Approach}

If we change the semantics of LSP in such a way that interference after
the last internal transition is not included in the eff-condition, 
the assignment-rule can be slightly simplified. The
reason is that the second occurrence of $R$ in the conclusions eff-condition
can be removed. Moreover, the await-rule can be changed in a similar way.
Unfortunately, the parallel-rule becomes more complicated:

\[\begin{array}{l}
\neg(W_1\And E_2|R_2)\And \neg(W_2\And E_1|R_1)\And \neg(W_1\And W_2)\\
z_1\satis (\vartheta,\alpha)::(P,R_1,W\Or W_1,G\And R_2,E_1)\\
z_2\satis(\vartheta,\alpha)::(P,R_2,W\Or W_2,G\And R_1,E_2)\\
\overline{\{z_1\parallel z_2\}\satis(\vartheta,\alpha)::
(P,R_1\And R_2,W,G,
(E_1|R_1\And E_2)\Or (E_1\And E_2|R_2))}
\end{array}\]

\noindent Redefining the eff-condition in this way can therefore not be said to result in any
overall simplification of the decomposition rules.

It may be argued that our approach, where the eff-condition covers interference 
both before the first and after the last internal transition, may lead to extra 
proof-work
when two statements are composed in sequence. The reason is that the interference
after the last internal transition of the first statement and before the first
internal transition of the second statement must be included in the eff-conditions of
both statements.

But then, what about pre-condition preservation?
If the eff-condition of the first statement has been determined, and it
is known that this eff-condition is preserved by the rely-condition, 
it is enough to prove
that the first statement's eff-condition implies the second statement's pre-condition
to make sure that even the latter assertion is preserved by the rely-condition.

Otherwise it would have been necessary firstly to prove that the first 
statement's eff-condition
implies the second statement's pre-condition, and then show that this pre-condition
is preserved by the rely-condition.

Moreover, the first alternative approach certainly results in more proof-work 
when it comes to parallel composition.

\subsection{Second Alternative Approach}

The arguments for and against this approach are similar to
the previous case.

\subsection{Third Alternative Approach}

In this case, the sequential-rule must be changed to

\[\begin{array}{l}
\overleftharpoon{P_2}\And R\Implies P_2\\
z_1\satis(\vartheta,\alpha)::(P_1,R,W,G,P_2\And E_1)\\
z_2\satis(\vartheta,\alpha)::({P_2}^R,R,W,G,E_2)\\
\overline{z_1;z_2\satis(\vartheta,\alpha)::(P_1,R,W,G,E_1|R|E_2).}
\end{array}\]

\noindent The occurrence of $R$ in the conclusion's eff-condition is needed
to cover interference between the last internal transition of the first statement
and the first internal transition of the second. 

Unfortunately, some of the other rules like the while-rule
and the parallel-rule become much more complicated, and the
approach has been rejected for that reason.

\section{Multi-State Assertions}

\subsection{Accessing the Initial State}

As pointed out in \cite{dg:note}, in some cases it is useful 
to refer to the initial state in
the rely- or guar-conditions in the same way that we can use hooked variables
to refer to the initial state in the eff-condition. The same is of course true for
the wait-condition.

For example, if primed\footnote{The combination of primes and hooks is not good.
The introduction of three- or four-state assertions would require
a better naming convention.} variables in the wait-, rely- and guar-conditions
refer to the initial state, then the specification

\begin{tuple}{\{S\}}{\emptyset
            }{\true
            }{S=\overleftharpoon{S}\Or (\#\overleftharpoon{S}=\#S' -  1\And \overleftharpoon{S}\subset S\And \#S=\#S')
            }{\#S=\#S' -  1
            }{S=\overleftharpoon{S}\Or (\#\overleftharpoon{S}=\#S'\And S\subset \overleftharpoon{S}\And \#S=\#S' -  1)
            }{\#S=\#\overleftharpoon{S},}
\end{tuple}

\noindent where $\#S$ denotes the number of elements in the set $S$,
describes a program which either deadlocks in a state such that the size of
$S$ has been reduced with one, or repeats the following procedure a finite number 
of times:
\begin{itemize}
\item Reduce the size of $S$ by one and wait until an element
has been added.
\end{itemize}

\subsection{Accessing the Final State}

If primed variables are employed to let the rely-, guar- and wait-conditions refer
to a computation's initial state, why not introduce a similar convention to permit
the same assertions to refer to the final state too? (Remember that a terminating
program has only finite computations with respect to a convergent
environment.) If for example double-primed variables refer to the final state, then 
an implementation of 

\begin{tuple}{\{S\}}{\emptyset
            }{\true
            }{S=\overleftharpoon{S}
            }{\false
            }{\#S\ge\#S''
            }{\true}
\end{tuple}

\noindent can only terminate in a state in which the number of elements in $S$ is less than
or equal to the sets size at any previous state.

\subsection{Where to Stop?}

It does not have to stop with assertions accessing four states. On some
occasions it could be useful to refer to any state between the current
state and the initial state. One might then for example express that a specific
action can only take place if a particular flag has not been switched off between
the initial state and the current state.

So the question is not only, should we allow the different assertions to refer to
the initial state, the final state, etc.\ , but also --- where shall we say `stop'?

\subsection{Conclusion}

It may be true that the introduction of three- and four-state assertions has
a simplifying effect on some specifications. 
The examples above seem to confirm this.

Such a restricted use of multi-state \index{multi-state assertions} 
assertions will not give us
the necessary expressive power to get rid of auxiliary variables altogether,
but this is of course no argument against employing them to
simplify specifications, since binary assertions are used for the same
purpose.

The reason, why we have decided not to introduce assertions of more
than two states, is the complicating effect they have on the decomposition-rules. 
Try for example to formulate the sequential-rule when the rely-, wait-
and guar-conditions can refer to both the initial and the final states.

\section{Atomicity}

\label{atomic:ref}
Both expressions and assignments have so far been required to be atomic. As 
explained above (see page \ref{overlap:ref}) this does not mean that the
execution of two expressions cannot overlap in time, but only that they
behave as if they were executed in sequential order.

Some may argue that since LSP constrains the environment with a rely-condition, we
can do better than that; more precisely it should be possible to reason on the level
of memory reference without any (other) atomicity constraint.

Unfortunately, this is more difficult than it sounds. First of all, it would lead to
to a more complicated semantics, and the change would certainly have a
similar effect on some of the decomposition-rules. But this is of course only what one
could expect. 

However, the main argument against such a modification is that it would necessitate
a fundamental change in the way we are dealing with auxiliary variables, because it
would no longer be possible to place an assignment-statement inside the body 
of an await-statement without changing the behaviour of the algorithm.

\section{Boolean Tests of If and While}

\label{aux:ref}

In the definition of our programming language the Boolean tests of if- and 
while-statements are prohibited from accessing global variables. As explained 
earlier (see page \pageref{booltest:text}) this constraint does not reduce the set of
possible algorithms, but has the obvious disadvantage of increasing the length of
programs. However, this is in our opinion outweighed by simpler
decomposition-rules, and more importantly, that it is easier to reason
with auxiliary variables.

When reasoning with auxiliary variables it is often of great importance that
the auxiliary structure is updated in the same internal transition as the
global structure is updated or read.

This is achieved in the case of the assignment-statement, because due to the
assignment-rule, the execution of an assignment-statement of the form

\begin{prg}
v:=r\\
\end{prg}

\noindent actually corresponds to the execution of a statement of the form

\begin{prg}
\nawait\ b\ \ndo\\
\qquad a_1:=u_1;\\
\qquad\quad\vdots\\
\qquad a_m:=u_m;\\
\qquad v:=r\\
\nod.\\
\end{prg}

\noindent Similarly, due to the await-rule the execution of an await-statement of the form

\begin{prg}
\nawait\ b\ \ndo\ z\ \nod\\
\end{prg}

\noindent corresponds to the execution of a statement of the form

\begin{prg}
\nawait\ b\ \ndo\\
\qquad z;\\
\qquad a_1:=u_1;\\
\qquad\qquad\vdots\\
\qquad a_m:=u_m\\
\nod.\\
\end{prg}

\noindent Moreover, due to the constraint on the Boolean tests of if- and while-statements, 
this is also achieved in their cases, since the execution of 
a statement of the form

\begin{prg}
\nbegin\\
\qquad \loc\ v;\\
\qquad v:=b;\\
\qquad \nif\ v\ \nthen\ z_1\ \nelse\ z_2\ \nfi\\
\nend\\
\end{prg}

\noindent corresponds to the execution of a statement of the form

\begin{prg}
\nbegin\\
\qquad \loc\ v;\\
\qquad \nawait\ \true\ \ndo\\
\qquad\qquad a_1:=v_1;\\
\qquad\qquad\quad \vdots\\
\qquad\qquad a_m:=v_m;\\
\qquad\qquad v:=b\\
\qquad \nod;\\
\qquad \nif\ v\ \nthen\ z_1'\ \nelse\ z_2'\ \nfi\\
\nend,\\
\end{prg}

\noindent where $z_1'$ and $z_2'$ are the results of adding auxiliary structure to
respectively $z_1$ and $z_2$, while the execution of a statement of the form

\begin{prg}
\nbegin\\
\qquad \loc\ v;\\
\qquad v:=b;\\
\qquad \nwhile\ v\ \ndo\ z; v:=b\ \nod\\
\nend\\
\end{prg}

\noindent corresponds to the execution of a statement of the form

\begin{prg}
\nbegin\\
\qquad \loc\ v;\\
\qquad \nawait\ \true\ \ndo\\
\qquad\qquad a_1:=v_1;\\
\qquad\qquad\quad \vdots\\
\qquad\qquad a_m:=v_m;\\
\qquad\qquad v:=b\\
\qquad \nod;\\
\qquad \nwhile\ v\ \ndo\\
\qquad\qquad z';\\
\qquad\qquad \nawait\ \true\ \ndo\\
\qquad\qquad\qquad a_1:=v_1;\\
\qquad\qquad\qquad\quad \vdots\\
\qquad\qquad\qquad a_m:=v_m;\\
\qquad\qquad\qquad v:=b\\
\qquad\qquad\nod\\
\qquad \nod\\
\nend,\\
\end{prg}

\noindent where $z'$ is the result of adding auxiliary structure to $z$.
The same is of course not true for if- and while-statements if the truth
values of their Boolean
tests are not maintained by the environment. In that case, one possibility is
to allow auxiliary structure of the form
\begin{eqnarray}
<ite>&::=&\nif<tst>:<prg>\nthen <prg>\nelse <prg> \nfi,\nonumber\\
<wd>&::=&\nwhile<tst>:<prg>\ndo <prg>\nod,\nonumber
\end{eqnarray}
where $<tst>:<prg>$ is a pair of a Boolean test and a program which is assumed to 
be executed in isolation; in other words, as one internal transition. 
However, this would lead to a more complicated and less intuitive semantics.

\section{Fairness}

\subsection{Introduction}

\label{fairness:ref}
When discussing fairness, it is usual to distinguish between 
weak fairness\index{weak fairness}; that an event will not be infinitely postponed provided that it 
remains continuously enabled, and strong fairness\index{strong fairness}; that an event
will not be infinitely postponed provided that it is enabled
infinitely often.
See \cite{nf:fairness} for a more detailed discussion.

The programming language discussed in this thesis is unfair. This means that
LSP is not complete with respect to a weakly-fair programming language, and even less so
if the language is strongly fair.

For example, as
explained earlier, if the language is unfair the parallel composition of the 
two programs

\begin{prg}
b:=\false,\\
\end{prg}

\noindent and

\begin{prg}
\nbegin\\
\qquad\loc\ v;\\
\qquad v:=b;\\
\qquad \nwhile\ v\ \ndo\\
\qquad\qquad x:=x+1;\\
\qquad\qquad v:=b\\
\qquad \nod\\
\nend\\
\end{prg}

\noindent is not guaranteed to terminate because the first may be 
infinitely overtaken by the second. In other words, LSP cannot be used to prove
properties of programs whose algorithms depend upon busy waiting.

\subsection{Two Alternative Systems}

Nevertheless, we believe that it is possible to transform the present system
into two new systems; one which deals with weak fairness, and another 
which is specially designed to handle
strong fairness. The decomposition-rules for the two systems
are closely related to the
rules given in LSP. 

The basic idea is that although a busy-waiting process never becomes blocked, it is
actually waiting for the environment to release it. Thus instead of only
using the wait-condition to characterise the set of states in which
the implementation may become blocked, we will also use the wait-condition to describe
the set of states in which a busy process needs help from its environment
to be released.

\section{Data Reification}

\subsection{Introduction}

Program development is often divided into two rather separate subfields:
\begin{itemize}
\item program decomposition,
\item data reification \index{data reification} (also called data refinement).
\end{itemize}
The first topic has been discussed in detail above, but so far nothing has 
been said about
data reification. 

The earliest suggestions for a reification method were given in  
\cite{rm:reification}, \cite{carh:reification} and \cite{cbj:reification}. 
These ideas have since then
been further developed and discussed by many authors, see for example 
\cite{cbj:80}, \cite{jh:refined}, \cite{tn:reification}, \cite{ews:ref} and
\cite{ma:reification}.

\subsection{Jones' System}

Jones did not give any reification-rule in \cite{cbj:thesis}, but stated informally
that in the same way as the VDM-rule allows the pre-condition to be weakened
and the post-condition to be strengthened, it is sound to weaken the pre- and
rely-conditions and strengthen the guar- and eff-conditions.

This rule is not complete.
The incompleteness has been discussed by several authors (see
\cite{dg:note}, \cite{jw:VDM+} and \cite{qx:note}), but so far no complete
rule has been proposed.  A reification-rule for
rely-guarantee specifications is suggested in \cite{pg:reification}, 
but as pointed out in that paper; the method is incomplete and deals only with
safety properties.

\subsection{LSP}

The reification-rule indicated by Jones can be generalised to LSP in an obvious way; 
by insisting that the wait-condition is also strengthened. How to
formulate a complete rule is an open question.

\section{Nondeterminacy}

\subsection{Introduction}

\label{nondet:ref}
So far only deterministic program constructs have been considered (although the 
parallel-statement gives rise to nondeterministic behaviour). By introducing two new
statements, selection \index{nondeterministic selection} ($<sl>$) and repetition 
\index{nondeterministic repetition} ($<rp>$):
\begin{eqnarray}
<sl>&::=&\{<gl>\},\nonumber\\
<rp>&::=&\ast \{<gl>\},\nonumber\\
<gl>&::=&<tst>\rightarrow<stm>|<tst>\rightarrow<stm>\sqcap<gl>,\nonumber
\end{eqnarray}
we will show how LSP can
be generalised to deal with 
nondeterminism in the style of Dijkstra \cite{ewd:guar}.

\subsection{Selection}

We will use

\begin{prg}
\{\sqcap_{j\in \{1,\ \ldots\ ,m\}} b_j\rightarrow z_j\}\\
\end{prg}

\noindent as short for

\begin{prg}
\{b_1\rightarrow z_1\sqcap\ \ldots\ \sqcap b_m\rightarrow z_m\}.\\
\end{prg}

\noindent For each $j$, $b_j\rightarrow z_j$ is
called a guarded command, while the Boolean test $b_j$ is referred to as a guard.
Basically the selection-statement selects a guarded command with a true guard
and executes its body. If no guard is true when the selection
takes place, the statement aborts. 

\label{abort:ref}
The problem with modeling this in the operational semantics is that
the statement may abort. Fortunately, since we are only concerned with converging 
programs, this may be dealt with in the same way as nonterminating await-bodies
were dealt with above; namely by employing computations with infinitely many
internal identity steps:

\begin{itemize}
\item $<\{\sqcap_{j\in A} b_j\rightarrow z_j\},s>\intern < z_k,s>$ if $s
\models_{\pi}b_k$ and $k\in A$,
\item $<\{\sqcap_{j\in A} b_j\rightarrow z_j\},s>\intern <\{\sqcap_{j\in A} 
b_j\rightarrow z_j\},s>$ if $s\models_{\pi}\bigwedge_{j\in A}\neg b_j$,
\end{itemize}

\noindent To see that this has the desired effect, let $z$ denote the program

\begin{prg}
\{x=1\rightarrow x:=10\sqcap x=2\rightarrow x:=10\}.\\
\end{prg}

\noindent Clearly if $s(x)=0$, then there is an infinite computation 
$\sigma$ of the form 

\begin{flist}
<z,s>\intern<z,s>\intern\ \ldots\ \intern<z,s>\intern\ \ldots\ ,
\end{flist}

\noindent Moreover, $\sigma$ is an element of $ext_{\pi}[\true,\true]$, but
not an element of 

\begin{flist}
int_{\pi}[\false,\true,\true]. 
\end{flist}

\noindent Thus

\begin{flist}
z\satis (\{x\},\emptyset)::(\true,\true,\false,\true,\true)
\end{flist}

\noindent is not valid.

\subsection{Repetition}

In the same way as above

\begin{prg}
\ast \{\sqcap_{j\in \{1\ ,\ldots,\ m\}} b_j\rightarrow z_j\}\\
\end{prg}

\noindent is short for

\begin{prg}
\ast \{b_1\rightarrow z_1\sqcap\ \ldots\ \sqcap b_m\rightarrow z_m\}.\\
\end{prg}

\noindent Basically, this statement can be viewed as a loop which iteratively
executes the selection-statement $\{\sqcap_{j\in A} b_j\rightarrow z_j\}$ until
the selection of the guarded command takes place in a state
where none of the guards is true, in which case the statement
terminates normally. 
This can be described in the operational semantics as follows:

\begin{itemize}
\item $<\ast \{\sqcap_{j\in A} b_j\rightarrow z_j\},s>\intern < z_k;\ast \{\sqcap_{j\in A} 
b_j\rightarrow z_j\},s>$ if 
$s\models_{\pi} b_k$ and $k\in A$,
\item $<\ast \{\sqcap_{j\in A} b_j\rightarrow z_j\},s>\intern <\epsilon,s>$ if 
$s\models_{\pi}\bigwedge_{j\in A}\neg b_j$.
\end{itemize}

\subsection{Decomposition-Rules}

As in the case of the if- and while-statements, the environment
is required to preserve the truth-values of the different guards. This means
the following two rules are sufficient:

\begin{flist}
P\Implies \bigvee_{j\in A} b_j\\
 z_j \satis  (\vartheta,\alpha):: (P\And b_j,R,W,G,E)_{j\in A}\\
\overline{ \{\sqcap_{j\in A} b_j\rightarrow z_j\} \satis  (\vartheta,\alpha)
:: (P,R,W,G,E)}
\end{flist}

\begin{flist}
{\rm\underline{wf}}\ Z\\
 z_j  \satis  (\vartheta,\alpha):: (P\And b_j,R,W,G,P\And Z)_{j\in A}\\
\overline{\ast \{\sqcap_{j\in A} b_j\rightarrow z_j\} \satis  (\vartheta,\alpha) 
:: (P,R,W,G,(Z^{\dagger}\Or  R)\And \bigwedge_{j\in A}\neg b_j)}
\end{flist}

\noindent The only `new' with respect to the if- and while-rules 
is that the selection-rule needs an extra premise to ensure that
at least one guard is true when the selection takes place.

\section{Partial Functions}

\subsection{Introduction}

\label{part:ref}
So far it has been assumed that all functions are total. This constraint is 
too strong if we want to apply LSP to `real' problems. Division
by 0 is for example one thing which can lead to program abortion.
It should therefore not be possible to
prove that

\begin{flist}
x:=x/x\satis(\{x\},\emptyset)::(\true,x=\overleftharpoon{x},\false,\true,\true).
\end{flist}

\noindent The reason is of course that the program aborts if the initial value of $x$ is
0. If we use LSP as it is, and ignore that all functions are required to
be total, this specified program is actually `provable'. Hence, to handle 
partial functions some modifications are necessary.

Several logics have been suggested to deal with partial functions. 
We will employ LPF \cite{hb:LPF} here, but this does not mean that we 
cannot give similar rules with respect to other `partial' logics, like
for example the `weak logic' described in \cite{oo:weak} or LCF \cite{mg:LCF}.
A discussion of LPF with respect to its usability in program
development can be found in \cite{jhc:comp}. 

\subsection{LPF}

In LPF the propositional operators are given the strongest monotone 
extensions of their two-valued interpretations, while quantifiers range only
over the `proper' values of their bound sets. The resulting logic employs only
the normal collection of operators and all theorems of LPF hold
in classical first-order predicate calculus as well. 
However, the opposite is not true;
because the `law of the excluded middle' no longer is valid, there are many
truths of classical first-order logic that do not hold in LPF.

We will use $\delta(n)$ to mean a Boolean expression which is true if and only if
$n$ denotes a proper element of its sort. For example, if $b$ is an expression
of sort $\Bool$, then $\delta(b)$ is equivalent to

\begin{flist}
b\Or \neg b.
\end{flist}

\noindent Observe, that $\delta(b)$ is undefined if $b$ is undefined. At semantic level, we
will use the nonmonotonic operator $\Delta$ to express if an expression is
undefined or not. This means that $\Delta(b)$ is true if $b$ is true 
or false; otherwise false.

\subsection{Changes to the Operational Semantics}

Because the right-hand expression of the assignment-statement, and the Boolean tests
of the if-, while- and await-statements may be undefined, it is necessary to
change the operational semantics of these statements. 
Abortion will be modeled in the same way as
in the case of the selection-statement:
\begin{itemize}
\item $<v:=r,s>\intern <\epsilon,s(^v_r)>$ if $s\models_{\pi} \Delta(r)$, 
and $s(^v_r)$ denotes
the state that
is obtained from $s$, by mapping the variable $v$ to the value of the term $r$,
determined by $\pi$ and $s$, and leaving
all other maplets unchanged,
\item $<v:=r,s>\intern <v:=r,s>$ if $s\models_{\pi}\neg \Delta(r)$,
\item $<\nif\ b\ \nthen\  z_1\ \nelse\  z_2\ \nfi,s>\intern < z_1,s>$ if 
$s\models_{\pi}b\And \Delta(b)$,
\item $<\nif\ b\ \nthen\  z_1\ \nelse\  z_2\ \nfi,s>\intern < z_2,s>$ if $s
\models_{\pi}\neg b\And \Delta(b)$,
\item $<\nif\ b\ \nthen\  z_1\ \nelse\  z_2\ \nfi,s>\intern <\nif\ b\ \nthen\  z_1\ \nelse\  z_2\ \nfi,s>$ if $s\models_{\pi}\neg\Delta(b)$,
\item $<\nwhile\ b\ \ndo\  z_1\ \nod,s>\intern < z_1;\nwhile\ b\ \ndo\  z_1\ 
\nod,s>$ if 
$s\models_{\pi} b\And \Delta(b)$,
\item $<\nwhile\ b\ \ndo\  z_1\ \nod,s>\intern <\epsilon,s>$ if $s\models_{\pi}
\neg b\And \Delta(b)$,
\item $<\nwhile\ b\ \ndo\  z_1\ \nod,s>\intern <\nwhile\ b\ \ndo\  z_1\ \nod,s>$ 
if $s\models_{\pi}\neg \Delta(b)$,
\item $<\nawait\ b\ \ndo\  z_1\ \nod,s_1>\intern < z_n,s_n>$ 
if $s_1\models_{\pi} b\And \Delta(b)$,
and 

\begin{itemize}
\item there is a list of configurations $< z_2,s_2>,
< z_3,s_3>,\ \ldots\ ,< z_{n - 
1},s_{n -  1}>$, such that for all $1<k\le n$,
$< z_{k -  1},s_{k -  1}>\intern < z_k,s_k>$ and
$ z_n=\epsilon$, 
\end{itemize}

\item $<\nawait\ b\ \ndo\  z_1\ \nod,s_1>\intern <\nawait\ b\ \ndo\  z_1\ \nod,s_1>$
if $s_1\models_{\pi}\neg\Delta(b)$, or
if $s_1\models_{\pi} b$ and 

\begin{itemize}
\item there is an infinite list of configurations $< z_2,s_2>,
< z_3,s_3>,\ \ldots\ ,< z_n,s_n>,\ \ldots\ $, such that for all $k> 1$, 
$< z_{k -  1},s_{k -  1}>\intern< z_k,s_k>$, or
\item there is a finite list of configurations $< z_2,s_2>,
< z_3,s_3>,\ \ldots\ ,< z_n,s_n>$, where $ z_n\neq \epsilon$, there is no
configuration $< z_{n+1},s_{n+1}>$ such that 
$< z_n,s_n>\intern< z_{n+1},s_{n+1}>$, and 
for all $1<k\le n$, 
$< z_{k -  1},s_{k -  1}>\intern< z_k,s_k>$.
\end{itemize}
\end{itemize}

\subsection{Modified Rules}

Since the assignment-statement aborts if the assigned values are undefined, it is
necessary to add an extra premise to the await-rule to make this impossible:

\begin{flist}
P^R\Implies \delta(r)\And\bigwedge_{a\in\alpha}\delta(u_a)\\
\underline{\overleftharpoon{P^R}\And v=\overleftharpoon{r}\And I_{\{v\}\cup\alpha}\And 
\bigwedge_{a\in\alpha}a=\overleftharpoon{u_a}\Rightarrow G\And E}\\
v:=r \satis  (\vartheta,\alpha):: (P,R,W,G,R|E|R)
\end{flist}

\noindent The if-statement aborts if the Boolean test is undefined when
it is evaluated. Since the truth-value of the 
Boolean test cannot be changed by the environment, it is
enough to give the conclusion's pre-condition an extra conjunct:

\begin{flist}
 z_1 \satis  (\vartheta,\alpha):: (P\And b,R,W,G,E)\\
 z_2 \satis  (\vartheta,\alpha):: (P\And \neg b,R,W,G,E)\\
\overline{\nif\ b\ \nthen\  z_1\ \nelse\  z_2\ \nfi \satis  (\vartheta,\alpha)
:: (P\And \delta(b),R,W,G,E)}
\end{flist}

\noindent In the case of the while-statement, the Boolean test may be evaluated any number of
times. Thus the second premise must be changed to make sure that if the
Boolean test evaluates to true, then it is also defined the next time it is
evaluated, in which case it is enough to strengthen the conclusion's pre-condition
in the same way as above:

\begin{flist}
{\rm\underline{wf}}\ Z\\
 z  \satis  (\vartheta,\alpha):: (P\And b,R,W,G,P\And \delta(b)\And Z)\\
\overline{\nwhile\ b\ \ndo\  z \ \nod \satis  (\vartheta,\alpha) 
:: (P\And \delta(b),R,W,G,(Z^{\dagger}\Or  R)\And \neg b)}
\end{flist}

\noindent It is a bit more difficult to formulate the await-rule:

\begin{flist}
P^R\Implies (W\And \neg b)\Or b\\
E_1|(\bigwedge_{a\in\alpha}a=\overleftharpoon{u_a}\And I_{\alpha})\Implies G\And E_2\\
 z  \satis  (\vartheta,\alpha):: (P^R\And b,I,\false,\true,E_1\And
\bigwedge_{a\in\alpha}\delta(u_a))\\
\overline{\nawait\ b\ \ndo\  z \ \nod \satis  (\vartheta,\alpha):: 
(P,R,W,G,R|E_2|R)}
\end{flist}

\noindent The first premise ensures that the statement can only become blocked in a state
which satisfies $W$, and that the Boolean test is defined when it is evaluated.
The third premise guarantees that the expressions assigned to the auxiliary
variables are defined.

\section{CSP}

In this thesis we are only concerned with shared-state concurrency. This does 
not mean that similar strategies cannot be employed to develop 
communication-based programs.

Levin and Gries have suggested
a proof-system \cite{gml:csp} which is closely related to 
the method
for shared-state concurrency described in \cite{so:await}. 
Since LSP can be seen as
a compositional version of the latter, it should be fairly obvious that our system
can be transformed into a compositional version of the Levin/Gries approach.

\section{Without Reflexivity and Transitivity Constraints}

\label{reftran:ref}

\subsection{Modifications}

The rely-condition has so far been constrained to be both reflexive and transitive,
while the guar-condition has been required to be reflexive. 
The reason why we introduced these requirements is 
that they have a slightly simplifying effect on both the decomposition-rules and
the meaning of a specification. 
On the other hand, the obvious disadvantage is that some specifications become
more complicated.

However, these constraints can easily be removed by 
redefining the rely-
and guar-conditions
to denote respectively any atomic state change by the environment and any
atomic state change by the implementation. This means that $ext$ and $int$
(see page \pageref{int:def}) must be reformulated as below:

\begin{definition}
Given a glo-set $\vartheta$, a pre-condition $P$, a rely-condition $R$, and a
structure $\pi$, then
$ext_{\pi}[\vartheta,P,R]$ denotes the set of all computations $\sigma$ in $\pi$,
such that:
\begin{itemize}

\item $\delta(\sigma_1)\models_{\pi} P$,

\item for all $1\le j< len(\sigma)$, if $\lambda(\sigma_j)=e$ and $\delta(\sigma_j)
\stackrel{\vartheta}{\neq}\delta(\sigma_{j+1})$ then

\begin{flist}
(\delta(\sigma_j),\delta(\sigma_{j+1}))\models_{\pi} R,
\end{flist}

\item if $len(\sigma)=\infty$, then 
for all $j\ge 1$, there is a $k\ge j$, such that $\lambda(\sigma_k)=i$.
\end{itemize}
\end{definition}

\begin{definition}
Given a glo-set $\vartheta$, 
a wait-condition $W$, a guar-condition $G$, an eff-condition $E$, 
 and a structure $\pi$, then
$int_{\pi}[\vartheta,W,G,E]$ denotes the set of all computations $\sigma$ in $\pi$,
such that:
\begin{itemize}
\item $len(\sigma)\neq \infty$,

\item for all $1\le j<len(\sigma)$, if $\lambda(\sigma_j)=i$ and $\delta(\sigma_j)
\stackrel{\vartheta}{\neq}\delta(\sigma_{j+1})$ then

\begin{flist}
(\delta(\sigma_j),\delta(\sigma_{j+1}))\models_{\pi} G, 
\end{flist}

\item if $\tau(\sigma_{len(\sigma)})\neq\epsilon$ then 
$\delta(\sigma_{len(\sigma)})\models_{\pi}W$,

\item if $\tau(\sigma_{len(\sigma)})=\epsilon$ then 
$(\delta(\sigma_1),\delta(\sigma_{len(\sigma)}))\models_{\pi}E$.

\end{itemize}
\end{definition}

\noindent Satisfaction can then be defined as earlier.

\subsection{Alternative Rules}

It is only necessary to change
the skip-, assignment-, while-, await- and effect-rules (observe that $P^R$ is
equivalent to $P^{R^*}$):

\begin{flist}
\\
\nskip \satis  (\vartheta,\alpha):: (P,R,W,G,R^*)
\end{flist}

\begin{flist}
\underline{\overleftharpoon{P^R}\And v=\overleftharpoon{r}\And I_{\{v\}\cup \alpha}\And 
\bigwedge_{a\in\alpha} a=\overleftharpoon{u_a}
\Rightarrow (G\Or I)\And E}\\
v:=r \satis  (\vartheta,\alpha):: (P,R,W,G,R^*|E|R^*)\\
\\
{where\ for\ all\ }a\in\alpha, var[u_a]\subseteq \vartheta\cup\{a\}
\end{flist}

\begin{flist}
{\rm\underline{wf}}\ Z\\
 z  \satis  (\vartheta,\alpha):: (P\And b,R,W,G,P\And Z)\\
\overline{\nwhile\ b\ \ndo\  z \ \nod \satis  (\vartheta,\alpha) 
:: (P,R,W,G,(Z^{\dagger}\Or  R^*)\And \neg b)}
\end{flist}

\begin{flist}
P^R\And \neg b\Rightarrow W\\
E_1|(\bigwedge_{a\in\alpha}a=\overleftharpoon{u_a}\And I_{\alpha})\Implies (G\Or I)\And E_2\\
 z  \satis  (\vartheta,\alpha):: (P^R\And b,I,\false,\true,E_1)\\
\overline{\nawait\ b\ \ndo\  z \ \nod \satis  (\vartheta,\alpha):: (P,R,W,G,R^*|E_2|R^*)}\\
\\
{where\ for\ all\ }a\in\alpha, var[u_a]\subseteq \vartheta\cup\{a\}
\end{flist}

\begin{flist}
 z  \satis  (\vartheta,\alpha):: (P,R,W,G,E)\\
\overline{ z  \satis  (\vartheta,\alpha):: (P,R,W,G,E\And (R\Or G)^*)}
\end{flist}

\section{Allowing Environments to Diverge}

\label{conver:ref}
As mentioned earlier, the assumption that the environment is convergent can
easily be removed. It is enough to change the definitions of $ext$ (see page
\pageref{ext:def}) and $int$ (see page \pageref{int:def}) to:

\begin{definition}
Given a pre-condition $P$, a rely-condition $R$, and a
structure $\pi$, then
$ext_{\pi}[P,R]$ denotes the set of all computations $\sigma$ in $\pi$,
such that:
\begin{itemize}

\item $\delta(\sigma_1)\models_{\pi} P$,

\item for all $1\le j< len(\sigma)$, if $\lambda(\sigma_j)=e$ then
$(\delta(\sigma_j),\delta(\sigma_{j+1}))\models_{\pi} R$.
\end{itemize}
\end{definition}

\begin{definition}
Given a wait-condition $W$, a guar-condition $G$, an eff-condition $E$, 
and a structure $\pi$, then
$int_{\pi}[W,G,E]$ denotes the set of all computations $\sigma$ in $\pi$,
such that:
\begin{itemize}
\item if $len(\sigma)=\infty$, then there is a $j\ge 1$, 
such that for all $k\ge j$,
$\lambda(\sigma_k)=e$,

\item for all $1\le j<len(\sigma)$, if $\lambda(\sigma_j)=i$ then
$(\delta(\sigma_j),\delta(\sigma_{j+1}))\models_{\pi} G$, 

\item if $len(\sigma)\neq\infty$ and $\tau(\sigma_{len(\sigma)})\neq\epsilon$ then 
$\delta(\sigma_{len(\sigma)})\models_{\pi}W$,

\item if $len(\sigma)\neq\infty$ and $\tau(\sigma_{len(\sigma)})=\epsilon$ then 
$(\delta(\sigma_1),\delta(\sigma_{len(\sigma)}))\models_{\pi}E$\footnote{This 
constraint is sufficient, because if there are an infinite computation $\sigma$ 
and $j\ge 1$, such that $\tau(\sigma_j)=\epsilon$, then $\sigma(1,\ldots,j)$ is a
computation.}.
\end{itemize}
\end{definition}

\noindent Observe, that $int$ is not constraining the infinite elements of $ext$ that have only
a finite number of internal transitions.  Thus the only difference from earlier is
that we can no longer use LSP to prove termination, but only that a program
terminates or is infinitely overtaken by the environment.
(This is related to the view taken in
\cite{qx:note}.)

\section{Wait as an Assumption, Not as a Commitment}

So far the wait-condition has characterised a commitment.
It is also possible to interpret the wait-condition 
as an assumption
on the environment.

It is enough to strengthen the definition of
a specification (see page \pageref{spec:def}) 
with another constraint\footnote{This constraint
can be dropped. It was introduced to
ensure that the fourth assumption in definition \ref{ketil:def} is consistent with
the second assumption. (This is related
to implementability --- see page \pageref{imple:ref}.)}; namely that the environment can
always reach a state which does not satisfy the wait-condition, and thereafter
redefine $ext$ and $int$ (see page \pageref{int:def}) 
as below:

\begin{definition}
A specification is a tuple of the form

\begin{flist}
(\vartheta,\alpha)::(P,R,W,G,E),
\end{flist}

\noindent where $R$, $G$ and $E$ are binary assertions, 
$P$ and $W$ are unary assertions, $\vartheta$
and $\alpha$ are finite, disjoint sets of variables, such that for
all structures $\pi$:
\begin{itemize}
\item $R$ is reflexive and transitive,
\item $G$ is reflexive,
\item for any state $s_1$,
 there is a state $s_2$, such that $(s_1,s_2)\models_{\pi} R$ and $s_2
 \models_{\pi}\neg W$,
\item the unhooked version of any free variable occurring in $(P,R,W,G,E)$ is an
element of $\vartheta\cup \alpha$.
\end{itemize}
\end{definition}

\begin{definition}\label{ketil:def}
Given a pre-condition $P$, a rely-condition $R$, a wait-condition $W$, and a
structure $\pi$, then
$ext_{\pi}[P,R,W]$ denotes the set of all computations $\sigma$ in $\pi$,
such that:
\begin{itemize}

\item $\delta(\sigma_1)\models_{\pi} P$,

\item for all $1\le j< len(\sigma)$, if $\lambda(\sigma_j)=e$ then
$(\delta(\sigma_j),\delta(\sigma_{j+1}))\models_{\pi} R$,
\item if $len(\sigma)=\infty$, then 
for all $j\ge 1$, there is a $k\ge j$, such that $\lambda(\sigma_k)=i$, 
\item if $len(\sigma)\neq \infty$ and $\tau(\sigma_{len(\sigma)})\neq \epsilon$, 
then $\delta(\sigma_{len(\sigma)})\models_{\pi}\neg W$.
\end{itemize}
\end{definition}

\begin{definition}
Given a guar-condition $G$, an eff-condition $E$, and a structure $\pi$, then
$int_{\pi}[G,E]$ denotes the set of all computations $\sigma$ in $\pi$,
such that:
\begin{itemize}
\item $len(\sigma)\neq \infty$,

\item $\tau(\sigma_{len(\sigma)})=\epsilon$,

\item for all $1\le j<len(\sigma)$, if $\lambda(\sigma_j)=i$ then
$(\delta(\sigma_j),\delta(\sigma_{j+1}))\models_{\pi} G$, 

\item $(\delta(\sigma_1),\delta(\sigma_{len(\sigma)}))\models_{\pi}E$.

\end{itemize}
\end{definition}

\noindent The decomposition-rules are the same as before, but due to the new constraint
on a specification, it is no longer possible for a program to become blocked
in a state in which it cannot be released by the environment. Moreover, due to
the new assumption on the environment, an implementation can be thought of as
totally correct even when the wait-condition is not equivalent to false.

This means that, in the same 
way as the wait-condition characterises the set of states in which it is safe for
the implementation to become blocked, it is safe for the environment to become
blocked in a state which neither satisfies the wait- nor the eff-condition (with
all free hooked variables bound by existential quantifiers).

In other words, the wait- and eff-conditions generate an assumption/com\-mit\-ment
relationship similar to the one between the rely- and guar-conditions.

\section{Proof-Obligations on Specifications}

\subsection{Implementability}

In VDM \cite{cbj:VDM} specifications are required to satisfy an obligation
called implementability\index{implementability}\footnote{In 
\cite{cbj:VDM2} the same obligation is
called satisfiability.}; namely that for any state which satisfies the pre-condition,
there is a state transition which satisfies the eff-condition. 
Since we are only interested in total correctness, it is sensible
to formulate a similar implementability constraint for our system.

More formally, given a specification $(\vartheta,\alpha)::(P,R,W,G,E)$,
this means that

\begin{itemize}
\item for all structures $\pi$ and all states $s_1$, $s_1\models P$ implies there is
a state $s_2$ such that $(s_1,s_2)\models_{\pi} E$.
\end{itemize}

\noindent By insisting that specifications satisfy this constraint we may eliminate
some inconsistencies, but far from all. Consider for example the following
specification:

\begin{tuple}{\{x\}}{\emptyset}{\true}{x=\overleftharpoon{x}}{\false}{x\ge \overleftharpoon{x}}{x<\overleftharpoon{x}.}
\end{tuple}

\noindent Clearly, the eff-condition is defined for any state which satisfies the 
pre-condition. Nevertheless, this specification does not make sense,
since none of the alternatives offered by the eff-condition is 
`reachable' by the transitive 
closure of the rely- and guar-conditions.

One possibility is instead to employ the constraint\footnote{
This was suggested in \cite{jw:VDM+} with respect to Jones' rely/guar-method.}:

\begin{itemize}
\item for all structures $\pi$ and all states $s_1$, $s_1\models P$ implies there is
a state $s_2$ such that $(s_1,s_2)\models_{\pi} E\And (G\Or R)^{\dagger}$.
\end{itemize}

\noindent This disqualifies the specification above, but it is not difficult to find cases
where a still stronger constraint is desirable.
There is for example 
no program which satisfies the specification

\begin{tuple}{\{x\}}{\emptyset}{x=0}{\true}{
\false}{x\ge \overleftharpoon{x}}{x=2,}
\end{tuple}

\noindent because the environment can update $x$ as it likes.

\label{imple:ref}

It may also be argued that since LSP is a method for the development of totally
correct programs, an implementability
constraint should restrict specifications in such a way 
that an implementation cannot become blocked in a state
in which it cannot be released by the environment.

It is still an open question how to formulate a sufficiently strong requirement, 
and how difficult it would be to prove that a specification satisfies
such a constraint. See \cite{dg:note} for a further discussion.

\section{Related Work}

\subsection{Owicki/Gries}

LSP can be understood as a compositional version of the proof system proposed by
Owicki and Gries in \cite{so:await}\footnote{The additional interference-freedom
requirement for total correctness proposed by Qwicki and Gries is not correct
\cite{kra:owicki}.}. The rely-, guar- and wait-conditions have been
introduced to avoid the final non-interference and freedom-from-deadlock proofs.

The handling of auxiliary variables has also been changed. Auxiliary variables do
not have to be implemented. Constraints on their use are built into the
decomposition-rules. Moreover, in the Owicki/Gries approach 
auxiliary variables can only be used as a verification tool, 
while in LSP auxiliary
variables can be employed both as verification and specification tools.

\subsection{Jones}

It is correct to consider Jones' system \cite{cbj:ifip} as a restricted version of our
system. There are two main differences. First of all, LSP has a wait-condition 
which makes it possible to deal with synchronisation. Secondly, because auxiliary
variables may be employed both as specification and verification tools, 
LSP is more expressive.

\subsection{Soundararajan}

One important advantage of LSP with respect to 
Soundararajan's method \cite{ns:oslo} is that
auxiliary variables can be of any sort. It is shown in \cite{oo:conc} how reasoning
with this system can be simplified by splitting up the original traces into
history variables. 
However, the use of auxiliary variables is still
more restricted than in LSP.

Another obvious disadvantage with the Soundararajan approach is that it only deals
with partial correctness.

\subsection{Barringer/Kuiper/Pnueli}

The system proposed by Barringer, Kuiper and Pnueli \cite{hb:class} is much more
general than ours. However, we believe that our approach is conceptually
simpler and therefore better suited in the area where it can be employed.

Another advantage with LSP is that it makes a clearer distinction between
assumptions (about the environment) and commitments (which the implementation 
is required to satisfy).

\subsection{Stirling}

We have already discussed some of the differences between Stirling's method 
\cite{cs:note} and
LSP, namely that the rely- and guar-conditions are sets of invariants and not
binary assertions, that the eff-condition is unary and not binary, and that 
interference after the last internal transition is treated differently.

Another important difference is that Stirling is only concerned with 
partial correctness. Moreover,
his system does not allow auxiliary variables to be employed as a specification
tool.